# A Process Model to Improve Information Security Governance in Organisations

**Chee Kong Wong**
ORCHID ID: 0000-0002-4088-9751

Submitted in total fulfilment of the requirements for the degree of
Doctor of Philosophy

May 2022

School of Computing and Information Systems
Melbourne School of Engineering
The University of Melbourne

# Abstract

Information security is an increasingly important topic among senior organisational stakeholders (i.e. the board and executive management) as organisations acknowledge the potential for operational disruption, reputational loss, impact to share value and financial penalties. As information resources are a strategic asset to organisations, there is an expectation that these stakeholders will demonstrate their fiduciary duty of care by implementing information security governance (ISG).

Compared to corporate governance, ISG is a relatively new and under-researched area. A review of the literature shows the lack of an ISG framework or model that: (1) incorporates the broad areas of ISG; (2) explains how to implement ISG; (3) is empirically grounded; and (4) identifies the processes required to be undertaken by various stakeholder groups involved in ISG.

The practical requirement for an ISG framework or model to help organisations improve their implementation of ISG and the research gaps have led to the following research question:

"How can ISG be implemented in organisations?"

To address the research question, this research has adopted an exploratory research approach. First, a conceptual ISG process model was proposed based on synthesis of extant literature and detailed review of relevant frameworks and models. The conceptual ISG process model was subsequently refined based on empirical data gathered from 3 case study organisations comprising one financial institution in Singapore and two financial institutions in Malaysia. The refined ISG process model was finally validated in 6 expert interviews.

This research addresses the aforementioned practice requirements and research gaps by introducing an empirically grounded ISG process model as a practical reference to facilitate the implementation of ISG in organisations.

Specifically, the research contributes by: (1) developing ISG process theory, as ISG is a series of events occurring within an organisational context; and (2) developing an information-processing perspective on ISG, as the process model identifies the information and communication flows, and the relationships among stakeholder groups. In addition, the research has: (3) empirically examined and validated the ISG process model based on how ISG is practised in real-world organisations; (4) examined corporate governance theories to provide additional perspectives to ensure that the ISG process model is aligned with corporate governance objectives; (5) identified additional factors that influence the implementation of ISG requiring further research; and finally (6) expanded existing seminal research by introducing an empirically grounded ISG process model that has been developed based on synthesis of cumulative knowledge from previous research and validated with empirical data.

This research is the most comprehensive study to date that has developed an empirically grounded ISG process model identifying stakeholder groups and explaining how core ISG processes and sub-processes interact. An ISG process model is easier to visualise for practitioners and easier to implement as it allows practitioners to structure their thinking according to the stages of the process model and change activities in their organisations.

# Declaration

This is to certify that:

i. the thesis comprises only my original work towards the PhD except where indicated in the preface;

ii. due acknowledgement has been made in the text to all other material used; and

iii. the thesis is less than 100,000 words in length, exclusive of tables, maps, bibliographies and appendices.

Signature:

Name: Chee Kong Wong

Date: 08 May 2022

# Preface

This thesis was edited and proofread by a nationally accredited professional editor, Mary-Jo O'Rourke AE, who is familiar with the limitation on editorial intervention in accordance with the *Australian standards for editing practice* and the *Guidelines for thesis editing* (Institute of Professional Editors, 2019). The editor has no specific knowledge in the academic discipline of the thesis. Compliance with these policies has ensured that the thesis retains its integrity as entirely the work of the student.

This section also includes the list of peer-reviewed articles that I have published during my PhD research. Elements of these articles are included in this thesis but no specific passages have been quoted verbatim from published papers where I am the primary author.

# Acknowledgements

First and foremost, I would like to express my sincere gratitude to my supervisors, Associate Professor Dr Atif Ahmad, Associate Professor Dr Sean Maynard and Dr Humza Naseer, for their tireless effort in guiding me from the start to the successful completion of this research. Specifically, Dr Atif Ahmad for his intellectual challenge on concepts and principles of information security, Dr Sean Maynard for his broad perspectives and intellectual ideas on the managerial aspects of information security research and Dr Humza Naseer on his interest in ensuring a robust research methodology. Their challenges in our regular meetings helped stimulated greater engagement with the content and strengthened the learning on the topics being researched.

I would also extend my appreciation to Professor Dr Tony Wirth, the chair of my advisory committee, for his comments and suggestions during my annual PhD progress reviews.

Not to be forgotten, many thanks to the Graduate Research team, Faculty of Engineering and Information Technology, which has provided guidance and support throughout my candidature as a research student at the University of Melbourne.

I would also like to record my thanks to the case study organisations and the interviewees who participated in this research, all of whom I cannot name due to anonymity and confidentiality concerns, but they know who they are. They provided invaluable insights and without these data the research would not have been successful.

I would like to acknowledge the Australian Government Research Training Program Scholarship that has made this research possible.

Last but not least, I would like to thank my wife and my two sons, who have provided me with emotional support and encouragement throughout my journey in completing this research. Over the last few years, my wife and sons have sacrificed their vacations and weekends with me as I have spent all these times working on my research and writing up my thesis. They have shown great understanding and shared my ambition in completing my research and thesis towards my PhD. No words can express my appreciation to them as they are my pillars of strength.

# Table of Contents

# List of Tables

# List of Figures

# Chapter 1
# Introduction

This thesis study develops an empirically grounded information security governance (ISG) process model that can help improve the implementation of ISG in organisations. This chapter provides an overview of the research and is structured into 7 sections. Section 1.1 provides a background on information security and ISG, and the importance of ISG in the dynamic security environment in organisations. Section 1.2 explains the motivation for the research based on the current challenges in practice and research, while Section 1.3 outlines the current ISG research and identifies the research gaps. Sections 1.4 and 1.5 present the research questions and research scope, respectively. Section 1.6 provides an overview of the exploratory research design adopted in this research and Section 1.7 concludes this chapter with an outline of the complete thesis.

## 1.1 Research Background

Information security risk has emerged as a systemic risk concern for organisations and this risk is treated as a key operational risk among other risks such as geopolitical, supply chain and climate risks that can severely impact on the operations of organisations. Information security-related breaches (e.g. data fraud/theft and cyberattacks) have been identified as the top 5 global risks in terms of likelihood from 2017 to 2019 (World Economic Forum, 2017, 2018, 2019), and the global costs of such breaches are estimated to increase to more than USD6 trillion by 2021 from USD400 billion in 2015 (Morgan, 2019). Moreover, information theft has been identified as the most expensive and fastest rising consequence of cybercrime (Accenture and Ponemon Institute, 2019; Ponemon Institute, 2019).

The recent pandemic situation has brought forward major digitalisation programs in government and private organisations at an alarming speed. The proliferation of online banking, online shopping and remote working has increased the exposure of information security vulnerabilities, while at the same time information security crime is becoming industrialised with targeted attacks that affect the profits of organisations (e.g. through ransomware attacks). These scenarios over the last few years have raised serious concerns among the boards of directors and senior executives of organisations (Deloitte, 2017; Ernst & Young, 2019a; McKinsey, 2021).

Leaders of organisations acknowledge that the impact of information security risk can be damaging as it may involve direct or indirect monetary losses. Such losses can be attributed to revenue loss due to operational disruptions, reputational loss and loss of customer trust, and drop in share price (Ahmad et al., 2014; Ahmad et al., 2019; Elyas et al., 2014; Schneier, 2013). In addition, the introduction of new regulations such as the General Data Protection Regulation (European Parliament and Council of the European Union, 2016), Australia's *Security of Critical Infrastructure Act 2018* (Commonwealth of Australia, 2018), the Singapore Cybersecurity Act (Cyber Security Agency of Singapore, 2018) and additional local regulatory requirements are holding leaders of organisations accountable for the protection of information assets. These regulations add to the financial impact of information security breaches as regulators impose hefty penalties and fines. It is also not uncommon to see the demand for leadership changes and class-action lawsuits as a result of information security breaches. These incidents have created increased awareness of the fiduciary duty of care of boards of directors and the expectations of executive management in protecting organisations. Some recent high-profile information security incidents (Center for Strategic and International Studies, 2021) include an attack on Singapore's largest healthcare institution leading to the leakage of personal information including that of the Prime Minister (Tham, 2018), state-sponsored hackers accessing the computer systems of the Australian Federal Parliament (Miller, 2019), an attack on Capital One stealing data on 100 million credit card applications and personal identification details (Flitter & Weise, 2019) and a USD50 million cyber ransom data leak at Saudi Aramco (Murphy & Sheppard, 2021).

It is interesting to note that there was a focus in the first half of the decade 2010-2020 on corporate risk management where major improvements in corporate risk governance were made. There were increased involvement of boards and enhancements of the roles of C-level (chief) executives, especially the role of chief risk officers (Deloitte, 2019; Ernst & Young and Institute of International Finance, 2019). This led to renewed emphasis on strong corporate risk culture to support corporate governance with the introduction of risk management, oversight and assurance processes, and a focus on clear roles and responsibilities of boards, executive and operational management. The second half of the decade saw increased focus on information security risk. However, while there were a lot of similar conversations on addressing information security risks through ISG, it has unfortunately continued to be an area that has had multiple interpretations and been relegated to a technical concern but needs further investigation by both information security practitioners and academicians.

Boards realise that their role is overseeing the long-term strategy and sustainable business of organisations. While boards and executive management are engaging more intimately with information security matters and becoming more conscious of their organisations' information security risks, they are still in search of guidance and models that can simplify their understanding and implementation of ISG (Hake, 2015; Lidster & Rahman, 2018; McMillan & Scholtz, 2013). At the same time, the proliferation of standards and frameworks in information security has caused confusion to organisations seeking guidance and implementation of ISG (Farrell, 2015; Lidster & Rahman, 2018; Westby, 2015). These have added to the difficulty in understanding information security risks while ensuring effective ISG and protection.

Therefore, this research develops a practical ISG process model that can help practitioners improve ISG implementation in organisations and subsequently improve the governance of information security.

## 1.2 Motivations for the Research

Corporate governance is critical to the functioning of an organisation in facilitating effective and prudent leadership. In corporate governance, boards and executive management have a fiduciary duty to protect the organisation's assets and value. This fiduciary duty used to be undertaken from the perspective of financial assets. However, as information is now a strategic asset for organisations, this fiduciary duty has extended to include the protection of such information (Holzinger, 2000; Thomson & Solms, 2003; Westby, 2015). As a result, the first motivation is to recognise that the importance of ISG has become paramount for boards and executive management, i.e. ISG is no longer a technical discipline but incorporates technical, organisational and managerial aspects, and organisations are looking at ways to improve the implementation of ISG ( Korhonen et al., 2012; Tan et al., 2017; Westby, 2015).

The literature review of information security topics (see Chapter 2) shows that research on information security frameworks, strategies, policies, risk management and compliance has been increasing over the last decade. For example, there is considerable interest in how organisations respond to cyber attack (Kotsias et al., 2022; Ahmad et al., 2020; Ahmad et al., 2021; Tan et al., 2003; Shedden et al., 2010; Shedden et al., 2012); how they can mitigate such risks (Abdul Hamid et al., 2022; Abdul Molok et al., 2010; Alshaikh et al., 2014; Alshaikh et al., 2021; Maynard et al., 2011); from strategic to operational activities (Ahmad & Ruighaver, 2005; Ahmad et al., 2002). If we consider these topics to be part of ISG research as per the IBM information security framework (Buecker et al., 2013), then we conclude that there has been increasing research interest in ISG. While there is increased interest in ISG in relation to both academia and professional practice, there continues to be a challenge in adopting a standardised definition of ISG which may be attributed to the different contexts, cultural and intellectual backgrounds and interests of scholars and practitioners (Koh et al., 2005; Moulton & Coles, 2003; Tan et al., 2017). Therefore, the second motivation is to develop a consistent interpretation of ISG to drive better understanding of ISG.

A detailed analysis of research on ISG frameworks and models shows that ISG frameworks and models have been developed to either explain the ISG phenomenon or facilitate the implementation of ISG. There are various frameworks and models which can be attributed to the varying definitions of ISG and models that have been developed based on specific emphasis on ISG principles such as principles of good governance (Kim, 2007; Ohki et al., 2009; von Solms & von Solms, 2006), risk management (Conner & Coviello, 2004; Posthumus & von Solms, 2004) and consolidation of best practices and standards (Alves et al., 2006; Da Veiga & Eloff, 2007; Sajko et al., 2011). In addition, existing frameworks and models are generally hypothetical conceptual models developed based on consolidated knowledge of concepts and standards, and were not developed based on empirically grounded research. On top of these frameworks and models found in the academic literature, various frameworks and models have also been introduced by professional bodies and practitioners (Gartner, 2010; Information Systems Audit and Control Association (ISACA), 2012; Institute of Internal Auditors, 2010) and standards bodies (International Organization for Standardization, 2013; National Institute of Standards and Technology, 2011) to help organisations implement ISG. Such proliferation of frameworks, models and standards has provided awareness of and guidance in the implementation of ISG, but has also added to the confusion for organisations in implementing ISG. In addition, information security is an applied discipline and therefore its research should have an applied orientation towards improving practice (Benbasat et al., 1987; Darke et al., 1998). Hence, the third motivation for this research is to develop a practical ISG framework or model for practitioners that encompasses all the areas of ISG, building on the cumulative knowledge of previous research and existing framework and models, and most importantly empirically grounded on real-world practices.

Compared to ISG, corporate governance is a well-researched area dating back to 1992 with the introduction of the *Report of the Committee on the Financial Aspects of Corporate Governance*, generally known as the *Cadbury report* (Cadbury, 1992). Furthermore, there is extensive research on corporate governance theories (Bajo Davó et al., 2019; Chambers & Cornforth, 2010; Donaldson & Davis, 1991; Freeman, 2010) that has studied the various roles of the board and management. The same cannot be said of ISG as information systems and specifically information security is a newer discipline that has picked up significant interest only recently due to the increased speed of digital adoption in today's business. This is the 4th and final motivation, where the research assesses how key corporate governance theories in relation to the board's role and governance models which are more mature can provide additional insights in the research on ISG. This helps in the development of an ISG model that complies with corporate governance requirements and is simple to understand and adopt by those who are not information systems or information security practitioners, i.e. by boards and non-technical executive management.

## 1.3 Research Gaps

Initial information security research was in the technical areas that involved computer security and access controls (Blake & Ayyagari, 2012; Dhillon & Backhouse, 2001). However, information security research has slowly expanded over the last two decades to cover research on the human and managerial aspects of information security strategies, governance, policies, risk management, compliance, education and awareness, and incident management, as well as the roles of boards and management in information security (Silic & Back, 2014; Tan et al., 2017; von Solms, 2006; Williams, 2007a). Specifically, research on ISG has gained significant traction beginning in the early 2000s when calls were made to consider information security a governance issue as business faced increased scrutiny (Conner et al., 2003; Conner & Coviello, 2004; von Solms & Strous, 2003). Since then, research has been conducted in various areas of ISG and papers have been published to cover many areas of ISG as the responsibility to research these newer areas falls onto information systems and information security researchers, as the topic of research is multidisciplinary rather than technically focused.

Research on ISG and specifically on ISG frameworks and models has been fragmented, not cumulative, as well as very diverse in interpretation. This situation in ISG research may be attributed to a lack of consistent interpretation of ISG and differing contexts, cultural and intellectual backgrounds and interests of scholars and practitioners (Alshaikh et al., 2014; Maynard et al., 2018; Moulton & Coles, 2003). This has resulted in research on ISG frameworks and models that has specific emphasis on areas such as principles of good governance (Ohki et al., 2009; von Solms & von Solms, 2006), risk management (Conner & Coviello, 2004; Posthumus & von Solms, 2004) and consolidation of best practices and standards (Da Veiga & Eloff, 2007; Sajko et al., 2011). This leads to the identification of the first gap in research, i.e. the lack of a holistic ISG framework or model that incorporates and brings together the many areas of ISG.

Research has focused on developing ISG frameworks and models to explain ISG and this includes frameworks and models that explain the need for checks and balances (Maynard et al., 2018; Mishra, 2007; von Solms & von Solms, 2006), explain specific areas of governance (Alves et al., 2006; Carcary et al., 2016; Conner & Coviello, 2004) and identify the critical components of ISG (Alqurashi et al., 2013; Da Veiga & Eloff, 2007; Park et al., 2006). The challenges in implementing ISG have motivated researchers to research and develop ISG frameworks and models that aim to facilitate the implementation of ISG, but these frameworks and models ended up focusing on “what” that is required to implement ISG (Conner et al., 2003; Maleh et al., 2018; Mathew, 2018; Ohki et al., 2009; Sajko et al., 2011). This leads to the second gap, as there continues to be a lack of ISG frameworks and models that provide guidance on “how” to implement ISG in organisations.

In addition to the above analysis, most of the developed ISG frameworks and models are conceptual models that have been developed based on theoretical analysis of information security and ISG requirements (Alshaikh et al., 2014; Siponen et al., 2008). Furthermore, the ISG frameworks introduced by standards and professional bodies remain abstract, providing little evidence of an empirically validated model. The need for a practical ISG framework or model that is empirically validated and so can act as a reliable source for organisations is the third gap in research on ISG frameworks and models.

While efforts have been made in ISG research to develop frameworks and models to explain the concepts of ISG and to facilitate implementation of ISG, there is still lacking an ISG framework or model that can easily identify the ISG processes required to be undertaken by various stakeholder groups in an organisation to implement ISG. Process models are under-represented in information systems research (Markus & Robey, 1988; Radeke, 2010; Shaw & Jarvenpaa, 1997). This represents the 4th and final gap that is identified in this research.

Figure 1-1 summarises the motivations for this research and research gaps that informed the research question for this study.

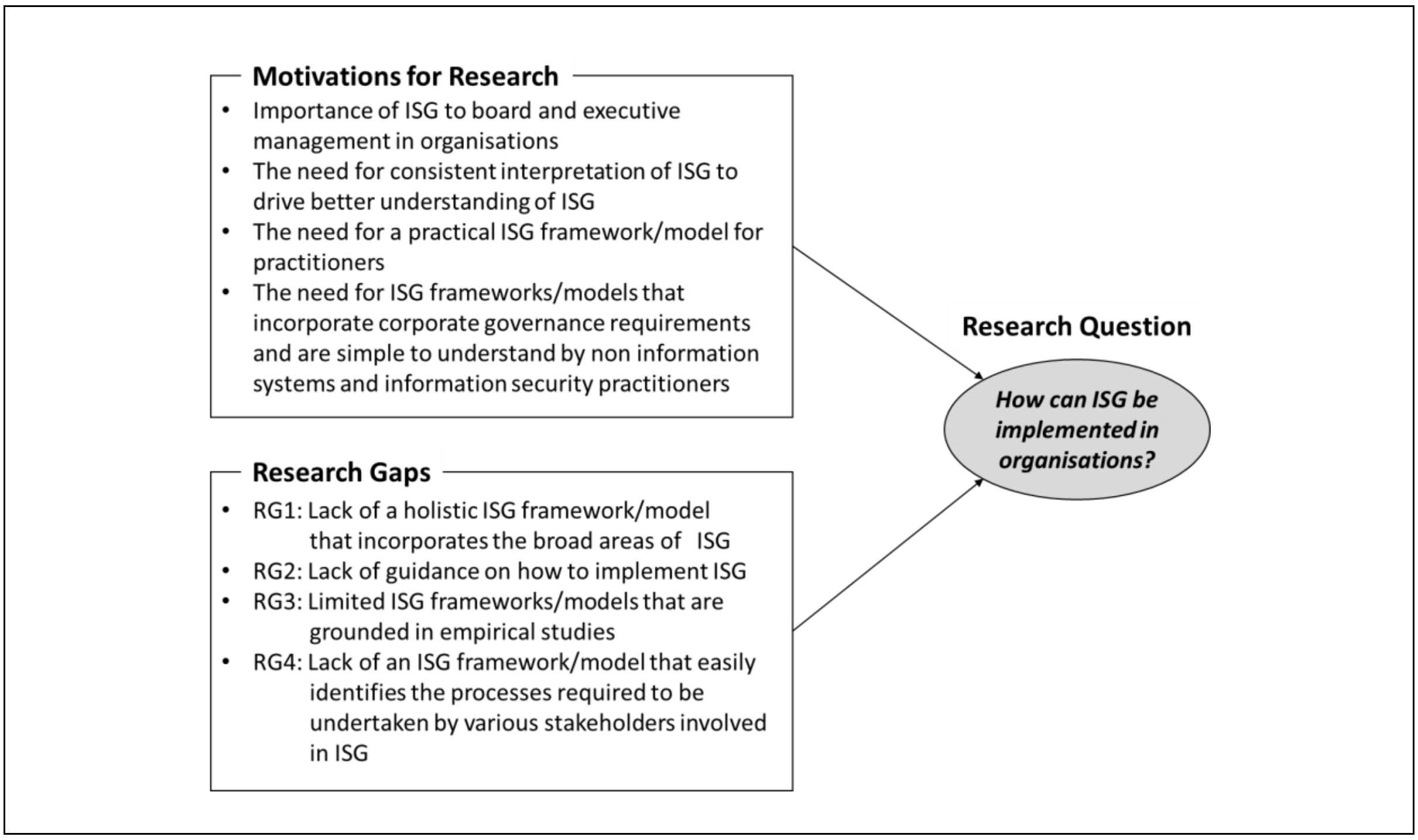

Figure 1-1: Motivations for research and research gaps that informed the research question

Understanding of the practical challenges in implementing ISG and the identified research gaps informed the main research question for this study:

"How can ISG be implemented in organisations?"

## 1.4 Research Question

This research focuses on answering the "how" to implement ISG in organisations, i.e. the following research question:

"How can ISG be implemented in organisations?"

In order to answer this research question, it is important to address 3 related sub-questions, as illustrated in Figure 1-2.

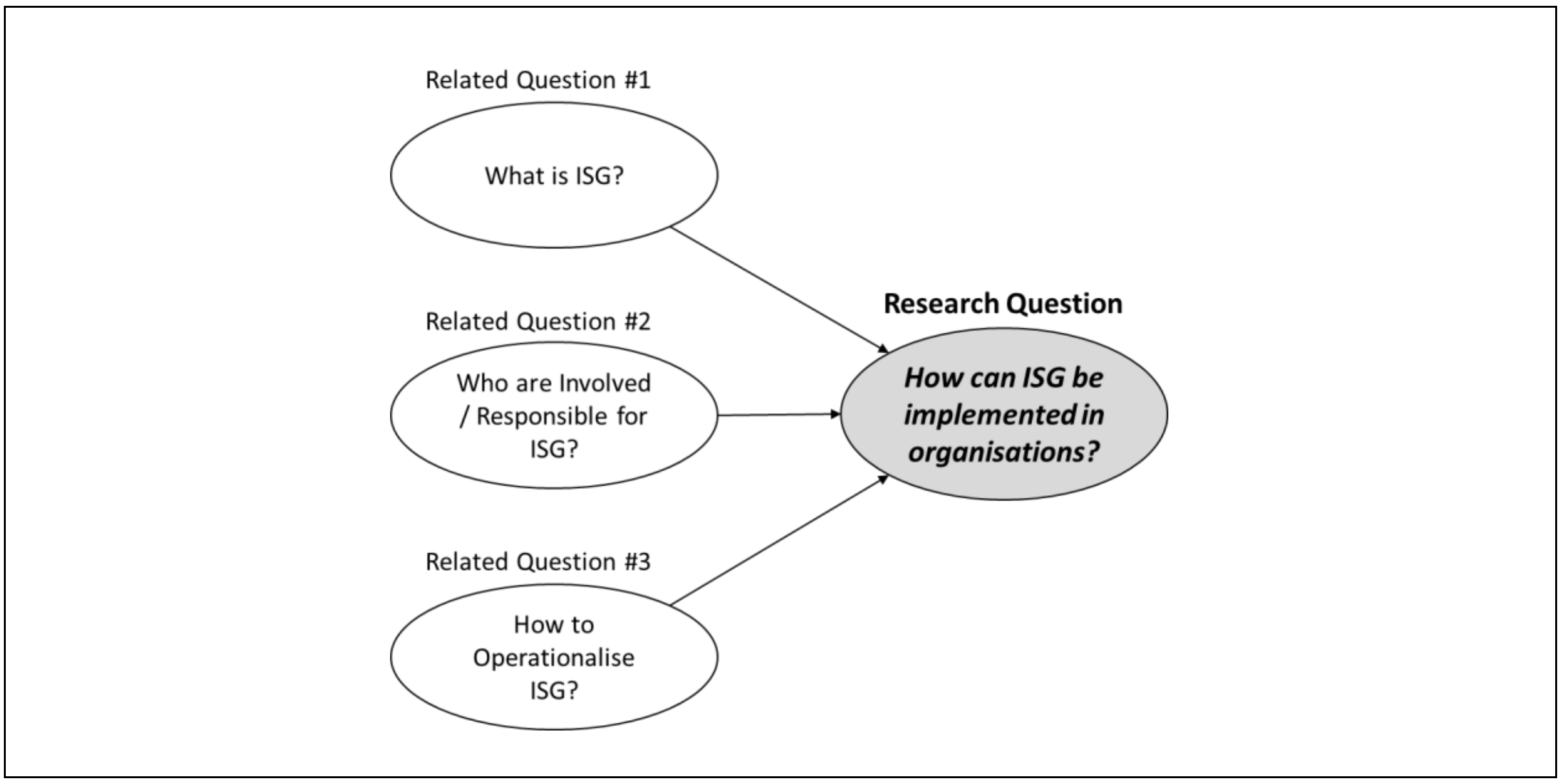

Figure 1-2: Research question and related sub-questions

First, what is ISG? There are many definitions of ISG, hence, it is imperative to understand what represents ISG so that the study can be focused on addressing the scope of ISG that is required in governing information security in organisations. Second, what are the activities involved in ISG? And 3rd, who are involved and responsible for the implementation of ISG?

When organisations know what activities are involved in ISG and who are involved and responsible for implementing ISG, organisations will be able to implement ISG to drive improvement in the governance of information security. Although the related sub-questions begin with the understanding and identification of the relevant concepts (i.e. the "what" and "who"), the focus of this study is on defining the underlying concepts, mechanisms and approach that are required to improve the implementation of ISG in organisations (i.e. the "how").

## 1.5 Scope of Research

ISG is a multidimensional discipline and has many interpretations. This research focuses on the development of an ISG model that can help organisations implement ISG. In this research, the definition of ISG incorporates the principles of good corporate governance, framework of rules, relationships, systems and processes, the value aspect, attaining objectives and monitoring performance, and concepts of information technology (IT) governance. Therefore, ISG is defined as follows:

> ISG is the framework of rules, relationships, systems and processes by which the security objectives of the organisation are set and the means of attaining those objectives and monitoring performance are determined.

This definition of ISG informs the scope of this research. Other definitions are:

a. Organisation: All organisations in both public and private sectors, as ISG is important and applicable to all organisations
b. Framework/model: An empirically grounded model that facilitates ISG implementation in organisations
c. Processes: All ISG processes as informed by various ISG research and ISG models, e.g. von Solms's direct-control model (2006) and ISO 27014 (2013)
d. Stakeholder groups: Identification of the stakeholder groups which are involved and responsible for ISG as compared to information security management

## 1.6 Research Design

An overview of the research design is given in Figure 1-3.

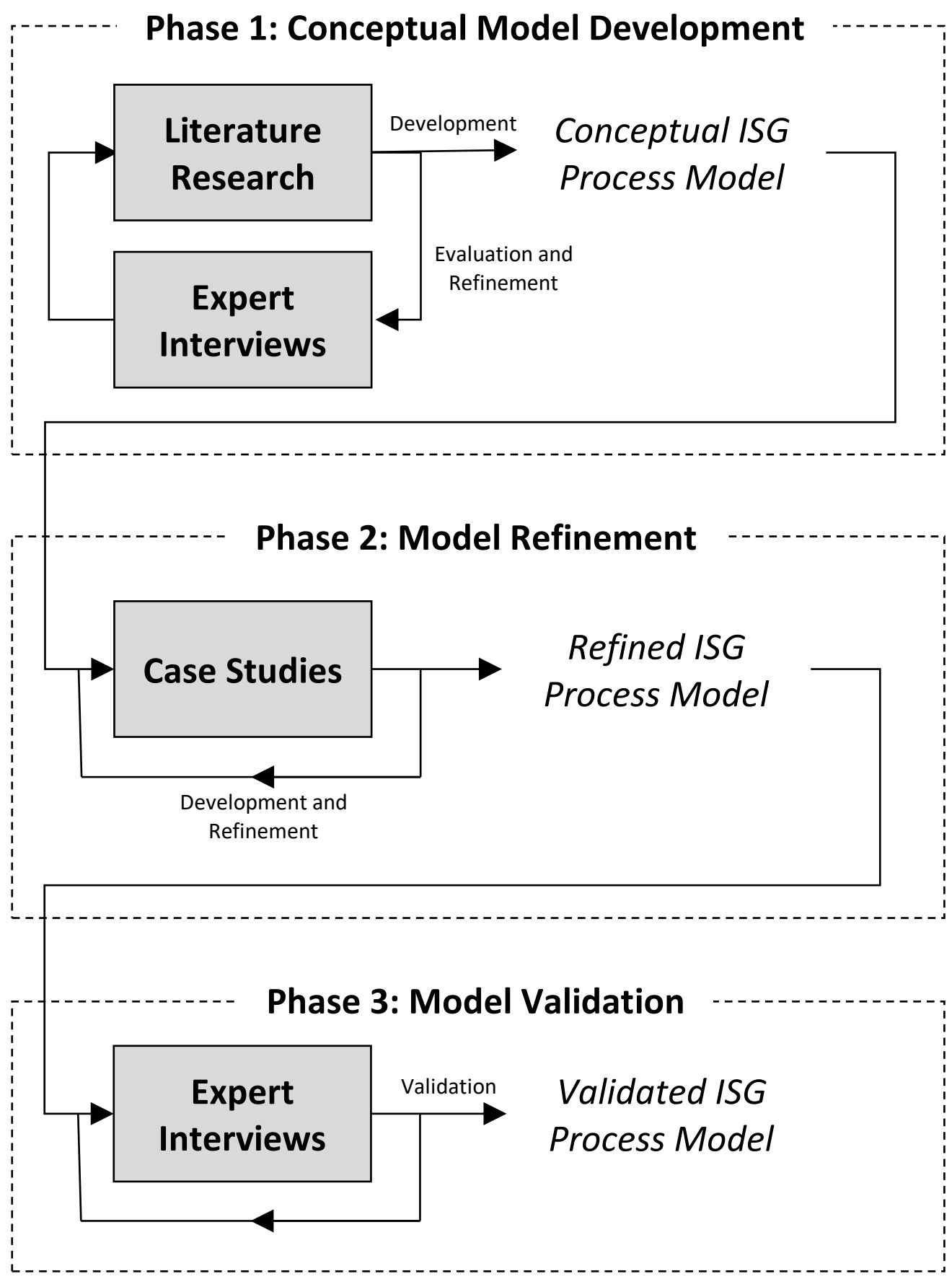

Figure 1-3: Research design

ISG straddles both information systems and business management, therefore the first step towards a better understanding of how to improve the implementation of ISG in organisations is a thorough review and analysis of interdisciplinary literature across information security management, ISG and corporate governance. A total of 129 articles on ISG and 43 articles on corporate governance from journals, conference proceedings and professional publications have been analysed. The objectives were to identify common themes, key features and governance processes that constitute ISG. Further, existing theoretical ISG frameworks and models were analysed and integrated with the findings from the literature review to develop the conceptual ISG process model. This conceptual ISG process model was then shown to 4 information systems and security practitioners to seek initial expert comments on the need and relevance of the model and to test the interview questionnaire that was used in subsequent empirical research. This conceptual ISG process model was also used to structure and support data collection and analysis in the next phase of model refinement. This represents Phase 1: Conceptual model development.

The constructivist paradigm drives the inquiry of this research and a qualitative research approach has been selected because the focus of the study is to address the “how” in implementing ISG in organisations. To answer the research question empirically, an exploratory field study has been conducted using multiple case study and expert interview methods. Qualitative research allows the development of a rich picture of the research phenomena and provides the opportunity to investigate aspects of the phenomena that may not be obvious at the outset of the research (Darke et al., 1998; Eisenhardt, 1989b; Miles et al., 2014; Yin, 2018).

In Phase 2, 3 financial institutions (one from Singapore and two from Malaysia) were selected for the multiple case study design where a total of 17 on-site interviews were conducted with participants across different management hierarchies within these financial institutions. As this research studies and incorporates concepts of ISG and corporate governance, financial institutions were selected because financial institutions are strictly regulated with established corporate governance processes, are known to have a mature security posture with in-house, well-resourced and permanent information security teams and are investing heavily in information security initiatives. The interview data together with other documentation collected from these case studies were analysed to identify emergent themes (known as second-order themes in this thesis) and aggregated dimensions (Creswell, 2013; Gioia et al., 2012; Yin, 2018) to further develop and refine the conceptual ISG process model. These empirical data from the case studies provided in-depth understanding of the roles of the various ISG stakeholders, and the ISG processes and sub-process that were practised in implementing ISG. The result of Phase 2 is the refined ISG process model.

This refined ISG process model from Phase 2 was then taken through validation and confirmation in Phase 3. Phase 3 utilised the expert interview research method (Bogner et al., 2009; Pfadenhauer, 2009) where 6 experts comprising information security consultants, a chief information security officer (CISO), a chief information officer (CIO) and a chief information risk officer were interviewed for their expert remarks on the refined ISG process model. The findings from these expert interviews assisted in validating and confirming the proposed ISG process model and provided further triangulation and supported the generalisation of the theories. The result of Phase 3 is the validated and therefore, the proposed ISG process model.

## 1.7 Thesis Outline

This thesis consists of 7 further chapters within 3 main parts, as illustrated in Figure 1-4.

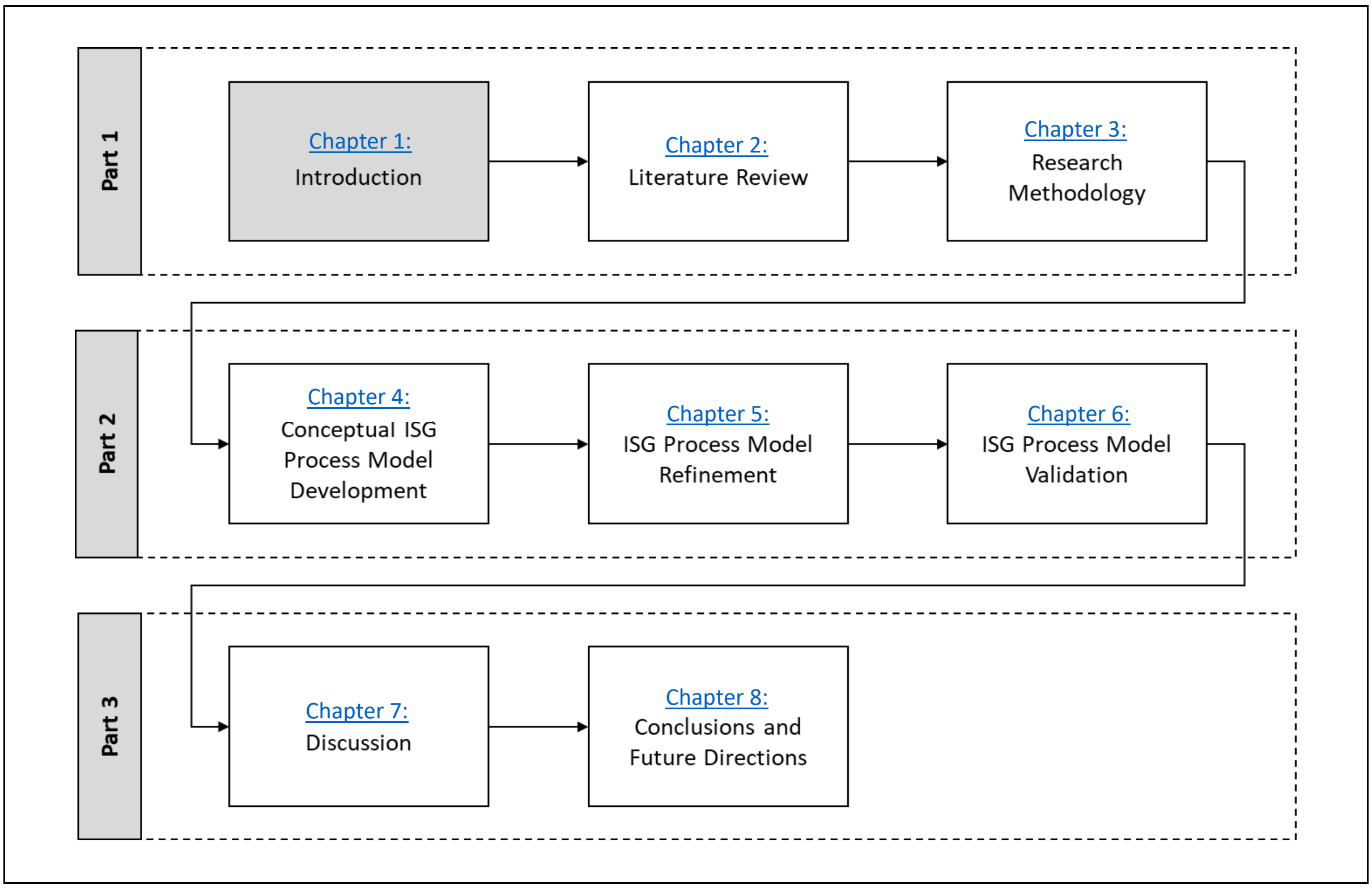

Figure 1-4: Thesis outline

In Part 1 (Chapters 2 and 3), the research and the research methodology are placed in perspective based on the study of extant literature. Chapter 2 informs this research with all relevant background information, providing a multidisciplinary review of information security, ISG and corporate governance literature that guides the inquiry. Based on the gaps identified in the literature review, the research objectives and research question were framed accordingly. Chapter 3 discusses the research methodology and research design, providing the philosophical foundations, the exploratory design approach which comprises case study and expert interviews research methods, and how the empirical data were gathered and analysed for theory generation.

In Part 2 (Chapters 4, 5 and 6), the development of the ISG process model is presented. Chapter 4 explains how the conceptual ISG process model was developed based on the analysis of existing ISG models and the synthesis of knowledge from the literature review. Chapter 5 explains the analysis of case study data for theory generation and the use of this analysis to refine the conceptual ISG process model. Then Chapter 6 discusses the validation and confirmation of the refined ISG process model through expert interviews to produce the final proposed ISG process model.

Part 3 (Chapters 7 and 8) contains the discussions, recommendations and conclusions of the current research study. Chapter 7 discusses the key findings, the theoretical integration with extant literature bringing in similarities and confirmations, and divergences. It also identifies additional findings such as factors that influence the implementation of ISG. Finally, Chapter 8 provides the conclusion and identifies the key contributions of the research to theory and practice. It also highlights the limitations of this study and identifies opportunities for future research.

# Chapter 2
# Literature Review

This chapter reviews the literature on information security, specifically on ISG, and the literature on corporate governance with the aim of understanding the current state of knowledge, building an argument and putting the research into context. Section 2.1 describes the literature review approach. Section 2.2 provides an overview of information security research, while Section 2.3 provides a detailed analysis of the research on ISG to build a body of knowledge on ISG and to understand the gap in ISG research. Section 2.4 reviews the literature on corporate governance and the theories behind corporate governance to understand the functions of boards in governing corporations in order to provide an additional theoretical lens for understanding the functions of governance. This additional study of corporate governance research and practices was conducted as this area has been extensively studied in attempts to understand the concept of governance (Sundaramurthy & Lewis, 2003; Turnbull, 1997). Section 2.5 analyses the gaps in ISG research and concludes that scholarly and empirical studies have been limited in guiding the implementation of ISG in organisations.

## 2.1 Literature Review Approach

A literature review is a critical step in creating new knowledge as it helps with the awareness of existing knowledge and the understanding of research undertaken by other researchers (Boell & Cecez-Kecmanovic, 2014; Kitchenham, 2004; Webster & Watson, 2002). The literature review for this thesis adopted the hermeneutic framework suggested by Boell and Cecez-Kecmanovic (2014), which proposed two major hermeneutic circles including a recursive search and acquisition of articles, and a wider recursive analysis and interpretation, steps that are mutually intertwined. The detailed approach included recursive steps of searching, reading, mapping and classifying, critical assessment, argument development and research question development and refinement until the researcher believed that a well-argued literature review had been achieved and a justified research question had been defined.

In this research, two literature reviews were conducted to build knowledge of ISG and of corporate governance, as illustrated in Figure 2-1.

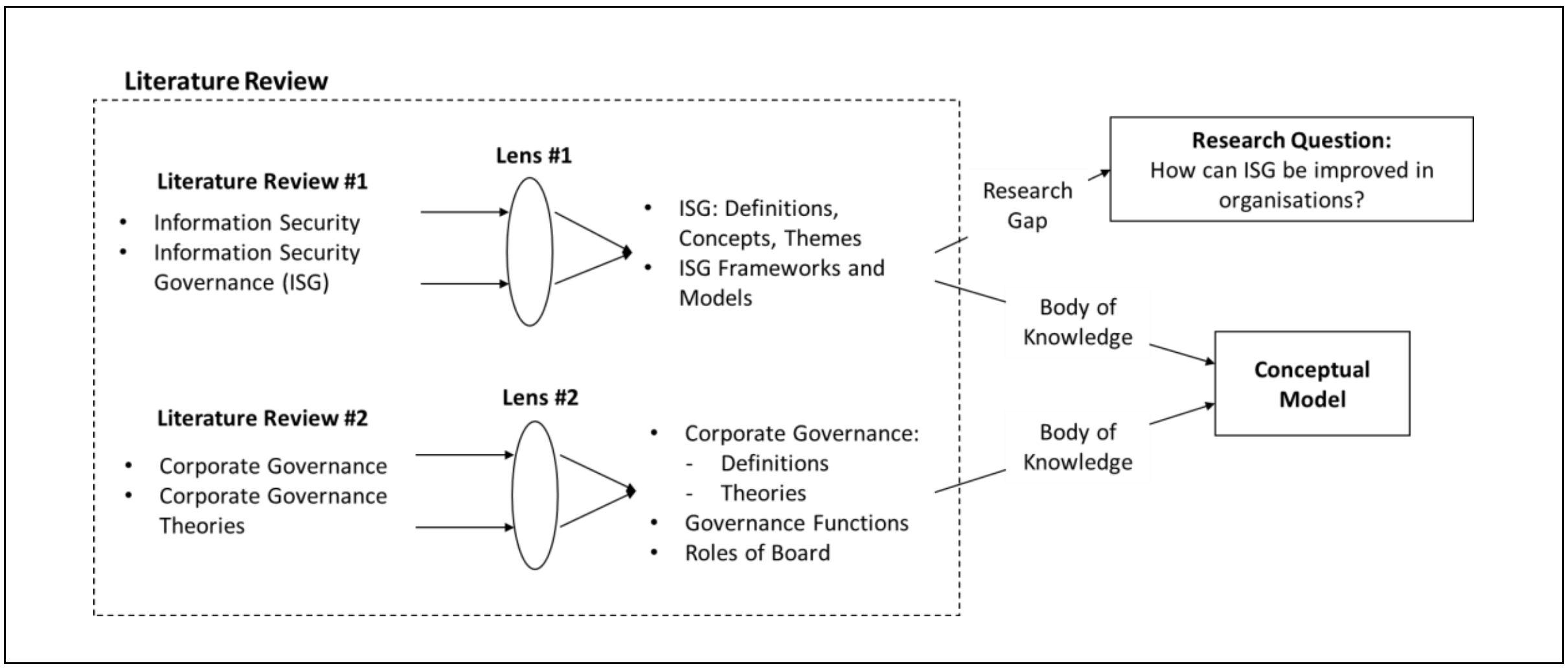

Figure 2-1: Literature review approach and purpose

The first literature review began with an initial review of 9 literature review or meta-analysis papers published between 2001 and 2019 on broader information security topics. This literature review or meta-analysis summarised existing research and provided a theoretical basis in guiding new directions and priority research areas for information security researchers. This meta-analysis also provided initial understanding of ISG research within the context of other areas of research on information security and set the scene for a detailed literature review of ISG.

For the detailed literature review of ISG, a literature search was conducted using the Scopus online database to search the relevant terms because of its good coverage of both academic and practitioner journals. Additional searches were done for peer-reviewed articles from key information journals and conferences using Google Scholar, Science Direct, ProQuest, JSTOR and AIS Electronic library. These searches adopted a search string i.e. using the terms "information security governance" OR "information security and corporate governance" OR "information technology security governance" yielded 688 articles. After eliminating duplicates and reading the abstracts, the total number of articles was reduced to 129 that directly concern ISG. These 129 articles formed the basis of the literature review for ISG. The articles on "information security governance" were read in detailed, examined, coded and mapped, and critically analysed to identify key research themes and research gaps in the area of ISG. The literature review on ISG established the following key themes which are discussed in detail in the following sections:

a. Lack of a consistent interpretation and definition of ISG
b. Various motivations, approaches and underlying principles adopted to develop ISG frameworks and models
c. Information security as a governance concern for the board and executive management
d. ISG comprising key principles of information security strategy and policy, risk management, compliance and assurance

A second, separate literature search was conducted with the following search string i.e. “corporate governance” OR “organisation governance” OR “business governance” OR “governance, risk and compliance” using Google Scholar, Business Source Complete, Scopus and Informit Business Collection to search for articles on corporate governance. The initial search yielded 447 articles, which were reduced to 95 after removing duplicates and reading the abstract, introduction and conclusion sections. These 95 articles were further reduced to 43 shortlisted articles for detailed review as these provided the foundation and theories behind the development of corporate governance. The purpose of this second literature review was to provide an additional theoretical lens to study the rationale for governance and how the theories of corporate governance could be applied to the governance of information security.

A specific literature review on IT governance was not done as most initial ISG research incorporated the principles of IT governance (Alqurashi et al., 2013; Da Veiga & Eloff, 2007; Lessing & von Solms, 2008). This research considered ISG as either a subset of IT governance or the intersection of IT governance and corporate governance. Moreover, the focus of this research was on information security, which is beyond IT security, hence the focus on using the knowledge of both ISG and corporate governance.

## 2.2 Information Security

Initial research on information security was largely focused on the technical areas that involved computer security, access controls, asset management and identity management (Blake & Ayyagari, 2012; Dhillon & Backhouse, 2001). However, during the last two decades published literature in the technical context has decreased, while interest in information security research in the human and managerial aspects covering the areas of information security strategies (Tan et al., 2017), governance (Holgate et al., 2012), policies (von Solms et al., 2011), risk management (Webb, Ahmad, et al., 2014), compliance (von Solms, 2005), education and awareness (Mishra & Dhillon, 2006) and incident management (Ahmad et al., 2015) has increased (Blake & Ayyagari, 2012; Silic & Back, 2014; Wu & Liu, 2019). Research explored management roles in information security covering board-level priorities (Bihari, 2008; Rothrock et al., 2018; Williams, 2007a), integration of technical and management processes (Bodin et al., 2005; Knapp et al., 2009; Straub et al., 2008), policy definition (Eloff & Eloff, 2005; Knapp et al., 2009) and risk management and compliance management (Straub et al., 2008; von Solms & von Solms, 2004, 2005; Webb, Ahmad, et al., 2014).

If we consider all research on information security frameworks, strategies, policies, risk management and compliance as part of the ISG research theme per the IBM information security framework (Buecker et al., 2013), then this research has shown increasing interest in ISG (Blake & Ayyagari, 2012; Zafar & Clark, 2009). This meta-analysis of the research literature confirms that in general, ISG is an area that continues to demand further research that contributes to theory and practice as digitalisation of businesses drives further integration of technical and managerial aspects of information security (Acuña, 2016; Wu & Liu, 2019).

The following section provides a detailed analysis of ISG research, i.e. research publications, that is focused on ISG as the key research area, and not managerial research literature as defined by Blake and Ayyagari (2012) and Zafar and Clark (2009), and ISG in practice based on professional information security publications.

## 2.3 Information Security Governance

Corporate governance is critical to facilitate effective and prudent leadership to deliver the strategic objectives of an organisation. Corporate governance became critical to the functioning of organisations after dramatic corporate failures such as that of Enron in 2001 and the associated demise of Arthur Andersen in 2002. As a result, new regulations and legislation such as the Sarbanes-Oxley Act (Sarbanes-Oxley Act, 2002), UK Corporate Governance Code (Financial Reporting Council, 2016), ASX Principles of Good Governance (ASX Corporate Governance Council, 2019) along with corporate governance codes in many countries have been introduced to improve monitoring and disclosure, thereby improving the governance of organisations. In corporate governance, the board of directors and executive management have a fiduciary duty to protect the organisation’s assets and value. Traditionally this was undertaken from the perspective of financial assets. However, as information is now a strategic asset for organisations, the fiduciary duty has extended to include the protection of such information. As a result, the importance of ISG has become paramount for boards of directors and executive management (Korhonen et al., 2012; Tan et al., 2017; Westby, 2015). In the early 2000s, there were already calls to consider information security as a governance issue as businesses faced increased scrutiny (Conner et al., 2003). In the latest Australian Prudential Regulation Authority (APRA) corporate plan for 2019–2023 (Australian Prudential Regulation Authority (APRA), 2019a), information security was identified as one of 4 key priority areas for the financial sector where the responsibility for information security falls squarely on the shoulders of the board of directors. A failure in this fiduciary duty of care can result in serious implications as evidenced in an information security incident at Equifax where the Chairman of the Board and Chief Executive Officer (CEO) were asked to step down (Sweet & Liedtke, 2017) and a breach at Yahoo that caused the CEO to lose her annual bonus and stock award (Goel, 2017). There is also an increasing trend of derivative claims and class-action lawsuits following information security breaches (Jones Day Publications, 2004; Romanosky et al., 2014; Talotta et al., 2015). These incidents have created increased awareness of the fiduciary duty of care of boards of directors and the serious need for the oversight function to review, monitor and govern information security.

### 2.3.1 Definition

Corporate governance is a well-researched area, but ISG has only gained significant interest from academics and professionals over the last two decades. A review of academic and professional literature on ISG has identified many diverse views (Moulton & Coles, 2003; Tan et al., 2017) and consensus is still not well established (Höne & Eloff, 2009). These diverse views may be attributed to the different contexts, cultural and intellectual backgrounds and interests of scholars and practitioners.

Researchers have defined ISG as systematic oversight and execution of information security functions (Conner et al., 2003; Rastogi & von Solms, 2006) and the establishment of a control environment where policies and procedures with defined roles and responsibilities are implemented (Conner et al., 2003; Conner & Coviello, 2004; Mishra & Dhillon, 2006; Saunders, 2011). ISG has also been defined as a set of policies and procedures that drives information security culture of awareness and accountability (Allen, 2005; Alves et al., 2006). Some researchers have further defined ISG to encompass the wider areas of information security strategy, objectives, organisation structure, risk management and monitoring of performance (Antoniou, 2018; Coetzee, 2012; Moulton & Coles, 2003; von Solms, 2005; von Solms & von Solms, 2009).

Analysis of the interpretation of ISG from professional publications and standards indicates similar diverse interpretations focusing on good practices, role of board of directors, strategic alignment and risk management. The *Global technology audit guide* on ISG published by the Institute of Internal Auditors (2010) does not provide a specific definition of ISG but depicts ISG as part of IT governance, which is defined as "consisting of leadership, organisational structures, and processes that ensure the enterprise's information technology sustains and supports the organisation's strategies and objectives". Gartner (2010) and the *Leading practices and guidelines for enterprise security of Australia* (Commonwealth of Australia, 2006) define ISG as "a set of principles, processes and actions required to protect organisation's information resources in pursuit of its business goals" while some publications focus on "the structure and role of the leadership in governing the processes that safeguard information" (Information Security Forum, 2011; IT Governance

Institute, 2006a). Another publication defines cybersecurity (rather than information security) governance as "the policies and processes required for risk management" (National Institute of Standards and Technology, 2018a). ISO/IEC 27014 (International Organization for Standardization, 2013) on governance of information security simply defines it as "the system by which an organization's information security activities are directed and controlled".

Table 2-1 provides a summary of the diverse definitions of ISG found in academic literature and professional publications.

Table 2-1: Definition of ISG from selected academic literature and professional publications.

| Source of Reference | Definition |
|---|---|
| Governing for enterprise security (Allen, 2005) | "Directing and controlling an organisation to establish and sustain a culture of security in the organisation's conduct (beliefs, behaviours, capabilities, and actions). Governing for enterprise security means viewing adequate security as a non-negotiable requirement of being in business." |
| Enterprise security governance: A practical guide to implement and control information security governance (Alves et al., 2006) | "Information security governance is the act of directing and controlling an organization aligned with the strategy and business objectives, establishing and retaining a culture of information security, optimizing the related processes (based on indicators and learned lessons), and assigning activities to the most competent people to perform the necessary actions." |
| A framework for the governance of information security: Can it be used in an organisation (Antoniou, 2018) | "Information security governance can be defined as the process of establishing and maintaining a framework and supporting management structure and processes to provide assurance that information security strategies are aligned with and support business objectives, are consistent with applicable laws and regulations through adherence to policies and internal controls, and provide assignment of responsibility, all in an effort to manage risk." |
| Towards a holistic information security governance framework for SOA (Coetzee, 2012) | "Information Security governance is a subset of corporate governance that provides strategic direction, ensures objectives are achieved, manages risk appropriately, uses organizational resources responsibility, and monitors the success or failure of the enterprise security programme." |
| Information security governance: Toward a framework for action (Conner et al., 2003) | "Governance entails the systematic oversight and execution of information security functions." |
| Information security governance: Business requirements and | "Information Security Governance is defined as the guidance and control of the information security activities of an organisation through the establishment of applicable policies, |

| Source of Reference | Definition |
|---|---|
| research directions (Höne & Eloff, 2009) | processes and procedures based on the risks faced by the information assets of the organisation.” |
| Improved Security through information security governance (Johnston & Hale, 2009) | “Information security governance is an essential element of enterprise governance and consists of the leadership, organizational structures, and processes involved in the protection of informational assets.” |
| Applying information security governance (Moulton & Coles, 2003) | “Our definition of information security governance is ‘the establishment and maintenance of the control environment to manage the risks relating to the confidentiality, integrity and availability of information and its supporting processes and systems’.” |
| Information security governance - a re-definition (Rastogi & von Solms, 2006) | “Information security governance consists of the frameworks for decision-making and performance measurement that boards of directors and executive management implement to fulfil their responsibility of providing oversight, as part of their overall responsibility for protecting stakeholder value, for effective implementation of information security in their organization.” |
| An information security governance framework (Da Veiga & Eloff, 2007) | “Information security governance can be described as the overall manner in which information security is deployed to mitigate risks.” |
| Information security governance - compliance management vs operational management (von Solms, 2005) | “Information security governance consists of the management commitment and leadership, organizational structures, user awareness and commitment, policies, procedures, processes, technologies and compliance enforcement mechanisms, all working together to ensure that the confidentiality, integrity and availability (CIA) of the company’s electronic assets (data, information, software, hardware, people, etc.) are maintained at all times.” |
| Information security governance: A risk assessment approach to health information systems protection (Williams, 2013) | “Information security governance is the set of responsibilities and practices exercised by the board and the executive management with the goal of providing strategic direction, ensuring that objectives are achieved, ascertaining that risks are managed appropriately and verifying that the enterprise’s resources are used responsibly.” |
| Leading practices and guidelines for enterprise security governance (Commonwealth of Australia, 2006) | “Leading practice dictates that security governance defines the core security principles, the accountabilities and actions of an organisation, to ensure that its objectives are achieved.” |
| The standard of good practice for information security (Information Security Forum, 2011) | “The framework by which policy and direction is set, providing executive management with assurance that security management activities are being performed correctly and consistently.” |
| Framework for improving critical infrastructure cybersecurity | “The policies, procedures, and processes to manage and monitor the organization’s regulatory, legal, risk, |

| Source of Reference | Definition |
|---|---|
| (National Institute of Standards and Technology, 2018a) | environmental, and operational requirements are understood and inform the management of cybersecurity risk.” |
| Information security governance guidance for boards of directors and executive management (IT Governance Institute, 2006a) | “Information security governance consists of the leadership, organisational structures and processes that safeguard information.” |
| Introducing the Gartner information security governance model (Gartner, 2010) | “Information security governance (ISG) is defined as "the processes that ensure the requisite actions are taken to protect the organization's information resources, in the most appropriate and efficient manner, in pursuit of its business goals." |
| ISO/IEC 27014 (International Organization for Standardization, 2013) | “System by which an organisation's information security activities are directed and controlled.” |

It can be concluded that there continue to be differing definitions and interpretations of ISG and there is a need to provide a more consistent interpretation to drive better understanding of ISG. While there are differing interpretations of ISG, there exist some common themes across the various definitions. The following sections explore these themes.

### 2.3.2 Information Security as a Governance Concern for the Board and Executive Management

Information security has found its way to becoming a key topic on the boardroom agenda (Anhal et al., 2003; Deloitte, 2018; Georg, 2017). Information systems play a key role in the continuous automation and digitalisation of organisation processes and information in both physical and digital formats is a strategic asset to organisations. The accessibility to information has made safeguarding the confidentiality, integrity and availability of information a high priority in all organisations. Fear of becoming the next victim of an information security breach, increasing investments and costs related to information security, potential financial and reputation losses, and regulatory and legal implications have kept boards and executive management on their toes (Allen, 2005; Georg, 2017). While there has been an increase in boards’ awareness attributed to widely reported information security breaches and proactive actions taken by organisations in educating their boards, research has highlighted challenges in the inadequate understanding of information security at the board level that hamper the effective governance of information security (Anhal et al., 2003; Georg, 2017).

There is extensive research on the roles and responsibilities of the board and executive management in corporate governance, but little is found on ISG (Alqurashi et al., 2013; Anhal et al., 2003; Georg, 2017). Information security research has identified the importance of the board and executive management in ISG and management, but very few studies have defined the roles and responsibilities in ISG expected of the board and executive management (Conner & Coviello, 2004).

Lindup (1996), Holzinger (2000) and von Solms (2001b) wrote about the need to incorporate information security as part of corporate governance and to consider security as part of business requirements and not solely as IT security. They opined that the board can provide effective leadership in driving an information security culture and ensure objective assurance of information security policies across the business due to its authority level in setting the right tone at the top and the independent position of the board. This call for the integration of ISG as part of corporate governance has been echoed by many researchers in papers that examined the relationship between corporate governance and the need for the board and executive management to protect information assets (Conner & Coviello, 2004; Fazlida & Said, 2015; Moulton & Coles, 2003; Thomson & Solms, 2003; Yngström, 2005). These researchers have highlighted the importance of the board and executive management in giving extra attention to information security as a strategic business concern, setting the right information security culture by communicating the appropriate message from the top to all levels of staff and ensuring effective information security management in the organisation. A meta-study was also done of various research on boards' involvement in information security and specifically in ISG from a sociological perspective to better understand the study of the contribution of the board and senior executives to information security (McFadzean et al., 2006).

With the introduction of corporate governance codes and legislation such as Sarbanes–Oxley on 30 July 2002 (Sarbanes-Oxley Act, 2002), the board and executive management are legally responsible to ensure the credibility of financial reporting by attesting to the effectiveness of internal controls, e.g. Sections 302 and 404 of Sarbanes–Oxley specifically mandate the CEO to have these responsibilities. Researchers have examined the responsibilities of the board and executive management required in governing information security by translating the needs and implications of Sarbanes–Oxley in ensuring information integrity and providing assurance on security control of information assets (Anand, 2008; Brown & Nasuti, 2005; Kim et al., 2008; Wallace et al., 2011; Westby, 2012; Williams, 2014). Bihari (2008) studied the roles and responsibilities of the board by drawing on the theoretical foundations of corporate governance in relation to individuals' rights, markets and ethical behaviour. In the latest Prudential Standard CPS 234 (Australian Prudential Regulation Authority (APRA), 2019c) published by APRA for regulated entities, the standard specifies the following:

> The Board of an APRA-regulated entity (Board) is ultimately responsible for the information security of the entity. The Board must ensure that the entity maintains information security in a manner commensurate with the size and extent of threats to its information assets, and which enables the continued sound operation of the entity.

The corresponding *Prudential practice guide* (Australian Prudential Regulation Authority (APRA), 2019b) provides the roles and responsibilities expected of the board, emphasising the importance of its role in ISG.

von Solms and von Solms (2009, 2006) developed a model for ISG. This model, which is popularly referred to as the direct-control model in the ISG literature, identifies 3 levels of management, i.e. strategic, tactical and operational, together with 3 distinct governance actions, i.e. direct, execute and control. The strategic level represents the board, which sets the direction on the protection of information assets which is then expanded into standards, policies and procedures at the tactical and operational levels. The board also undertakes control through reviewing compliance and conformance to original directives as strategic management. Expanding on the von Solms direct-control model, Korhonen et al. (2012) developed a framework that breaks down the organisation's management level into the decision-making levels of strategic steering, strategic implementation, tactical level, operational level and real-time level, together with the key governance processes of design, development, operations and monitoring. The strategic steering level comprises the executive management, which provides strategic direction and is accountable for the 4 processes from design to monitoring. However, the framework did not define the roles and responsibilities of the board.

Effective governance requires the board and executive management to make the right decisions to bring the right results, i.e. oversight in decision-making. This oversight is dependent on well-informed decision-making, as well as the engagement of the board and executive management (Rastogi & von Solms, 2006; Shaw, 2004; Whitman & Mattord, 2012). Adopting the practice of board committees in corporate governance, various committees are also proposed to facilitate better involvement and engagement between the board and executive management (Georg, 2017; Holzinger, 2000; Koh et al., 2005; Sajko et al., 2011). Information security discussions are then incorporated into various committees such as audit, risk management, IT and dedicated information security committees.

There is always a challenge in implementing ISG, especially at the board level, due to lack of understanding of information security. Anhal et al. (2003) suggested various actions to incentivise the board and executive management to discharge their ISG roles. They suggested aligning information security initiatives directly with the benefits of business in driving increased shareholder value, marketing differentiation, reducing insurance premiums, limiting legal liability and regulatory penalties, and adopting tools and methods such as a process model to simplify the design, implementation and monitoring.

### 2.3.3 Information Security Strategy and Policy

Governance involves the development of strategic directions and the translation of these directives into standards and policies to drive an effective information security environment (Ahmad et al., 2014; Horne et al., 2016; Zafar & Clark, 2009). Mishra (2015) conducted 52 interviews across 9 organisations in a study that resulted in the definition of 23 organisation security governance objectives. This study found that security controls that are aligned with ISG objectives which are also aligned with an organisation's strategy improve the effectiveness of information security controls. This is consistent with other studies which have argued that information security policies must be aligned with ISG objectives which are defined based on organisations' strategy (Eloff & Eloff, 2005; Lindup, 1996). Studies have also shown that external legal and regulatory requirements impact on an organisation's business strategy. Therefore, ISG needs to be aligned with an organisation's business strategy and the related legal and regulatory requirements impacting the organisation (Antoniou, 2018; Georg, 2017; Williams, 2013). In another study that identified the top governance practices and critical success factors for information security, it was found that alignment with strategy was one of the top 20 governance practices (Bobbert & Mulder, 2015).

More recent ISG research that catered for specific requirements such as ISG for cloud computing (Rebollo et al., 2014, 2015) also identified alignment with strategic business goals as a critical component of ISG as security policies that are defined must adapt to the use of emerging technologies. In another study on information security knowledge-sharing, the exploratory findings suggested that information security strategy within ISG needs to be aligned with business strategy so that security standards and policies can be developed to meet users' requirements (Flores et al., 2014).

As discussed in Section 2.3.2, effective governance requires the board and executive management to make the right decisions to bring the right results and this is dependent on well-informed decision-making. An information security strategy that is aligned with the business strategy facilitates well-informed decision-making as the strategy defines the directions and requirements that are required to reduce the security risk of an organisation, answering questions such as "What needs to be protected? Why does it need to be protected? How much do we need to invest? What is the impact if we don't?" (Allen, 2005; Allen & Westby, 2007b). The importance of strategic business alignment in defining information security strategy as part of ISG was further emphasised in an empirically grounded study (Maynard et al., 2018; Tan et al., 2017). In both the studies, the researchers argued that a clear business-aligned information security strategy that is shared across the organisation facilitates improved decision-making, enabling all stakeholders to coordinate their information security activities and adapt to a dynamic security environment. Standards and policies developed this way can enable decision-makers to understand the rationale for the controls, rather than just for the sake of compliance. When viewed from the opposite perspective, information security activities impede an organisation from achieving its business objectives when the information security strategy is not aligned with the organisation's business (Lidster & Rahman, 2018; Park et al., 2006; Webb, Maynard, et al., 2014).

Yaokumah and Brown (2014a) studied the stakeholder theory of corporate governance where the objective of an organisation is to create value for all its stakeholders by aligning the organisation's business goals with the objectives of its various stakeholders. Translating this to an ISG perspective, this involves defining an information security strategy that is aligned with the business strategy so that information security standards and policies are defined and implemented to protect the strategic information assets. The study, which was based on a web-based survey of 360 respondents from 120 organisations, showed that strategic business alignment is important and improves information security risk management, performance measurement, resource management and value delivery.

Discussions of ISG that emphasise information security strategy alignment with an organisation's business strategy lead naturally into the next section on information security risk management.

### 2.3.4 Information Security Risk Management

In corporate governance, risk management is a key principle for protecting the assets of an organisation to ensure the continuity of its operations (ASX Corporate Governance Council, 2019; van Manen & de Groot, 2009; Zabihollah, 2007). Risk management became an important topic of information security research from 2003 onwards where researchers proposed that risk management is key to the protection of information assets (Allen, 2005;

Anhal et al., 2003; Bobbert & Mulder, 2015; Conner & Coviello, 2004; Posthumus & von Solms, 2004; Straub et al., 2008; Webb, Maynard, et al., 2014).

Information security risk can be described as the probability of an unwanted occurrence such as an adverse event or loss of information assets (Whitman & Mattord, 2017) and information security risk management is defined as the process of identifying the risk, assessing the risk's relative magnitude and taking action to control the risk to an acceptable level (Giordano, 2010; Whitman & Mattord, 2017; Yaokumah & Brown, 2014b). While researchers have studied risk management extensively, it has been difficult to segregate the governance component within information security risk management as it has been treated as an integral component of ISG (Williams, 2013).

Information risk management straddles the governance and the management of information security. The ISG component of information security risk management generally refers to the governance functions of decision-making and oversight. This translates to the role of deciding the risk appetite, i.e. the relative risk that an organisation is willing to accept after the implementation of appropriate risk mitigation controls, and the oversight of the end-to-end information security risk management process ensuring that appropriate processes are adopted for risk identifications, risk assessment, risk mitigation and continuous risk monitoring (Georg, 2017; von Solms, 2006; Webb, Maynard, et al., 2014). Risk management has also been identified as a key practice in ISG in defining appropriate information security policies and selecting countermeasures for managing information security threats (Höne & Eloff, 2009). The objective of effective ISG has also been identified as guiding the understanding and evaluation of information security risk in order to implement appropriate controls within information security programs (Lidster & Rahman, 2018).

Williams (2013) developed an ISG model based on a risk assessment approach for the healthcare industry. In her research, she found that risk management was a focus for many organisations as it was seen to be more actionable and involved the mitigation of risk by early detection through risk assessment, monitoring and reporting. A number of studies on ISG adopted the definition of the IT Governance Institute (2006a) where IT risk management was identified as one of the 5 key focus areas of IT governance (Allen & Westby, 2007a; Asgarkhani et al., 2017; Miller et al., 2009; Yaokumah & Brown, 2014b; Zia, 2010). In a study conducted by Zia (2010), it was found that IT security risk management is a critical area of IT governance and is more mature in Australian non-government organisations compared to government organisations. Non-government organisations were found to have more documented processes and more defined security policies and risk management strategies which were aligned with international practices. Yaokumah and Brown (2014b) studied the impact of ISG components on information security risk management and found that organisations which align their information security strategy with business strategy have better risk management approaches and organisations which have invested in information security awareness and training better respond to risks. In the literature on effective governance of enterprise security, information security is considered a non-negotiable requirement in the running of a business, therefore information security risk management must be aligned with an organisation's strategic goals and the regulatory and legislation requirements (Allen & Westby, 2007b).

In corporate governance, the board of directors as part of the governing body of organisations is expected to provide active oversight including approving a risk management framework and risk appetite, and actively challenging management decisions and recommendations where appropriate. Since information security risk is one of the risk factors, the same is expected of boards of directors in the area of information security risk (Deloitte University Press, 2017; Georg, 2017). As information security risk management is such a high-priority area in practice, it is noteworthy that risk management is a key topic within ISG in most of the professional publications (Deloitte, 2018; Ernst & Young, 2013, 2015; Ferrillo, 2015; Wedutenko, 2015) and risk management is a key service that is being provided by companies offering information security consulting. Information security risk management is a critical practice area as continuous regulatory requirements are defined and updated such as new technology risk management guidelines (Australian Prudential Regulation Authority (APRA), 2019b; Bank Negara Malaysia, 2018; Monetary Authority of Singapore, 2013) to align with new information security risks for financial institutions in many countries.

In summary, the governance component of information security risk management involves oversight in deciding on a suitable risk framework, the risk appetite and the alignment of risk with business strategies, ensuring information security risk is managed.

### 2.3.5 Information Security Compliance

In the direct-control ISG model proposed by von Solms and von Solms (2006), the “control” action forms the 3rd action, which is to measure, monitor and report on the compliance with the directives that have been defined. Compliance should cover both internal compliance with the organisation’s standards and policies, and external compliance with international information security standards, codes of practice and legal and regulatory requirements (Da Veiga & Eloff, 2007). Compliance is considered a key component of a continuous governance process. In a literature review conducted by Höne and Eloff (2009), they found that compliance monitoring was one of the most important topics in ISG among computer security and computer fraud publications. Höne and Eloff believed that the high coverage of compliance and monitoring was attributable to the fact that stricter regulatory requirements and legislations have come into effect to penalise and prosecute offenders. These findings were confirmed with further research showing that information security compliance is intended to ensure an organisation adheres to defined standards, policies, regulatory and legal requirements to avoid breaches of these requirements. Researchers concurred that compliance has become a high priority when more legislation and regulatory requirements are introduced, as organisations have to avoid breaches which could lead to serious legal implications (Brown & Nasuti, 2005; Georg, 2017; Wallace et al., 2011).

In driving the sustainable governance of information security in organisations, Allen (2005) recommended that leaders of organisations, i.e. the board and executive management, who have the authority and accountability, must act to enforce compliance. Compliance is a part of governance in ensuring organisations comply with legislative and regulatory requirements. This is further supported by the ISG model developed by Williams (2013) where information security monitoring and compliance were identified as key components that drive the evaluation of metrics and external validation of compliance monitoring and regulatory requirements. By employing a strict compliance regime that is practised by all levels in an organisation, compliance with information security standards and policies helps prioritise the importance of information security that eventually becomes integral to the organisational culture (Corriss, 2010) and adds value to the organisation by enhancing its corporate reputation (Yaokumah & Brown, 2014b). Corriss (2010) proposed a progressive approach in driving compliance which starts with rewarding compliance in the early stages of introduction but eventually punishes noncompliance when an organisation's implementation of information security matures.

Expanding this research, Tan et al. (2010) conducted in-depth case studies evaluating the implementation of ISG enterprise-wide as compared to centrally driven corporate security governance. The research found that centrally driven corporate security governance promotes a compliance culture where compliance with corporate guidelines can become more important than improving security. Besides compliance, it is important to ensure a holistic ISG process that emphasises alignment of strategic directions and understanding of risk management across all levels of the organisation to enable better decision-making.

A review of the literature has found that compliance, together with a business strategy–driven risk agenda, and appropriate board and executive management oversight in decision-making, is required to drive effective ISG.

### 2.3.6 Information Security Assurance

Assurance can be categorised into internal and external audits in providing assurance on the information security of organisations and is conducted to provide assurance to management that relevant controls and responsibilities over information assets are being met (Anhal et al., 2003; Da Veiga & Eloff, 2007; Havelka & Merhout, 2013; Information Systems Audit and Control Association (ISACA), 2012). In addition, audits are done to provide assurance that all legislation and regulatory requirements are being complied with (Havelka & Merhout, 2013; Holzinger, 2000).

Assurance in general has been identified as an important governance and risk management mechanism (Busco et al., 2006; Carcello et al., 2011; Gramling et al., 2004; Steinbart et al., 2018) and has been extensively discussed in the literature in the accounting and corporate governance research domains. Assurance, i.e. the audit function in information security has more coverage in accounting journals than information systems security journals. This may be attributable to the focus of such journals where researchers' interests are in the areas of assurance, audit and accounting where the role of assurance has started to extend beyond financial assurances. This extended scope of assurance is evidenced in professional publications (Anhal et al., 2003; Institute of Internal Auditors, 2010, 2013; IT Governance Institute, 2006b; Pathak, 2004).
Mishra (2007) proposed a process model to conceptualise the audit function in an organisation. The study argued that information security audits strengthen ISG as the activities of audits such as internal control assessment, process standardisation, risk mitigation and training provide assurance that necessary actions have been implemented to protect information assets. Similar literature has confirmed that auditing as part of ISG provides assurance of the information security posture of organisations to stakeholders (Pathak, 2004).

A good working relationship between the internal audit function and the information security function enhances the overall effectiveness of ISG and information security as the openness between audits and the information security function allows better understanding of risks and improves access to evidence in recommending suitable controls to address these risks. This positive relationship between assurance and the governance of information security is evidenced in an empirical study conducted by Steinbart, Raschke, Gal and Dilla (2018) and has been confirmed by other studies on such relationships between the internal audit function and organisation governance (Busco et al., 2006; Gramling et al., 2004; Havelka & Merhout, 2013; Stoel et al., 2012).

In addition to assurance on information security posture, audits are also conducted to provide independent assurance that organisations comply with all required legislation and regulatory requirements. Working together with the board, internal and external audit functions through the audit committee provide the required governance oversight (Allen, 2005; Georg, 2017; Wallace et al., 2011). In a “call to action” paper (Holzinger, 2000), it was recommended that the board and senior management have to be responsible for governing the management of information security risk and this can be effectively done via the audit function with an audit committee. The board is also responsible for working with an external auditor that provides an independent assurance on whether the organisation is managing the information security risk with sufficient information security controls. The role of the board through audit committees has been emphasised in various studies that all believed the governance and assurance of information security reside with the board and senior management (Allen & Westby, 2007b, 2007a; Anhal et al., 2003; Georg, 2017; Williams, 2007a).

Certification as a form of assurance has also been discussed in a number of studies (Hall et al., 2015; Humphreys, 2008; von Solms, 2001a). Certification aims to evaluate the level of information security in an organisation against established standards such as ISO 27000s information security standards jointly published by the International Organisation for Standardisation and the International Electrotechnical Commission, the standards by the National Institute of Standards and Technology in the US (NIST) and the COBIT 5 business framework for the governance and management of enterprise IT to provide assurance on information security controls. As certification is provided by an external and independent third party, it is normally used as validation of assurance to increase the confidence and trust of customers and stakeholders. However, the effectiveness of certification continues to be an interest of research (Hall et al., 2015).

### 2.3.7 ISG Frameworks and Models

ISG frameworks and models are a research area that aims to define the ISG components, processes and areas of responsibilities. While these frameworks may have different underlying design principles, they all aim to facilitate education, decision-making and implementation of ISG (Kim, 2007; Williams et al., 2013). In addition, the majority of these frameworks are conceptual models developed based on consolidated knowledge of concepts and standards, and were not developed based on empirically grounded research.

Moreover, many of these frameworks appear to have been developed as independent initiatives based on the researchers' motivation, with very few that build on cumulative knowledge of previous research.

A total of 27 articles on ISG frameworks and models have been analysed to understand the motivation of the research on ISG frameworks and models, the development methods and the underlying design principles of the various ISG frameworks and models. Figure 2-2 provides an overview of the analysis showing the number (n) and total (N) of articles in each area of analysis. The details are discussed in the following subsections.

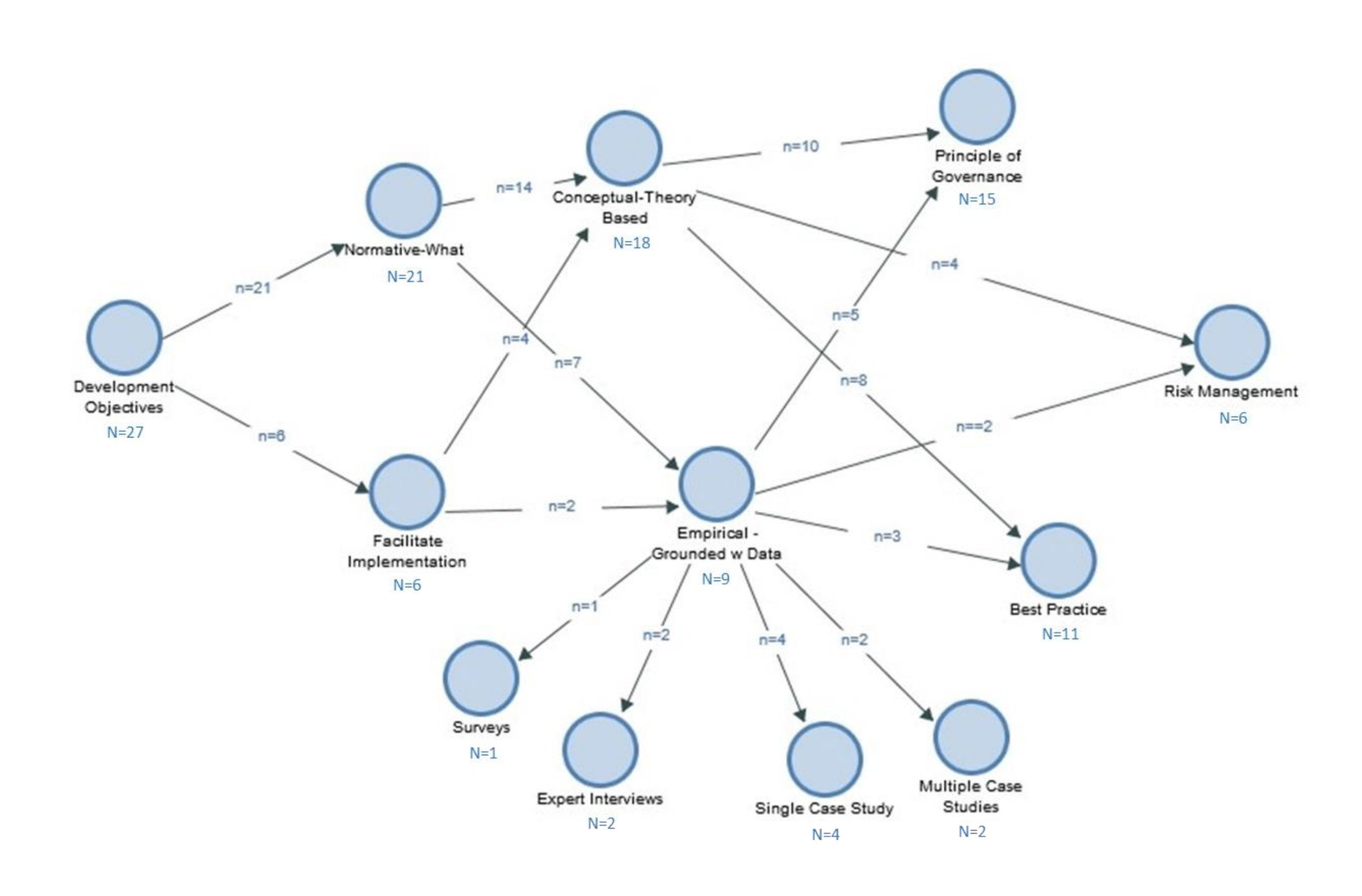

Figure 2-2: Coding of articles on ISG models/frameworks (as analysed with NVivo 11)

### *2.3.7.1 ISG Model Development Objectives*

The analysis of the literature shows that there are two main reasons that motivate researchers to develop ISG models: (1) models that are developed to explain the ISG phenomenon, as found in 21 studies; and (2) ISG models that aim to facilitate the implementation of ISG, as found in 6 studies.

Most researchers developed ISG models to explain the ISG phenomenon. As information security has become more critical to the operations of organisations and evolved from a technical to a human and managerial focus (Da Veiga & Eloff, 2007; Knapp et al., 2009; von Solms & von Solms, 2005; Webb, Maynard, et al., 2014), researchers have become motivated to develop ISG models to explain the concepts of governance, which comprises accountability and the need for checks and balances (Maynard et al., 2018; Mishra, 2007; von Solms & von Solms, 2006), and to define ISG by emphasising the critical success factors for ISG (Alqurashi et al., 2013; Da Veiga & Eloff, 2007; Gashgari et al., 2017; Kim, 2007; Lessing & von Solms, 2008; Park et al., 2006). All these ISG models define ISG and explain the difference between ISG and information security management.

As ISG has gained more interest among researchers, more have been motivated to study the challenges in implementing ISG, as identified by Fazlida and Said (2015), Holgate et al. (2012) and Yaokumah (2014a). These challenges in implementing ISG have motivated researchers to develop ISG models and frameworks to facilitate the implementation of ISG. These ISG models (Conner et al., 2003; Conner & Coviello, 2004; Maleh et al., 2018; Mathew, 2018; Ohki et al., 2009; Sajko et al., 2011; Williams, 2007b) focus primarily on identifying “what” is required to implement ISG. Some of the requirements identified in these models are:

a. Inclusion of clear roles and responsibilities between strategic and management processes
b. Alignment of strategic business objectives with information security policies
c. Continuous evaluation and monitoring to ensure compliance with security policies

In addition to ISG models developed by academic researchers, ISG models have also been introduced by information security professionals and standards organisations (Commonwealth of Australia, 2006; Gartner, 2010; Institute of Internal Auditors, 2010; International Organization for Standardization, 2013; National Institute of Standards and Technology, 2018b) to help explain ISG and to facilitate implementation of ISG.

The analysis of the literature has indicated that the bigger challenge to the implementation of ISG remains in answering "how" to implement ISG, i.e. a model that identifies the processes that can guide organisations to implement ISG in a practical manner.

#### *2.3.7.2 ISG Model Development Method*

The literature analysis has identified a total of 27 articles that developed ISG models and frameworks. The majority (18 articles) of these ISG models are conceptual models that have been developed based on theoretical analysis of information security and ISG requirements. Only 9 articles have introduced ISG models and frameworks that are grounded on empirical data. The 18 articles that have introduced conceptual ISG models adopted one or more of the following approaches:

a. Incorporation of various concepts such as corporate governance, strategic alignment of organisation objectives and risk management (Conner et al., 2003; Conner & Coviello, 2004; Fink et al., 2008; McDermid et al., 2010; Moreira et al., 2008; Ohki et al., 2009; Sajko et al., 2011; von Solms & von Solms, 2006)
b. Identification of critical success factors based analysis of information security and ISG extant literature and theoretical analysis of ISG requirements (Bobbert & Mulder, 2015; Da Veiga & Eloff, 2007; Kim, 2007; Park et al., 2006)
c. Analysis and aggregation of best practices and guidelines from professional bodies and standards such as ISO 27000 (2016), ISO 27014 (2013), COBIT 5 (2012) and Global Internal Audit Guide (2010) (Alves et al., 2006; Coetzee, 2012; Da Veiga & Eloff, 2007; Lessing & von Solms, 2008; Rebollo et al., 2014; Ula et al., 2011)

Our analysis shows that only 9 articles have introduced ISG models and frameworks that are grounded on empirical data. In developing these empirically tested ISG models, researchers adopted different methods of empirical testing which include the following.

a. Surveys (Maleh et al., 2018)

Maleh et al. (2018) developed an ISG model that incorporates ISG practices of organisations via a statistical and econometric analysis of data from a survey of 1000 participants from a mix of 500 large and medium-sized organisations across various industries.

b. Expert interviews (Mathew, 2018; Williams, 2007b)

Mathew (2018) developed an ISG conceptual model based on the plan-do-check-act cycle model of Deming (Walton, 1988) which was tested through 5 expert interviews from 5 different organisations in the UAE. Similarly, Williams (2007b) developed an ISG model for medical practice that was grounded on expert interviews with 5 general practitioners and 3 information security experts.

c. Single case study (Alqurashi et al., 2013; Antoniou, 2018; Tan & Ruighaver, 2004; Vinnakota, 2011)

Alqurashi et al. (2013) developed an ISG model based on the viable systems model that was tested through data simulation from a single case study undertaken by HP Laboratories. Both Tan and Ruighaver (2004) and Vinnakota (2011) conducted a single in-depth case study with qualitative methods to validate their ISG models. Tan and Ruighaver (2004) tested a model that consisted of inter-related ISG processes through stakeholder interviews at a research and development service organisation in Australia, while Vinnakota (2011) validated the practical use of the ISG model at a large telecommunications enterprise through a set of questionnaires. In a more recent study, Antoniou (2018) conducted an interpretive pilot case study to explore information security processes in a US organisation. Single case study method was the most common empirical method adopted by the previous researchers.

d. Multiple case studies (Mishra, 2015; Musa, 2018)

Mishra (2015) developed an ISG model that is based on a multiple case studies approach where 52 interviews were conducted across 9 organisations in different industries, while Musa (2018) introduced an ISG model that was validated with 8 publicly listed organisations in Malaysia.

#### *2.3.7.3 ISG Model Design Principles*

The detailed analysis of the ISG literature also shows that ISG models and frameworks have been developed based on one or more of the following design principles, i.e. principles of good governance, risk management and best practices and standards, as shown in Figure 2-2 and Table 2-2. There are 15 articles on frameworks based on the principles of good governance, 6 articles on frameworks based on a risk management approach and 11 articles on frameworks based on best practices and standards.

Table 2-2: Primary design principles driving ISG framework development.

| Primary Design Principles | Information Security Framework |
|---|---|
| Principles of good governance | (Bobbert & Mulder, 2015; Carcary et al., 2016; Conner et al., 2003; Conner & Coviello, 2004; Fink et al., 2008; International Organization for Standardization, 2013; Kim, 2007; Mathew, 2018; Mishra, 2015; Moreira et al., 2008; Ohki et al., 2009; Park et al., 2006; Tan & Ruighaver, 2004; Vinnakota, 2011) |
| Risk management approach | (Alves et al., 2006; Conner et al., 2003; Conner & Coviello, 2004; Mathew, 2018; Posthumus & von Solms, 2004) |
| Best practices and standards | (Alves et al., 2006; Coetzee, 2012; Conner et al., 2003; Conner & Coviello, 2004; Da Veiga & Eloff, 2007; Gashgari et al., 2017; Ibrahim et al., 2018; Lessing & von Solms, 2008; McDermid et al., 2010; Musa, 2018; Sajko et al., 2011; Ula et al., 2011) |

a. Principles of good governance

A key principle of corporate governance is to provide a solid foundation for management and oversight, i.e. clearly defining the division of responsibilities between the board and management (ASX Corporate Governance Council, 2019). Similarly, for ISG there should be a governing body that is responsible for setting information security strategy and direction, while management is responsible for implementing the information security strategy and directives to ensure security objectives are met (Conner et al., 2003; International Organization for Standardization, 2013; Mathew, 2018; Ohki et al., 2009). Another principle of good governance is the “direct and control” principle where directing and controlling of an organisation are critical (von Solms & von Solms, 2006). This is consistent with the “prescribes and checks” concept in IT governance principles which are identified as “evaluate, direct and monitor” in ISO 38500 (International Organization for Standardization, 2015).

Kim (2007) studied the principles of governance in various domains such as corporate governance, IT governance and ISG to develop an ISG framework that incorporates strategies, decision-making, accountability controls and performance monitoring. Similar approaches that incorporate key governance practices and critical success factors of governance practices across these various domains are found in the framework research by Bobbert and Mulder (2015) and Vinnakota (2011). Extending the “direct and control” principle of von Solms and von Solms (2006), Ohki et al. (2009) studied the need to improve the effectiveness of information security controls that are aligned with corporate strategies and the coordination between corporate governance and management efforts. They developed an information security framework that incorporates the functions of corporate governance and ISG, which is unique and different from other corporate activities and information security management and control mechanisms. Based on the ISO/IEC 38500 governance of information technology framework, Ohki et al. extended the components “direct, monitor and evaluate” to include two additional governance components of “oversee” and “report” in the proposed ISG framework.

b. Risk management approach

Risk management is fundamental to the corporate governance of an organisation (ASX Corporate Governance Council, 2019) as it helps protect organisational assets. As discussed in Section 2.3.4, risk management became an important topic of information security research from 2003 onwards where researchers proposed that risk management is key to the protection of information assets. A number of ISG frameworks developed during this period were driven by the principle of risk management (Alves et al., 2006; Conner et al., 2003; Conner & Coviello, 2004; Posthumus & von Solms, 2004).

*Information security governance: A call to action* (Conner et al., 2003; Conner & Coviello, 2004) developed and promoted a governance framework that aims to effectively implement ISG in organisations. The framework focuses on implementation of information security risk management and oversight, defining the roles and responsibilities of the various functional groups, i.e. the board, executives, management and all employees required to implement controls to manage information security risk. The framework aims to be prescriptive as it identifies the ISG functions and responsibilities of various functional groups to facilitate implementation. Posthumus and von Solms (2004) developed an ISG framework that aims to integrate ISG into corporate governance in protecting information assets. The framework was developed to address the risks in ensuring the confidentiality, integrity and availability of information assets. The research highlighted that a framework must address both internal risks (business and internal IT infrastructure) and external risks (legal and regulatory together with standards and best practices) to information assets and must include everyone in an organisation, i.e. governance and management, to be effective. While the research proposed key areas required for an ISG framework, it did not define processes and actions that need to be implemented, making the model generally descriptive.

c. Best practices and standards

Some researchers anchored their ISG frameworks on best practices and standards as they extracted relevant best practices from IT and information security standards and corporate governance documents. These models aim to incorporate salient elements extracted from relevant best practices and standards so that organisations can achieve more comprehensive ISG coverage than is provided with one single best practice or standard (Alves et al., 2006; Coetzee, 2012; Conner et al., 2003; Conner & Coviello, 2004; Da Veiga & Eloff, 2007; Gashgari et al., 2017; Lessing & von Solms, 2008; Musa, 2018; Sajko et al., 2011; Ula et al., 2011).

In the “call to action” papers (Conner et al., 2003; Conner & Coviello, 2004), a governance framework was developed to facilitate the understanding and implementation of ISG by identifying the roles and responsibilities of the board and management. This framework was built on the approach reflected in ISO 17799 and the US Federal Information Security Management Act. Da Veiga and Eloff (2007) argued that a comprehensive ISG framework helps organisations to develop a strong information security culture in protecting information assets. In their study, they developed a framework that incorporates key components extracted from ISO/IEC 177995 and ISO/IEC 27001, the capability maturity model and PROTECT (Eloff & Eloff, 2005). In a similar approach, an information security framework was defined by incorporating key features from COBIT 4.1, ISO 27000 family of security standards, ISO/IEC 38500, Information Security Governance: Guidance for Information Security Managers, NIST SP 800-53 and other standards and guidelines by Sajko, Hadjina and Sedinić (2011) and Lessing and von Solms (2008). Alves, Carmo and Almeida (2006) presented a framework that was developed by incorporating both strategic (balanced scorecard; political-economic-social-technology (PEST) and strengths-weakness-opportunities-threats (SWOT)) and technical objectives (COBIT and ISO/IEC 17799) with the aim of providing a balanced business and technical approach.

This section has provided a detailed literature review of ISG. While the context of this research is the governance of information security in organisations, it is also important and appropriate to study the concept of governance from the perspective of how organisations are governed (Bihari, 2008; Yaokumah & Brown, 2014a). The following section provides a review of relevant literature on corporate governance.

## 2.4 Corporate Governance

The term "governance" has become an important concept in a variety of disciplines including business and management, public administration, public policy and information systems (Cornforth, 2002). This section explores the main theories behind corporate governance as theories are statements of concepts and their inter-relationships illustrate the occurrence of the phenomenon (Corley & Gioia, 2011). A strong understanding of these theories and their constraints helped the researcher frame the investigation from an additional corporate governance perspective in investigating ISG.

Corporate governance can be defined in its simplest form as "the system by which companies are directed and controlled" as in the *Report of the Committee on the Financial Aspects of Corporate Governance*, popularly known as the *Cadbury report* (Cadbury, 1992). Similar to the discussion on ISG, there are wide variations in the definitions of corporate governance from academic scholarship and corporate governance codes that may be related to different cultural contexts, intellectual backgrounds and reflection of the changing attitudes over time (du Plessis et al., 2011; Gericke, 2018; Klettner, 2017; Turnbull, 1997). Some of these definitions are shown in Table 2-3.

Table 2-3: Definition of corporate governance, adapted from (Klettner, 2017).

| Source of Reference | Definition |
|---|---|
| Report of the Committee on The Financial Aspects of Corporate Governance (Cadbury, 1992) | "Corporate governance is the system by which companies are directed and controlled." |
| G20/OECD Principles of Corporate Governance (OECD, 2015) | "Corporate governance involves a set of relationships between a company's management, its board, its shareholders and other stakeholders. Corporate governance also provides the structure through which the objectives of the company are set, and the means of attaining those objectives and monitoring performance are determined." |
| Australian Corporate Governance Principles and Recommendations (ASX Corporate Governance Council, 2007) | "The phrase 'corporate governance' describes "the framework of rules, relationships, systems and processes within and by which authority is exercised and controlled within corporations. It encompasses the mechanisms by which companies, and those in control, are held to account." |
| UK Corporate Governance Code (Financial Reporting Council, 2016) | "Corporate governance is the system by which companies are directed and controlled. Boards of directors are responsible for the governance of their companies. The shareholders' role in governance is to appoint the directors and the auditors and to satisfy themselves that an appropriate governance structure is in place. The responsibilities of the board include setting the company's strategic aims, providing the leadership to put them into effect, supervising the management of the business and reporting to shareholders on their stewardship. The board's actions are subject to laws, regulations and the shareholders in general meeting." |
| Singapore Code of Corporate Governance (Monetary Authority of Singapore, 2018) | "Corporate governance refers to having the appropriate people, processes and structures to direct and manage the business and affairs of the company to enhance long-term shareholder value, whilst taking into account the interests of other stakeholders." |
| Malaysian Code on Corporate Governance (Securities Commission Malaysia, 2017) | "Corporate governance is defined as the process and structure used to direct and manage the business and affairs of the company towards promoting business prosperity and corporate accountability with the ultimate objective of realising long-term shareholder value while taking into account the interest of other stakeholders." |
| The Corporate Board: Confronting the Paradoxes (Demb & Neubauer, 1992) | "Corporate governance is the process by which corporations are made responsive to the rights and wishes of stakeholders." |
| A Survey of Corporate Governance (Shleifer & Vishny, 1997) | "Corporate governance deals with the ways in which suppliers of finance to corporations assure themselves of getting a return on their investment." |

| Source of Reference | Definition |
|---|---|
| Corporate Governance: Its scope, concerns and theories (Turnbull, 1997) | "Corporate governance describes all the influences affecting the institutional processes, including those for appointing the controllers and/or regulators, involved in organizing the production and sale of goods and services." |
| Corporate Governance: Decades of Dialogue and Data (Daily et al., 2003) | "We define governance as the determination of the broad uses to which organisational resources will be deployed and the resolution of conflicts among the myriad participants in organisations." |
| Corporate Governance (Monks & Minow, 2011) | "In essence, corporate governance is the structure that is intended to make sure: (1) that the right questions get asked and (2) that checks and balances are in place to make sure that the answers reflect what is best for the creation of long-term, sustainable, renewable value." |
| Principles of Contemporary Corporate Governance (du Plessis et al., 2011) | "The system of regulating and overseeing corporate conduct and of balancing the interests of all internal stakeholders and other parties (external stakeholders, governments and local communities who can be affected by the corporation's conduct, in order to ensure responsible behaviour by corporations and to achieve the maximum level of efficiency and profitability for a corporation." |

While the definitions of corporate governance are broad and varied, the definitions cover key areas such as the structures, roles and responsibilities of the board and management, processes encompassing directing, controlling and reporting, and relationships with shareholders and other stakeholders comprising employees, regulators, auditors and suppliers.

### 2.4.1 Corporate Governance Theories

The research into the functions, roles and effectiveness of corporate governance has extended over the years to include economic and finance, management, behavioural and regulatory theories (Chhotray & Stoker, 2009; Hung, 1998; Jensen & Meckling, 1976). In the context of this research, 3 dominant theories of corporate governance: agency theory, stakeholder theory and stewardship theory, are examined. These 3 theories are by no means the only options, but these different theoretical perspectives have been analysed to understand their issues, the functions of governance and the roles of the board and management.

#### 2.4.1.1 Agency Theory

Agency theory helps to explain a corporation's relations to its shareholders, but does not consider management issues or other external and internal stakeholders of a corporation (Nix & Chen, 2013). Agency theory has been very influential in corporate governance and has its origin in finance and economics. According to one of the most cited studies in agency theory (Jensen & Meckling, 1976), the principal–agent relationship is defined as follows:

> Agency relationship as a contract under which one or more persons (the principal(s) engage another person (the agent) to perform some service on their behalf which involves delegating some decision-making authority to the agent.

In agency theory, corporate governance is the mechanism that aims to address the shortcomings attributed to the differences and conflicts in desires and goals of the owner (principal) and the management (agent). Agency theory is also concerned with the problem of risk sharing that arises when the principal and agent have different attitudes towards risk as they have different risk preferences (Eisenhardt, 1989a). The focus on the principal-agent problem gives rise to the compliance model where agents have to be monitored to ensure that they comply with the objectives that are set by the principal. In addition, a prudent risk management model is required to manage the differing attitudes towards risk.

From an agency theory perspective, boards can be used as monitoring mechanisms for shareholders to control the agents driving the control and conformance functions (Eisenhardt, 1989a; Fama & Jensen, 1983).

#### 2.4.1.2 Stakeholder Theory

Stakeholder theory believes that a corporation needs more than just its shareholders to survive. Besides shareholders, there are other stakeholders which include employees, customers, suppliers, regulators and other groups who can affect or are affected by activities of the corporation (Hung, 1998; Shailer, 2018) and the interests of these stakeholders are very important and have to be taken care of for the survival of the corporation (Freeman, 2010). According to Hung (1998), stakeholder theory adopts a pluralistic approach to organisations and the corporate governance mechanism is based on the notion that there are many groups in society besides shareholders and employees to whom the corporation is responsible. The objectives of a corporation can only be achieved by balancing the often conflicting interests of these stakeholders.

In stakeholder theory, the board continues with the monitoring of management (agents) in driving control and conformance functions, but the board has added responsibilities in ensuring that the management meets the interests of not just the shareholders but all other stakeholders. Therefore, the board must monitor management on behalf of many stakeholders and engage in communications with a wide range of stakeholders when setting the strategy and advising management (Klettner, 2017).

Stakeholder theory is a theory of organisation management and is a broader theory compared to agency theory.

#### *2.4.1.3 Stewardship Theory*

Stewardship theory looks at a corporation and the behaviour of its management very differently from agency theory. Agency theory suggests that management focuses on self-interest and will pursue its own interests rather than the interests of the shareholders if not monitored. However, stewardship theory assumes that management (agent) wants to do a good job in achieving the goals of the owner (principal) (Donaldson & Davis, 1991).

Stewardship theory is derived from the disciplines of sociology and psychology, and is based on collaboration, mentoring and empowerment with the belief that management works in the best interest of the corporation (Nix & Chen, 2013). Stewardship theory promotes a partnership model (Cornforth, 2002) where the board actively participates in the strategy and policy formulation and implementation (Klettner, 2017; Nix & Chen, 2013). The main function of the board is to work together with management to improve the corporation's performance, and not to ensure managerial compliance or conformance.

Table 2-4 summarises the key features of the different perspectives of these corporate governance theories.

Table 2-4: Comparison of theoretical perspectives on corporate governance, adapted from (Bajo Davó et al., 2019; Cornforth, 2002).

| Theory | Interests | Board Members | Board Role | Model | Reference Literature |
|---|---|---|---|---|---|
| Agency theory | Owners and managers have different and potential conflicting interests | Owners' representatives | Compliance / conformance and risk management: Safeguard owners' interests, monitor and control managers ensuring compliance | Compliance / risk management model | (Bajo Davó et al., 2019; Cornforth, 2002; Eisenhardt, 1989a; Fama & Jensen, 1983; Klettner, 2017) |
| Stakeholder theory | Stakeholders have different interests | Stakeholder representatives | Balancing stakeholder needs: Strategic alignment, coordinating with management | Stakeholder model | (Bajo Davó et al., 2019; Cornforth, 2002; Freeman, 2010; Hung, 1998) |
| Stewardship theory | Owners and managers share similar interests | Experts | Improve performance: Strategic alignment, add value to strategic decisions, work closely with management | Partnership model | (Bajo Davó et al., 2019; Cornforth, 2002; Donaldson & Davis, 1991) |

### 2.4.2 Corporate Governance Mechanisms

Nearly all modern governance research on governance mechanisms conceptualises them as deterrents to managerial self-interest (Daily et al., 2003). Corporate governance mechanisms have been designed to provide shareholders with assurance that managers are striving to achieve the interests of the shareholders (Daily et al., 2003; Shleifer & Vishny, 1997).

Boards are one of the corporate governance mechanisms. As specified in the *Cadbury report* (Cadbury, 1992), the board of directors is responsible for the governance of its company. The board receives its authority from the shareholders and has the responsibility to control, monitor and advise the management on behalf of the shareholders (Aguilera et al., 2011; Bajo Davó et al., 2019). The control and monitoring roles of the board are aligned with agency theory, where the primary concern of the board is to curb the self-serving interests of management (agent) that may work against the interests of the shareholders (principal) (Aguilera et al., 2011; Fama & Jensen, 1983; Klettner, 2017). However, from the perspective of stewardship theory, boards are expected to work closely with management to provide strategic direction and support to achieve the goals of the corporation and shareholders. In addition, from the stakeholder theory it is important to ensure that boards manage the various stakeholders' interests, e.g. meeting regulators' requirements and ongoing communications in updating the performance to other stakeholders (Cornforth, 2002; Donaldson & Davis, 1991).

Many academics agree that none of these theories can independently provide a full explanation of the role of the board and there is a need for a multi-theoretical approach (Klettner, 2017). Therefore, the development of modern corporate governance has incorporated the various features from these theories to drive a more inclusive mechanism where boards are responsible for the various functions of directing, controlling and monitoring the corporation, as well as communicating to shareholders and other stakeholders the results of their performance (Bajo Davó et al., 2019). Academic scholars have also identified these functions in the control role of the board (oversight, monitoring and compliance), the service role of the board (strategy and policy) and the institutional role of the board (networking and stakeholder communication) to better understand the value-adding functions of boards in governing corporations (Klettner, 2017). The board functions are consistent with the roles of overseeing, directing, reviewing and monitoring, as mentioned repeatedly in the OECD Principles of Corporate Governance (du Plessis et al., 2011; OECD, 2015).

Corporate governance in practice has also been subjected to major reforms over the years and changes to the codes and regulations among others have included board configuration, separation of the position of board chair and CEO, imposing age and term limits for boards of directors and executive compensation (Calder, 2008; Daily et al., 2003; Klettner, 2017). However, for the purpose of this research the literature review has focused only on the corporate governance theories and the role of the board as a governance mechanism to help understand and develop the ISG process model.

#### *2.4.2.1 Board's Corporate Governance Roles*

As discussed in the previous sections, various theories have been used to identify the functions of the board. In a number of these studies, the function of the board in corporate governance is a multifaceted function and may involve overseeing, directing, reviewing and monitoring to ensure the corporation meets the interests of the shareholders and wider stakeholders (Klettner, 2017). Demb and Neubauer (1992) supported this approach that embraces the simultaneous need for control and collaboration in meeting the current corporation needs.

Three key board corporate governance roles are discussed below.

a. Control role

The board's control role is a central element from agency theory where control is essential in ensuring that management is aligned with the shareholders' interests. This control role is required in stakeholder theory to ensure all stakeholders' interests are managed. The components of control comprise activities that involve board oversight, monitoring and ensuring compliance with policies that have been defined. In practice, the board's role in oversight, monitoring and compliance has become increasingly important due to shareholder activism (Daily et al., 2003; Turnbull, 1997) which has demanded more stringent requirements in corporate monitoring. In accordance with this corporate control role, the board aims to reduce agency costs arising from the noncompliance of management through involvement in driving shareholders' objectives through strategic decision-making and control (Zahra & Pearce, 1989). Zahra and Pearce (1989) argued that agency theory places a premium on the board's strategic contribution through the board's involvement in the development of the corporation's strategic direction and the setting of guidelines for implementation and effective control of the strategy. Hence, even in a primarily control role, the board is indirectly involved in strategic decisions.

b. Service role

Stewardship theory stresses service and promotes active board participation in guiding and advising management in achieving improved performance of a corporation (Sundaramurthy & Lewis, 2003). The board is involved in setting the strategy of the corporation with management and is continuously aware of the corporation's strategic direction in driving effective corporate governance (Bart & Bontis, 2003; Klettner, 2017). As stewardship theory encourages partnership and collaboration between the board and management, the board is actively involved in guiding policy formulation and implementation to achieve the corporation's goals.

c. Institutional role

Networking and stakeholder communication are the other key roles of the board. These roles are suggested by stakeholder theory, where the board and management need to balance the interests of all stakeholders (Chambers & Cornforth, 2010; Freeman, 2010). Stakeholder communication ensures transparency in updating all impacted stakeholders.

Borrowing from the extensive corporate governance literature, ISG processes can be examined and defined through an additional theoretical lens based on the understanding of the functions and roles of the board in corporate governance. In fact, many researchers (Allen & Westby, 2007a; Antoniou, 2018; von Solms, 2001b; Whitman & Mattord, 2012) have described ISG as the application of the principles of corporate governance to the information security function.

## 2.5 Research Gap

There are many interpretations of the term "ISG" as discussed in Section 2.3.1. First and foremost, it is critical to provide a consistent and holistic interpretation that covers the broad areas of governance to define the scope of this research. For the purpose of this research, the definition of ISG incorporates the principles of good corporate governance, framework of rules, relationships, systems and processes, the value aspect, attaining objectives and monitoring performance, and concepts of IT governance.

Therefore, ISG is defined as follows:

> ISG is the framework of rules, relationships, systems and processes by which the security objectives of the organisation are set and the means of attaining those objectives and monitoring performance are determined.

The literature review has concluded that there are common themes, various frameworks and critical success factors that have been identified to represent ISG that could lead to improved ISG but may be difficult to implement in practice. The literature review has also concluded that practitioners do not have guidance on how to implement ISG (Lidster & Rahman, 2018; Maleh et al., 2018; Williams, 2013) and the proposed governance models and methods are not capable of adapting to the dynamic security environment (Lidster & Rahman, 2018; Maynard et al., 2018; Williams, 2013) as most of these models are conceptual models.

A further assessment of these situations as discussed in this chapter has identified shortcomings that point to the need to develop models of governance that are grounded in empirical studies of organisation practices that can improve ISG and can be practically implemented in organisations. These shortcomings in ISG research are summarised in Table 2-5, which forms the basis of the research discussed in this thesis.

Table 2-5: ISG research gaps.

| Research Gap | Description |
|---|---|
| RG1: Lack of a holistic ISG framework or model that incorporates the broad areas of ISG | ISG frameworks and models have been presented based on the various concepts and principles of corporate governance, board's responsibilities, risk management, best practices and assurance. There is a lack of ISG frameworks or models that bring all the concepts of ISG together. |
| RG2: Lack of guidance on how to implement ISG | Literature review indicated that most (21) ISG frameworks and models have been developed to |

| Research Gap | Description |
|---|---|
| | explain ISG. There are few (6) ISG frameworks and models that have been developed to assist the implementation of ISG. Even these ISG frameworks and models that were developed to facilitate the implementation of ISG only highlighted what is required to implement ISG and do not address “how” to implement ISG. |
| RG3: Limited ISG frameworks and models that are grounded in empirical studies | Literature review has identified only 9 out of a total of 27 ISG frameworks and models that have been developed grounded on empirical studies. Only two among the 9 have been developed based on multiple case studies methodology. |
| RG4: Lack of an ISG framework or model that easily identifies the processes required to be undertaken by various stakeholder groups involved in ISG | Among the 6 ISG frameworks and models that have been developed to facilitate the implementation of ISG, these frameworks and models only identify “what” that must be implemented covering key principles and critical success factors. Only one ISG framework (Ohki et al., 2009) highlighted at a very high level the core ISG process in ISG, while the other key ISG processes are found in professional publications (Gartner, 2010; Institute of Internal Auditors, 2010; International Organization for Standardization, 2013). However, these are still normative models. |

## 2.6 Summary

This chapter has critically analysed the literature relating to ISG to build the body of knowledge required to identify the key themes that shaped the research questions and research design and methodology. This chapter has also analysed the 3 dominant theories of corporate governance and the functions of governance together with the role of the board and executive management, which has provided an additional perspective on the governance of information security. The next chapter will describe the research design and methodology used in this study.

# Chapter 3
# Research Methodology

The purpose of this chapter is to describe the research methodology adopted for this study. A methodology is defined as the research process from the identification of research problems and the selection of a research strategy through to the analysis, interpretation of findings and reporting of research results (O'Connor, 2012). The nature of the research questions, the researcher's experience and the nature of the study influence the selection of a suitable research methodology (Creswell, 2013).

Sections 3.1 to 3.4 describe the research objectives, the researcher's philosophical assumptions, the process of inquiry and the scope of research that have been adopted in addressing this research. Then Sections 3.5 and 3.6 discuss the specific research methods, while Sections 3.7 and 3.8 explain the data collection and data analysis processes. Section 3.9 states the limitations and biases of the research and Section 3.10 concludes by highlighting the actions taken to ensure the validity and reliability of this research.

## 3.1 Research Objectives

The aim of this study is to develop an ISG process model that explains how ISG can be implemented in organisations to improve ISG. The model aims to identify the governance processes, the relationships between the processes, the stakeholder groups and their roles in ISG. Specifically, the research aims to answer the following research question:

"How can ISG be implemented in organisations?"

The objective of the ISG process model is to provide organisations with an understanding of the processes involved in ISG, the stakeholder groups who are involved with specific roles and responsibilities, and how these ISG processes work together in driving improved ISG in organisations. With the help of this ISG process model, organisations can then improve their implementation of ISG. Consequently, this may also improve the overall information security posture of the organisation.

## 3.2 Research Paradigm

There are 4 philosophical worldviews or paradigms, i.e. postpositivism, constructivism (interpretivist), transformativism and pragmatism (Creswell, 2013). Based on these philosophical assumptions, researchers decide on their preferred research method, which may be quantitative, qualitative or a mixed methods approach (Burrell & Morgan, 2016; Creswell, 2013). The choice of research paradigm is dependent on the nature of the research questions that need to be addressed and the paradigms that have previously been embraced in the field of research. In the context of this study, the field is information systems.

The constructivist/interpretivist research paradigm has been selected for this study as the research aims to explore and understand the research problem through interpreting the meanings assigned by individuals or groups of individuals working in a social environment and relies on the participants' views of the situation being studied to develop a theory or pattern of meaning (Creswell, 2013; Walsham, 1995).

In this research, the main objective is to "make sense" of the ISG practices of organisations. Therefore, qualitative research is most suitable. This approach allows researchers to "see and understand" the context within which decisions and actions take place (Orlikowski & Baroudi, 2002).

### 3.2.1 Researcher's Role

In qualitative research, the researcher plays many roles as they are involved in the gathering of data, analysis of the data, interpreting of the data and finally writing up the findings of the research (Creswell, 2013; Yin, 2018). Therefore, in this research the researcher has taken the necessary actions to ensure the validity and reliability of the research.

As data gathering involved working with people, ethics approval from the Human Ethics Advisory Group (HEAG) was required prior to data collection for this study conducted in the School of Computing and Information Systems at the University of Melbourne. An ethics application was submitted and approved (Ethics ID: 1749890) for this study before any research was undertaken. The researcher has taken relevant steps to comply with the ethics guidelines of the HEAG and the ethics approval was renewed annually throughout the duration of this research. A copy of the ethics approval is found in Appendix A.

The researcher also ensured that the necessary approvals were obtained from the interviewees prior to any data gathering. Consent letters were signed by all interviewees and interviewees were also informed about the safeguarding of the confidentiality of information and that all data gathered would be anonymised to protect the identity of the interviewees. The Consent Form and Plain Language Statement (PLS) are found in Appendix B.

As the project involved interpretive research and is based on both expert interviews and case study methods, the findings are heavily dependent on the integrative powers of the researcher (Benbasat et al., 1987). It was therefore critical for the researcher to be aware of and to minimise bias in shaping and influencing the direction of the research (Creswell, 2013; Yin, 2018). Further actions taken for this research are discussed in Section 3.10.

## 3.3 Research Design

The aim of this research is to explore how ISG can be implemented and improved in organisations. As highlighted in Section 3.1, the research needs to identify the governance processes, the relationships between the processes, the stakeholder groups and their roles in ISG in the proposed ISG process model.

The constructivist paradigm drives the inquiry of this research. The researcher has selected a qualitative research approach, which is best suited to address the research question of "how?". The heart of the research is semi-structured interviews to obtain both retrospective and real-world accounts from those people experiencing the phenomenon of theoretical interest (Gioia et al., 2012). As this is qualitative research, the following are key characteristics of the research (Creswell, 2013; Williamson, 2002; Yin, 2018):

a. The research was conducted on-site.

b. The research employed multiple data sources such as semi-structured interviews, site process walk-throughs, documents and reports.
c. The researcher played a primary role in data collection through examination of documentation, attending and observing process walk-throughs, interviewing participants and transcribing interviews.
d. The researcher was involved in all data analysis, coding of data and identifying of the emerging themes and theories.

This exploratory research was structured into 3 phases which took the initially developed conceptual process model through to empirical analysis for further development, evaluation and refinement, and then to final validation with expert interviews. The research design consisted of different methodologies and techniques to collect, analyse and validate empirical data in the different phases. The research design is illustrated in Figure 3.1.

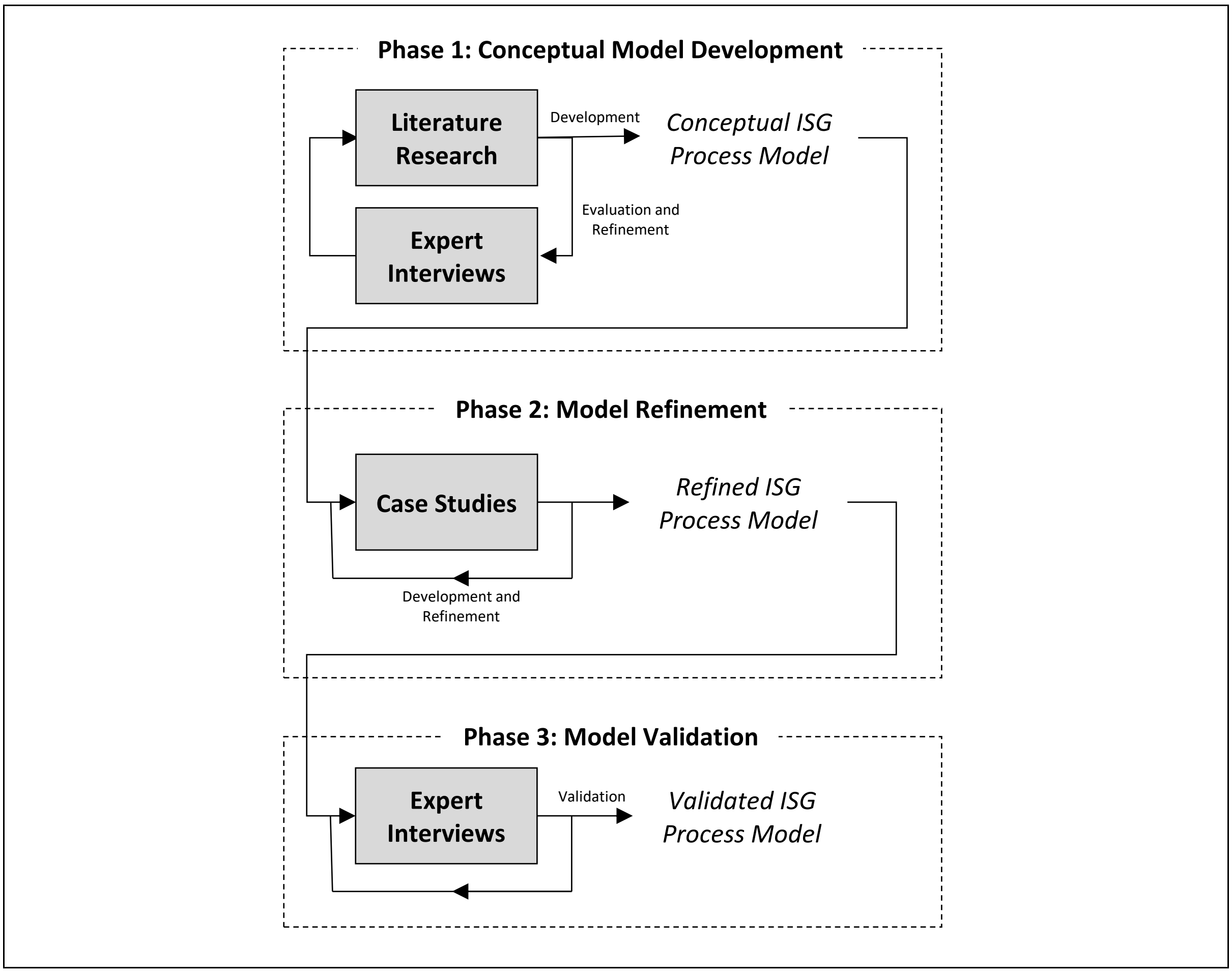

Figure 3-1: Exploratory research design

a. Phase 1: Conceptual model development

Phase 1 focused on the development of a conceptual model. This phase consisted of two stages, i.e. the literature research and the expert interviews. During the literature research, the area of ISG and associated domains, and corporate governance theories were investigated from a theoretical perspective. A detailed analysis of various information covering existing literature, journal articles, conference proceedings, professional publications, security vendor whitepapers and consultant reports was undertaken to frame and develop the conceptual ISG process model. This conceptual ISG process model identifies the core ISG processes, stakeholder groups and their relationships with each other.

The second stage was where expert interviews were used as preparation for data collection, i.e. an approach to test the interview questionnaire and to obtain an initial understanding from an empirical perspective of the subject of research (Yin, 2018). These experts were asked to share their understanding and expert views on ISG. Subsequently, the experts were shown the conceptual model and were asked to comment on the need and relevance of the model in facilitating understanding and implementation of ISG. Their comments and knowledge were analysed to refine the interview questionnaire and to evaluate and improve the conceptual model.

A detailed discussion of the conceptual model development is found in Chapter 4.

b. Phase 2: Model refinement

Phase 2 was exploratory research aimed at developing theories based on detailed analysis of case study interviews. In this phase, 3 case study organisations were selected and a total of 17 interviews across the various hierarchies within the organisations were conducted. Open-ended questions in a semi-structured manner were used in the interviews to elicit ideas, comments and knowledge relating to ISG. A copy of the interview guide is provided for reference in Appendix C. These interviews were transcribed, coded and analysed accordingly to identify second-order themes and aggregated dimensions, which were used to further develop and refine the ISG process model. Follow-up interviews were arranged with interviewees when additional clarifications were required or information needed to be verified.

A detailed discussion of the case study research method for this exploratory research is given in Section 3.6 and the analysis of the case studies in generating the refined ISG process model is discussed in detail in Chapter 5.

c. Phase 3: Model validation

Phase 3 focused on validating and confirming the proposed theory, i.e. the refined ISG model that was developed from Phase 2. Phase 3 utilised the expert interview research method where 6 experts were interviewed about their expert practices in the field of ISG. The experts were subsequently shown the refined ISG process model developed from Phase 2 to solicit their views and feedback. The expert interview data were transcribed, coded and analysed to identify new and validate existing second-order themes, aggregated dimensions and therefore the ISG process model.

An evaluation was done on whether to use the focus group or expert interview method in validating the refined ISG process model. One of the objectives of the validation was to share the refined ISG process model with the expert interviewees to seek their knowledge and feedback in validating and confirming the ISG process model. As all the selected interviewees were individuals involved in governance, they were also individuals who held senior roles in organisations and it would have been extremely difficult to get groups of them together at the same place and same time for a focus group discussion. Therefore, it was more feasible to arrange one-to-one interviews with the selected interviewees. In addition, the expert interview method was more suitable for the specific purpose of Phase 3 as it allowed more time and flexibility for the interviewees (Bogner & Menz, 2009; Meuser & Nagel, 2009) to share their expert knowledge and views in confirming the ISG process model without being afraid of unintentionally sharing competitive or organisation-sensitive information.

A description of the expert interview research method is found in Section 3.5 and detailed analysis of the expert interviews and validation of the model to finalise the validated ISG process model is found in Chapter 6.

## 3.4 Scope of Study

In designing the research and deciding on the industry which would provide the most relevant data for the topic of information security and ISG, several critical industries such as financial services, telecommunications and energy were considered. For the purpose of this research, the financial services industry has been selected. The financial services industry was selected for the research as information security is a critical area of concern for the industry because the cost of information breaches is extremely high, with significant impacts on both customers and organisational operations (Bouveret, 2018; European Central Bank, 2018; IBM, 2019; Ponemon Institute, 2016, 2018). Financial services organisations have also invested heavily in information security and are leading the way in information security practices. Therefore, conducting research in financial services organisations provided the opportunity to understand the actual situational challenges and capture the knowledge of practitioners to identify potential best practices for theory development.

## 3.5 Expert Interviews

This section describes the expert interview which has been used as one of the research methods in this study. Expert interviews as a method of qualitative empirical research have been developed and used since the early 1990s (Bogner et al., 2009) and have been further discussed and evaluated as a method of theory generation (Bogner & Menz, 2009; Meuser & Nagel, 2009; Pfadenhauer, 2009).

Based on the definition provided by Bogner and Menz (2009), an expert is an individual who possesses technical, process and interpretative knowledge in a specific field of action by virtue of the fact that the individual practises in a particular professional area, which in the case of this research is in one or all of the areas of information security, ISG or corporate governance.

Bogner and Menz (2009) developed a typology of the expert interview in which 3 forms of expert interviews were identified, i.e. the exploratory interview, systematising interview and theory-generating interview. The first form is the exploratory interview, which serves to establish initial orientation in a field that is new or poorly defined. The interviews are generally conducted as openly as possible with the aim of sounding out the subject under investigation. The second form is the systematising interview, which has a different focus on gaining access to the expert's daily routines and practices to obtain systematic and complete information. These interviews are normally done with a structured questionnaire such as via a survey as data need to be comparable in relation to the subject matter and they are therefore often used in multi-method research approaches. The 3rd form is the theory-generating interview, where the goal is to obtain interpretive knowledge. These interviews are semi-structured interviews that allow the expert flexibility in sharing their knowledge but in relation to the topic under investigation so that the researcher is able interpret the knowledge during data analysis.

In this research, expert interviews were used in two phases for two different purposes.

a. The conceptual ISG model in Phase 1 was developed purely based on theoretical literature and the researcher's understanding of the problem area. Four expert interviews were used in Phase 1 as pilot interviews and as an exploratory research tool to establish an initial orientation in the target problem area, to structure the area under investigation and to refine the interview guide that was used in subsequent data gathering in the case study research in Phase 2 and further expert interviews in Phase 3. The initial feedback from the expert interviews was also used to evaluate and refine the conceptual ISG process model from a practical perspective (Bogner & Menz, 2009). Therefore, the first form of exploratory interview was used in Phase 1.
b. In Phase 3, 6 expert interviews were used to validate and confirm the refined ISG process model developed in Phase 2. Therefore, the 3rd form of expert interview, i.e. the theory-generating interview, was used (Bogner & Menz, 2009). A semi-structured interview guide was developed for the interviews to guide the experts on the areas that were required to be validated, but the experts were allowed flexibility and were free to express their views based on their experience, knowledge and what was practised in their organisations. In addition, the refined ISG process model was also used as a catalyst to drive the latter part of the interview as the objective of the expert interviews was to obtain relevant feedback and knowledge confirmation from the experts. The findings from these interviews assisted in further triangulation and supported the generalisation of the theories.

### 3.5.1 Selecting Expert Interviewees

With the specific research objectives in mind, the experts were selected based on the following criteria:

a. Individuals working in the financial services industry
b. Individuals from different levels of the hierarchy of the organisations that mapped against the proposed conceptual model definition comprising members of the board and executive, tactical and operational management
c. Senior executives in organisations who understood governance or were involved in some governance processes; they could be involved in corporate governance, IT governance or ISG
d. Senior executives who were involved directly or indirectly with information security in the organisation
e. Information security risk consultants or information security professionals

Experts were identified through personal contacts and business networks, which meant some of the interviewees were direct contacts of the researcher while others were contacts of colleagues or friends. Due to the nature of the research, which is on the topic of governance, all experts who were identified held high positions in their organisations and were generally very busy with their work, and some had busy travel schedules. Therefore, it was critical to elaborate on the importance of the research topic and demonstrate the value of their participation to get them agree to invest their time for the interviews.

A total of 15 experts were contacted, but a final 10 experts were interviewed. Five of the contacted experts were not interviewed as they either declined the interviews upfront due to the sensitivity of the information security topic or had their interviews rescheduled too many times due to their busy schedules which resulted in the interviews ultimately being cancelled. Of the 10 experts, 4 experts were interviewed for Phase 1 and 6 experts were interviewed for Phase 3. There was no specific target number of expert interviews that was set in advance; however, the final number of expert interviews was reached when saturation was achieved as no additional information was being obtained from additional interviews. All these experts were sourced from organisations other than the case study organisations. The profiles of these experts are shown in Table 3-1.

Table 3-1: Expert interviewee profiles.

| **Interviewee** | **Organisation** | **Location & Role Coverage** | **Job Role** | **Involvement in Information Security** | **No. of Years in Industry** |
|---|---|---|---|---|---|
| **Expert Interviews in Phase 1** | | | | | |
| Expert #1 (CIO-MY_Bank) | Financial institution (bank) | Malaysia | CIO | Direct | 20 years+ |
| Expert #2 (CIO-SG_Bank) | Financial institution (bank) | Singapore | CIO | Direct | 30 years+ |
| Expert #3 (GCCO-MY_Bank) | Financial institution (bank) | Malaysia & South East Asia | Group CCO | Indirect | 20 years+ |
| Expert #4 (GCOO-MY_Bank) | Financial institution (bank) | Malaysia & South East Asia | Group COO | Indirect | 20 years+ |
| **Expert Interviews in Phase 3** | | | | | |
| Expert #5 (IS_ConsultingDirector) | Consulting firm | Singapore & South East Asia | Information Security Consultant | Direct | 20 years+ |
| Expert #6 (IS_Consultant) | Consulting firm | Singapore & South East Asia | Information Security Consultant | Direct | 10 years+ |
| Expert #7 (CISO-MY_Bank) | Financial institution (bank) | Malaysia & South East Asia | CISO | Direct | 25 years+ |
| Expert #8 (CIO-SG_InvestmentCo) | Financial institution (investment company) | Singapore | CIO | Direct | 25 years+ |

| Interviewee | Organisation | Location & Role Coverage | Job Role | Involvement in Information Security | No. of Years in Industry |
|---|---|---|---|---|---|
| Expert #9 (ChiefInfoRiskOfficer-APAC_InsuranceCo) | Financial institution (insurance company) | Malaysia & Asia Pacific | Regional Chief Information Risk Officer | Direct | 35 years+ |
| Expert #10 (IS_SoftwareConsultant) | Information security software company | Singapore | Information Security Consultant | Direct | 25 years+ |
| | | | | **Total number of expert interviewees** | **10** |

## 3.6 Case Study Research Method

The case study research method is a qualitative approach in which the researcher studies one case or multiple cases through detailed, in-depth data collection involving multiple sources of information that can cover participant interviews, process walk-throughs, documents and reports. The researcher conducts a detailed analysis to uncover new theory (Creswell, 2013; Eisenhardt, 1989b; Yin, 2018). In this study, the case study research method has been selected as the preferred method. Case studies can be used to study situational processes as they allow contextual factors and process elements to be considered in a real-life situation (O'Connor, 2012) and ISG is more of an organisational rather than technical field that involves many actors in a practical environment (Benbasat et al., 1987; Darke et al., 1998). In addition, the case study research method is well-suited to ISG research, which is a knowledge domain that continues to develop and few empirical studies have been undertaken in this area. Finally, the case study method is an ideal mode of inquiry for addressing research questions regarding "how" things occur such as the focus of this research, i.e. how ISG can be implemented in organisations and how ISG is operationalised in organisations - the investigation of situational processes (Orr & Scott, 2008; Yin, 2018).

A multiple case study design was adopted for this research to enhance the generalisability of the findings and multiple cases were carefully selected based on theoretical sampling to replicate or extend emergent theory (Eisenhardt, 1989b; Eisenhardt & Graebner, 2007; Gioia et al., 2012; Yin, 2018). Each case study is a distinct "experiment" that stands on its own as an analytic unit and multiple cases are discrete analytic units that serve as replications and extensions of the emerging theory (Eisenhardt, 1989b; Yin, 2018).

### 3.6.1 Setting

The case study research was conducted on-site at the locations of the case study organisations. In a few situations, the case study interviews were conducted at a location outside of the case study organisations where the interviewees were able to share information more freely without work interruptions. Where the case study interviews were done outside of the case study organisation, the data gathering process through the interviews was not impacted in any manner. When additional evidence was required, the researcher arranged follow-up interview sessions at the case study locations to collect the additional evidence such as organisation charts, reports and sample security frameworks, as well as policy documents. Further data collection such as process walk-throughs was also conducted on-site for all 3 case study organisations.

### 3.6.2 Selecting Cases

The selection of multiple case study organisations was done based on theoretical sampling with the purpose of replicating or extending emergent theory. While a fixed number of case study organisations was not established at the beginning of the research, the research ended with 3 case study organisations when saturation of the findings was achieved. The case study organisations were carefully selected with the intention of literal replication (Yin, 2018).

The selection of case study organisations for this research was difficult and limited due to financial institutions being very protective of their information security stature and initiatives, particularly when the industry is strictly regulated and there are strict regulations on information protection. In this research, 3 financial institutions were selected for the case studies based on the following reasons:

a. The organisations are strictly regulated with established corporate governance processes covering board oversight, risk management and compliance processes as described in their annual reports or corporate profiles.
b. The organisations are known to have mature information security postures and are investing heavily in information security initiatives.

Based on the above selection criteria and the interest from the organisations in participating, one financial institution in Singapore and two financial institutions in Malaysia were selected for the multiple case study research. Table 3-2 provides high-level profiles of the organisations. A more detailed description of the organisations together with the findings and analysis is found in Chapter 5.

Table 3-2: Case study organisation profiles.

| | FinServices_SG | FinServices_SEA | FinServices_MY |
|---|---|---|---|
| **Business Area** | Specialist financial service provider | Regional commercial bank | Local commercial bank |
| **Country of Operations** | Singapore and small regional operations across South-East Asia | Malaysia, Singapore, Vietnam, Hong Kong and China | Malaysia |
| **Number of Employees** | ~ 1000 | ~ 7000 | ~ 3500 |
| **Separate CISO Office** | Yes | No | No |
| **Illustrative Organisation Structure with Information Security Function** | 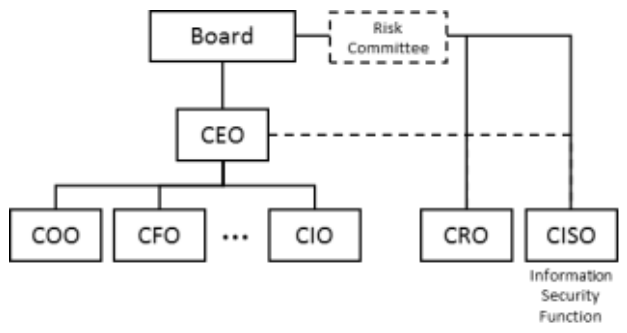  | 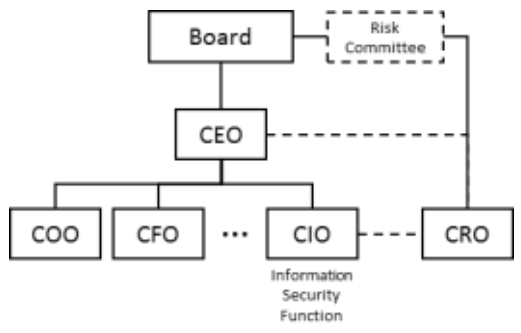  | 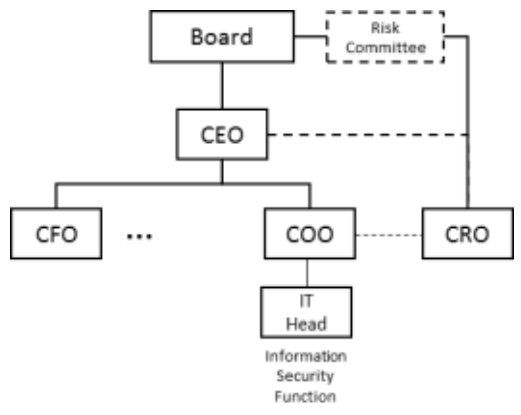  |

### 3.6.3 Selecting Case Study Interviewees

The case studies have been carefully selected to aid in answering the research questions on how ISG can be implemented and improved in organisations. To answer the research questions and to ensure that the relevant information was collected, the aim was to interview individuals in these organisations who were involved directly or indirectly in information security, IT security or governance processes covering decision oversight, risk and compliance, and auditing. The individuals were generally C-level executives holding the positions of CIO, CISO, chief operating officer (COO), chief risk officer (CRO), chief compliance officer (CCO) or board members of the organisations.

It is difficult to secure interviews with C-level executives as they tend to be busy and are generally not interested in participating in research unless the topic is of special interest to them. As information security is a high-priority topic for executives in the financial services industry, many of the individuals contacted for the study were willing to participate if an available time slot could be identified. Therefore, the challenge was to confirm the schedules for the interviews as there were normally several requests for rescheduling.

The approach adopted by the researcher was to contact one or more of the potential interviewees in the selected organisations who also happened to be a personal or business contact to discuss the purpose of the research and for the first interview. The CIO of FinServices_SG, a board member of FinServices_SEA and the COO of FinServices_MY were the first individuals who were interviewed in each of the organisations. Leveraging the snowball sampling technique, the researcher worked with these first interviewees to identify additional interviewees within the organisations whom the researcher believed were relevant to the study. It was easier for someone within the organisations to refer their colleagues and facilitate these additional invitations.

A total of 25 were approached, but 17 interviewees participated in the 3 case studies. Table 3-3 shows the case study interviewee profiles.

Table 3-3: Case study interviewee profiles.

| **Participant** | **Job Role** | **Involvement in Information Security** | **No. of Years in Industry** |
|---|---|---|---|
| **FinServices_SG** | | | |
| Interviewee #1: FinServices_SG_CIO | CIO | Direct | 20 years+ |
| Interviewee #2: FinServices_SG_DeputyCIO | Deputy Head of Technology (Deputy CIO) | Indirect | 25 years+ |
| Interviewee #3: FinServices_SG_CISO | CISO | Direct | 20 years+ |
| Interviewee #4: FinServices_SG_Director-InfoSecOfficer | Director, Information Security Officer | Direct | 20 years+ |
| Interviewee #5: FinServices_SG_Head-IT-Security (During Process Walk-through) | Head of IT Security | Direct | 15 years+ |
| Interviewee #6: FinServices_SG_IT-SecOfficer (During Process Walk-through) | IT Security Officer | Direct | 5 years+ |
| **FinServices_SEA** | | | |
| Interviewee #1: FinServices_SEA_Board | Non-Executive Board Member | Indirect | 35 years+ |
| Interviewee #2: FinServices_SEA_CFO | CFO | Indirect | 25 years+ |
| Interviewee #3: FinServices_SEA_CIO | CIO | Direct | 20 years+ |
| Interviewee #4: FinServices_SEA_IT-Architect | Head, Group IT Architect | Direct | 15 years+ |
| Interviewee #5: FinServices_SEA_Head-IT-Infra (During Process Walk-through) | Head of IT Infrastructure | Direct | 10 years+ |
| Interviewee #6: FinServices_SEA_IT-SecOfficer (During Process Walk-through) | IT Security Officer | Direct | 10 years+ |
| **FinServices_MY** | | | |
| Interviewee #1: FinServices_MY_COO | COO | Indirect | 35 years+ |
| Interviewee #2: FinServices_MY_Board | Non-Executive Board Member | Direct | 25 years+ |
| Interviewee #3: FinServices_MY_CRO | CRO | Indirect | 20 years+ |
| Interviewee #4: FinServices_MY_Head-IT (During Process Walk-through) | Head of IT | Direct | 20 years+ |
| Interviewee #5: FinServices_MY_IT-AppSecOfficer (During Process Walk-through) | IT Application Security Officer | Direct | 10 years+ |
| | | **Total number of case study interviewees** | **17** |

## 3.7 Data Collection Strategies

In qualitative research, researchers combine multiple data collection methods to enable triangulation to strengthen theory building (Eisenhardt, 1989b; Yin, 2018). Yin (2018) argued that case study evidence can come from 6 potential sources covering documents, archival records, interviews, direct observations, participant-observation and physical artefacts.

In both the expert interviews and case study research, interviews were the primary data source. During these interviews, additional documents were collected as references. Some organisation-specific documents were also shared under a non-disclosure agreement to provide a better understanding of the topics that were discussed during the interviews. These additional documents were normally shared during the process walk-throughs or during less formal discussions with the information security teams. The following sections describe the data gathering processes that were adopted in the research.

### 3.7.1 Interviews

Interviews were the main source of data for both expert interviews and case study research. All interviews followed a structured process to increase the comparability and reliability of the interviews across different interviews and across case studies (Schlegel, 2015). While the interview process was structured, the interview questions were semi-structured with a list of open-ended questions prepared as a guide for spontaneous discussion around second-order themes. Interviewees were provided the flexibility to elaborate on ideas. By adopting semi-structured questions, the researcher could also follow up leads provided by the interviewees. The interview guide was adapted based on feedback and experiences gained from the initial expert interviews in Phase 1. The semi-structured interview guide is shown in Appendix C.

The researcher also recognised that the interview questions could be modified with the progression of the research when required to drive specific insights as highlighted by the interviewees. These adjustments allowed the researcher to probe second-order themes with exiting interviewees in subsequent interviews and with subsequent interviewees (Eisenhardt, 1989b).

Each interview was structured as follows:

a. Welcome

Each interview started with small talk and a short discussion of the interviewee's background and current role. For interviewees who were personal and business contacts, discussions included some background on my objective in pursuing my PhD. These initial discussions helped to build rapport and a personal relationship, which are crucial for successful interviews. In addition, according to Schlegel (2015) gaining more information on the background and role of the participant helps to provide contextual information in the formulation of questions and the interpretation of the data collected.

b. Introduction to research

As part of the introduction, the researcher explained the purpose of the research as provided in the PLS, which was emailed earlier to the interviewees as part of the invitation email. The researcher also provided an introduction to the research topics to ensure a common understanding of the research project and avoid misunderstandings. Once a brief introduction was done, the researcher went through the key points on the Consent Form covering areas in ethics, confidentiality and consent to audio recording. The interview commenced only after agreement with the interviewee and the Consent Form was signed. A copy of the PLS and Consent Form are found in Appendix B.

c. Actual interview

The actual interview covered a few major areas introduced in the form of open-ended questions. First, the researcher sought the interviewee's understanding of information security, then a more specific understanding of ISG. The second area covered the implementation of information security and ISG specifically in the case study organisation. The 3rd area concerned the interviewee's role and other different roles within the organisation who were involved in ISG. Finally, for the 4th area the interviewee was asked their views on the future development of ISG and its challenges.

Although there was a conceptual ISG model that had been developed as a guide for fieldwork, the conceptual ISG model was not referenced in any part of the interview until the interview was near completion. Every attempt was made to begin all interviews with a *tabula rasa* – an open mind (Glaser & Strauss, 1967) – or with "the suspension of belief" as proposed by Gioia, Corley and Hamilton (2012) to minimise researcher bias. Only at the conclusion of each interview when all data gathering had been completed was the interviewee shown the conceptual ISG model as part of validation and asked about its suitability and its usefulness in helping organisations implement ISG. This generally started another round of discussion that provided further insights into the topic under research.

d. Closing

At the end of the interview, the interviewee was thanked for the participation. In some instances, interviewees were asked to introduce additional colleagues whom they believed were suitable to be interviewed for the research. The researcher also requested permission from the interviewee on potential follow-up interviews should there be the need for further clarification.

#### *3.7.1.1 Case Study Interviews*

Most of the case study interviews lasted between 45 and 90 minutes and were conducted between June 2018 and September 2018. Four of the interviews lasted about two hours as there were follow-up discussions with information security team members that were held on location at the case study organisations. Some less formal exchanges occurred with members of the team covering various areas of information security policies and procedures administration, and reviewing of sample information security team monthly meeting minutes and information security frameworks.

All interviews were conducted in the English language. Twelve of the 17 interviews were audio recorded. Audio-recorded interviews were transcribed within 7 days and transcriptions were done with Express Scribe V8.06 software distributed by NCH Software. For the 5 interviews that were not audio recorded, detailed interview notes were taken and special attention was paid to writing down specific quotes and comments verbatim. This was done to ensure the essence of the discussions was captured as interviewees' quotes are important for subsequent theory-generating data analysis (Gioia et al., 2012). These interviews generally took longer, lasting about 90 minutes, as extra effort was required for note-taking. Non-audio-recorded interviews were written up within 24 hours to ensure that the notes and impressions were fresh in the researcher's mind.

#### 3.7.1.2 Expert Interviews

Four expert interviews were conducted in Phase 1 during the development of the conceptual ISG model between December 2017 and June 2018, and 6 expert interviews were conducted during Phase 3 between August 2018 and March 2019 to validate and confirm the refined ISG process model that was developed in Phase 2. All these interviewees held senior positions in different organisations other than the case study organisations. As these interviewees were often constrained by the availability of their time for the interviews, all these expert interviews were completed within 60 minutes with the exception of two which took about 90 minutes as the interviewees spent more effort in elaborating on the discussions with illustrations and supporting documents.

As with the case study interviews, all expert interviews were conducted in the English language. Six of the 10 expert interviews were audio recorded. Similarly, audio-recorded interviews were transcribed within 7 days and transcriptions were done with Express Scribe V8.06 software distributed by NCH Software. For the 4 non-audio-recorded interviews, detailed interview notes were taken and written up within 24 hours to ensure that the notes and impressions were fresh in the researcher's mind.

### 3.7.2 Process Walk-throughs

In addition to the interviews, the researcher attended process walk-throughs at all 3 case study organisations. During these process walk-throughs, the researcher was taken through a description of the overall ISG and management processes adopted by the organisations. With the exception of one financial institution where the researcher was taken to visit and speak to the various departments, these process walk-throughs were normally done with the help of PowerPoint slide presentations. These allowed the researcher to have additional informal discussions with the information security team to better understand how information security was governed and managed. As these sessions covered the processes, the information was used to verify the data collected via interviews.

### 3.7.3 Documentation

Documentation formed another source of data that was gathered for this research. Documentation collected for this research included publicly available information such as annual reports of the case study organisations, press releases and investor information. These documents were collected prior to the interviews to provide the researcher with some background information. Specific documentation which was relevant for data analysis included published corporate governance statements and risk management processes of organisations, and regulatory guidelines which were found on regulators' websites.

In addition, the researcher was shown organisations' specific documentation which included information security frameworks, policies and procedures. While most of these documents were private and confidential, and were not allowed to be removed from the organisations, the researcher was allowed to copy down some relevant information (masked and anonymised) for the research.

In achieving strong reliability of the research, all data gathered for this research are stored in a research database. Sources of data covering interview transcripts, documentation, interview memos and notes are labelled and stored accordingly. All data sources that were required for analysis, i.e. coding, were labelled and imported electronically into NVivo 11 software as a centralised research database, as shown in Figure 3-2.

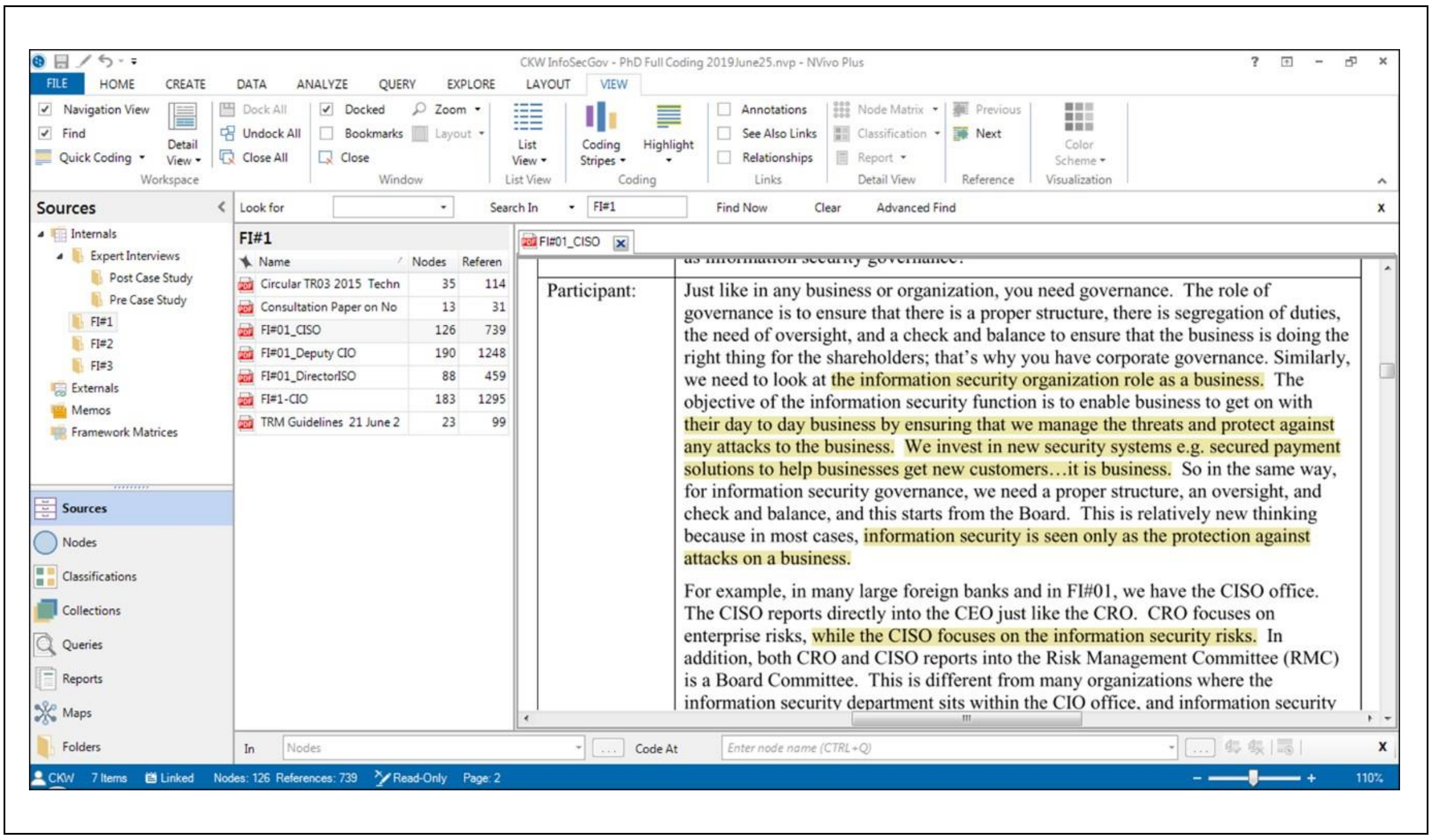

Figure 3-2: Research database of electronic data sources (NVivo 11)

## 3.8 Data Analysis Processes

The same data analysis processes were used to analyse the data gathered from expert interviews and from case studies as both datasets collected were qualitative data and primarily texts. A number of qualitative analysis approaches were considered for this research, namely the systematic approaches proposed by Eisenhardt (1989b) and Gioia et al. (2012) and the data displays to identify emergent theories from Miles, Huberman and Saldaňa (2014). However, the Gioia method was the primary method used for analysis of the raw data to generate theories in this research as the emergent theories were interpreted and grounded on empirical data with a priority on the voices of the interviewees (Gehman et al., 2018).

NVivo 11 software was used as the CAQDAS (Computer Assisted Qualitative Data Analysis Software) tool to facilitate the analysis. Specifically, NVivo 11 was used for the consolidation of all documentation as a centralised research database, capturing of research journals and theoretical memos, and coding to facilitate the analysis of emerging themes. Various queries and visual display features of NVivo 11 were leveraged to identify second-order themes and theories.

The data analysis processes are summarised and illustrated in Figure 3-3.

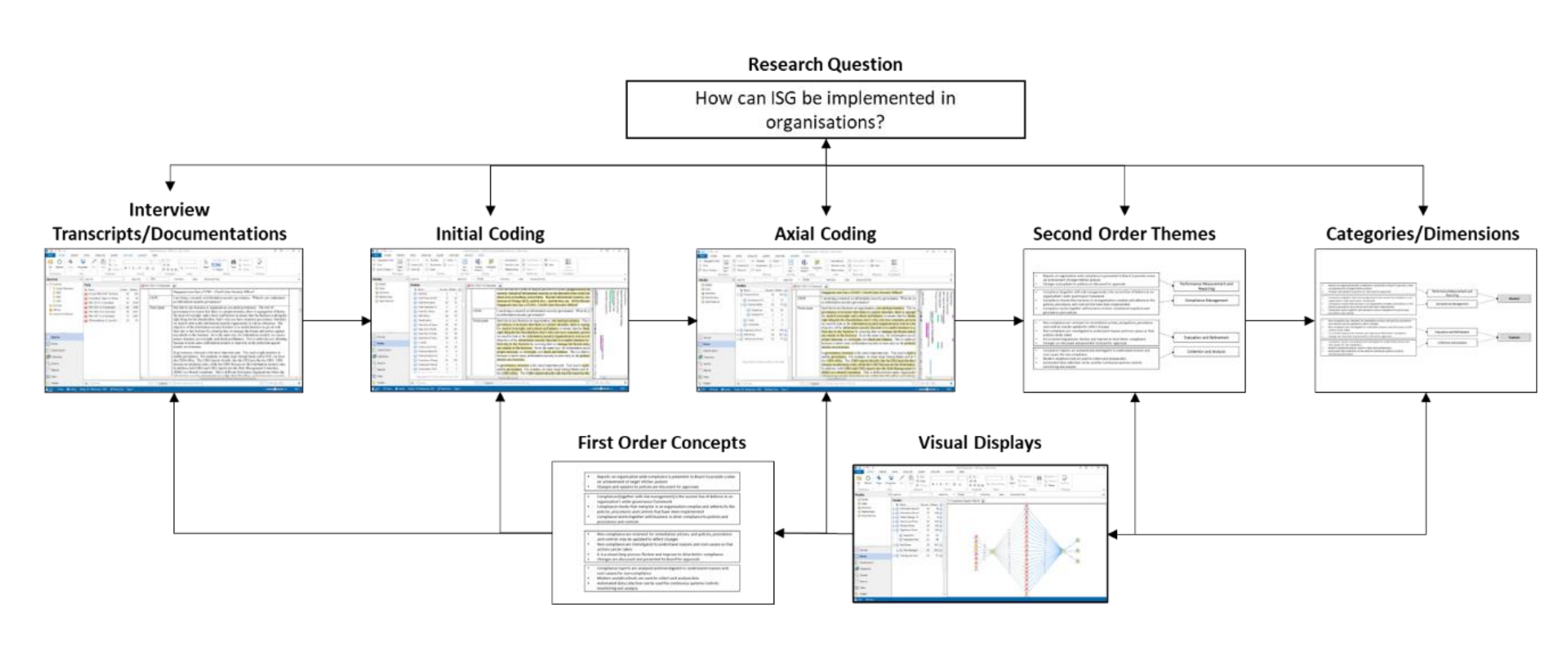

Figure 3-3: Overview of data analysis approach (adapted from Corbin & Strauss, 1990 and Gioia et al., 2012)

Data analysis started as soon as data gathering commenced and it was an iterative process. Once interviews were transcribed, the interview transcripts were read to capture initial concepts and relationships between different elements in the transcripts. Theoretical memos were written throughout the data analysis process as this facilitated reflection on the interview discussions and on the collected data (Charmaz, 2006; Miles et al., 2014). Moreover, the cycling back and forth between thinking about the existing data and generating strategies for collecting new and better data avoided the issue of collecting too much irrelevant data and missing out on opportunities to gather relevant data (Miles et al., 2014).

Data analysis began with the identification of initial concepts and terms thorough initial coding or open coding (Corbin & Strauss, 1990). “In-vivo” codes that captured ideas and verbatim quotes from interviewees were coded; otherwise, simple descriptive words/phrases were also used for the initial coding. A total of 194 codes were identified in the initial coding. The interview transcripts were re-read and re-coded continuously to refine the codes, forming the first-order concepts. Next, axial coding was conducted to identify relationships (similarities and differences) between and among the initial codes, and the number of initial codes was condensed, using the interviewees’ terms whenever possible (Gioia et al., 2012). Visual display features in NVivo 11 (e.g. node comparison charts) were also used to facilitate identification of similarities and differences in codes. These coding processes were done iteratively, always with the objective of answering the research question: "How can ISG be implemented in organisations?".

It was the continuous iterative coding and analysis that drove the abstraction of these codes to become higher level concepts and categories forming the second-level themes. As a result, a total of 20 second-order themes were identified, which were further condensed to 9 aggregated categories or dimensions. During the abstraction of second-order themes and the identification of dimensions, extant literature and existing theories were examined for possible explanations. This constant comparison of theories and data was carried out for iterating towards a theory that closely fit the empirical data. Throughout the coding and analysis processes, further questions were raised and these were addressed in subsequent interviews and case studies, which led to a more focused attempt to answer the research questions. These adjustments allowed the probing of second-order themes with existing interviewees or via the addition of new interviewees (Cavaye, 1996; Eisenhardt, 1989b).

These processes were done for all interview transcripts and documentation within a single case, and extended to other cases until no new information and no new second-order themes and dimensions were identified. The data gathering and analysis processes were deemed completed when theoretical saturation was achieved. Cross-case analysis was conducted with the aim of extending emerging theories and confirming replications (Yin, 2018).

Individual expert interview transcripts from Phase 3 were analysed in a similar way, but the coding and analysis adopted a template analysis approach (King et al., 2004, 2018) where the initial codes and second-order themes discovered during case study analysis ("a priori" themes) were used as the initial template. Additional codes, second-order themes and dimensions discovered were used to update the initial template. The codes, themes and dimensions were constantly compared against the codes, themes and dimensions that were identified. This continued across all 6 expert interviews that were used to validate the emergent theories.

The process across multiple case studies and expert interviews is illustrated in Figure 3-4.

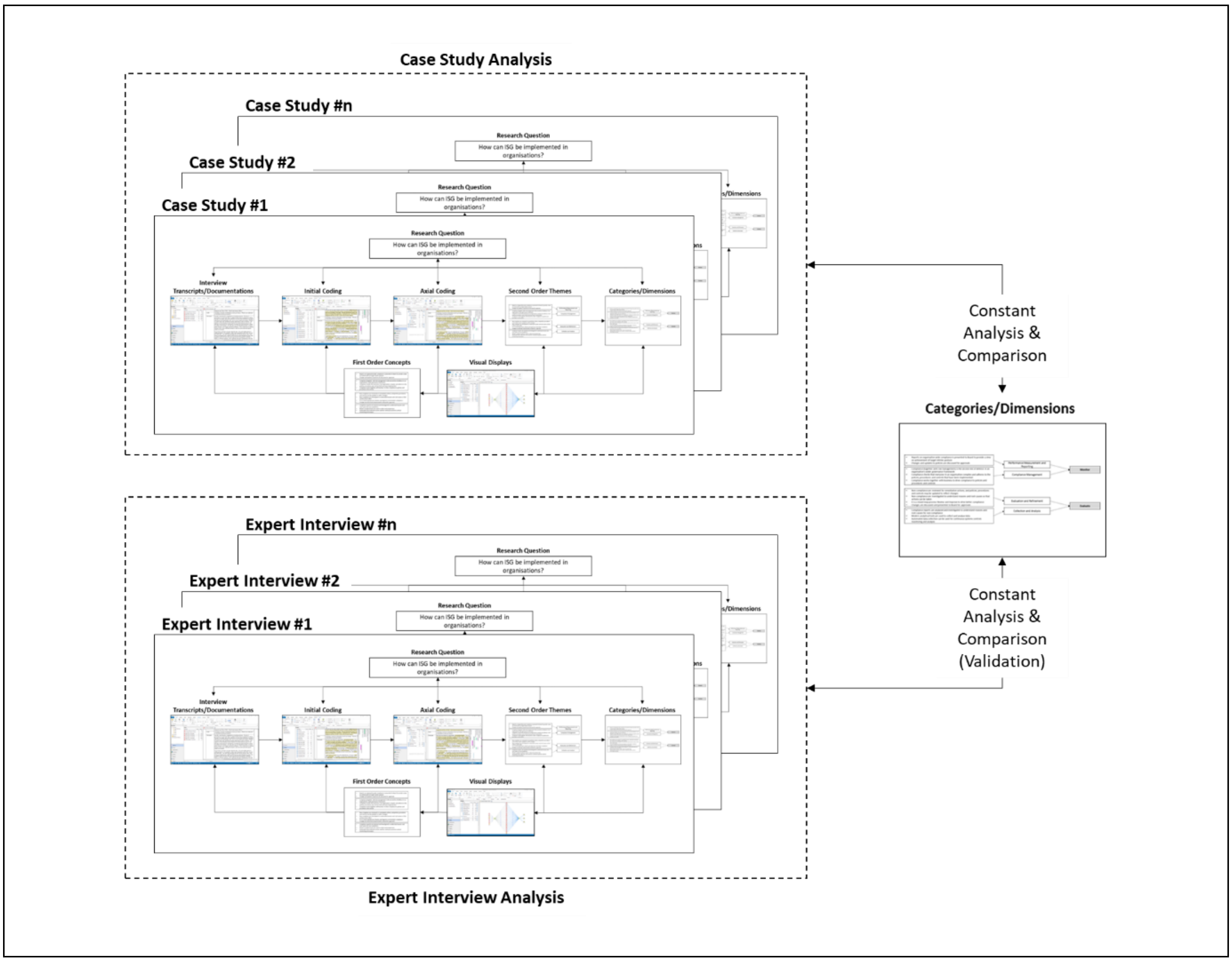

Figure 3-4: Overview of cross-case and expert interview analysis to validate emergent theories

## 3.9 Limitations and Bias

As with most research, this research has its limitations. Therefore, it is important in any research report to identify and disclose possible limitations and biases so that the boundaries of the research can be assessed (Creswell, 2013; Mruck & Mey, 2007; Yin, 2018):

a. The nature of information security as a topic area is very sensitive. Due to the sensitivities and confidentialities of the topic, not many organisations were willing to discuss it openly with external parties and this may have limited the information that was provided for the research.

b. The nature of the research in covering ISG involved participants/interviewees who were part of their governing bodies and this generally meant people on the board and C-level executives. It was difficult to get access to them as their schedules were very tight. Some interviews were rescheduled multiple times, prolonging the case study research in an organisation over a long period or interviews being subsequently cancelled.
c. In addition to the limited availability of board members and C-level executives mentioned above, these senior executives were generally very careful about disclosing information in research interviews as they were concerned about disclosing organisation-sensitive information.
d. The research focuses on the financial services industry. Although the research focuses on one industry, it should be applicable to other industries because the discussion of governance is similar across different industries.
e. One of the inherent biases in qualitative research is researcher bias, i.e. the researcher's personal views and perspectives. It is noted that the researcher was working in the IT consulting and financial services industry and therefore several of the interviewees were the researcher's business contacts. While some of the interviewees were known to the researcher, the topic under investigation was new to the researcher and it had little influence on the interpretation of the findings. In addition, the researcher has regularly discussed the research methodology and findings with the research supervisors.

The research design is cognisant of the potential limitations and biases, and therefore has adopted a series of approaches to assess rigour throughout the entire research process. The following section describes the techniques that have been adopted in ensuring the validity and reliability of the research.

## 3.10 Validity and Reliability

It is important to ensure the reliability and validity of a research study, especially in qualitative research. In order to evaluate the rigour of this qualitative research, specifically an exploratory research design, this study has adopted the following criteria proposed by Yin (2018) and Creswell (2013):

a. Construct validity, which demonstrates that the research was conducted in such a way that the subject under investigation was correctly identified
b. Internal validity, which assesses whether the interpretation of causal explanations has considered various effects that may not have been apparent during the research; this criterion is mainly a concern for explanatory research and not applicable to exploratory research (Yin, 2018)
c. External validity, which shows whether the research findings can be generalised to other situations
d. Reliability, which shows that the research has adopted a structured process that can be repeated to generate the same results

Table 3-4 shows the actions that have been adopted in the research to address the required criteria.

Table 3-4: Validity and reliability criteria (adapted from Yin 2018 & Creswell 2013).

| Quality of Research Design | Actions Taken in Research |
|---|---|
| a. Construct validity | • Multiple sources of evidence have been utilised to achieve data triangulation.<br>• Follow-up interviews with interviewees were conducted to clarify and confirm findings.<br>• Some of the interview transcripts were discussed with the interviewees to confirm their accuracy.<br>• Some of the insights from case studies were shared with selected interviewees to assess completeness and accuracy. |
| b. Internal validity | • According to Yin (2018), internal validity is not for descriptive or exploratory research. |
| c. External validity | • Replication logic was used across all 3 case studies that generated consistent case study findings achieving saturation.<br>• Further validation of case study findings with expert interviews was carried out in confirming analytic generalisation.<br>• Any biases the researcher may have brought to the research were clarified. |
| d. Reliability | • A case study database was created which contained all source data (i.e. interview transcripts, documents), coded documents and codebook. In addition, all working documents were stored as a consolidated working file in NVivo 11.<br>• All research data have been collected and stored in accordance with the University's ethics procedures. |

| Quality of Research Design | Actions Taken in Research |
|---|---|
| | • A case study protocol was developed consisting of a standard invitation letter, Consent Form, PLS and standard interview guide.<br>• The case study protocol has been used consistently across all 3 case studies.<br>• A research journal was maintained that contained a collection of memos, notes, preliminary ideas and concepts at different stages of the research.<br>• A chain of evidence was established with detailed description of the case studies and documentation from data gathering to analysis and final theory generation, as described in the various sections of this thesis. |

## 3.11 Summary

This chapter has explained the research design and methodology adopted in this research. It has justified the use of the constructivist/interpretivist research paradigm and the qualitative research method as the focus is to address the “how” in improving ISG in organisations. Subsequently, this chapter has explained the 3-phase exploratory research approach where the conceptual ISG process model was taken through to refinement and validation via case study and expert interview research methods. The chapter has covered the details of the case study and interview research methods, and discussed the selection of the case study organisations and the interviewees for both case study and expert interviews, followed by an explanation of the data collection and data analysis processes. Finally, this chapter has closed with a discussion of the limitations and biases of the research and the actions taken to ensure the reliability and validity of the research. The next chapter will discuss the development of the conceptual ISG process model.

# Chapter 4
# Conceptual ISG Process Model Development

This chapter focuses on the development of the conceptual ISG process model informed by the literature review and based on a selected set of models that had been developed by previous researchers. This is Phase 1 of the research design, as shown again in Figure 4-1.

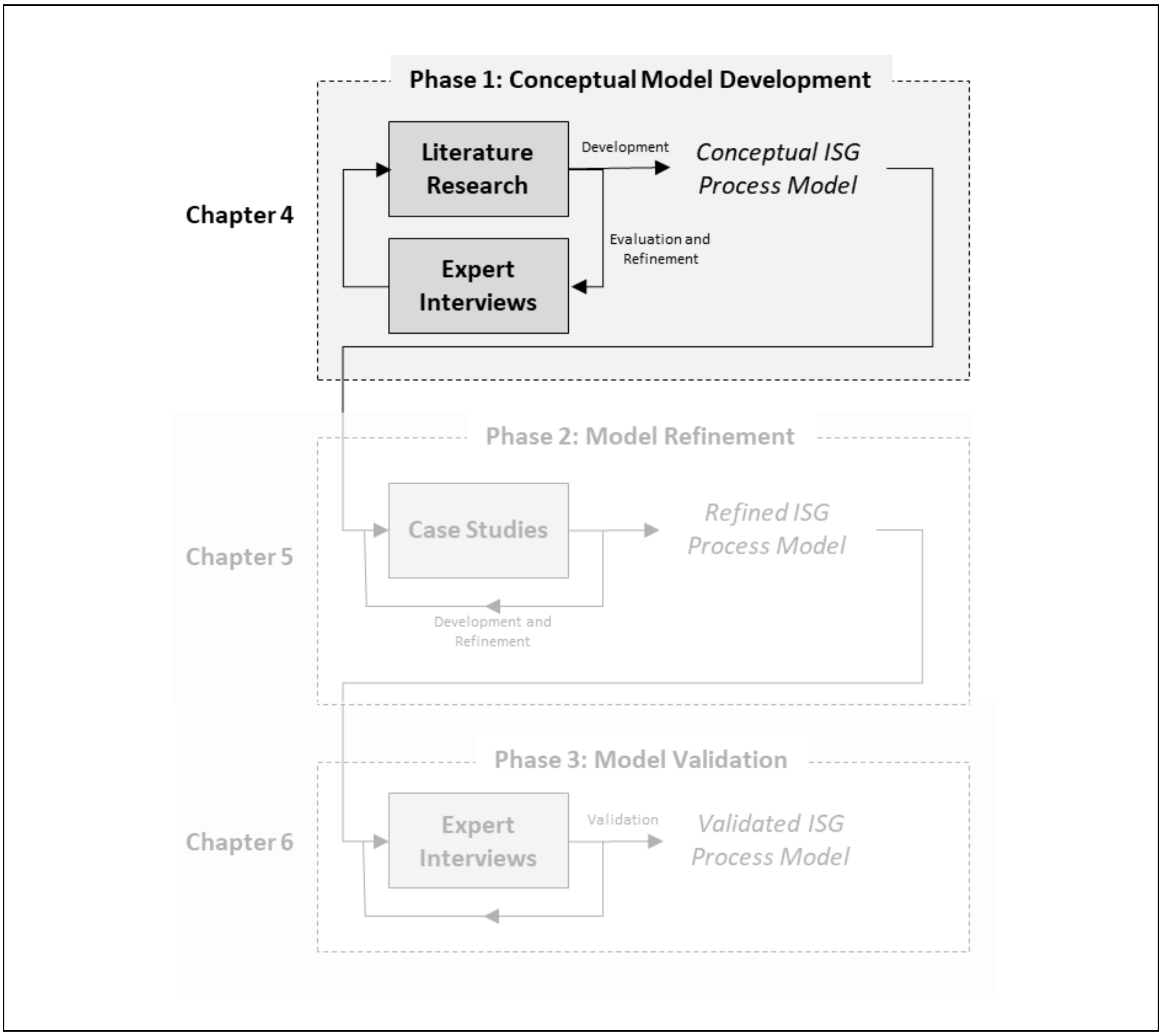

Figure 4-1: Phase 1: Conceptual model development

Section 4.1 provides an overview of the approach adopted to develop the conceptual ISG process model, which includes the identification of key ISG components from extant literature as discussed in Section 4.2. Section 4.3 provides insights into selected information security frameworks and models that have been studied to develop the conceptual model.

Section 4.4 describes the development of the initial conceptual process model from a synthesis of selected work conducted by various researchers and the different models proposed by the many professional and consultant reports and published standards. It also describes the approach where experts were consulted to provide initial feedback on the relevance of the process model in facilitating the understanding and implementation of ISG and to test the interview questionnaire which was used in subsequent case study interviews. The chapter concludes with Section 4.5 which shows the proposed conceptual ISG process model.

## 4.1 Conceptual Process Model Development Approach

A process model is a graphical representation of processes, i.e. business activities and their relationships, and is normally depicted as a set or series of inter-related and structured activities or tasks across an organisation that produces a specific service, product or outcome (Burgelman, 1996; Damelio, 2011; Jacka & Keller, 2009; Kalman, 2002). The processes depicted in a process model capture both formal and tacit knowledge and the model is a useful knowledge repository that helps to provide operational consistency in an organisation (Kmetz, 2012; White & Cicmil, 2016) and how an organisation gets its work done (Davenport, 1993; Kalman, 2002). Therefore, the development of the proposed conceptual ISG process model required the identification of the stakeholders involved and the related processes, i.e. business activities, their relationships and the workflow across an organisation involved in implementing ISG.

The definition of ISG as in Section 2.5, i.e.

> ISG is the framework of rules, relationships, systems and processes by which the security objectives of the organisation are set and the means of attaining those objectives and monitoring performance are determined.

and the requirements of a process model provided the basis to guide the identification of key components from extant literature and existing ISG framework and models to facilitate the abstraction of key components for the development of the conceptual process model. The process adopted in developing the conceptual ISG process model is shown in Figure 4-2.

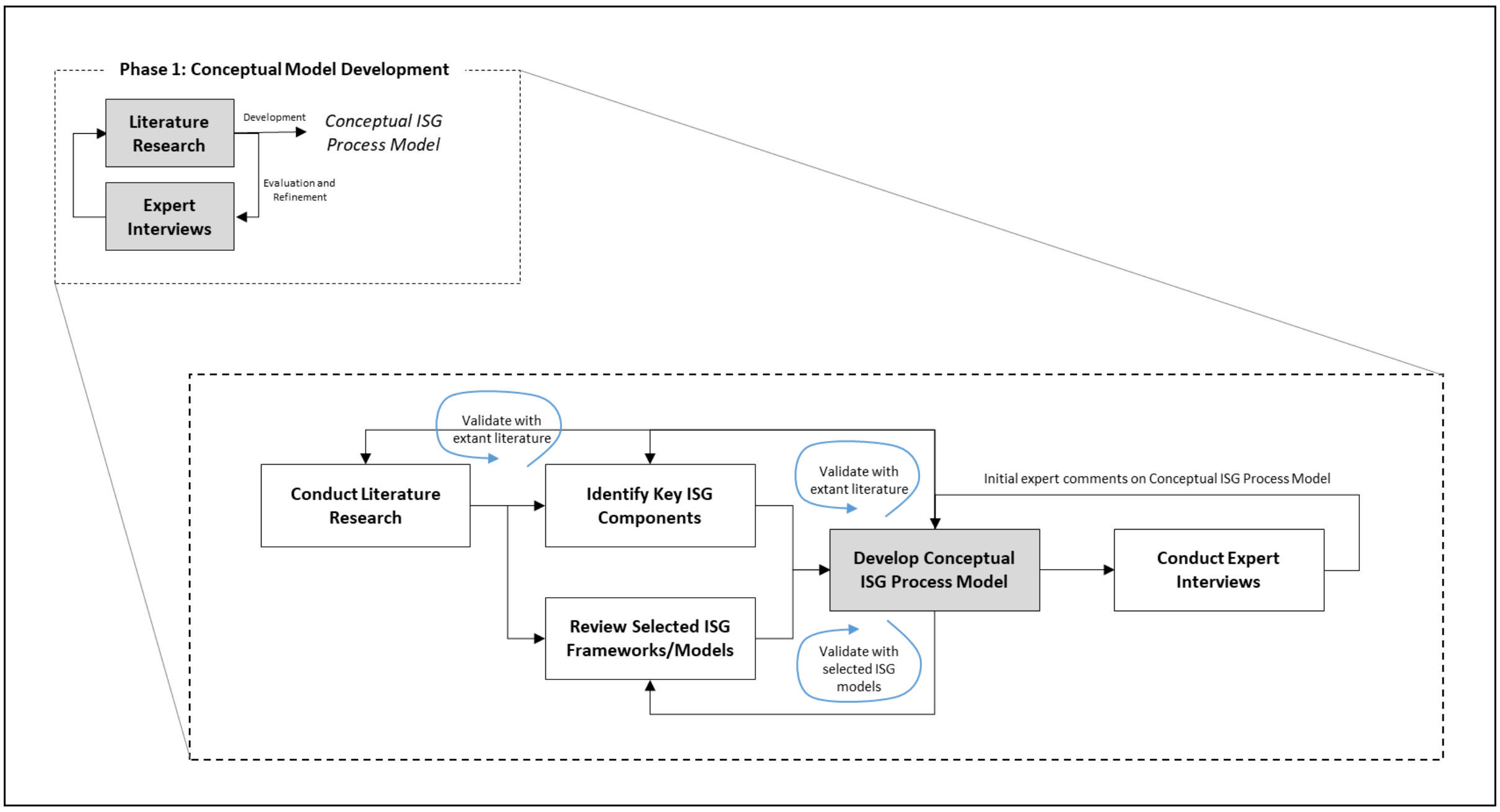

Figure 4-2: Conceptual ISG process model development approach

## 4.2 Identification of Key ISG Components

While Section 2.3 has provided a detailed literature review justifying further research in ISG and identifying the key concerns, needs and importance of ISG, this section uses the insights from the literature review to identify the key components required in the development of an ISG process model. For the purpose of this thesis, the term “components” refers to the elements that make up ISG, comprising of the ISG principles, stakeholders, processes and the interactions.

Figure 4-3 illustrates the key drivers that justify and influence the need for an ISG process model and the processes/functions that have been identified to be associated with ISG in extant literature.

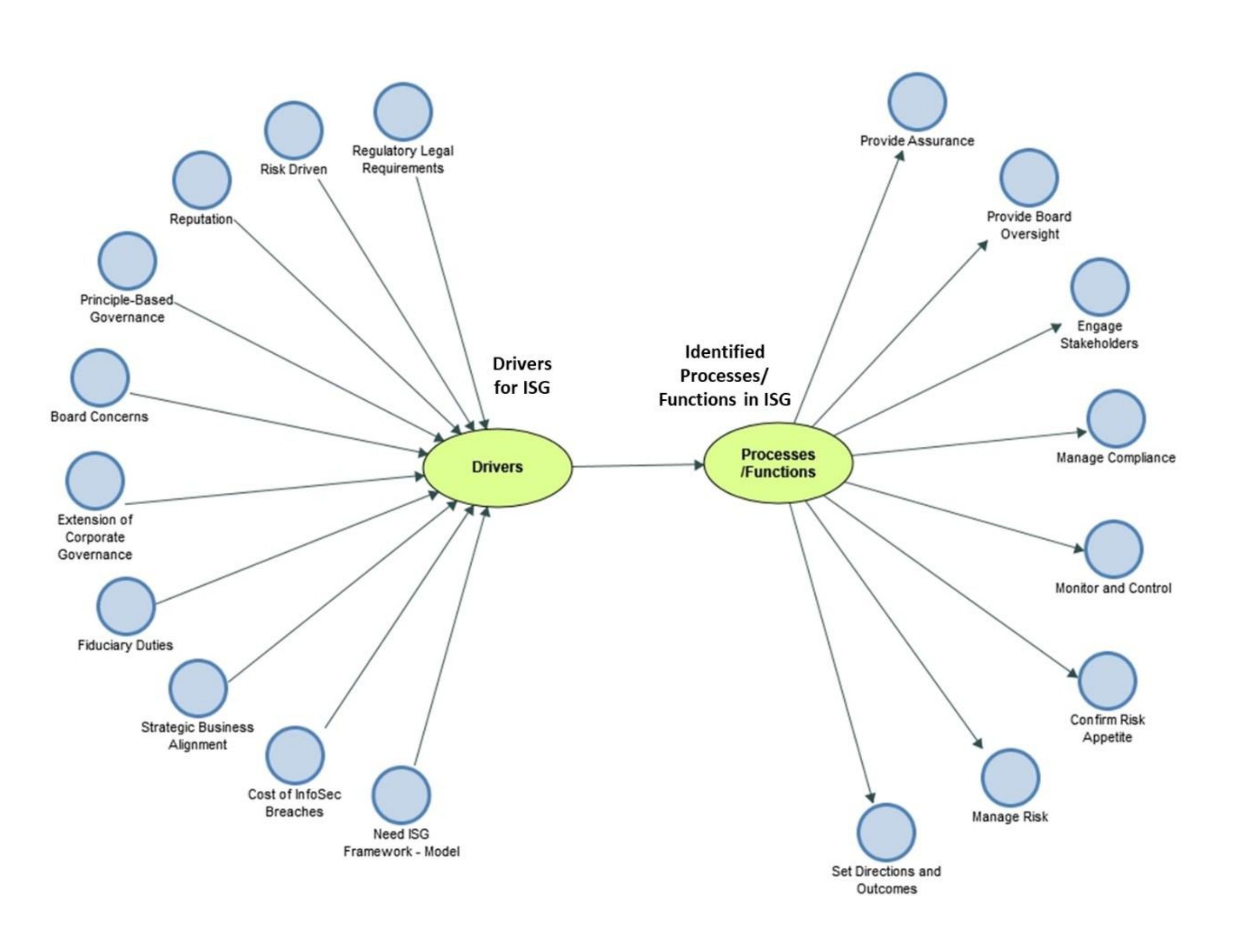

Figure 4-3: Key drivers and processes/functions of ISG from literature review (extracted from NVivo 11 analysis)

As discussed in detail in Section 2.3.2, Chapter 2, information security is a governance concern for the board and executive management, where board members have to discharge their fiduciary duty in protecting an organisation's information assets. Moreover, there is an increased requirement for organisations to comply with increased regulatory requirements and better appreciation of the risks associated with protecting information assets in organisations. These factors drive the need for better ISG and the need for an ISG process model that helps in the implementation of ISG.

The literature review also identified the following core process components that should be included in an ISG process model:

a. Provide assurance (Anhal et al., 2003; Fitzgerald, 2012; Holzinger, 2000)

b. Provide board oversight (Allen, 2005; Bihari, 2008; Georg, 2017; von Solms, 2001b; Westby, 2012)
c. Engage stakeholders (Allen, 2005; Gramling et al., 2004; Yaokumah & Brown, 2014b)
d. Manage compliance (Allen, 2005; Anhal et al., 2003; Antoniou, 2018; McFadzean et al., 2006; von Solms & von Solms, 2006; Williams et al., 2013)
e. Monitor and control (Allen & Westby, 2007b; von Solms & von Solms, 2006)
f. Confirm risk appetite (Allen, 2005; Bobbert & Mulder, 2015; Carcary et al., 2016; Höne & Eloff, 2009)
g. Manage risk (Allen, 2005; Anand, 2008; Anhal et al., 2003; Antoniou, 2018; Bobbert & Mulder, 2015; Maynard et al., 2018; von Solms & von Solms, 2009)
h. Set directions and outcomes (Holgate et al., 2012; Maynard et al., 2018; Rastogi & von Solms, 2006; Yaokumah & Brown, 2014a; Young, 2005)

The following section provides a review of 6 ISG frameworks/models that have been carefully selected to facilitate the development of the conceptual ISG process model.

## 4.3 Review of Selected ISG Frameworks and Models

Many researchers have developed frameworks and models for explaining or implementing ISG in organisations. As discussed in Subsection 2.3.7.3, these frameworks and models can be categorised by 3 design principles, i.e. principles of good governance, risk management approach, and best practices and standards. In this chapter, 4 frameworks and models have been selected from academic literature and two models from standards and professional publications for detailed analysis to identify salient features to be used for the development of the conceptual ISG model. These frameworks and models have been selected based on the following reasons:

a. They contain key ISG components such as functions/processes, stakeholders and the inter-relationships between components to facilitate ease of understanding.
b. They build cumulatively on knowledge of prior research and are grounded on extant literature, e.g. build upon early research on ISG models proposed by von Solms and von Solms (2006).

c. They propose an intra-organisational model that transverses strategic, tactical and operational management.
d. They are based on key governance principles.

The two additional models that were selected from standards and professional publications were analysed to provide an additional perspective on ISG models as these models were developed to facilitate the implementation of ISG. Key similarities and differences were identified as inputs in developing the conceptual ISG model.

The following sections provide analyses of the 6 selected ISG frameworks and models.

#### 4.3.1 ISG Based on Direct-Control Cycle by von Solms and von Solms (2009, 2006)

In 2006, von Solms and von Solms (2006) proposed an ISG model that closely reflects key principles of corporate governance because ISG is an integral component of corporate governance. Prior to this, most information security research focused on the technical aspect of information security (Conner et al., 2003; von Solms, 2006). According to von Solms (2006), corporate governance includes the two important aspects of directing and controlling an organisation. In directing the organisation, the board provides strategic direction, which is expanded into policies, standards and procedures which the next level of the organisation can operationalise. Similarly, in controlling the organisation, the board ensures that the organisation complies with all required regulatory requirements and all organisational directives, policies, standards and procedures. This direct-control cycle happens across all management levels of an organisation and is a core principle of corporate governance. The direct-control cycle forms the basis of the proposed ISG model, as shown in Figure 4-4.

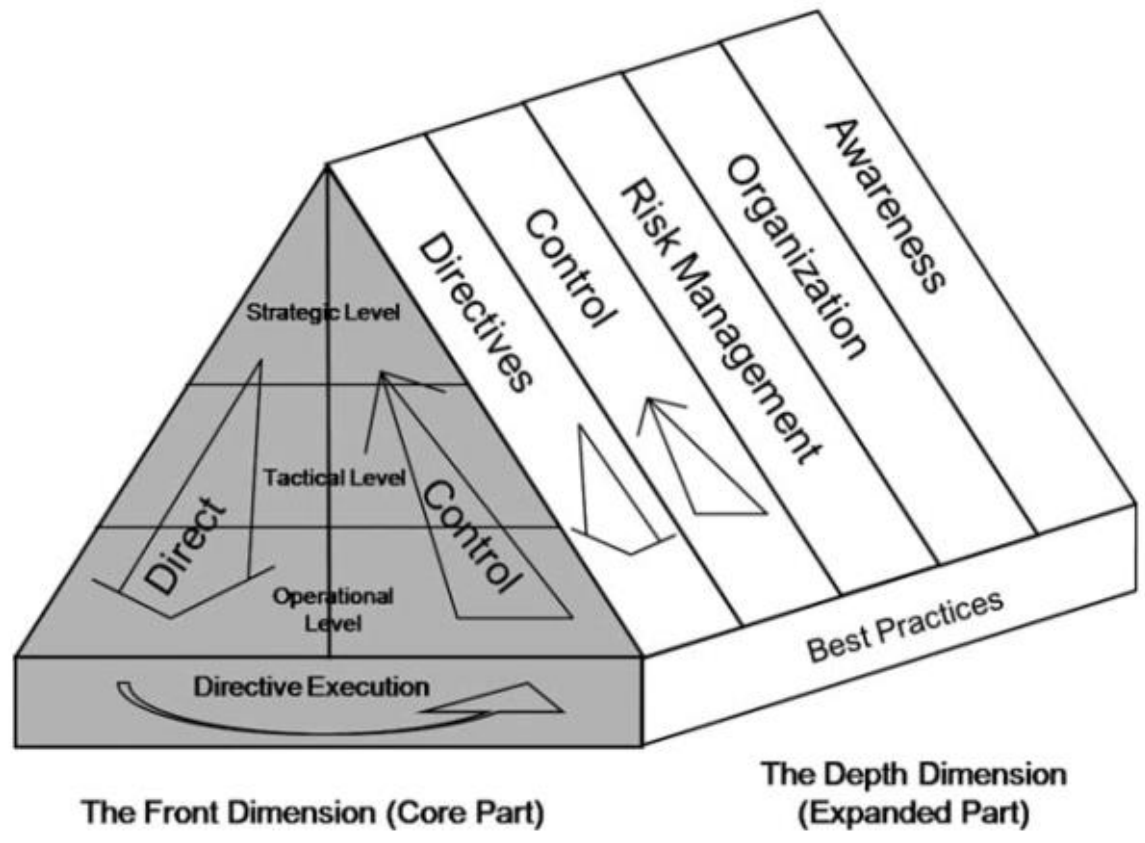

Figure 4-4: ISG based on direct-control cycle (von Solms & von Solms, 2009, 2006)

There are two core principles in this proposed ISG model:

a. Principle 1

The model covers 3 levels of management, i.e. “strategic”, “tactical” and “operational” as ISG involves activities across all levels in an organisation. The “strategic” level refers to the board and executive management, the “tactical” level is the senior and middle management, while the “operational” level comprises lower management and administration.

b. Principle 2

The model comprises 3 distinct actions, i.e. “direct”, “execute” and “control”, which are consistent with the principles of corporate governance. “Direct” specifies what must be done, indicated by the arrow pointing downward, “execute” implements the directives, as indicated by the left-to-right arrow, and “control” ensures the compliance of the execution against the directives.

In addition to the above, the model was expanded by von Solms (2009) to cover several information security-related dimensions, as shown in the expanded part of the model. The “directives” specify how directives are expanded into policies, standards and procedures for implementation, while the “control” ensures relevant information is collected for monitoring of the compliance against the directives. Additional dimensions encompass areas of risk management, clear organisational functions of ISG and management, and the importance of information security awareness across an organisation. Underpinning these dimensions is "best practices", which are collections of lessons learned, experiences and practices that guide the implementation of ISG.

The von Solms direct-control ISG model was developed based on the concept of corporate governance and incorporated core ISG processes involved across the various stakeholders within an organisation.

### 4.3.2 ISG Framework by Da Veiga and Eloff (2007)

In 2007, Da Veiga and Eloff (2007) developed an ISG framework by synthesising a list of information security components extracted from 4 approaches towards ISG, namely ISO 17799, components of PROTECT, the capability maturity model and information security architecture. The purpose of the framework, shown in Figure 4-5, is to assist management in implementing effective and comprehensive ISG that addresses technical, procedural and human components.

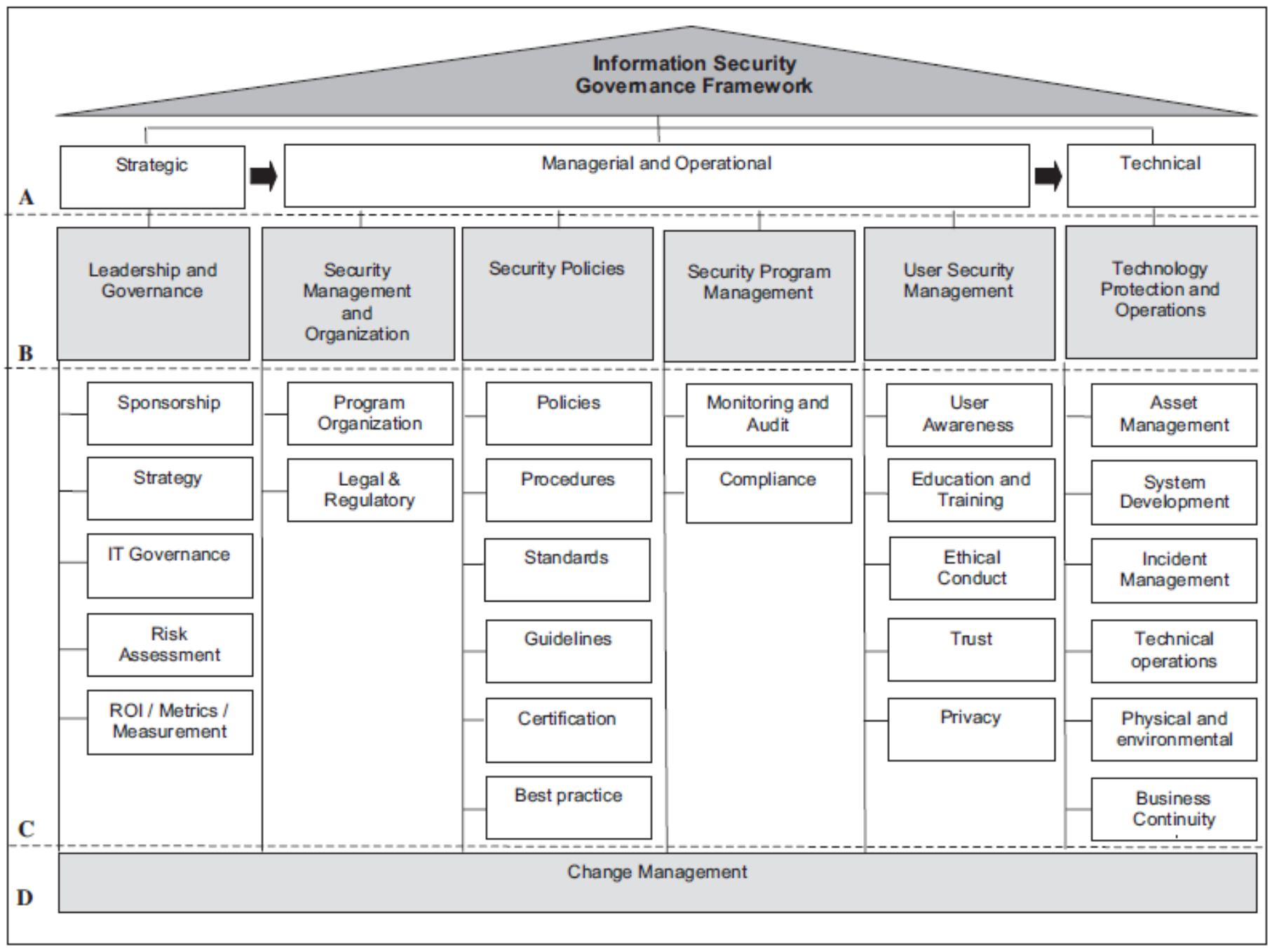

Figure 4-5: ISG framework (Da Veiga & Eloff, 2007)

In the proposed framework, the list of ISG components is grouped into 4 levels:

a. Level A: Strategic, management and technical protection components
b. Level B: 6 categories, namely, leadership and governance, security management and organisation, security policies, security program management, user security management, and technology protection and operations
c. Level C: A comprehensive list of information security components that help in the implementation of ISG
d. Level D: Change management, the foundation level that drives adoption and information security culture across the organisation

The Da Veiga and Eloff (2007) framework expanded beyond a technical framework to cover procedural and human behavioural components to provide a more comprehensive ISG framework. They were of the opinion that the implementation of an ISG framework is the first step in driving an effective information security culture to protect an organisation's information assets. The framework did not show all ISG processes, but identified key ISG functions and components to be undertaken by the various stakeholders within an organisation that are required to drive an acceptable information security culture.

### 4.3.3 ISG Framework by Ohki et al. (2009)

In 2009, Ohki et al. (2009) developed an ISG framework with the objective of combining the many information security guidelines and compliance requirements that were introduced in Japan to assist company executives to direct, monitor and evaluate information security-related activities in a unified manner. The proposed framework also helps company executives to overcome difficulties around information security management and differentiate ISG from related management processes and tools. In summary, it is a tool for corporate executives to explain their information security policy and related activities to stakeholders.

As ISG is recognised as a key component of corporate governance, it is important for an ISG framework to incorporate key principles of corporate governance (Ohki et al., 2009). The proposed framework shown in Figure 4-6 was designed to address 3 key requirements:

a. It must be aligned with the corporate governance framework, namely, corporate risk management, as information security is a major corporate risk.
b. It must be able to handle the unique characteristics of information security risk as information security risk can be very different from other corporate risks (e.g. speed of proliferation of risk, significant impact, organisation-wide being caused by anyone in an organisation, etc.).
c. It must be able to incorporate existing information security management and control mechanisms.

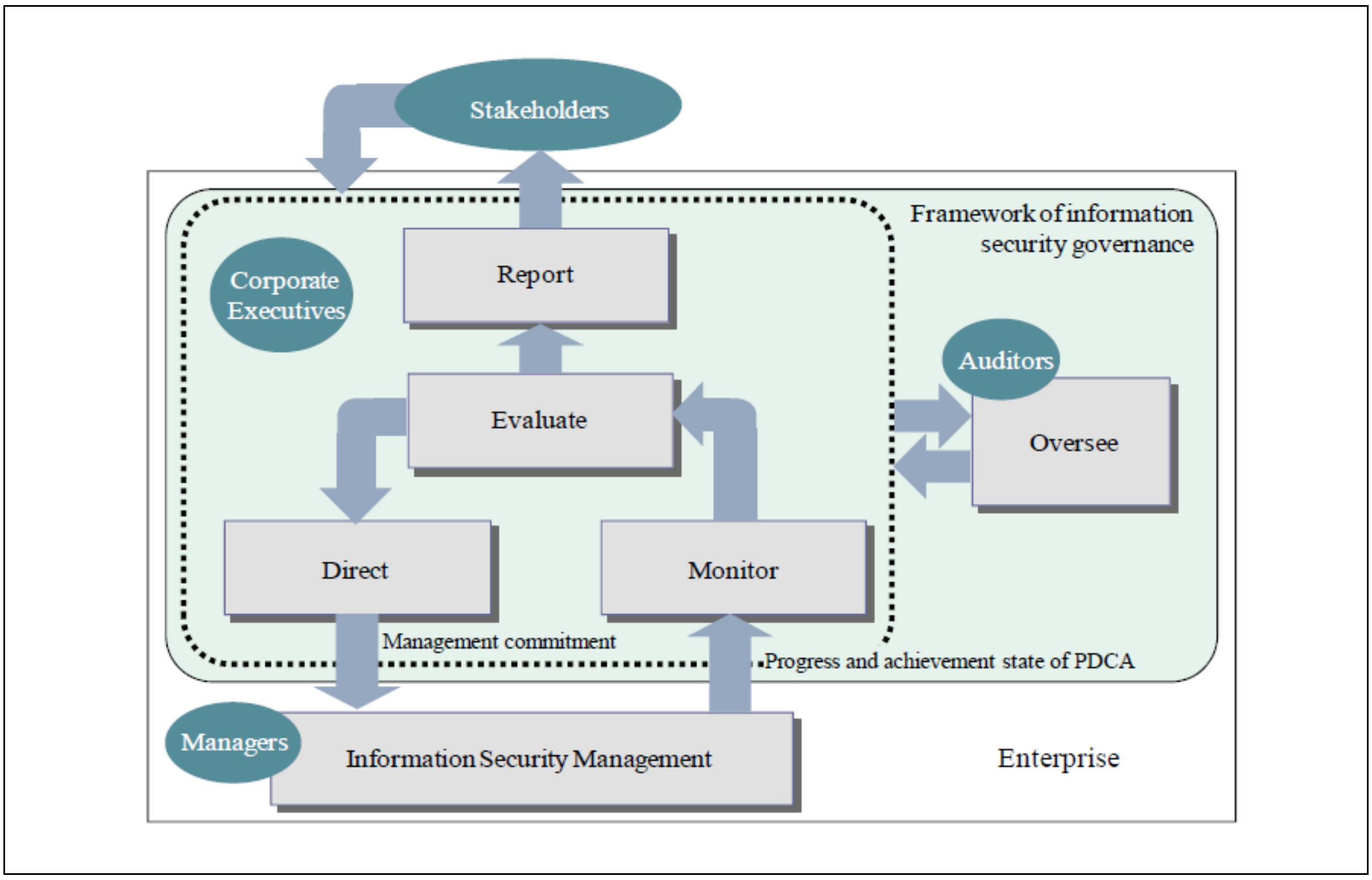

Figure 4-6: ISG framework (Ohki et al., 2009)

The ISG framework consists of 5 components with 3 key components, i.e. "direct", "monitor" and "evaluate" adopted from ISO/IEC 38500 Governance of Information Technology for the organisation. "Direct" provides guidance in business strategies and risk management, while "monitor" ensures governance activities are monitored through measurable indicators and "evaluate" assesses and verifies the results. These form the governing cycle of the information security management process. Two additional components were added, i.e. "oversee" and "report", to complete the governance functions. “Oversee” represents the auditing function to check and validate the corporate executive’s information security-related activities as it is important to obtain assurance from a third party. “Report” is proposed as the additional component to provide transparency and accountability of information security to stakeholders. This function helps build trust in the organisation as it is necessary to report and disclose activities relating to information security. The ISG framework consists of all 5 ISG components together with arrows that shows the relationships among these components. These components together with their relationships make it easy for organisations to adopt and check existing governance functions to identify improvements that are required to implement ISG.

In addition to the key governance components, the framework identifies key stakeholders who are involved in the various governance components. Corporate executives are responsible for the governing cycle of “direct”, “control” and “evaluate”, while managers are responsible for information security management. Auditors are responsible for the “oversee” component.

The Ohki et al. (2009) ISG framework was developed based on corporate governance principles and deemed to have incorporated all the core governance processes that are undertaken by the various stakeholders within an organisation. In addition, the framework shows the relationships and interactions among these processes.

### 4.3.4 Integrated Framework for Security Governance by Kim (2007) and Park et al. (2006)

The integrated framework for security governance developed by Kim (2007) and Park et al. (2006) was based on analysis of the limitations of information security management and IT governance that can be extended to cover broader ISG. The objective of the integrated framework for security governance was to facilitate better understanding among corporate executives in implementing ISG.

The proposed integrated framework for security governance integrates information security strategies, security controls and performance management across various internal stakeholders (i.e. employees and management) and external stakeholders (i.e. government, shareholders, customers, media and suppliers). The integrated framework for security governance consists of 4 domains and two relationship categories, as shown in Figure 4-7. The 4 domains are “community”, “security”, “performance” and “information”, which represent the holistic responsibility of all organisation members in governing information security, while the two relationship categories of “harmonisation” and “flywheel” govern the relationships among the domains and deal with the virtuous cycle of corporate security, respectively.

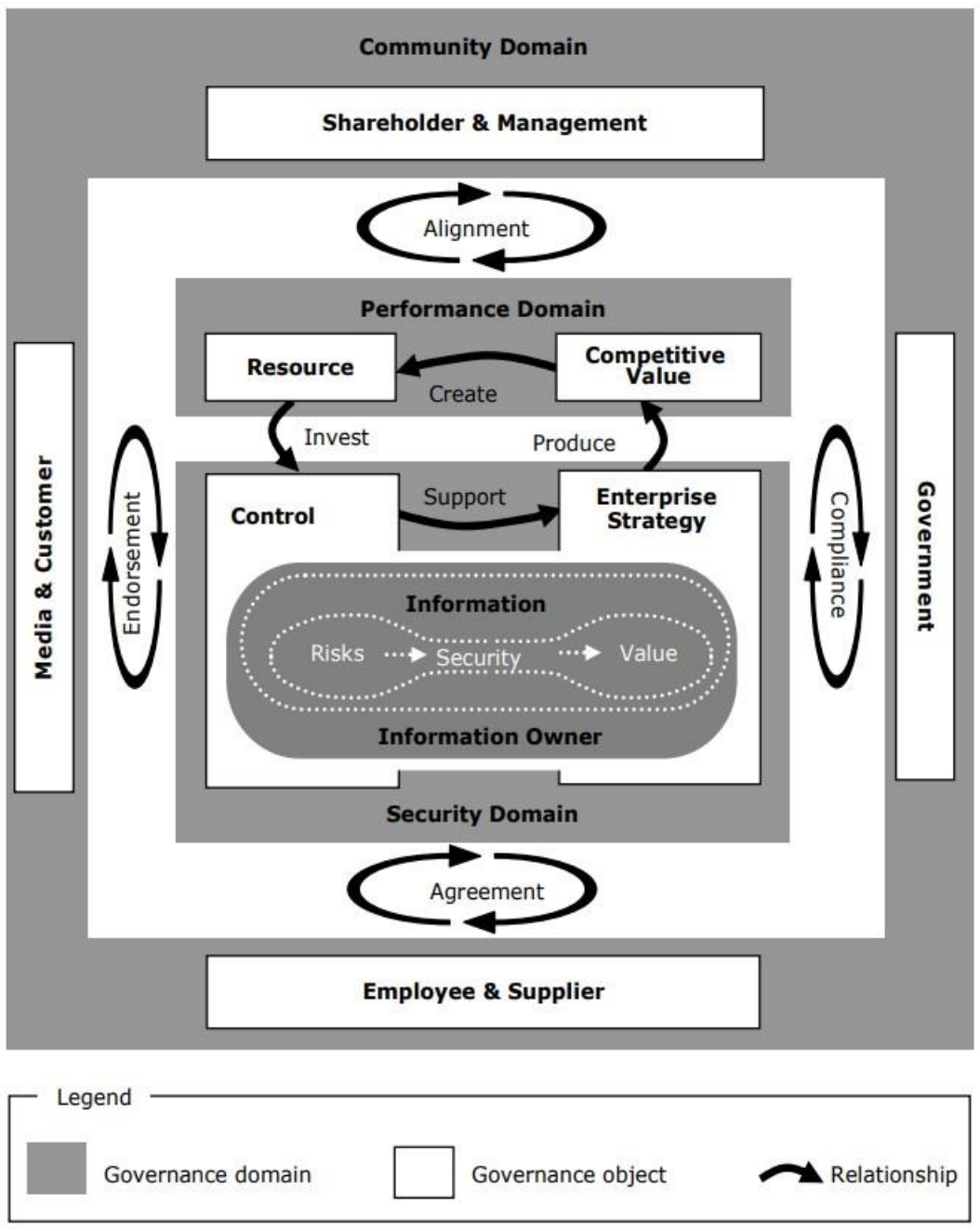

Figure 4-7: Integrated framework for security governance (Kim, 2007; Park et al., 2006)

This integrated framework for security governance was driven by the need to expand corporate governance and IT governance to ISG and aimed to integrate various stakeholders impacted by information security (depicted by the “harmonisation” relationship category) to achieve an improved information security outcome (depicted by the “flywheel” relationship category).

### 4.3.5 ISO/IEC 27014 ISG Model (International Organization for Standardization, 2013)

The ISO/IEC 27014 Information technology - Security techniques - Governance of information security technologies was published on 15 May 2013 (International

Organization for Standardization, 2013). This ISO standard is part of the ISO/IEC 27000 series of standards on information security and aims to provide a framework of ISG components where organisations can assess and implement ISG.

The scope of ISO/IEC 27014 is as stated below:

> provides guidance on concepts and principles for the governance of information security, by which organisations can valuate, direct, monitor and communicate the information security related activities within the organisation… This International Standard is applicable to all types and sizes of organisations.

The ISO 27014 model is shown in Figure 4-8.

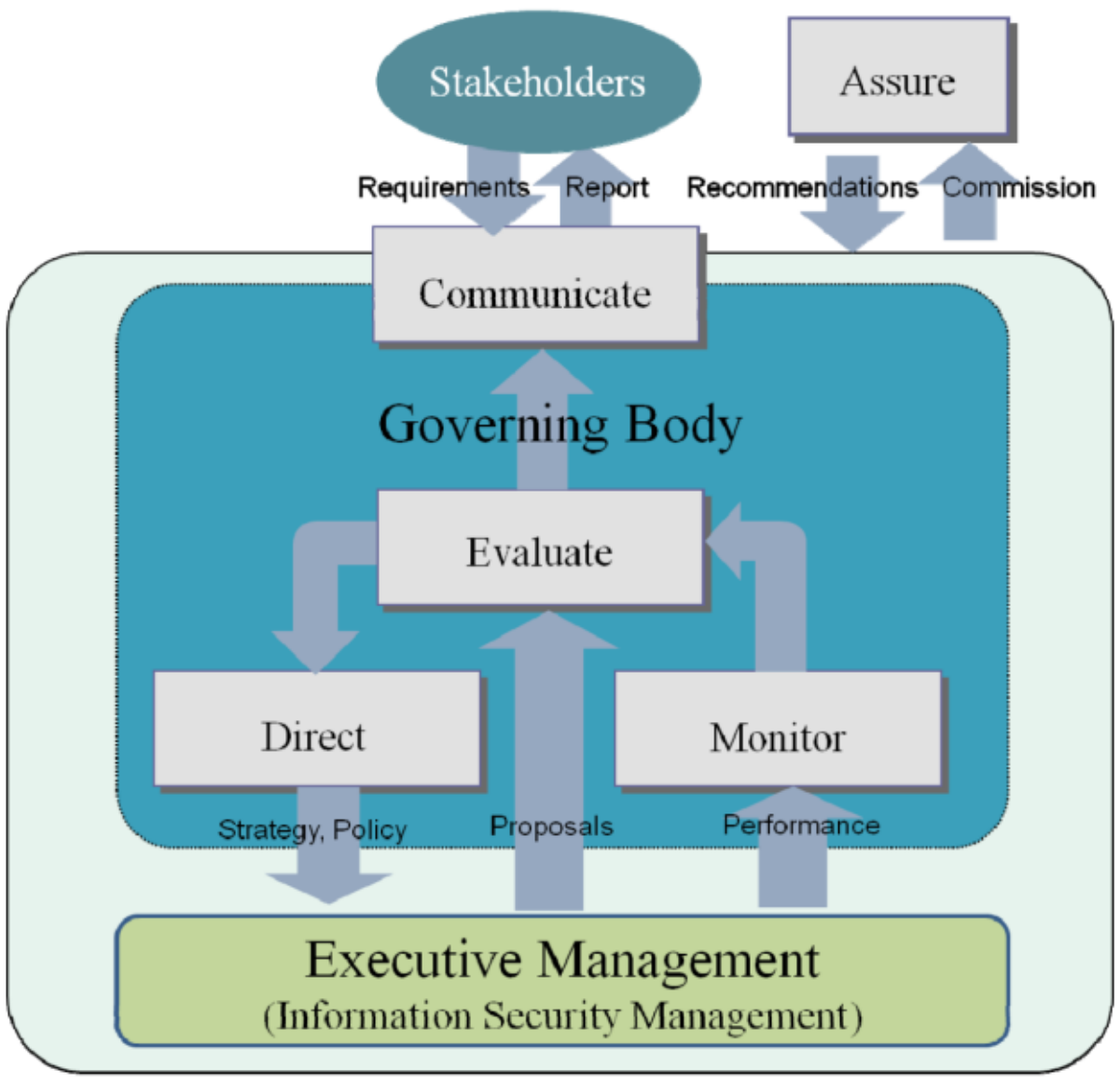

Figure 4-8: ISO 27014 ISG model (International Organization for Standardization, 2013)

This standard identified 6 action-oriented ISG principles that formed the rules for governance actions and conduct in implementing ISG. These 6 principles are as follows:

a. Principle 1: Establish organisation-wide information security

Information security is an organisation-wide concern that considers organisation strategy, business objectives and all relevant aspects of an organisation, and should be the responsibility of everyone across an organisation.

b. Principle 2: Adopt a risk-based approach

Information security is one of the many risks in an organisation. Therefore, information security risk should be managed consistent with an organisation's overall risk management approach.

c. Principle 3: Set the direction of investment decisions

Investment in information security initiatives should be based on business requirements and risk management directions.

d. Principle 4: Ensure conformance with internal and external requirements

Information security standards, policies and practices should be implemented in accordance with an organisation's operating environment, in compliance with legal and regulatory requirements.

e. Principle 5: Foster a security-positive environment

Information security initiatives should be promoted and coordinated across the organisation and relevant external stakeholders to drive an effective information security culture.

f. Principle 6: Review performance in relation to business outcomes

Information security initiatives should be monitored to ensure that they are effective in meeting the information security objectives which are aligned to an organisation's business objectives.

Based on these objectives, the ISO 27014 ISG model specifies 5 core ISG processes that provide a powerful link between the various stakeholders responsible for the governance of information security. The key stakeholders in the ISO 27014 model are the "governing body", "executive management" and "stakeholder". The "governing body" is a person or group of people who are accountable for the organisation and is not the executive management, while the “executive management” generally comprises the C-level executives in an organisation who have been delegated responsibility from the governing body (International Organization for Standardization, 2013). The standard defines a “stakeholder” as any person or organisation that interacts with the organisation.

In the ISO 27014 model, the “governing body” performs the “direct”, “monitor”, “evaluate” and “communicate” processes in governing information security. The “assure” process provides an independent opinion on the governance of information security. “Direct” provides the direction on information security strategies and objectives, “monitor” assesses the achievement of strategic objectives and “evaluate” assesses the requirements and determines the adjustments that are required. “Communicate” enables the exchange of information between the “governing body” and “stakeholders”. “Assure” is an independent process that assesses and validates the achievement of the governance of information security and is normally conducted by an independent party such as an external auditor.

The ISO 27014 ISG model is based on governance principles and identified all key ISG components which include core ISG processes, key stakeholders and the relationships and interactions with these processes.

### 4.3.6 ISG Model by Gartner (2010)

Gartner is a leading technology research and advisory company. Gartner introduced an ISG model in 2010 with the aims of formalising a common definition of ISG and identifying specific ISG functions and processes for information security practitioners (Gartner, 2010). Gartner’s ISG model is based on two key principles:

a. Principle 1: Governance processes are decision-making and oversight processes, not execution processes.
b. Principle 2: The objective is the attainment of business goals, not IT goals.

According to the Gartner model (2010), ISG is a set of processes and functions within the major processes of “plan”, “implement”, “manage” and “monitor”. This model is another informative model as it specifies the processes and functions that should be done but does not explain how it should be done or by whom in the organisation. This model also does not cover the organisational perspective, i.e. the responsibilities of the different persons or group of persons in an organisation.

While Gartner introduced this model as an ISG model, it is interesting to note that the processes of “plan”, “implement”, “manage” and “monitor” are actually defined as management processes rather than governance processes when compared against the definition stated in COBIT 5 (Information Systems Audit and Control Association (ISACA), 2012). This is another example of a non-consistent definition of ISG and continuous use of the terms “governance” and “management” interchangeably in the information security field.

The Gartner model is shown in Figure 4-9.

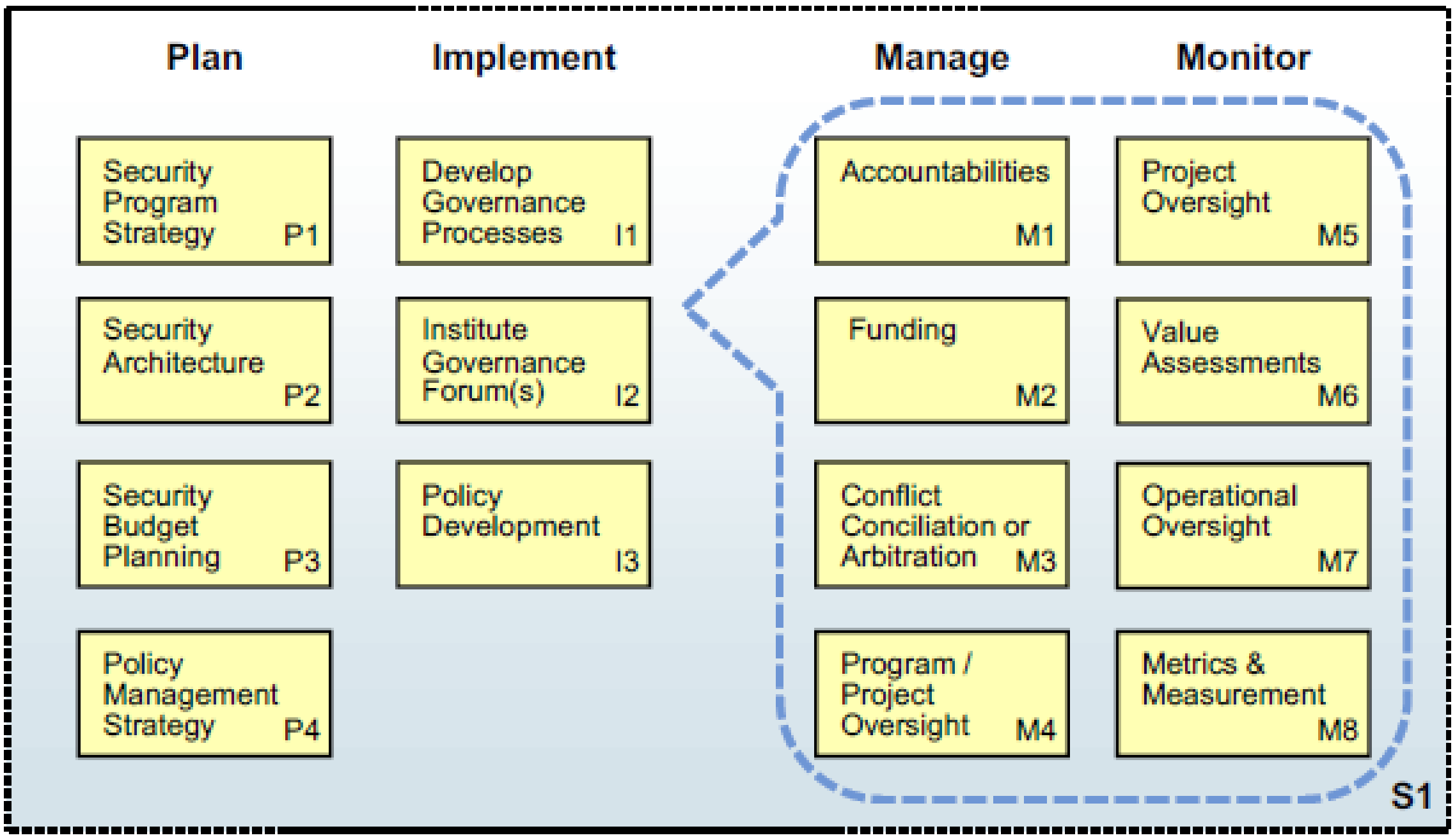

Figure 4-9: Gartner ISG model (Gartner, 2010)

Similar to the other ISG models, the Gartner ISG model identifies the key ISG components that are required to govern information security. While the model defines the core processes of plan, implement, manage and monitor, it does not identify the processes involved in ISG, but merely identifies the required ISG components.

### 4.3.7 Summary of ISG Frameworks and Models Review

The review of the 6 ISG frameworks and models in the previous sections allows for the identification of key features, similarities, differences and gaps that facilitates the abstraction of the key components required in developing a conceptual ISG process model.

Table 4-1 shows a summary of the 6 ISG frameworks and models reviewed.

Table 4-1: Summary of ISG frameworks and models reviewed.

| ISG Components | von Solms | Da Veiga | Ohki | Kim/Park | ISO 27014 | Gartner |
|---|---|---|---|---|---|---|
| Informative or normative model that identifies the "what" in ISG | ● | ● | ● | ● | ● | ● |
| Objective to facilitate the implementation of ISG | | | ● | ● | ● | ● |
| Based on corporate governance principles | ● | ● | ● | ● | ● | |
| Process driven | ● | | ● | | ● | ● |
| Closed-loop process | ● | | ● | | ● | ● |
| Identifies core ISG processes/functions/activities | | | | | | |
| a. Provide assurance | | ● | ● | | ● | |
| b. Provide board oversight | ● | ● | | | ● | ● |
| c. Engage shareholders | ● | ● | ● | ● | ● | |
| d. Manage compliance | ● | ● | ● | ● | ● | ● |
| e. Monitor and control | ● | ● | ● | ● | ● | ● |
| f. Manage risk | ● | ● | | | | |
| g. Set directions and outcomes | ● | ● | ● | ● | ● | ● |
| Demonstrates relationships across ISG Components | ● | ● | ● | ● | ● | ● |

| ISG Components | von Solms | Da Veiga | Ohki | Kim/Park | ISO 27014 | Gartner |
|---|---|---|---|---|---|---|
| Identifies key stakeholders' involvement | ● | ● | ● | ● | ● | |

All 6 frameworks and models that have been analysed are either informative or normative frameworks and models that identify key components of ISG and what should and must be done to implement ISG. Four of the frameworks, i.e. the ISG framework by Ohki et al. (2009), the integrated framework for security governance by Kim and Park et al. (2007; 2006), the ISO/IEC 27014 ISG model (2013) and the Gartner ISG model (2010), aim to facilitate the implementation of ISG by identifying the ISG components that must be considered in ISG but do not identify "how" to implement ISG. Therefore, the purpose of the following section is to develop a prescriptive ISG model that specifies the processes which can facilitate organisations to implement ISG. The proposed conceptual ISG model is a process-driven model as it identifies all governance processes, illustrates clear relationships among these processes between management and corporate governance mechanisms, and supports cross-functional information flow throughout the organisation.

## 4.4 Towards a Conceptual ISG Process Model

This conceptual ISG process model has been developed by incorporating the body of knowledge on ISG and associated information security domains, and corporate governance theories built from detailed literature reviews in Sections 2.3 and 2.4, key ISG processes and functions identified in Section 4.2 and analysis of the 6 selected ISG models discussed in Section 4.3. The purpose of a process model is to provide a common framework, i.e. a shared understanding of how work gets done (Davenport, 1993; Kalman, 2002).

This research has adopted a process-mapping approach to define and map out the core ISG processes that form the ISG process model required to facilitate ISG implementation. Process mapping enables the identified stakeholders and core processes to be graphically documented to show the simplified workflow and interactions of the processes and stakeholders in a cross-functional process diagram (Damelio, 2011; Madison, 2005; Wang et al., 2009; White & Cicmil, 2016). Based on the identification of ISG stakeholders and discovery of core ISG processes from the literature review and ISG frameworks and models analysis, the following process-mapping steps have been carried out to develop the conceptual ISG process model:

a. Step 1: Confirm the key design principles that guide the development of the ISG process model.
b. Step 2: Map the key stakeholders who are involved in ISG.
c. Step 3: Map all core ISG processes and the process owners (stakeholders) required to implement ISG.
d. Step 4: Identify the relationships and interactions between the ISG processes.

This conceptual ISG model was than then shown to 4 information security systems and security practitioners to seek initial expert comments on the need for and relevance of the model. These expert comments were incorporated into the development of the conceptual ISG process model and used as input to design the questionnaire for subsequent case study research to test and validate the conceptual ISG process model empirically.

The following sections describe the development of the conceptual ISG process model.

### 4.4.1 Confirmation of Key Design Principles

All ISG frameworks and models that were analysed had adopted some key principles. Additional reviews of related ISG literature also revealed a certain set of principles. Drawing on these many principles, the following are the 4 design principles that have been identified to guide the development of the conceptual ISG process model.

a. Principle 1: ISG is organisation-wide and is business driven

As information security is an organisation-wide concern, the governance of information security should be organisation-wide and consider organisation strategy, business objectives and all relevant aspects of an organisation.

b. Principle 2: Risk management is fundamental to ISG just as to corporate governance

As information security is one of the major risks to an organisation, ISG should adopt a risk management framework that is consistent with corporate governance.

c. Principle 3: The ISG model must clearly identify the governance processes together with clear roles and responsibilities of relevant stakeholders

ISG must have clearly identified governance processes. To be effective, clearly identified stakeholders must be assigned to be responsible for these governance processes.

d. Principle 4: ISG consists of closed-loop processes that drive continuous improvements in meeting its information security objectives

ISG sets the direction for all required information security objectives. In addition, it is equally important to regularly obtain feedback to ensure actions are taken or changed accordingly to meet the required information security objectives.

### 4.4.2 Clear Identification of Stakeholders

The conceptual model uses the von Solms and von Solms (2006) 3 management levels as the basis for the identification of stakeholders. In addition, drawing on the concepts of corporate governance (ASX Corporate Governance Council, 2019; Monetary Authority of Singapore, 2018; OECD, 2015), the strategic management level is expanded into two separate groups, i.e. “strategic - board” and “strategic - executive” as it is important to recognise the different roles performed by these groups. While both groups are responsible for the overall strategic direction of an organisation, the “strategic - board” are generally non-executive board members and are hired to provide oversight management of the “strategic - executive” group (International Organization for Standardization, 2013). In the proposed ISG process model, the “strategic - board” and “strategic - executive” are collectively known as the “governing body” i.e. the group of people that is accountable for the performance and conformance of the organisation. The other two management levels are “management - tactical” and “management - operational” (Korhonen et al., 2012; von Solms & von Solms, 2006), who have delegated responsibilities from the “governing body” for implementation of strategies and policies.

These 4 management levels, as shown in Figure 4-10, are identified as the stakeholder groups in the conceptual ISG process model.

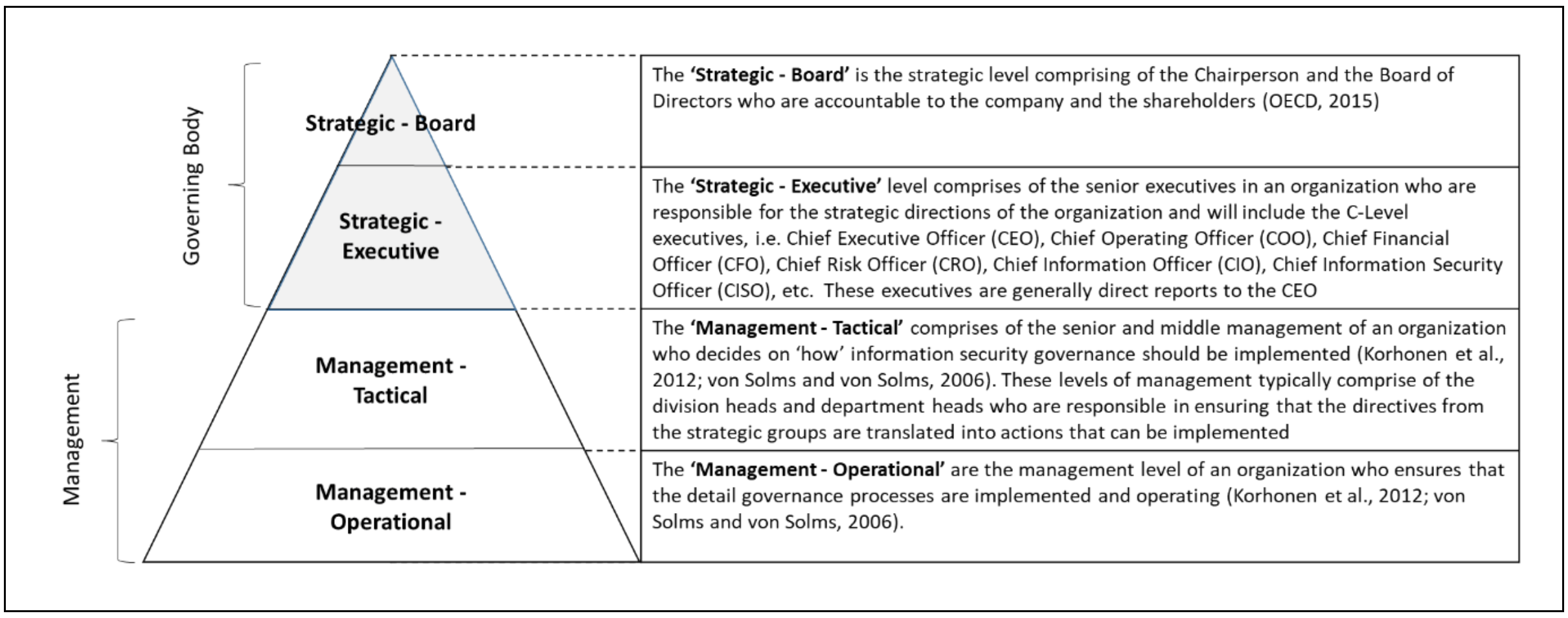

Figure 4-10: Proposed ISG stakeholder groups in an organisation, adapted from (International Organization for Standardization, 2013; Ohki et al., 2009; von Solms & von Solms, 2006)

The clear roles and responsibilities between the governing body and management, and across the various stakeholder groups, encourage segregation of duties, which is a key principle of governance structure and supports cross-functional information flow throughout the organisation.

Drawing on the selected ISG frameworks and models and the importance of auditing and assurance (Allen & Westby, 2007a; Institute of Internal Auditors, 2010, 2013; International Organization for Standardization, 2013; Ohki et al., 2009), one additional stakeholder group, i.e. "external" has been added to represent the persons or entities outside the organisation that affect or are affected by the information security activities of the organisation. In a regulated environment, this is the regulator or authority that grants the operating licence of the organisation. This "external" stakeholder group can also be external entities that conduct independent audits or certifications on the information security stature of an organisation.

These 5 stakeholder groups are mapped as 5 columns in a cross-functional process map in developing the conceptual ISG process model (Figure 4-11).

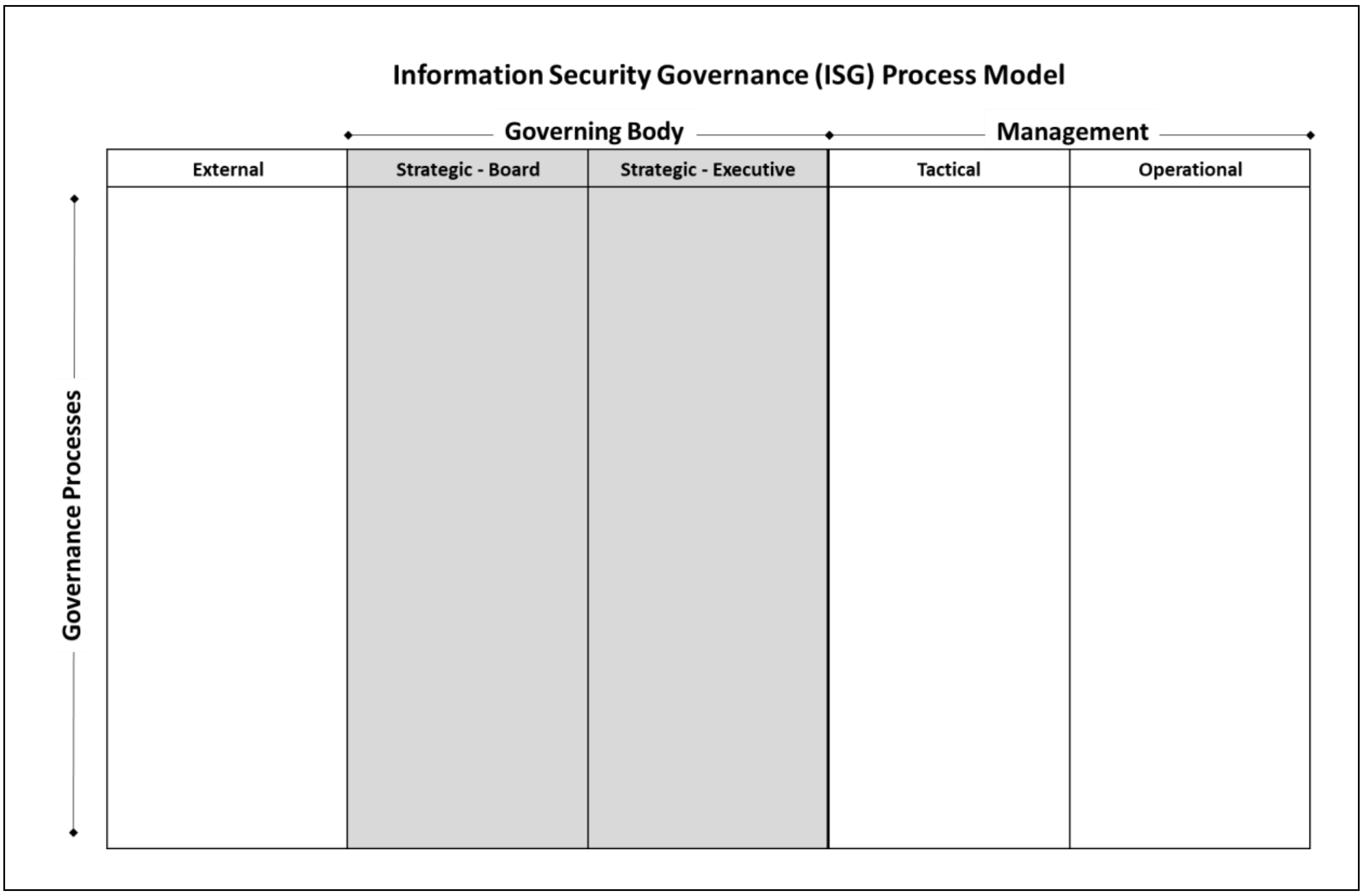

Figure 4-11: Stakeholder groups in a cross-functional process map (conceptual ISG process model)

### 4.4.3 Definition of Core Governance Processes and Sub-processes

Core ISG processes have been identified based on the body of knowledge from literature reviews and the selected ISG frameworks. While the names of these core ISG processes differ in the selected models, the meanings and roles of these processes are consistent and aim to achieve similar objectives. These core processes from the selected ISG frameworks and models are documented in Table 4-2.

Table 4-2: Core ISG processes in various ISG frameworks and models.

| von Solms & von Solms | Ohki et al. | ISO 27014 | Objectives | Conceptual ISG Process Model (Proposed Name) |
|---|---|---|---|---|
| Direct | Direct | Direct | Provides overall guidance and directions (directives) so that management can implement information security principles. | Direct |
| Control | Monitor | Monitor | Assesses the achievement/progress of information security objectives as defined in the directives. | Monitor |
| – | Evaluate | Evaluate | Undertakes evaluations and comparisons to determine changes and adjustments required to meet current and future information security objectives. | Evaluate |
| – | Report | Communicate | Demonstrates accountability and transparency through reporting and communication regarding the information security program undertaken to protect the organisation and respond to security incidents. | Communicate |
| – | Oversee | Assure | Conducts checks and validations by independent parties (e.g. reviews, audits and certifications) to ensure compliance with desired level of information security. | Assure |

As a cross-functional process diagram is used as part of process mapping to illustrate these core ISG processes, these core ISG processes are mapped on the left column in the conceptual ISG process model, as shown in Figure 4-12. The process flows among the 4 core processes as interpreted from both the von Solms direct-control cycle (2006) and Ohki et al.'s (2009) ISG framework together with ISO 27014 (2013) are shown as the connecting arrows. Further sub-processes that are identified from literature review are documented in process profile worksheets which will then be mapped onto this cross-functional process diagram.

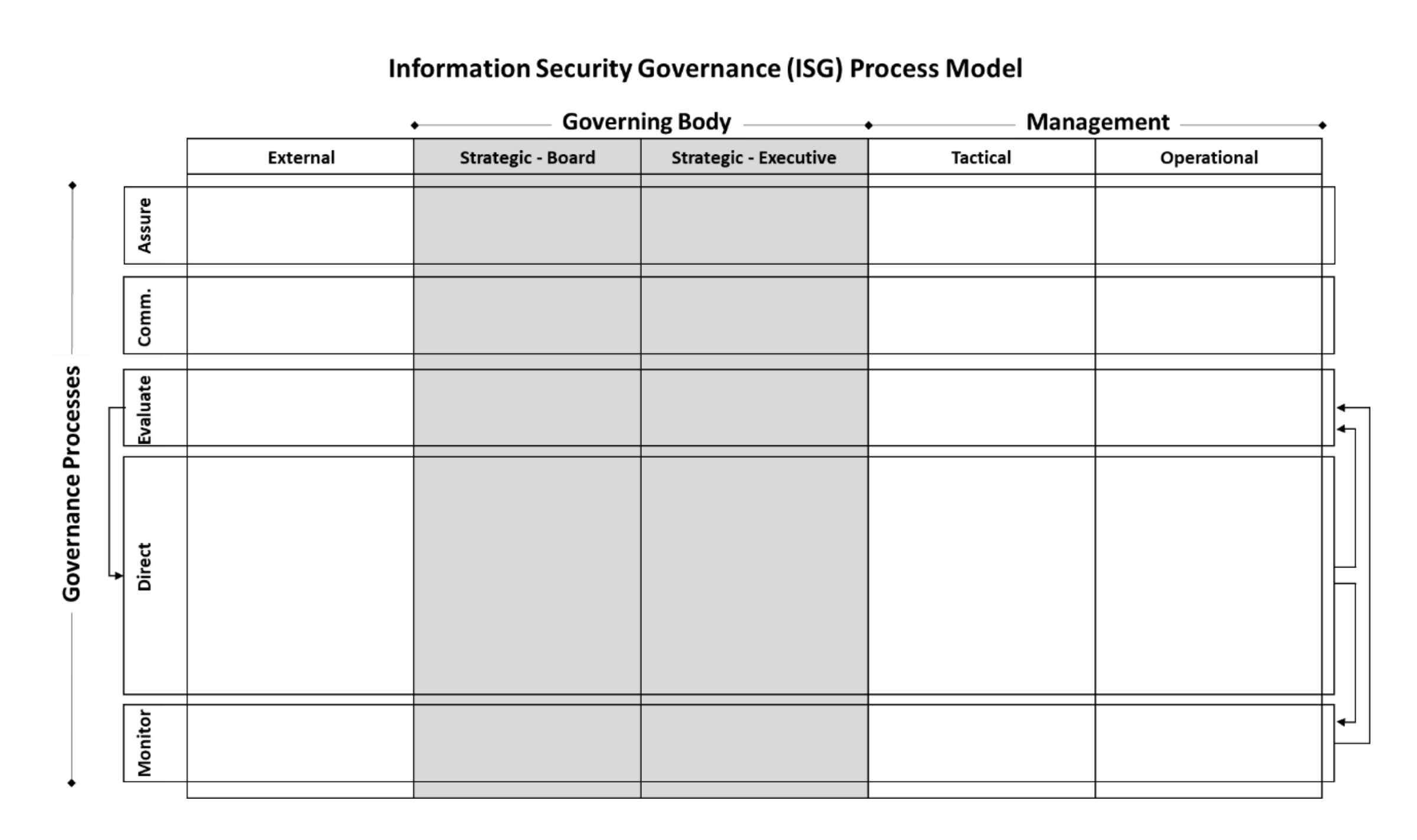

Figure 4-12: Core ISG processes in a cross-functional process map (conceptual ISG process model)

### 4.4.3.1 Direct, Monitor and Evaluate

Consistent with all 3 frameworks and models, the key governance principles involve the “direct and control” or “direct and monitor” processes. A third process, i.e. “evaluate” is not found in the von Solms model but was introduced by Ohki et al. and ISO 27014, although the function of evaluation is described in von Solms’s “monitor” process. “Evaluate” is the checks and balances process that analyses the information from “control” or “monitor” against “direct” to determine changes or adjustments that are required to meet current and future information security objectives (International Organization for Standardization, 2013). The “direct”, “monitor” and “evaluate” processes ensure a closed-loop process where the ISG processes can be regularly adjusted to meet the changing requirements of information security as business requirements and information security threats change. This closed-loop process is akin to the Deming cycle in the continuous quality improvement model (Gartner, 2010; Walton, 1988).

#### *4.4.3.2 Direct*

As ISG is related to the achievement of business goals (Conner & Coviello, 2004; Gartner, 2010; Korhonen et al., 2012; Tan et al., 2017), the “direct” process includes processes that understand an organisation’s vision, mission and business strategies so that information security initiatives can be identified and aligned. The second key component of ISG is the alignment with an organisation’s corporate risk management approach. Hence, ISG comprises processes that aim to understand the information security risk appetite that an organisation is willing to accept, so that relevant directives are provided for risk management (Da Veiga & Eloff, 2007; Webb, Maynard, et al., 2014; Yaokumah & Brown, 2014b). These business directions together with the risk appetite drive all the information security standards, policies and subsequently the procedures and initiatives for an organisation. Table 4-3 to Table 4-8 show the 6 sub-processes identified for the “direct” process:

Table 4-3: Process profile worksheet - align information security objectives with business strategy.

| **Core Process Name: Direct** | |
|---|---|
| **Sub-Process Name** | **Sub-Process Owner** |
| 1. Align information security objectives with business strategy (Allen, 2005; Antoniou, 2018; Holzinger, 2000; Lidster & Rahman, 2018; Ohki et al., 2009; Tan et al., 2017; Yaokumah & Brown, 2014a). | • Strategic - board<br>• Strategic - executive |
| **Process Description** | |
| Ensure alignment of information security objectives with organisation’s vision, mission and objectives, and corporate governance. | |
| **Input** | |
| • Organisation’s business strategy<br>• Information security objectives | |
| **Output** | |
| • Information security objectives aligned with organisation’s business strategy | |

Table 4-4: Process profile worksheet - confirm risk appetite.

| **Core Process Name: Direct** | |
|---|---|
| **Sub-Process Name** | **Sub-Process Owner** |
| 2. Confirm risk appetite (Allen, 2005; Bobbert & Mulder, 2015; Carcary et al., 2016; Da Veiga & Eloff, 2007; Gashgari et al., 2017; Georg, 2017; von Solms et al., 2011; Webb, Maynard, et al., 2014). | • Strategic - board<br>• Strategic - executive |
| **Process Description** | |
| Confirm the level of risk an organisation is willing to accept. | |
| **Input** | |
| • Information security objectives aligned with organisation's business strategy<br>• Risk assessment findings | |
| **Output** | |
| • Agreed risk appetite | |

Table 4-5: Process profile worksheet - manage risk.

| **Core Process Name: Direct** | |
|---|---|
| **Sub-Process Name** | **Sub-Process Owner** |
| 3. Manage risk (Allen, 2005; Anhal et al., 2003; Antoniou, 2018; Bobbert & Mulder, 2015; Da Veiga & Eloff, 2007; Lidster & Rahman, 2018; von Solms & von Solms, 2006; Webb, Ahmad, et al., 2014). | • Management - tactical<br>• Management - operational |
| **Process Description** | |
| Execute appropriate measures to manage and mitigate risks. | |
| **Input** | |
| • Agreed risk appetite | |
| **Output** | |
| • Risk assessment findings/results<br>• Risk management initiatives | |

Table 4-6: Process profile worksheet - define board directives.

| Core Process Name: Direct | |
|---|---|
| **Sub-Process Name** | **Sub-Process Owner** |
| 4. Define board directives (Allen, 2005; Gashgari et al., 2017; Ohki et al., 2009; Sajko et al., 2011; Tan et al., 2010; von Solms & von Solms, 2006; Williams et al., 2013). | • Strategic - board |
| **Process Description** | |
| Set the information security directions. | |
| **Input** | |
| • Information security objectives aligned with organisation's business strategy<br>• Agreed risk appetite | |
| **Output** | |
| • Information security directives | |

Table 4-7: Process profile worksheet - define security policies and standards.

| Core Process Name: Direct | |
|---|---|
| **Sub-Process Name** | **Sub-Process Owner** |
| 5. Define security policies and standards (Sajko et al., 2011; Tan et al., 2010, 2017; von Solms, 2001a). | • Strategic - executive<br>• Management - tactical |
| **Process Description** | |
| Develop organisation's information security policies and standards based on defined information security directions. | |
| **Input** | |
| • Information security directions | |
| **Output** | |
| • Information security policies and standards | |

Table 4-8: Process profile worksheet - define information security procedures.

| **Core Process Name: Direct** | |
|---|---|
| **Sub-Process Name** | **Sub-Process Owner** |
| 6. Define information security procedures (Allen & Westby, 2007a; Alves et al., 2006; Conner & Coviello, 2004; Mishra & Dhillon, 2006; von Solms & von Solms, 2006). | • Management - operational |
| **Process Description** | |
| Develop information security procedures based on defined information security policies and standards. | |
| **Input** | |
| • Information security policies and standards | |
| **Output** | |
| • Information security procedures | |

#### *4.4.3.3 Monitor*

The "monitor" process assesses the directives that have been set to ensure that these directives are implemented and followed. This is also a compliance process to ensure that an organisation complies with the information security policies, standards and procedures to achieve the intended information security objectives. Table 4-9 and Table 4-10 show the 2 sub-processes identified for the "monitor" process:

Table 4-9: Process profile worksheet - measure performance.

| **Core Process Name: Monitor** | |
|---|---|
| **Sub-Process Name** | **Sub-Process Owner** |
| 1. Measure performance (Allen, 2005; Anhal et al., 2003; Giordano, 2010; Höne & Eloff, 2009; Lidster & Rahman, 2018). | • Strategic - board |
| **Process Description** | |
| Review results on compliance to determine performance of compliance with information security objectives. | |
| **Input** | |
| • Results on compliance with information security directives, policies and standards and procedures | |
| **Output** | |
| • Recommendations and decisions on next actions based on results of compliance with information security directives, policies and standards and procedures | |

Table 4-10: Process profile worksheet - manage compliance.

| **Core Process Name: Monitor** | |
|---|---|
| **Sub-Process Name** | **Sub-Process Owner** |
| 2. Manage compliance (Allen, 2005; Anhal et al., 2003; Antoniou, 2018; Höne & Eloff, 2009; Lidster & Rahman, 2018; Ohki et al., 2009). | • Strategic - executive<br>• Management - tactical<br>• Management - operational |
| **Process Description** | |
| Conduct regular compliance tests on actual implementation of information security directives, policies and standards and procedures in practice, and provide report on compliance. | |
| **Input** | |
| • Information security directives, policies and standards and procedures<br>• Actual implementation of information security directives, policies and standards and procedures in practice | |
| **Output** | |
| • Results on compliance with information security directives, policies and standards and procedures | |

#### 4.4.3.4 Evaluate

In relation to "monitor", the "evaluate" process assesses the results of compliance and evaluates and determines the necessary changes and adjustments that may be required to meet current and future information security requirements. Table 4-11 and Table 4-12 show the 2 sub-processes identified for the "evaluate" process:

Table 4-11: Process profile worksheet - evaluate and refine.

| **Core Process Name: Evaluate** | |
|---|---|
| **Sub-Process Name** | **Sub-Process Owner** |
| 1. Evaluate and refine (Alves et al., 2006; Flores et al., 2011; Huang & Farn, 2016; Ohki et al., 2009; von Solms et al., 2011; von Solms & von Solms, 2006). | • Strategic - board<br>• Strategic - executive |
| **Process Description** | |
| Consider current and future to determine changes required to meet information security objectives. | |
| **Input** | |
| • Information security directives, policies and standards and procedures<br>• Results on compliance, i.e. gaps in meeting information security directives, policies and standards and procedures | |
| **Output** | |
| • Recommendations on changes on information security directives, polices, standards and procedures taking into consideration compliance results and the changing requirements of business strategies | |

Table 4-12: Process profile worksheet - collect and compare.

| **Core Process Name: Evaluate** | |
|---|---|
| **Sub-Process Name** | **Sub-Process Owner** |
| 2. Collect and compare (Asgarkhani et al., 2017; Huang & Farn, 2016; Kim et al., 2008; Miller et al., 2009; Ohki et al., 2009; von Solms et al., 2011; von Solms & von Solms, 2006). | • Management - tactical<br>• Management - operational |
| **Process Description** | |
| Collect information from "control" and compare with directives from "direct" to determine gaps in information security objectives. | |
| **Input** | |
| • Information security directives, policies and standards and procedures<br>• Results on compliance with information security directives, policies and standards and procedures | |
| **Output** | |
| • Results on compliance, i.e. gaps in meeting information security directives, policies and standards and procedures | |

#### *4.4.3.5 Communicate*

Disclosure and transparency are other key principles in corporate governance frameworks where timely and accurate disclosure is made on all material matters (ASX Corporate Governance Council, 2019; OECD, 2015). Accordingly, clear communication is a good governance principle that has been included in the models proposed by Ohki et al. (2009) and ISO 27014 (International Organization for Standardization, 2013), known as the "report" and "communicate" processes, respectively. In addition, the "communicate" process is bidirectional where communications include the recognition of regulatory obligations and stakeholders' expectations of information security (International Organization for Standardization, 2013). Table 4-13 shows the sub-processes identified for the "communicate" process:

Table 4-13: Process profile worksheet - engage stakeholders.

| **Core Process Name: Communicate** | |
|---|---|
| **Sub-Process Name** | **Sub-Process Owner** |
| 1. Engage stakeholders (Allen, 2005; Allen & Westby, 2007a; De Bruin & von Solms, 2016; Mahncke, 2013; Ohki et al., 2009). | • Strategic - board<br>• Strategic - executive |
| **Process Description** | |
| Report to shareholders and regulators on information security performance and compliance status. | |
| **Input** | |
| • Information security directives, policies and standards and procedures<br>• Results on compliance, i.e. gaps in meeting information security directives, policies and standards and procedures<br>• Recommendations on changes on information security directives, polices, standards and procedures taking into consideration compliance results and the changing requirements of business strategies | |
| **Output** | |
| • Communications to stakeholders and regulators, i.e. in the form of meetings, briefing sessions, reports and newsletters | |

#### *4.4.3.6 Assure*

The final process is “oversee” or “assure” i.e. the governance process where checks and validations are carried out by an independent party such as audits, reviews and certifications. This can help to ensure that the organisation is complying with its accountability to achieve the desired level of information security standards (Institute of Internal Auditors, 2010, 2013; International Organization for Standardization, 2013). Table 4-14 shows the sub-processes identified for the “assure” process:

Table 4-14: Process profile worksheet - conduct audits and certifications.

| **Core Process Name: Assure** | |
|---|---|
| **Sub-Process Name** | **Sub-Process Owner** |
| 1. Conduct audits and certifications (Anhal et al., 2003; Conner & Coviello, 2004; Farrell, 2010; Holzinger, 2000; Mishra, 2007; Ohki et al., 2009; Steinbart et al., 2018). | • External<br>• Strategic - board<br>• Strategic - executive |
| **Process Description** | |
| Conduct periodic audits and certifications. | |
| **Input** | |
| • Findings from on-site review of ISG processes and interviews with employees and stakeholders of organisation<br>• Information security directives, policies and standards and procedures<br>• Results on compliance, i.e. gaps in meeting information security directives, policies and standards and procedures<br>• Recommendations on changes on information security directives, polices, standards and procedures taking into consideration compliance results and the changing requirements of business strategies | |
| **Output** | |
| • Independent audit reports<br>• Independent external certifications of information security status of organisation | |

Based on the details on the sub-process profile worksheets as described above, these identified ISG processes and sub-processes are mapped against the stakeholder groups, as shown in a cross-functional process map in Figure 4-13.

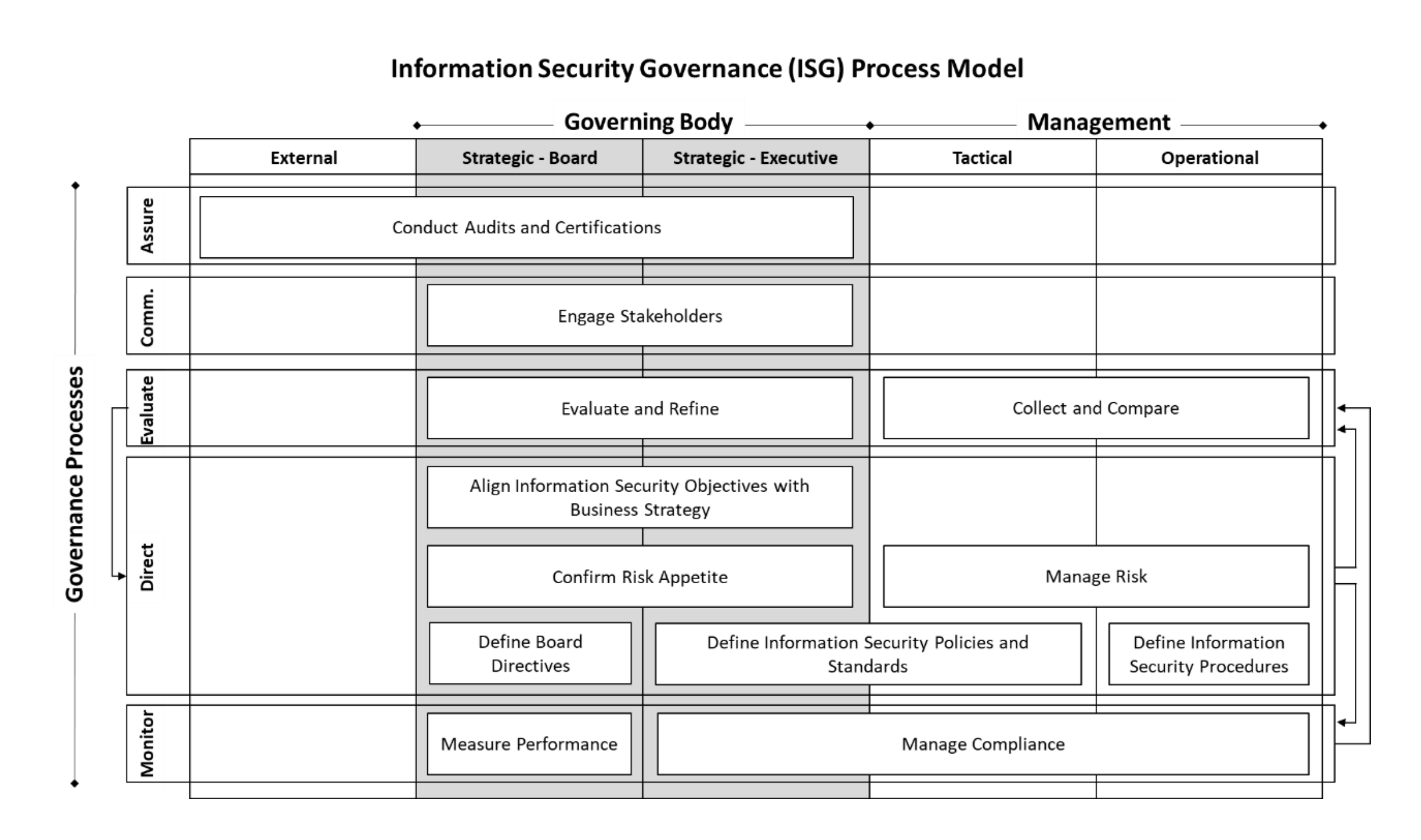

Figure 4-13: Stakeholder groups and processes in a cross-functional process map (conceptual ISG process model)

The next step in developing the conceptual ISG process model was to translate the interactions among the stakeholder groups and the dependencies of the processes and sub-processes as identified in the sub-process profile worksheets.

### 4.4.4 Interactions Among Stakeholders Groups and Across Processes/Sub-processes

An effective ISG model needs to have clear definitions of the governance processes and their interactions with each other to provide a framework that can guide an organisation in ISG implementation. The “direct”, “monitor” and “evaluate” processes provide a closed-loop process where ISG can be regularly adjusted to meet the changing requirements of information security (International Organization for Standardization, 2013; Ohki et al., 2009; von Solms & von Solms, 2006). This closed-loop process is illustrated in the conceptual model as arrows connecting the “direct”, “monitor” and “evaluate” processes. In addition to the interactions across the core governance processes, the interactions among the stakeholder groups are critical. These interactions together with the process flows across the various stakeholder groups provide clear segregation of roles and responsibilities. These are illustrated by the process flow arrows and the grouping of processes within the stakeholder groups in the conceptual model. These processes are consistent with the process flows and interactions as found in the selected ISG models from Ohki et al. (2009) and ISO 27014 (2013).

In addition to the interactions of the core ISG processes, there are also interactions and interdependencies between the ISG sub-processes as identified in the sub-process profile worksheets in Section 4.4.3. These interactions and process flows are shown as process flow arrows in the conceptual ISG process model in Figure 4-14.

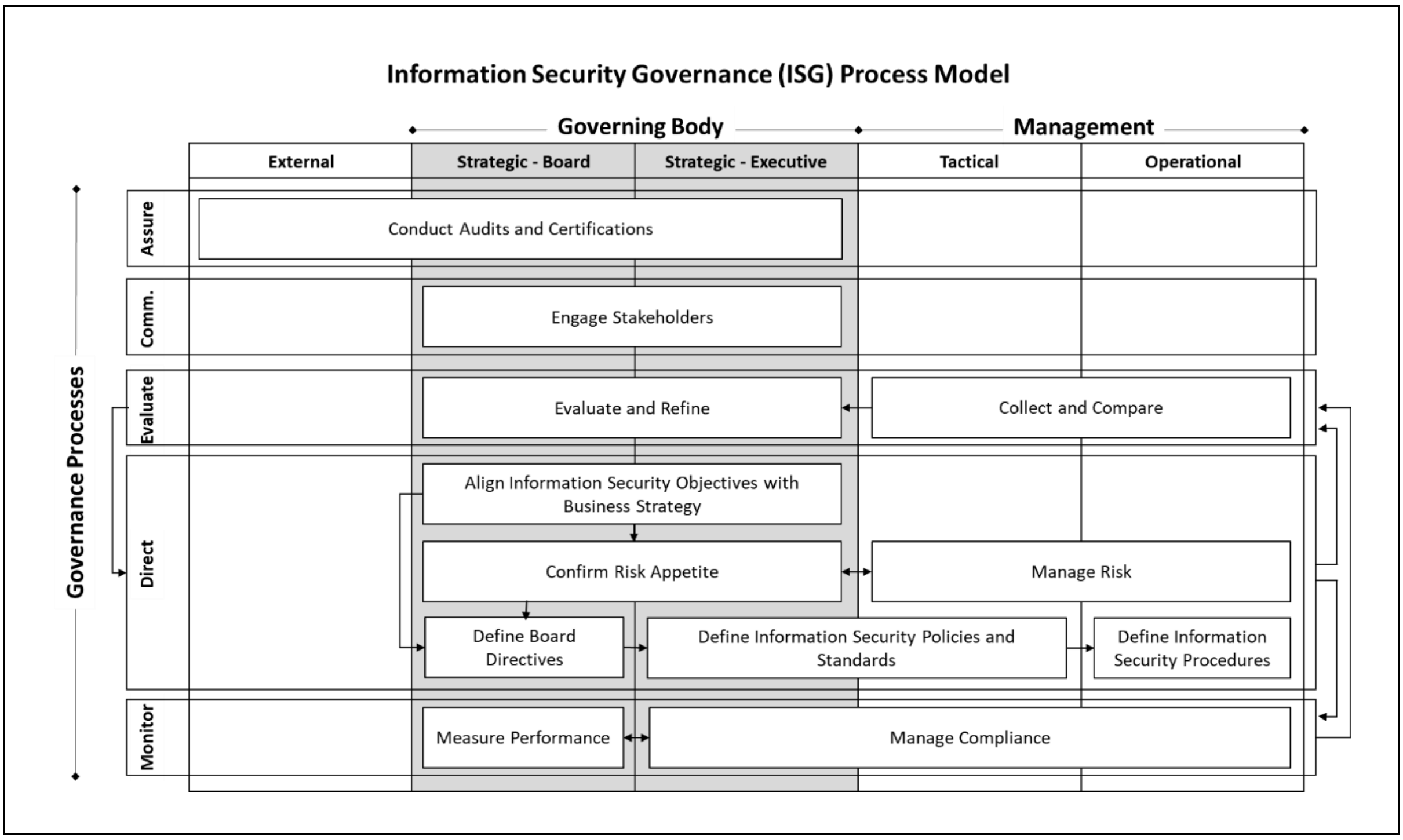

Figure 4-14: Interactions between stakeholders and processes/sub-processes in a cross-functional process map (conceptual ISG process model)

## 4.5 Proposed Conceptual ISG Process Model

The conceptual ISG process model was developed by adopting the process-mapping approach where the stakeholder groups and the ISG core processes and sub-processes were mapped onto a cross-functional diagram, which is shown again in Figure 4-15.

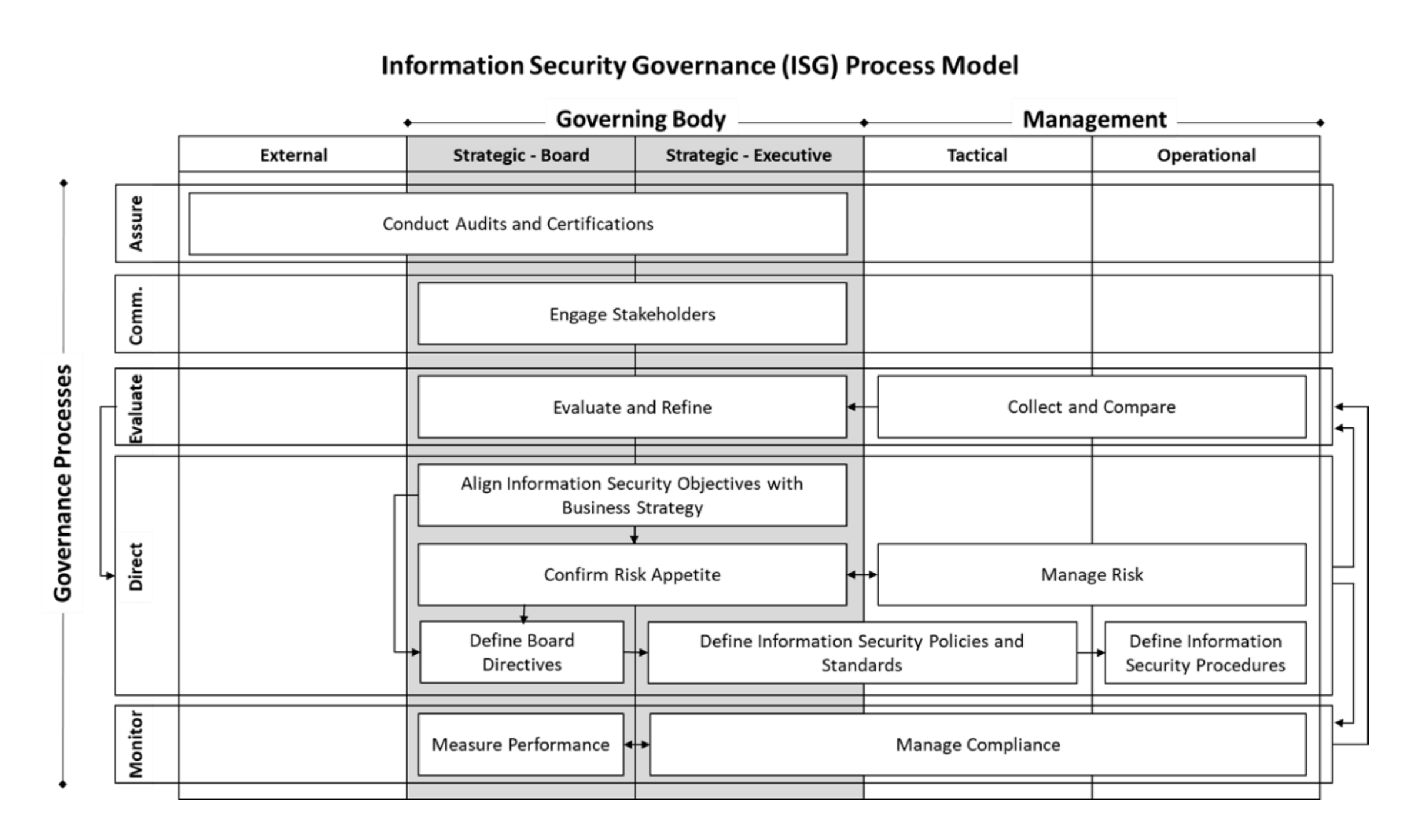

Figure 4-15: Conceptual ISG process model

The conceptual ISG process model shows the core ISG processes together with the sub-processes against the 5 stakeholder groups. The stakeholder groups (i.e. external, strategic - board, strategic - management, tactical and operational) are depicted on the top horizontal row, while the core processes (i.e. assure, communicate, evaluate, direct and monitor) are shown on the left side of the conceptual model. As indicated in Section 4.4.4, the interactions among the stakeholders and the process flows are shown as arrows connecting the processes and sub-processes.

This conceptual ISG process model was shown to initial 4 information security systems and security practitioners to seek their expert comments on the need for and relevance of the model and to help develop the questionnaire for subsequent case study interviews. Table 4-15 shows a summary of these 4 expert interviewees' profiles.

Table 4-15: Expert interviewee profiles (extracted from Table 3-1).

| Interviewee | Organisation | Location & Role Coverage | Job Role | Involvement in Information Security | No. of Years in Industry |
|---|---|---|---|---|---|
| **Expert Interviews in Phase 1** | | | | | |
| Expert #1 (CIO-MY_Bank) | Financial institution (bank) | Malaysia | CIO | Direct | 20 years+ |
| Expert #2 (CIO-SG_Bank) | Financial institution (bank) | Singapore | CIO | Direct | 30 years+ |
| Expert #3 (GCCO-MY_Bank) | Financial institution (bank) | Malaysia & South East Asia | Group CCO | Indirect | 20 years+ |
| Expert #4 (GCOO-MY_Bank) | Financial institution (bank) | Malaysia & South East Asia | Group COO | Indirect | 20 years+ |
| | | | | **Total number of expert interviewees** | **4** |

Some initial comments and feedback on the conceptual ISG model are shown in Table 4-16.

Table 4-16: Initial comments on conceptual ISG process model.

| | Initial Comments |
|---|---|
| Clear identification of stakeholders | "have a structure in placed. I believe roles and responsibilities are often the next issue … there must be accountability and oversight at the top." (GCCO-MY_Bank)<br><br>"just like corporate governance it begins at the top, i.e. the board. In my view, information security governance needs a proper structure where the board provides the overall oversight in the governance." (GCOO-MY_Bank)<br><br>"you have Line three, the auditors who will provide the assurance that all are in placed. We have both the internal and external auditors who will cover information security as part of their assurance scope of work." (CIO-MY_Bank)<br><br>"all banks, we have external auditors, regulator MAS auditors, etc." (CIO-SG_Bank) |
| Definition of core governance processes and sub-processes | |
| • Direct | "the bank still drives our information security initiatives and investment from a business strategy … both MAS requirements and |

| | Initial Comments |
|---|---|
| | business will drive our information security budget and prioritisation of information security projects." (CIO-SG_Bank)<br><br>"We start with the type of customer, we want to have a risk appetite, so if we do bank in higher risk customers, what would be our risk assessment methodology, all the controls, and … all the other stuff." (GCCO-MY_Bank)<br><br>"a lot of it is driven from business needs, regulatory requirements or risk management related." (GCCO-MY_Bank)<br><br>"The board should be responsible for oversight in overall decision-making maybe to include approving the investment for information security, approving the risk profile and facilitating meeting to ensure that the bank is focused and has a priority in information security." (GCOO-MY_Bank) |
| • Monitor | "you have the Line two, where risk and compliance, and IT is involved in making sure that the information security policies are implemented and complied with." (CIO-MY_Bank)<br><br>"We are now working on a proactive compliance process." (GCCO-MY_Bank)<br><br>"we have compliance department who will work closely with O&M [Organisation & Method Department] and my IT security team to drive compliance, making sure everyone follows the procedures. Compliance department will conduct their compliance audit … ah … I think annually, some departments maybe every six months to ensure that they comply. Any noncompliance will be raised in our management or Exco meetings. We will investigate the noncompliance and take necessary actions." (GCOO-MY_Bank) |
| • Evaluate | "The risk and compliance team will check to ensure compliance and act on noncompliance to put them back on the proper track." (CIO-MY_Bank)<br><br>"the chief risk officer … one of the key role is to ensure all risk areas must have proper policies and framework established, reviewed, refreshed." (CIO-SG_Bank)<br><br>"Any noncompliance will be raised in our management or Exco meetings. We will investigate the noncompliance and take necessary actions." (GCOO-MY_Bank) |
| • Communicate | "You need to keep the board updated, keeping them aware of the latest progress in the bank and in the industry … This also act as an education session to the board to keep them aware of the activities in the bank." (CIO-MY_Bank) |

| | Initial Comments |
|---|---|
| | “Communication to ensure constant updates to board and regulators.” (CIO-SG_Bank)<br>“a process where board is kept updated and aware and making sure board is involved in the decision-making process.” (GCCO-MY_Bank)<br><br>“We the senior management updates the board so the board can have an oversight of what is happening in the bank.” (GCOO-MY_Bank) |
| • Assure | “you have Leve three, the auditors who will provide the assurance that all are in place. We have both the internal and external auditors who will cover information security as part of their assurance scope of work.” (CIO-MY_Bank)<br><br>“Assurance is given … all banks, we have external auditors, regulator MAS auditors, etc.” (CIO-SG_Bank)<br><br>“Just like corporate governance, there should be the three lines of defence, the user at the front line, risk management and compliance function at the second line, and finally through to internal audit or assurance … information security governance adopts a similar line of defence approach.” (GCOO-MY_Bank) |

## 4.6 Summary

This chapter has analysed 4 selected ISG models that were developed by previous researchers and two models from the ISO standard and a professional publication. A detailed analysis of these 6 models and a detailed literature review discovered key ISG principles, the ISG stakeholder groups and processes as well as sub-processes and their interactions. Based on this body of knowledge, a proposed conceptual ISG process model was developed to incorporate key ISG principles that comprise good governance, risk management and best practices from international standards. The conceptual ISG process model is a process-driven model as it identifies the core ISG processes and sub-processes, the process flows and relationships. The model also maps these processes against the various stakeholder groups to show the responsibilities of these stakeholder groups in relation to these processes. This conceptual process model will facilitate the implementation of ISG in organisations which will lead to improved ISG in organisations.

As most previous models are hypothetical models developed based on key principles and theoretical frameworks, it is imperative to develop a model that is grounded on empirical data. The following chapter will describe the approach adopted to further develop and refine this conceptual ISG process model with actual data gathered from 3 case study organisations.

# Chapter 5
# ISG Process Model Refinement

The previous chapter has explained how the conceptual ISG process model was developed based on literature research and the analysis of 6 selected ISG frameworks and models, as well as initial feedback from expert interviews. This chapter presents the results and key findings of Phase 2: Model refinement of the research design, as shown in Figure 5-1, where the refined ISG process model was developed. Phase 2 focused on the development of the ISG process model through a multiple case study method where 3 case study organisations in the financial services industry were analysed. A total of 17 interviews were conducted with individuals who were primarily senior executives across different roles and were directly or indirectly involved in ISG.

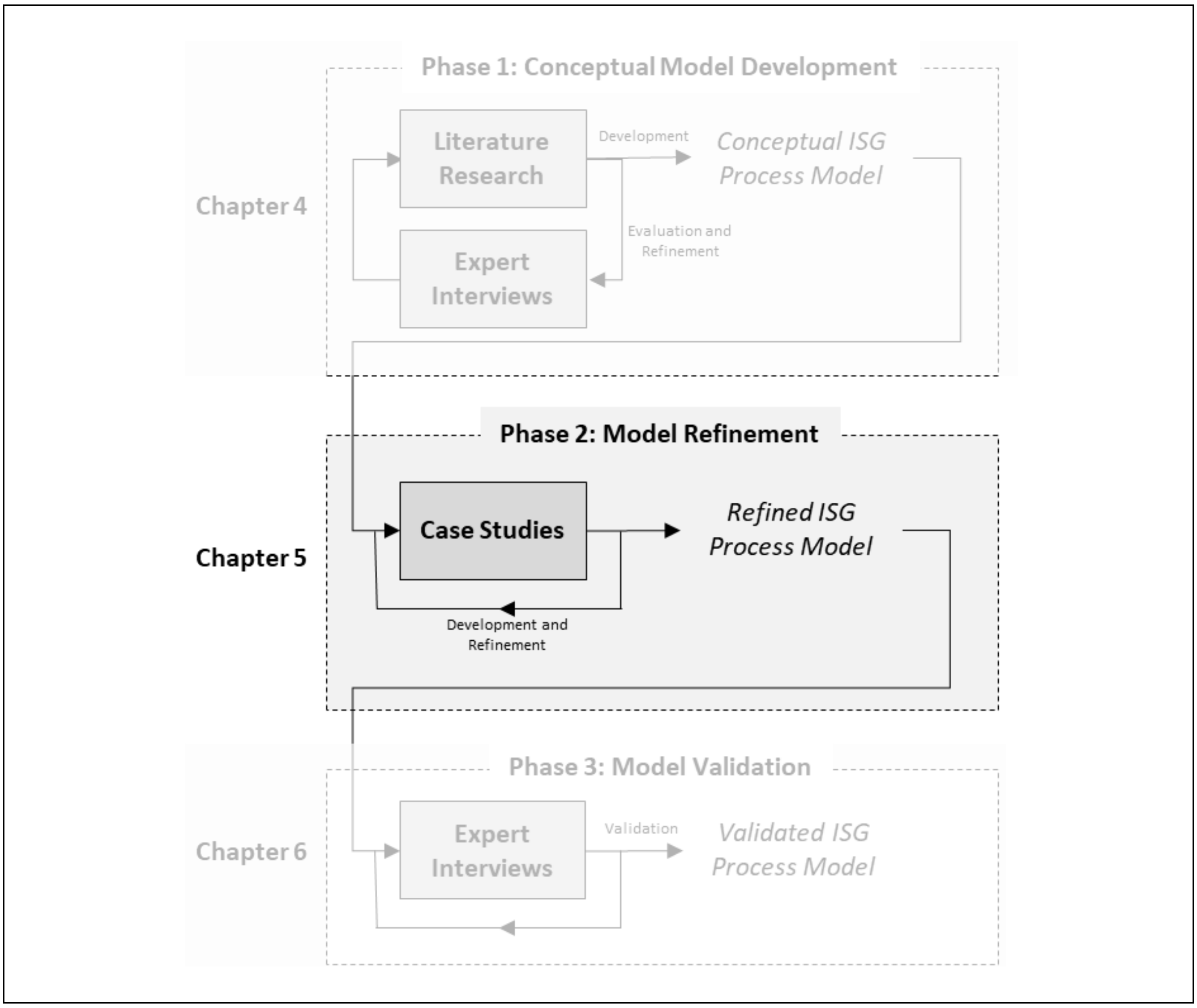

Figure 5-1: Phase 2: Model refinement

Section 5.1 introduces the 3 case study organisations. Section 5.2 summarises the coding and analysis of case study data to discover the second-order themes and aggregated dimensions, while Section 5.3 discusses in detail the discovery of the second-order themes and aggregated dimensions that led to the identification of the core components of ISG, i.e. the stakeholder groups, ISG processes and sub-processes required to develop the ISG process model. Section 5.4 compares the ISG process model discovered from the case study data with the conceptual ISG process model to confirm the theories and identify the differences. This section also highlights the refinements made to the conceptual model that led to the refined ISG process model. This chapter concludes with Section 5.5 that shows the refined ISG process model.

## 5.1 Case Study Organisations

Case study analysis allowed the researcher to understand the organisation and to position the research within the setting of the organisation, extracting best practices adopted in the organisation. As discussed in Section 3.6.2, financial institutions have been selected as the researcher believes that they will provide the best practices in how ISG has been implemented and thus provide insightful data for analysis. Financial institutions are strictly regulated and therefore also have established corporate governance processes covering board oversight, risk management and compliance processes that extend to cover information security. In addition, financial institutions have invested heavily in information security initiatives to protect themselves from information security risks (AustCyber, 2018; Ernst & Young, 2017; Ponemon Institute, 2019).

Three financial institutions have been identified for the multiple case study research. These 3 financial institutions have the following key characteristics:

a. The financial institutions have designated in-house information security teams with specific knowledge, skills and abilities, typically led by a CISO or dedicated IT security team within the CIO or COO functions.

b. The financial institutions have adopted information security processes which are compliant with the respective regulators, the Monetary Authority of Singapore (2013) and Bank Negara Malaysia (2018).
c. The financial institutions have strong corporate governance processes which are basic requirements for continuing business operations as they are in a strictly regulated industry. Financial institutions in Singapore are regulated by the Monetary Authority of Singapore, while financial institutions in Malaysia are regulated by Bank Negara Malaysia.

The following sections provide descriptions of the financial institutions.

### 5.1.1 FinServices_SG

FinServices_SG is an innovative technology-driven financial services organisation in Singapore providing specialist services such as electronic payments and fund transfers. It is headquartered in Singapore and has a small regional operational footprint. As it is a technology-driven financial services organisation, it has adopted leading-edge technology solutions in providing its services across the entire payments value chains of both retail and corporate customers, and has embarked on various digital transformation initiatives to drive improved customer experience and convenience. FinServices_SG is a financial institution and operates in the financial services industry, and therefore, FinServices_SG has to comply with the regulatory requirements of the Monetary Authority of Singapore. This means that FinServices_SG has adopted information security and corporate governance processes that are compliant with these regulatory requirements.

FinServices_SG has demonstrated leading practices in information security and has an independent information security function within the organisation that is headed by the CISO. The CISO has a direct reporting line to the Board and the Board Risk Committee, as well as a dotted reporting line to the CEO for administrative functions only. He views information security as a business and looks at his information security organisation as a business unit, rather than a cost centre:

> Similarly, we need to look at the information security organisation or role as a business. The objective of the information security function is to enable business to get on with their day-to-day business by ensuring that we manage the threats and protect against any attacks to the business. We invest in new security systems, e.g. secure payment solutions to help businesses get new customers … it's business. So in the same way, for information security, we need a proper structure, an oversight, and checks and balances. This is relatively new thinking because in most cases, information security is seen only as the protection against attacks on a business. (FinServices_SG_CISO)

With this new paradigm thinking, FinServices_SG has redefined its information security framework that standardises the overall direction for all FinServices_SG companies with consistent standards, policies and procedures within the group and embarked on leading information security initiatives that drive proactive information security threats surveillance:

> We have policies and procedures for everything, i.e. for user access controls, network security, hardware hardening, data encryption, information life-cycle management, document management, and training and awareness. In addition, we have strict governance which is our "third eye" to provide oversight, making sure information security initiatives are done right. Oversight include monitoring, compliance checks to ensure that we are actually doing what we are set out to do according to our policies and procedures. (FinServices_SG_Director-InfoSecOfficer)

In addition to the CISO (FinServices_SG_CISO), 3 additional executives were interviewed from FinServices_SG, i.e. the CIO (FinServices_SG_CIO), Deputy Head of Technology/Deputy CIO (FinServices_SG_DeputyCIO) and Director – Information Security Officer (FinServices_SG_Director-InfoSecOfficer). Both the Head of IT Security (FinServices_SG_Head-IT-Security) and IT Security Officer (FinServices_SG_IT-SecOfficer) also attended a discussion session where the CISO shared his vision of his information security organisation.

### 5.1.2 FinServices_SEA

FinServices_SEA is a commercial bank in Malaysia providing services in retail, corporate and institutional banking. It is headquartered in Malaysia with regional operations in selected countries in South-East Asia, Hong Kong and China, and employs more than 7000 employees. FinServices_SEA is a financial institution operating in a heavily regulated environment as it must comply with various financial regulators across all operating countries including Bank Negara Malaysia (Central Bank of Malaysia), Monetary Authority of Singapore, Otoritas Jasa Keuangan of Indonesia and Hong Kong Monetary Authority among others.

Corporate governance is a major area for all banks and similarly in FinServices_SEA. Information security is considered part of the risk management framework within the corporate governance risk management framework and has been identified as a specific risk component. Specifically, on ISG, one of the Board of Directors had the following to say:

> The bank is big in corporate governance. Corporate governance is the process and structure used to direct and manage the business and affairs of the bank to ensure business prosperity and corporate accountability, realising shareholder value and protecting all stakeholders' interests. The board is responsible for effective stewardship and control of the bank. Key responsibilities cover formulation of corporate policies and strategies, overseeing and evaluating the conduct of the bank's businesses, identifying principal risks and ensuring that the risks are managed, and reviewing and approving strategic business decisions. As risk is a key area of responsibility, information security governance falls within corporate governance as information security risk is a major component of the bank's risk. (FinServices_SEA_Board)

The bank's Board of Directors plays a critical role in corporate governance. Since ISG is considered a key component of corporate governance from the risk perspective, the bank has placed a high priority on ISG which can demonstrated by the continuous awareness and training sessions for their board members through regular information security and cybersecurity training, and thought leadership sessions conducted by both the internal information security team and external consultants.

In FinServices_SEA, there is a dedicated information security/IT security team that is part of the CIO's functions. The dedicated information security team is structured within the CIO's functions as FinServices_SEA believes that information security is a key responsibility of the CIO in maintaining the confidentiality, integrity and availability of all information within FinServices_SEA. The CIO also has a dotted reporting line to the CRO with respect to all information security risk matters. FinServices_SEA has a rolling 3-year security strategy which includes the implementation of intelligent anti-fraud services, a Security Operations Centre (SOC), and a common security architecture standard.

For this research, 4 individuals from FinServices_SEA were interviewed. They were a non-executive member of the Board of Directors (FinServices_SEA_Board), the Chief Financial Officer (FinServices_SEA_CFO), CIO (FinServices_SEA_CIO) and IT Architect (FinServices_SEA_IT-Architect). Two additional individuals, i.e. the Head of IT Infrastructure (FinServices_SEA_Head-IT-Infra) and IT Security Officer (FinServices_SEA_IT-SecOfficer) participated in less formal discussions during a process walk-through session where the researcher had the opportunity to examine the information security framework, policies and procedures.

### 5.1.3 FinServices_MY

FinServices_MY is a small commercial bank in Malaysia with a focus on providing efficient services to both retail customers and small and medium enterprises in Malaysia. FinServices_MY is well known for providing excellent customer service and quick turn-around in processing of business transactions, e.g. mortgage and trade finance processing. A number of innovations have been introduced in the areas of customer on-boarding, anti-fraud checking and mobile application leveraging on various mobile devices.

Similarly, FinServices_MY has a strong emphasis on compliance with regulatory requirements as the financial services industry is heavily regulated by Bank Negara Malaysia.

When FinServices_MY was asked about ISG, FinServices_MY compared ISG with corporate governance and emphasised its importance in the financial services industry:

> Information is your strategic asset whether you are with a bank or a retailer. Information is the lifeblood of an organisation. Information comes in many forms, both physical documents or digital. You must be aware of the risks associated with the capturing, handling, processing and storing of this information. Since we are looking at it from the information security risk angle, we have to treat this as part of the bank's risk. Information security risk is part of operational risks and information technology risk or cybersecurity risk that falls within our bank-wide risk management framework, which is part of the bank's corporate governance framework. Therefore, information security governance should be treated as part of corporate governance framework. (FinServices_MY_Board)

FinServices_MY had a small security incident a few years ago that caused interruption to the online banking system for a short time. While it was a small incident that had no financial implications, it was a wake-up call for FinServices_MY and created awareness for everyone in the bank from the Board of Directors to the bank tellers at the branches. Since then, FinServices_MY has improved the information security posture of the bank, engaged external consultants and implemented new security solutions, revised security standards and processes, and implemented improved policies and procedures complying with the regulatory and international standards (Bank Negara Malaysia, 2018; National Institute of Standards and Technology, 2018b):

> Operational risk includes any other non-financial risk and this will include systems down due to IT issues, flooding at branches, etc. Obviously this include cyber attacks. In this area of information security risk, my team work closely with the COO and IT as they have specialist resources that can help me … As I am the CRO, I will make sure that the IT team conduct a detail information security assessment. They [referring to the IT Team] will work with all business departments and back office departments to

> understand their information security risk and identify actions needed to be taken to manage these risks. (FinServices_MY_CRO)

In FinServices_MY, information security is the responsibility of the COO, where a dedicated team is responsible for information security functions. The COO has a dotted reporting line to the CRO with respect to all information security risk matters.

In FinServices_MY, a non-executive member of the Board of Directors (FinServices_MY_Board), the COO (FinServices_MY_COO) and the CRO (FinServices_MY_CRO) were interviewed. Two additional information security team members, i.e. the Head of IT (FinServices_MY_Head-IT) and IT Applications Security Officer (FinServices_MY_IT-AppSecOfficer) were invited to join the interview with FinServices_MY_COO in the latter part of the interview session where they shared the operations aspect of information security.

## 5.2 Coding and Analysis

The coding and analysis of all case study data followed the data analysis processes that were discussed in Section 3.8. Data analysis began with the identification of initial concepts and terms through initial coding or open coding (Corbin & Strauss, 1990) where "In-vivo" codes were created, e.g. to label the data that captured the organisational activities that explain the ISG process. These initial codes were continuously refined by re-reading and re-coding the interview scripts until they formed the first-order concepts. The next step of data analysis was to link the similar concepts from the first-order concepts while retaining the interviewees' terms whenever possible to develop the second-level themes (axial coding). This iterative coding and analysis facilitated the abstraction of first-order concepts to higher level concepts and categories forming the second-level themes. Further analysis of these second-order themes helped to develop the final dimensions (Gioia et al., 2012). Table 5-1 and Table 5-2 show a summary of the aggregated dimensions and themes (axial codes) against the case study organisations. The "intensity" represents the total number of individual statements of all interviews that relate to the particular dimension and theme, and provided the evidence for the discovery of the themes and dimensions from the case study interviews.

Table 5-1: Intensity by aggregated dimensions and themes (axial codes) against case studies for core governance stakeholder groups (extracted from analysis in NVivo 11).

| **Aggregated Dimension** | **Theme (Axial Code)** | **Intensity#** | | | |
|---|---|---|---|---|---|
| | | **FinServices_SG** | **FinServices_SEA** | **FinServices_MY** | **Total** |
| **External** | External | 14 | 17 | 13 | **44** |
| **Governing Body** | Strategic – board | 34 | 25 | 27 | **86** |
| | Strategic – executive | 21 | 11 | 14 | **46** |
| **Management** | Management | 20 | 12 | 11 | **43** |

*#Intensity represents the total number of individual statements of all interviews that relate to a particular dimension and theme (or axial code)*

Table 5-2: Intensity by aggregated dimensions and themes (axial codes) against case studies for core ISG processes (extracted from analysis in NVivo 11).

| **Aggregated Dimension** | **Theme (Axial Code)** | **Intensity#** | | | |
|---|---|---|---|---|---|
| | | **FinServices_SG** | **FinServices_SEA** | **FinServices_MY** | **Total** |
| **Assure** | Conduct external audits and certifications | 19 | 18 | 19 | **56** |
| | Provide oversight | 35 | 17 | 15 | **67** |
| | Conduct internal audit | 16 | 14 | 5 | **35** |
| **Communicate** | Engage stakeholders | 31 | 25 | 6 | **62** |
| **Evaluate** | Evaluate and refine | 16 | 12 | 5 | **33** |
| | Collect and analyse | 3 | 7 | 0 | **10** |
| **Monitor** | Measure and report performance | 10 | 10 | 1 | **21** |
| | Manage compliance | 35 | 43 | 19 | **97** |
| **Direct** | Define information security objectives to comply with regulatory requirements | 16 | 14 | 14 | **44** |
| | Define information security objectives to support business strategy | 41 | 45 | 20 | **106** |
| | Confirm risk appetite | 39 | 27 | 16 | **82** |
| | Manage risk | 51 | 40 | 30 | **121** |
| | Confirm information security strategy and objectives | 18 | 12 | 3 | **33** |
| | Implement information security standards, policies and controls | 40 | 24 | 13 | **77** |

*#Intensity represents the total number of individual statements of all interviews that relate to a particular dimension and theme (or axial code)*

## 5.3 Refinement of ISG Process Model through Multiple Case Study

This section describes how the refined ISG process model as shown in Figure 5-2 was developed.

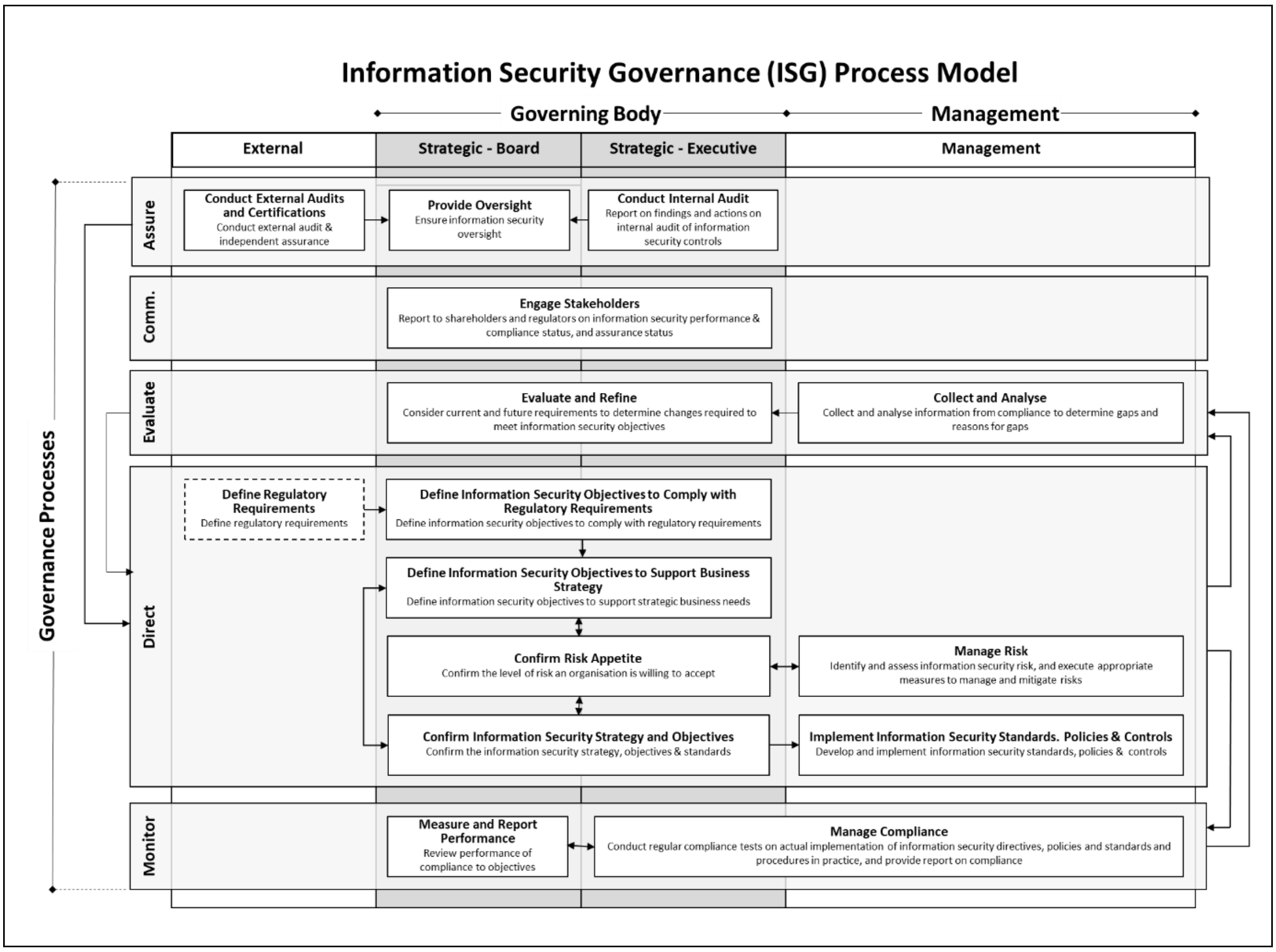

Figure 5-2: Refined ISG process model

The next section starts with describing the case study analysis that suggested the existence of the key stakeholder groups who are involved in implementing ISG. This is followed by subsequent sections that discuss the analysis which facilitated the identification of the core ISG processes. Throughout the analysis, comparisons were made against extant literature and the conceptual ISG model, iterating towards the refined ISG process model. The analysis helped to confirm the ISG processes and sub-processes as identified in the conceptual ISG process model and also identified new sub-processes practised in the case study organisations which are reflected in the refined ISG process model.

### 5.3.1 ISG Stakeholder Groups and Structure

The purpose of this section is to identify the stakeholders and the structure required to implement ISG. In the data analysis, first-order concepts and subsequent higher order themes and dimensions in relation to the ISG stakeholders and structure were identified to build a data structure, as shown in Figure 5-3.

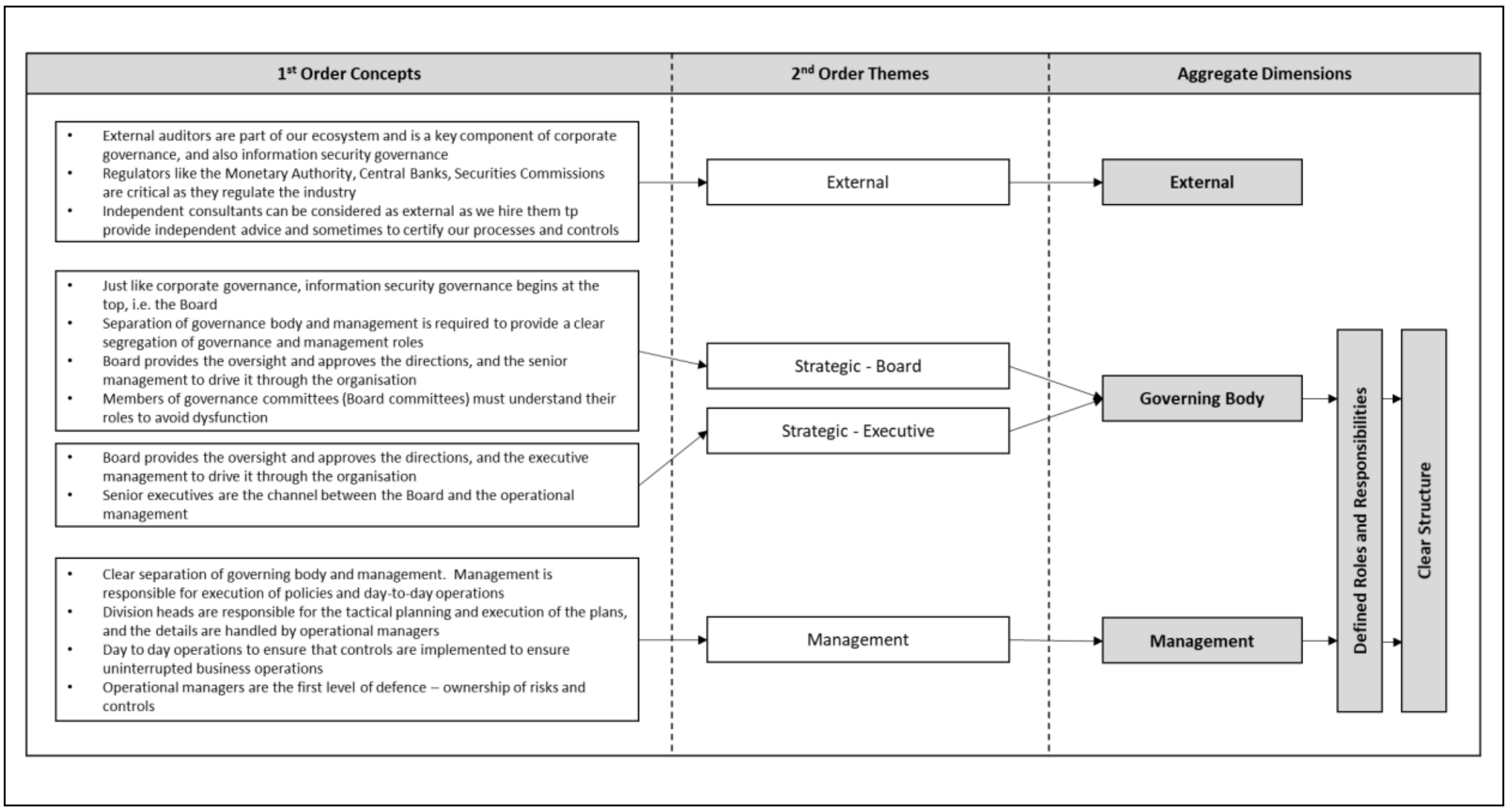

Figure 5-3: Data structure – ISG structure

The dominant theme that emerged from the interviews in relation to good ISG was the importance of a clear structure that drives defined roles and responsibilities. This theme resonates with the principles of good corporate governance which emphasise the requirement of a structure that drives clear roles and responsibilities:

> should establish and disclose the respective roles and responsibilities of its board and management, and how their performance is monitored and evaluated. (ASX Corporate Governance Council, 2007)

> Corporate governance also provides the structure which the objectives of the company are set, and the means of attaining those objectives and monitoring performance are determined. (OECD, 2015)

> Corporate governance refers to having the appropriate people, processes and structures to direct and manage the business. (Monetary Authority of Singapore, 2018)

The following subsections provide the analysis and representative quotes that were gathered from the case studies to identify the second-order themes and aggregated dimensions.

#### *5.3.1.1 External*

"External" refers to external parties, i.e. parties not within the same organisation who are involved in ISG processes. Data from case study interviews identified "external" to include external auditors, external consultants and any independent parties such as regulators who provide independent assurance or conduct independent reviews or assessments of the information security posture of an organisation. External auditors provide independent assurance, while regulators ensure supervision and enforce compliance with regulatory requirements. Table 5-3 shows representative quotes from the interviewees in relation to external parties' involvement in ISG. In addition, the involvement of external parties is also consistent with the adherence of these case study organisations to corporate governance guidelines that mandate the use of external auditors in assurance.

Table 5-3: Data supporting "external" (ISG structure/stakeholder - external).

| Dimension: **External**; Theme: **External** | |
|---|---|
| **Case Study** | **Representative Quotes** |
| FinServices_SG | "In addition, we have an external assurance function where we engage external consultants like you to conduct a review annually just to ensure that we are in compliance with our internal policies and procedures. In addition, the external assurance ensures we are kept aligned with industry best practices and standards. Assurance helps to ensure that we are adopting the right approach and that our policies and procedures are aligned to best practices in the industry. External assurance also helps us with findings to justify to the board should we need additional investments to bring our compliance to industry standards and new regulatory or future requirements." (FinServices_SG_CISO)<br><br>"While we are the CISO office and is an independent function, we are still part of the organisation. We need an external party who is outside of the organisation like the external auditor to provide an independent assessment of our internal controls." (FinServices_SG_CISO)<br><br>"Remember the various regulators, they are part and parcel of the governance, and outside of the organisation." (FinServices_SG_CIO) |
| FinServices_SEA | "Just like corporate governance, you need the external auditors to provide an independent view. While the conduct their financial audit, the external auditors |

| Dimension: **External**; Theme: **External** | |
|---|---|
| **Case Study** | **Representative Quotes** |
| | will also review our internal controls. This includes our security access controls." (FinServices_SEA_Board)<br>"Oh yeah! Forgotten, there is also regulators doing assurance, our central bank. They [referring to regulators] will conduct scheduled audit, maybe every two years, but they can also conduct a surprise audit! Normally, they will focus on financial audits, but with the internet banking and digital here and digital there, central bank comes to conduct information security audit and they are very detailed." (FinServices_SEA_Board)<br>"I mentioned about the board and our senior management. Actually, you also have the external auditor who is involved. They [referring to the external auditor] play a critical part in providing the external independent assurance of our internal controls and this includes information security or specifically IT controls. Then, you have central bank and securities commission, the regulators. Together, they play a critical role in the overall governance of our information security." (FinServices_SEA_CFO)<br>"Talking about regulators, don't forget the central bank. They are the regulatory body that are involved too. While they are external parties to the organisation, they are very involved in ensuring regulatory compliance." (FinServices_SEA_CIO) |
| FinServices_MY | "External auditors and regulators are external parties that provide the assurance and I believe they are critical to the overall governance process. These independent assurance are critical for banks and public listed organisations as they provide the confidence to the public." (FinServices_MY_COO)<br>"We hire people like you to undertake our independent checks to ensure that we are doing the right thing. We also hire external consultants to help us put in place proper processes and learn from best practices adopted by banks globally and regionally." (FinServices_MY_Board)<br>"Finally, you have the check and compliance where you have the audit and compliance people to ensure that the policies and procedures are complied. In addition, you also have the external auditors that conduct the independent checks." (FinServices_MY_Board) |

#### *5.3.1.2 Governing Body*

The analysis of case study data facilitated the identification of two second-order themes, i.e. "strategic - board" and "strategic - executive" which are responsible for providing the overall leadership in ISG. The "strategic - board" is represented by the board of directors, which is responsible for ensuring effective stewardship and control of ISG in the organisation, which includes oversight, direction setting and approval of information security policies and strategies; confirmation of the information security risk appetite; and approval of budget required for information security initiatives. The “strategic - executive” i.e. the C-level executives are entrusted by the board of directors to implement the strategies defined and approved by the board of directors. The C-level executives work with the management team to implement and execute the information security strategies and to ensure the day-to-day running of the organisation. With a clear structure, there is defined division of responsibilities between the board of directors, C-level executives and management. The collective stakeholder groups of “strategic - board” and “strategic - executive” are identified as the aggregated dimension of the “governing body”. This “governing body” is responsible for the overall ISG in an organisation, as highlighted by the CIO of FinServices_SG and the COO of FinServices_MY:

> information security governance is the oversight that is required by senior management, especially the C-levels and the board in ensuring that the management is doing all the right things in managing information security. By oversight, I mean the responsibilities to ask the right questions, check and get updates on the information security situation, and to guide the organisation in the right direction. As you know, the board and the C-levels have onerous responsibilities on information security, therefore they better do a good job in oversight. (FinServices_SG_CIO)
>
> The board together with the Exco/senior management provides an oversight of the overall information security activities undertaken by the bank, by the operational management team. (FinServices_MY_COO)

Additional representative quotes provided by the interviewees that facilitated the identification of these themes and dimension are found in Table 5-4 and Table 5-5.

Table 5-4: Data supporting “strategic - board” (ISG structure/stakeholder - governing body).

| Dimension: **Governing Body**; Theme: **Strategic-Board** | |
|---|---|
| **Case Study** | **Representative Quotes** |
| FinServices_SG | “In governance, structure is the most important part. You need a right structure to enable governance. Corporate governance, IT governance or information |

| Dimension: **Governing Body**; Theme: **Strategic-Board** | |
|---|---|
| **Case Study** | **Representative Quotes** |
| | security governance. Structure drives clear roles and responsibilities. It starts at the board as board has an oversight responsibility in governance." (FinServices_SG_CISO)<br>"In my view, for information security governance, we need clear structure and responsibilities of the board. They [referring to the board] approve the risk appetite, the budget required for the security investments. The executive management execute it and the operations is done by the line management." (FinServices_SG_CIO)<br>"I believe good governance starts from the top, especially the board who sets the tone. They have a duty to make sure the organisation is doing the right thing for information security." (FinServices_SG_Director-InfoSecOfficer)<br>"We differentiate the roles and processes between the board and executive management as the board are normally only involved in approvals and endorsements while the executive management actually work on the direction, get it executed." (FinServices_SG_Director-InfoSecOfficer) |
| FinServices_SEA | "Just like corporate governance, process and structure are critical to direct and manage the business. There is a clear division of responsibilities between the board and the management, the chairman and the managing director/CEO." (FinServices_SEA_Board)<br>"I strongly believe good governance starts from the board as we have a fiduciary duty and is responsible for the overall governance." (FinServices_SEA_Board)<br>"Decisions on information security are tabled to the board for approval." (FinServices_SEA_CFO) |
| FinServices_MY | "Board delegates the independent oversight over to the board committees to work with the executive management. However, the ultimate responsibility and the final decision rest with the board." (FinServices_MY_Board)<br>"Board takes their responsibilities seriously in information security governance through the various board committees – IT Committee and Risk Committee. These committees are chaired by a board member." (FinServices_MY_COO)<br>"My role is to consolidate all the risk profile and present it to the RMC, then for the board to agree on the risk appetite that the bank is OK to accept." (FinServices_MY_CRO) |

Table 5-5: Data supporting "strategic - executive" (ISG structure/stakeholder - governing body).

| Dimension: **Governing Body**; Theme: **Strategic-Executive** | |
|---|---|
| **Case Study** | **Representative Quotes** |
| FinServices_SG | "We in the CISO team develops the information security standards, policies and procedures based on the directions provided by the board. Specifically, the CISO is responsible to make sure we execute what is approved by the board." (FinServices_SG_CISO) |

| Dimension: **Governing Body**; Theme: **Strategic-Executive** | |
|---|---|
| **Case Study** | **Representative Quotes** |
| | "I as the CIO will be responsible to implement and comply to the standards." (FinServices_SG_CIO)<br>"Executive management, i.e. the C-levels are responsible to take the board's direction to define the standards and policies, and ensure that it is implemented throughout the organisations." (FinServices_SG_DeputyCIO) |
| FinServices_SEA | "All C-levels executive are part of executive management." (FinServices_SEA_CFO)<br>"For example, the Risk Management Committee is chaired by a board member, but the team on the committee are executives who will make sure that the bank has the proper standards and policies for implementation." (FinServices_SEA_IT-Architect)<br>"Clear structure and segregation of duties are important. Board has approval oversight and management is responsible to ensure proper execution and compliance." (FinServices_SEA_Board) |
| FinServices_MY | "They are the ones [referring to the C-level executives] who make sure that policies and procedures are defined and cascaded down the other levels for execution." (FinServices_MY_Board)<br>"In our context, responsible at C-level executives means someone will lose their job, the banks will get fined by the central bank/regulators, there are liabilities to the customers and the public." (FinServices_MY_COO)<br>"All executive management, i.e. the C-level executives are involved in reviewing the risk profile and approving their various department profiles and defining the risk mitigation plan and initiatives. So, all the executive management is responsible for risk management." (FinServices_MY_CRO) |

#### *5.3.1.3 Management*

"Management" is the other second-order theme and aggregate dimension discovered in the case study data analysis relating to ISG structure. It is found that management is responsible for the execution and ongoing implementation and operation of information security programs to ensure uninterrupted business operations. Table 5-6 shows the evidence for the required structure of "management".

Table 5-6: Data supporting “management” (ISG process model - management).

| Dimension: **Management**; Theme: **Management** | |
|---|---|
| **Case Study** | **Representative Quotes** |
| FinServices_SG | “We will assign this to the various business for execution. The various business management will be responsible to ensure execution.” (FinServices_SG_CISO)<br>“Management is like what I do, keep the lights on, making sure we have the right access controls that we are using, we have the latest version of anti-virus running on all the machines, we have the latest operating system patches on all the servers, etc … Line management will be responsible to ensure all information security policies are implemented and procedures are adhered in our day-to-day operations.” (FinServices_SG_DeputyCIO) |
| FinServices_SEA | “The RMC will work with the C-level management to make sure that the bank has proper standards and policies for implementation. Then the management and operational managers will be tasked to implement on the ground.” (FinServices_SEA_IT-Architect) |
| FinServices_MY | “The operations management ensure that the policies and standards are implemented and operationalised.” (FinServices_MY_Board)<br>“Then, we have the operational management that enforces the implementation. We have IT, Risks and Compliance involved in ensuring process, policies and procedures are implemented and adhered.” (FinServices_MY_COO) |

Figure 5-4 shows the second-order themes and aggregated dimensions for the stakeholder groups who are involved in ISG in an organisation as mapped onto a cross-functional process map.

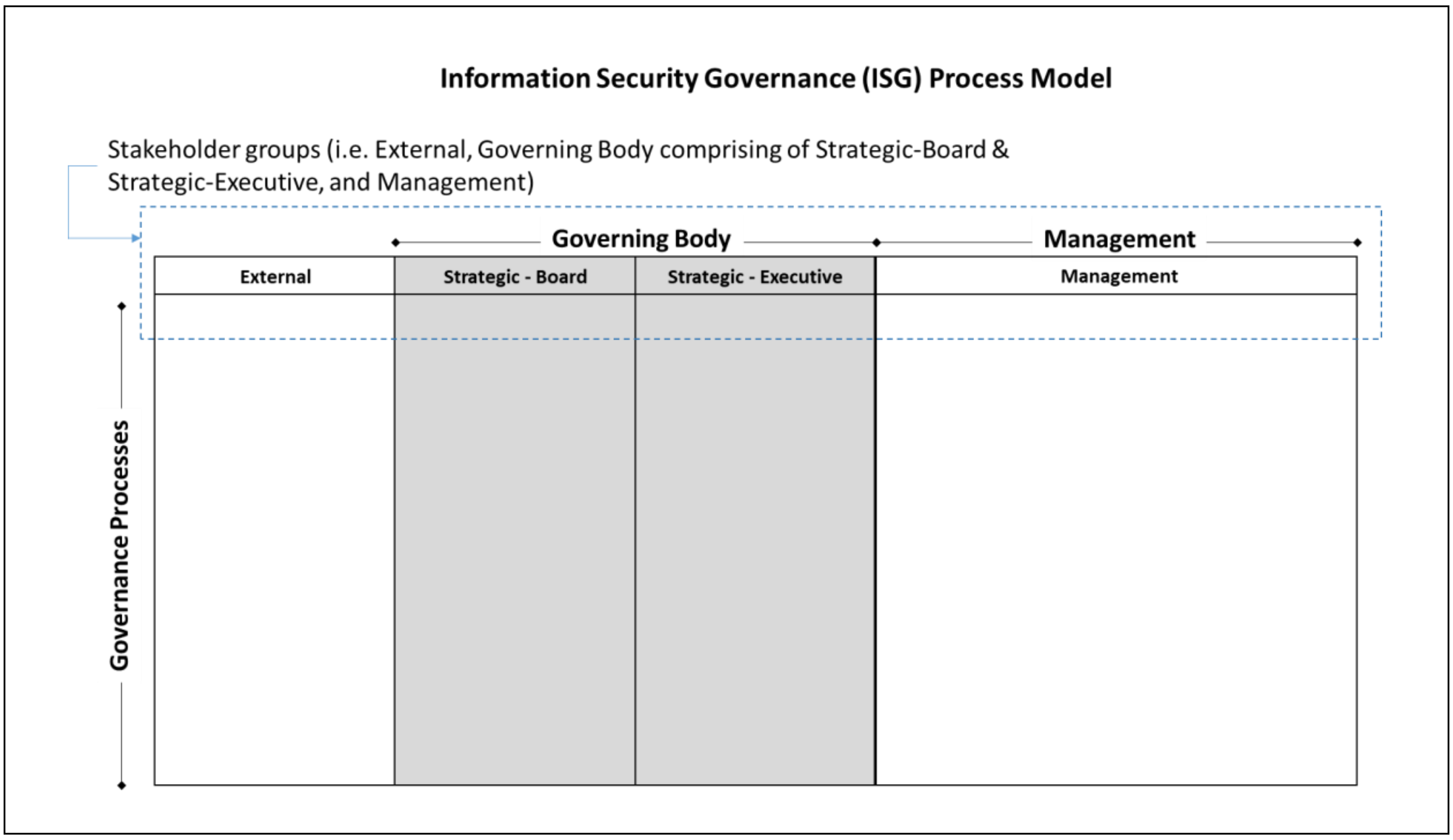

Figure 5-4: ISG – Stakeholder groups as discovered from case study data

### 5.3.2 ISG Processes

In all the case studies, it was apparent that the interviewees were discussing the processes involved in the governance process. Interviewees consistently agreed that governance involves a set of processes and it is similar for ISG.

Interview data was analysed and codes were defined and further analysed to identify second-order themes which confirmed that ISG is implemented as a set of processes and can be represented by a process model. Once the codes and second-order theme were identified, a data structure for ISG processes as a process model was defined, as shown in Figure 5-5.

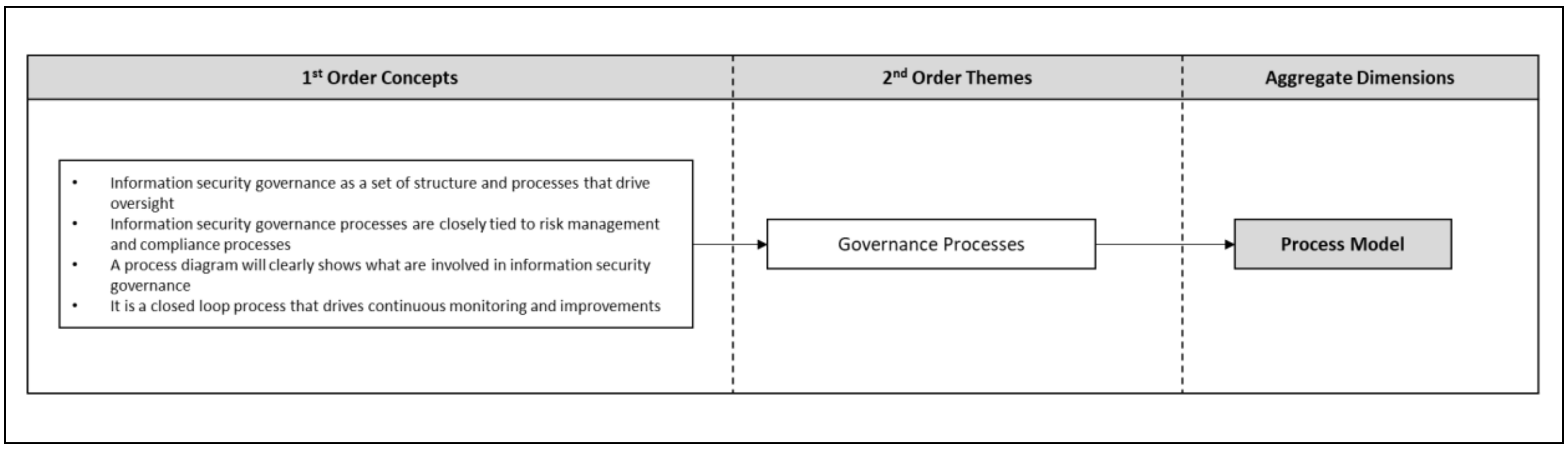

Figure 5-5: Data structure - ISG process model

Table 5-7 shows representative quotes from the interviewees across the case studies that informed that ISG involves a set of processes.

Table 5-7: ISG - data supporting governance processes (ISG process model).

| Dimension: **ISG Process Model**; Theme: **Governance Processes** | |
|---|---|
| **Case Study** | **Representative Quotes** |
| FinServices_SG | "We have a process where the board is involved through the Risk Management Committee (RMC)." (FinServices_SG_CIO)<br>"The CISO team is responsible for driving the information security risk assessment process." (FinServices_SG_CIO)<br>"this type of model will be very useful to guide them towards a proper governance process." (FinServices_SG_CISO)<br>"compliance becomes a monitoring and feedback process for us." (FinServices_SG_DeputyCIO) |
| FinServices_SEA | "Just like corporate governance, process and structure are critical to direct and manage the business. There is a clear division of responsibilities between the board and the management, the chairman and the managing director/CEO" (FinServices_SEA_Board)<br>"So we have processes that the board is involved in setting the direction, agreeing on the targets (business, risks, etc.) and the management will take it to the next level of work and implementation." (FinServices_SEA_Board)<br>"we have a proper process to derive the allocated budget based on the agreed risk profile." (FinServices_SEA_CFO)<br>"The self-declaration/dispensation and audit process is a well-practised closed-loop process." (FinServices_SEA_CIO) |
| FinServices_MY | "Clear structure, with clear roles and responsibilities is the basis of good governance … Superimposed this with a clear process model will make the implementation easy." (FinServices_MY_Board)<br>"After the structure is in place, you need a proper process." (FinServices_MY_CIO) |

| Dimension: **ISG Process Model**; Theme: **Governance Processes** | |
|---|---|
| **Case Study** | **Representative Quotes** |
| | "These committee defines the directions, the risk appetite and what should be done at the bank. These will then be cascaded to the Exco, i.e. the executive committee or the senior management team who is responsible to make it happens on the ground. The board together with the Exco/senior management provides an oversight of the overall information security activities undertaken by the bank, by the operational management team. It is like a process from the board to the operational management and then a feedback loop back to the board." (FinServices_MY_COO) |

The following sections identify the ISG processes and sub-processes.

#### *5.3.2.1 Direct*

"Direct" in the ISG model refers to the provision of overall guidance and direction in governance. From the analysis of the data gathered from the case studies, it can be shown that ISG is primarily regulatory, risk and business driven. "Regulatory driven" surfaced prominently in all the interviewees' comments. Regulatory requirements have significant influence on information security as these requirements are dictated by the regulators (Bank Negara Malaysia, 2018; Monetary Authority of Singapore, 2013). Aside from regulatory requirements, risk and business requirements are key inputs into the "direct" sub-processes.

The open coding and first-order concepts, and subsequent aggregation of second-order themes for the "direct" dimension are shown in the data structure in Figure 5-6. Six second-order themes support the "direct" dimension, i.e. "define information security objectives to comply with regulatory requirements", "define information security objectives to support business strategy", "confirm risk appetite", "manage risk", "confirm information security strategy and objectives" and "implement information security standards, policies and controls".

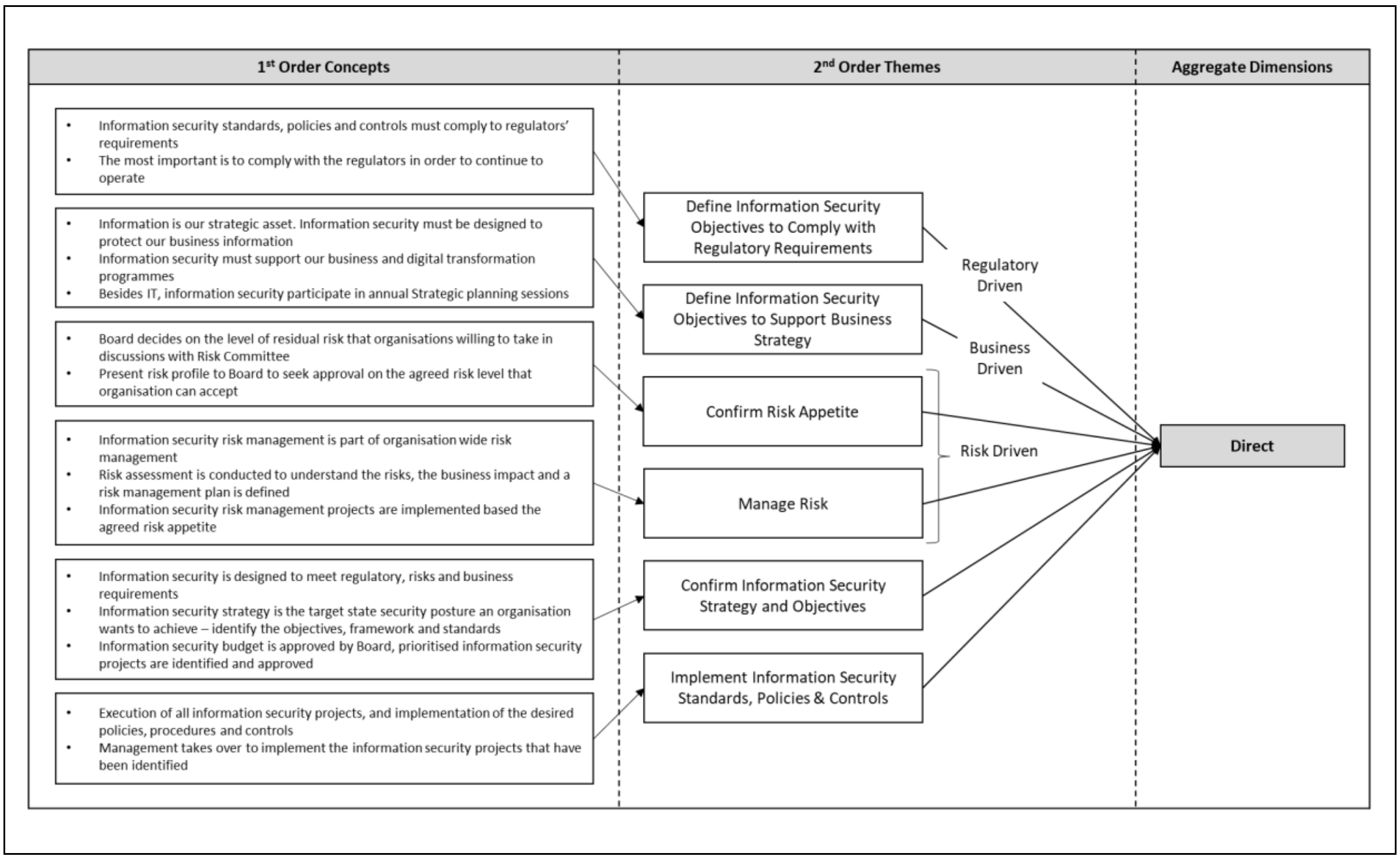

Figure 5-6: Data structure – ISG processes (direct)

The following subsections provide analysis of the case study data in identifying the sub-processes (i.e. second-order themes) in "direct".

#### 5.3.2.1.1 Define Information Security Objectives to Comply with Regulatory Requirements

Regulatory requirements surfaced as the most important component that drives ISG. All case studies concluded that the regulators, while being an external stakeholder group in ISG, are responsible for defining the requirements. Organisations have no option but to comply with the requirements that are defined by the regulators, which are generally the central bank or monetary authority responsible in governing the operations of financial institutions.

The board, which has fiduciary responsibilities towards the regulators and shareholders, is responsible for ensuring that the organisation complies with regulatory requirements. These were confirmed consistently by the FinServices_SG_CIO, FinServices_SG_CISO, FinServices_SG_DeputyCIO, FinServices_SEA_Board, FinServices_SEA_CFO, FinServices_SEA_CIO, FinServices_MY_Board and FinServices_MY_COO and their representative quotes are shown in Table 5-8. These regulatory requirements are key inputs into the strategic business direction and risk management approach of the organisations which are ultimately used to confirm the information security strategy and objectives.

Table 5-8: Data supporting “define information security objectives to comply with regulatory requirements” (direct).

| Dimension: **Direct**; Theme: **Define Information Security Objectives to Comply with Regulatory Requirements** | |
|---|---|
| **Case Study** | **Representative Quotes** |
| FinServices_SG | “we are in the financial industry that is so regulated by the monetary authority.” (FinServices_SG_CIO)<br>“compliance to industry standards and new regulatory or future requirements.” (FinServices_SG_CISO)<br>“Financial institutions are heavily regulated … proper processes for compliance and audit, etc … strict regulations and heavy fines make the difference.” (FinServices_SG_DeputyCIO) |
| FinServices_SEA | “making sure we adhere and comply to policies and regulations … regulators doing assurance, our central bank … scheduled audit, maybe every two years, but they also conduct a surprise audit!” (FinServices_SEA_Board)<br>“central bank that goes after the board and the bank’s management.” (FinServices_SEA_Board)<br>“central bank issues some new regulations that warrants the implementation of some new software or systems, this will be allocated as out-of-cycle budgeting items.” (FinServices_SEA_CFO)<br>“An easy way to determine the budget is based on the regulatory compliance requirements. Compliance is a non-negotiation matter in a bank.” (FinServices_SEA_CIO) |
| FinServices_MY | “We need to comply, comply with all the regulations and we have a long list of GP (guidelines) to comply as a bank. I believe central bank as the regulator is driving the strict adherence to the information security guidelines and banks have to spend. No choice.” (FinServices_MY_Board)<br>“banks spend money in information security, especially in cybersecurity because banks need to comply to central bank regulations and the fines that central bank imposes.” (FinServices_MY_Board)<br>“Banks are heavily regulated and most banks focus on getting things done to comply with regulations.” (FinServices_MY_COO)<br>“The risk management process is very well documented and practised as we are strictly regulated by BNM.” (FinServices_MY_CRO) |

#### 5.3.2.1.2 Define Information Security Objectives to Support Business Strategy

"Define information security objectives to support business strategy" is the process where information security strategy, objectives and standards are defined based on an organisation's business vision, mission and business objectives. The CISO of FinServices_SG had this to say about how the information security function should support business:

> The objective of the information security function is to enable business to get on with their day-to-day business by ensuring that we manage the threats and protect against any attacks to the business. (FinServices_SG_CISO)

Other interviewees echoed the need for information security requirements to meet the strategic business objectives of an organisation:

> we are able to plan out our investments and information security projects to align with our business strategy. For example, we are embarking on digital banking initiatives and these require enhanced security protection. So, we need to spend more in specific areas. (FinServices_MY_COO)

Additional representative quotes from the case studies are found in Table 5-9.

Table 5-9: Data supporting "define information security objectives to support business strategy" (direct).

| Dimension: **Direct**; Theme: **Define Information Security Objectives to Support Business Strategy** | |
|---|---|
| **Case Study** | **Representative Quotes** |
| FinServices_SG | "When we conduct the risk assessment, it is assessed against the business. In this way it is done with the business in mind and it is indirectly driven by strategy. For example, when we drive our strategy towards more digital payments, our information security risk is assessed against the digital risks. Automatically, our information security initiatives, policies and procedures are driven by these risks associated with the digital strategy." (FinServices_SG_CISO)<br>"We do bring risk and specifically information security risks in during our strategic planning process." (FinServices_SG_DeputyCIO) |
| FinServices_SEA | "We look at the business plans, what is our business targets, the marketing and sales initiatives, the IT projects that are required etc." (FinServices_SEA_Board)<br>"The risks are related to the strategic business objectives of the bank. Then, everything flows from here. With the risks agreed, you start to develop all the |

| Dimension: **Direct**; Theme: **Define Information Security Objectives to Support Business Strategy** | |
|---|---|
| **Case Study** | **Representative Quotes** |
| | risk mitigation strategies, the projects required, the budget required and the resources involved.” (FinServices_SEA_Board)<br>“the information security team will meet with me and my team to understand my business, and work to identify the potential information security threats that can affect my finance function and the operations of the bank.” (FinServices_SEA_CFO)<br>“new digital bank initiatives, new mobile payment solution offerings, cloud solutions, outsourcing strategies, etc. In every of these initiatives, my information security architecture team is involved working together with the other information security teams to identify potential security risks and look at ways to address or mitigate these risks … information security is aligned to the strategic business projects.” (FinServices_SEA_IT-Architect) |
| FinServices_MY | “always look at information as the asset of an organisation, then you classify the different type of information and consider the different level of protection to ensure the security of these information.” (FinServices_MY_Board)<br>“information security should start with the business strategy. Everything should be driven from the business strategy. Based on your strategy, you consider the risks, and then the investment should be determined to manage the risks ... the bank is embarking on major initiative in digitalisation to enhance customer experience by providing new customer touch points, automation of processes with RPA (robotic process automation), AI (artificial intelligence), etc.” (FinServices_MY_Board) |

##### 5.3.2.1.3 Confirm Risk Appetite

As discussed in the earlier sections, information security in an organisation is driven by regulatory requirements, strategic business objectives and the level of risk that the organisation is willing to accept. Managing information security risk is a key component of ISG and management. To effectively govern and manage information security risk, an organisation is required to understand the risks, accept the risk profile to work with and identify the actions required to manage and mitigate the risks.

“Confirm risk appetite” is a theme discovered within “direct”. The board, which is responsible for effective stewardship and control of an organisation, is responsible for approving the risk appetite within which the organisation operates and this is consistent with the overall code of good corporate governance. Risk appetite is defined by Committee of Sponsoring Organisations (COSO) as the amount of risk that an organisation is willing to accept in pursuit of its business objectives (Moeller, 2011). Excerpts from the code of corporate governance are shown below:

> The Board determines the nature and extent of the significant risks which the company is willing to take in achieving its strategic objectives and value creation. The Board sets up a Board Risk Committee to specifically address this, if appropriate. (Monetary Authority of Singapore, 2018)

> set the risk appetite within which the board expects management to operate and ensure that there is an appropriate risk management framework to identify, analyse, evaluate, manage and monitor significant financial and non-financial risks. (Securities Commission Malaysia, 2017)

As information security risk is a key risk in an organisation, ISG adopts a similar approach to corporate governance which includes confirmation of the risk appetite by the board. The C-level executives, i.e. the CRO, CISO or CIO work closely with the board in providing a risk profile and confirming the risk appetite. This approach was confirmed and shared by all interviewees in the case studies, as shown in the representative codes in Table 5-10.

Table 5-10: Data supporting “confirm risk appetite” (direct).

| Dimension: **Direct**; Theme: **Confirm Risk Appetite** | |
|---|---|
| **Case Study** | **Representative Quotes** |
| FinServices_SG | “we have an organisation wide enterprise risk profile where the board is involved in the discussions and final approval. The enterprise risk profile documents the various risks of the organisation that cover strategic risks, operational risks, financial risks and information security risks. The Chief Risk Officer (CRO) will present all the risks, the risks profile and the proposed residual risks, risk appetite to the Board. As information security risks is a specialised area, the Chief Information Security Officer (CISO) will present the information security risks profile. We discuss the details with the board and seek for the board’s approval on the accepted risk appetite.” (FinServices_SG_CIO)<br>“We then take this to the RMC and subsequently present to board for agreement and approval on the agreed level of risk acceptance.” (FinServices_SG_CISO) |
| FinServices_SEA | “The board acknowledges its overall responsibility for the risk management and internal control environment and its effectiveness in safeguarding |

| Dimension: **Direct**; Theme: **Confirm Risk Appetite** | |
|---|---|
| **Case Study** | **Representative Quotes** |
| | shareholders' interests and the bank's assets. This include information security risks." (FinServices_SEA_Board)<br>"The CRO will present and seek approval from the RMC, then the RMC will update and get the final approval from the board." (FinServices_SEA_Board)<br>"The board will deliberate on it and will decide on a suitable risk profile for the bank." (FinServices_SEA_CFO) |
| FinServices_MY | "the board should provide the guidance on the level of risk acceptable to the bank." (FinServices_MY_Board)<br>"These (board) committees define the directions, the risk appetite and what should be done at the bank." (FinServices_MY_COO)<br>"At the start of the financial year, we will present our risk assessment results, present our risk profile to the RMC. We will recommend a proposed risk profile and the actions that will be undertaken to protect the risk profile. We seek the board to approve our recommendations." (FinServices_MY_CRO) |

##### 5.3.2.1.4 Manage Risk

From the data gathered through case study interviews, it is apparent that a lot of effort is spent in organisations on information security risk management–related activities which cover risk identification, risk assessment, risk monitoring, risk mitigation and risk reporting.

The coding and interpretation of first-order concepts highlight the activities in risk identification and assessment, where organisations undertake an assessment of the potential threats and attacks, and the impact on the ongoing business operations, and risk mitigation activities, where actions are taken to address the information security risks through changes in processes, implementation of controls or introduction of technology solutions. These were highlighted by the various interviewees and are consistent across the case studies. While "confirm risk appetite" is the responsibility of the board together with the C-level executives, these risk identification, assessment and mitigation activities are owned and managed by everyone in management, primarily the tactical and operational management. For example, the CRO and CISO work with their management teams to undertake the assessments and propose a risk profile for the board to approve as part of "manage risk" activities. FinServices_SG_CIO, FinServices_SEA_CFO and FinServices_MY_Board had the following to say:

> The Chief Risk Officer (CRO) will present all the risks, the risks profile and the proposed residual risks, risk appetite to the board … the Chief Information Security Officer (CISO) will present the information security risks profile. (FinServices_SG_CIO)

> The IT and the risk divisions, i.e. CIO and CRO will do some bank-wide risk assessment and decide on a risk profile, together with a set of risk mitigation initiatives. (FinServices_SEA_CFO)

> the bank conducts a detailed risk assessment, understanding the various risk areas, its chances of happening and the impact of the loss … Based on the risk assessment and how much risk we are willing to accept, we define the actions to mitigate or manage the risks. (FinServices_MY_Board)

The detailed analysis of the coding and first-order concepts from the interviewees' quotes was aggregated to the "manage risk" dimension, which covers "identify and assess information security risk" and "implement risk management and mitigation" activities. Table 5-11 shows representative quotes selected from detailed case study data. As the research has identified a larger scope of monitoring and reporting that covers areas beyond risk, risk monitoring and reporting have been aggregated into the "monitor" dimension, as discussed in Subsection 5.3.2.2.

Table 5-11: Data supporting "manage risk" (direct).

| Dimension: **Direct**; Theme: **Manage Risk** | |
|---|---|
| **Case Study** | **Representative Quotes** |
| FinServices_SG | "CISO team is responsible for driving the information security risk assessment process." (FinServices_SG_CIO)<br>"The CISO takes the lead to conduct an organisation-wide information security risk assessment with the business to identify the risk profile and potential threats and the impact." (FinServices_SG_CISO) |
| FinServices_SEA | "based on the risk assessment and the risk appetite, the budget and the information security risks projects will be decided." (FinServices_SEA_Board) |

| Dimension: **Direct**; Theme: **Manage Risk** | |
|---|---|
| **Case Study** | **Representative Quotes** |
| | "the information security team will meet with me and my team to understand my business and work to identify the potential information security threats that can affect my finance function and the operations of the bank, i.e. potential loss of financial information, stolen financial information by hackers, financial information systems going down, potential ransomware, etc." (FinServices_SEA_CFO)<br>"we will be reporting on information security incidents (if any), what happened, what was the risks, how we have managed the incident and how we prevent it from recurring." (FinServices_SEA_CIO)<br>"The CRO team owns the risk assessment process and works with the CIO team on the information security risk assessment." (FinServices_SEA_IT-Architect) |
| FinServices_MY | "IT team conducts a detailed information security risk assessment. They [referring to the IT team] will work with all business departments and back office departments to understand their information security risks and identify actions that need to be taken to protect these risks." (FinServices_MY_CRO) |

#### 5.3.2.1.5 Confirm Information Security Strategy and Objectives

"Confirm information security strategy and objectives" is another core sub-process within "direct" that has been identified from the data analysis. Just like business strategy or IT strategy, which need to be defined clearly and approved by the board, information security strategy and objectives need to be confirmed and approved by the board. Information security strategy and objectives are defined to support the requirements of regulators, to meet the needs of strategic business objectives and to support the mitigation and management of information security risks. Once the information security strategy and objectives have been approved, the management can be assigned to develop and implement the supporting standards, policies, procedures and controls.

The following quotations facilitated discovery of the "confirm information security strategy and objectives" theme which is performed by the board to provide the direction for subsequent implementation:

> the board approves our information security risk profile when they [referring to the board] approve the overall organisation's risk profile. Based on the risk profile, the CISO develops the security strategy and framework. These strategy and framework need to be presented and approved by the board. (FinServices_SG_CIO)

Just like IT strategy, you need to present the information security strategy together with the proposed initiatives and budget for board to approve. Once you get the approval, the management will take it to implementation. Then you can define the policies and procedures to support the strategy that was approved. (FinServices_SEA_CIO)

The central bank mandated that all major enhancements and new internet-based systems to go through an external and independent review and attestation by a board member responsible for security before banks can launch such systems. This has put the emphasis that board are ultimately responsible for the information security of the bank. (FinServices_MY_COO)

The CISO or CIO who is responsible for information security will develop the information security strategy, framework and objectives, and seek approval from the board. The approved information security strategy will be the basis to drive implementation of the relevant information security initiatives. Further quotations extracted from the case study interviews are provided in Table 5-12.

Table 5-12: Data supporting “confirm information security strategy and objectives” (direct).

| Dimension: **Direct**; Theme: **Confirm Information Security Strategy and Objectives** | |
|---|---|
| **Case Study** | **Representative Quotes** |
| FinServices_SG | “Also, using the information from the risk assessment, we develop the information security framework based on the NIST model to define the roadmap for all the information security projects. This will then be presented to the board for approval so that the details can be developed and rolled out across the organisation.” (FinServices_SG_CISO) |
| FinServices_SEA | “For information security, this is what we do in the bank. Every year, as part of our planning and budgeting cycle, we review our strategy, our business plans and the budget required to run the bank for the financial year … have a process that the board is involved in setting the direction, agreeing on the targets |

| Dimension: **Direct**; Theme: **Confirm Information Security Strategy and Objectives** | |
|---|---|
| **Case Study** | **Representative Quotes** |
| | (business, risks, etc.) and the management will take it to the next level of work and implementation." (FinServices_SEA_Board) |
| FinServices_MY | "Based on the risk assessment and the willingness to accept the level of risks, we can define the required information security roadmap." (FinServices_MY_COO) |

#### 5.3.2.1.6 Implement Information Security Standards, Policies & Controls

"Implement Information security standards, policies and controls" is the sub-process that implements the approved information security strategy framework and objectives. It is critical that the relevant standards and the correct policies and procedures are implemented to ensure a secure environment to protect the business from potential information security threats and attacks. As highlighted in earlier subsections, these are implemented to meet regulatory requirements, to support strategic business needs and to mitigate identified risks. FinServices_SG_Director-InfoSecOfficer mentioned the following:

> Besides the risk assessment and risk management that I have discussed, we are also responsible in the development and the implementation of security standards, policies and procedures for the organisation. (FinServices_SG_Director-InfoSecOfficer)

The statement that the management team works together to ensure proper implementation was echoed by FinServices_MY_COO:

> These will then be cascaded to the Exco, i.e. the executive committee or the senior management team who is responsible to make it happens on the ground. (FinServices_MY_COO)

Further quotations providing evidence that facilitated the emergence of this dimension are shown in Table 5-13.

Table 5-13: Data supporting “implement information security standards, policies and controls” (direct).

| Dimension: **Direct**; Theme: **Implement Information Security Standards, Policies & Controls** | |
|---|---|
| **Case Study** | **Representative Quotes** |
| FinServices_SG | “the CISO team develops the information security standards, policies and procedures. I as the CIO will be responsible to implement and comply to the standards from the information systems perspective.” (FinServices_SG_CIO)<br>“using the information from the risk assessment, we develop the information security framework.” (FinServices_SG_CISO)<br>“everyone is very good in implementing information security initiatives. We have policies for user access controls, network security, hardware hardening, data encryption, training and awareness. We do this day in, day out and we make sure that we do this well so that operations run without any issue.” (FinServices_SG_Director-InfoSecOfficer)<br>“The executive management, i.e. the C-levels are responsible to take the board’s direction to define the standards and policies, and ensure that it is implemented throughout the organisation … line management will be responsible to ensure all information security policies are implemented and procedures are adhered in our day-to-day operations.” (FinServices_SG_DeputyCIO) |
| FinServices_SEA | “IT would like to hire consultants like you to conduct information security review or develop a framework or policy.” (FinServices_SEA_CFO)<br>“In my bank, the risk and prevention defines the policies and procedures, and the various department executes it accordingly.” (FinServices_SEA_CIO)<br>“The information security team helps in ensuring that the policies are implemented across the bank.” (FinServices_SEA_IT-Architect)<br>“In every of these initiatives, my information security architecture team is involved working together with the other information security teams to identify potential security risks and look at ways to address or mitigate these risks.” (FinServices_SEA_IT-Architect) |
| FinServices_MY | “Then, there is the management and executive committees (Exco) that is run by the senior management. They are the ones who make sure that policies and procedures are defined and cascaded down the other levels for execution.” (FinServices_MY_Board)<br>“the IT team will define specific controls that are required to be implemented. It could also be changes on policies, procedures and new training etc.” (FinServices_MY_CRO) |

As discussed in the previous subsections, the “direct” dimension represents one of the core ISG process. Within “direct” the research has identified 6 key second-order themes that represent the sub-processes within the “direct” process. These sub-process are mapped against various stakeholder groups in a cross-functional process map shown in Figure 5-7.

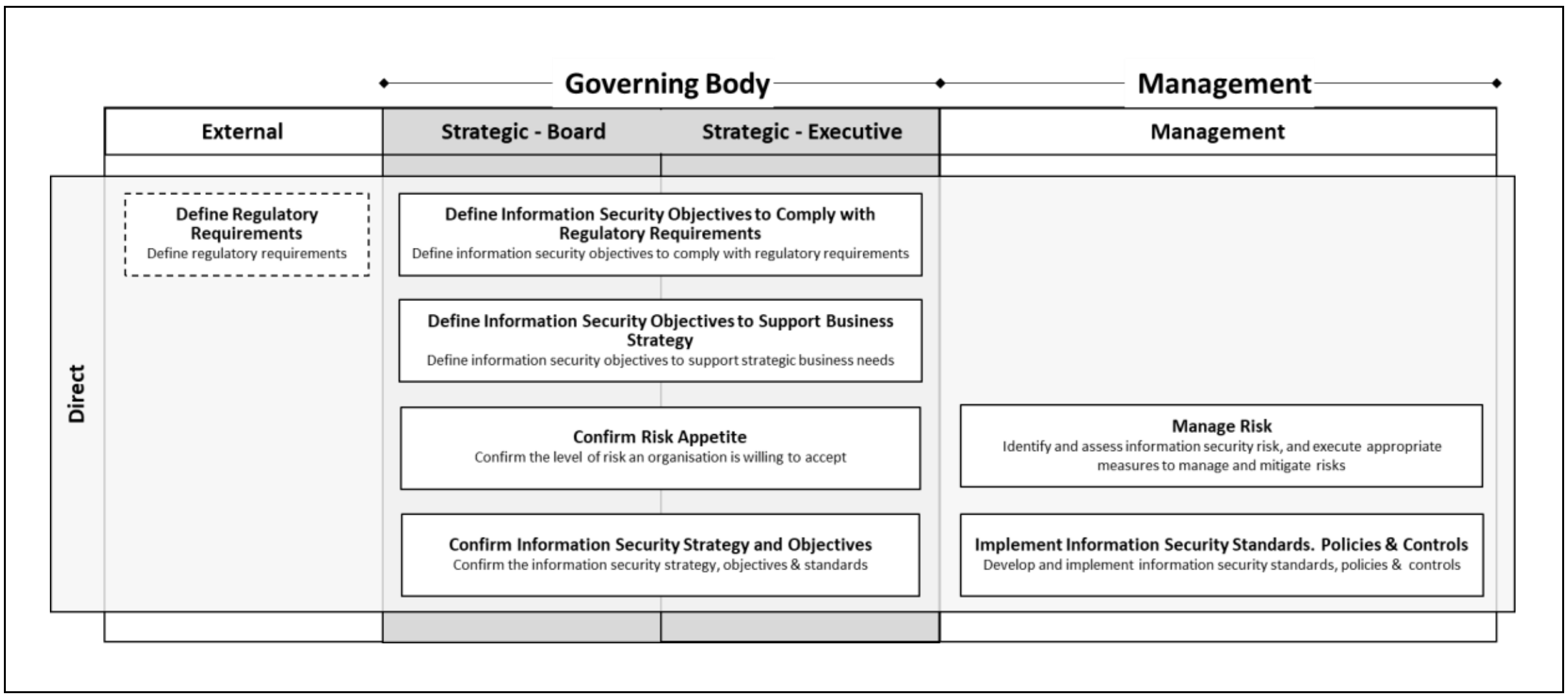

Figure 5-7: ISG second-order themes/sub-processes (direct)

The following subsection discusses the sub-process flows and interactions among these sub-processes.

#### 5.3.2.1.7 Sub-Process Flows and Interactions within Direct

This subsection identifies the sub-process flows and interactions among these sub-processes as discovered from the case study data. Figure 5-8 shows the sub-process flows and interactions among the sub-processes.

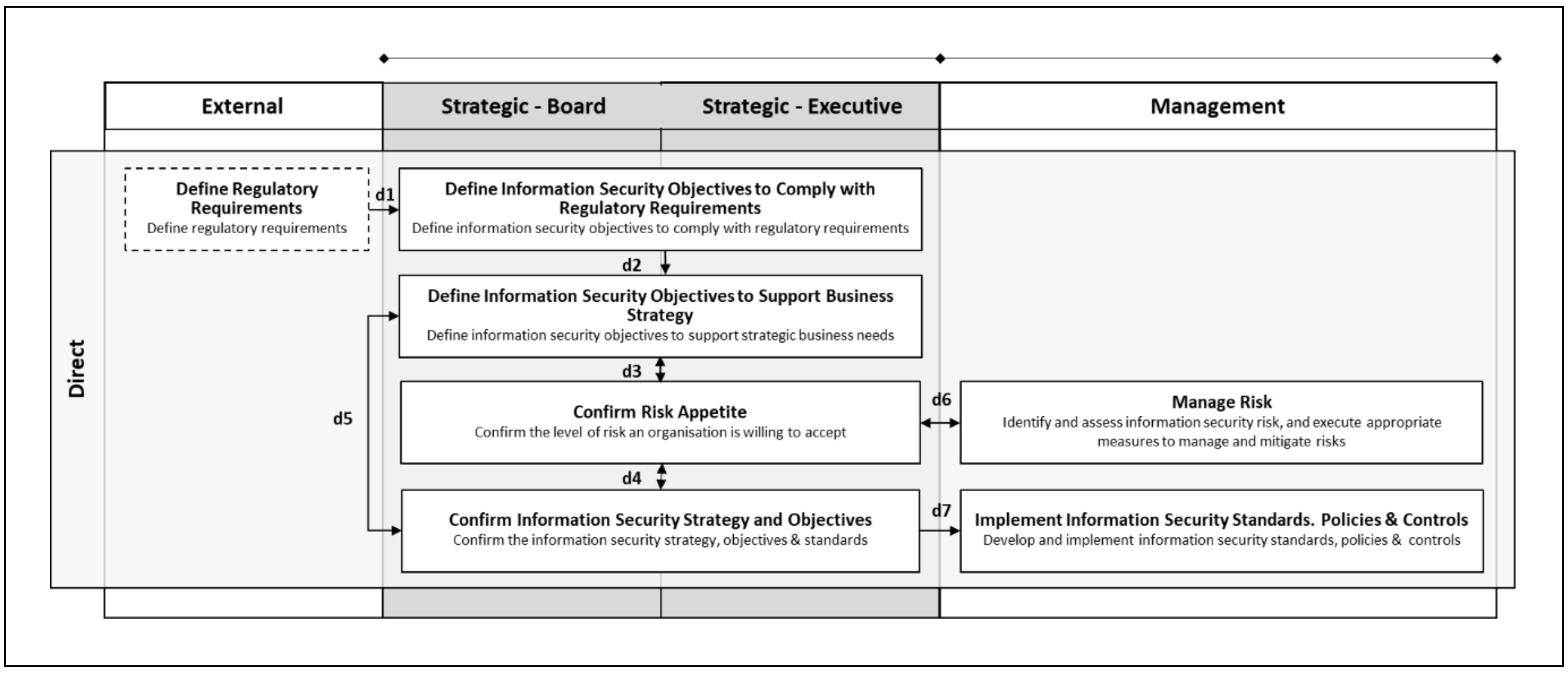

Figure 5-8: ISG sub-process flows and interactions (direct)

“Define regulatory requirements” has been identified as a sub-process that is outside of the organisation i.e. where regulatory requirements are defined and set by the regulators.

These regulatory requirements are important inputs into the "define information security objectives to comply with requirement" sub-process as evidenced from the case study data. The input is shown as a one-directional arrow (→), d1:

> we are in the financial industry that is so regulated by the monetary authority. (FinServices_SG_CIO)
>
> compliance to industry standards and new regulatory or future requirements. (FinServices_SG_CISO)
>
> In the recent monetary authority (MAS) technology risk audit, we assigned a team to work closely with MAS in supporting their audit. (FinServices_SG_Director-InfoSecOfficer)
>
> central bank issue some new regulations that warrants the implementation of some new software or systems. (FinServices_SEA_CFO)
>
>> Priority for banks is to ensure compliance to regulatory, that is not negotiable, and board members will approve as long as it is for compliance to regulators! (FinServices_MY_Board)

The output from "define information security objectives to comply with regulatory requirements" is used as input to "define information security objectives to support business strategy" as shown by a one-directional arrow (→), d2. Both d1 and d2 are one-directional arrows as the need to comply with regulatory requirements is non-negotiable:

> central bank issues some new regulations that warrants the implementation of some new software or systems, this will be allocated as out-of-cycle budgeting items. (FinServices_SEA_CFO)

> An easy way to determine the budget is based on the regulatory compliance requirements. Compliance is a non-negotiation matter in a bank. (FinServices_SEA_CIO)
>
> Priority for banks is to ensure compliance to regulatory, that is not negotiable, and board members will approve as long as it is for compliance to regulators! (FinServices_MY_Board)
>
> banks spends money in information security, especially in cybersecurity, because banks need to comply to central bank regulations and the fines that central bank imposes. (FinServices_MY_Board)

The next interaction is between the "define information security objectives to support business strategy" and "confirm risk appetite" sub-processes, which is shown as a bidirectional arrow ( ↔ ), d3. This interaction shows that the output from the information security objectives defined is validated with the risk appetite and this is an iterative process where discussions are held to finalise suitable information security objectives to both support the business strategy and meet the required risk appetite. This is evidenced in the following representative quotes:

> information security should start with the business strategy. Everything should be driven from the business strategy. Based on your strategy, you consider the risks and the investment should be determined to manage the risk. (FinServices_MY_Board)
>
> The risks are related to the strategic business objectives of the bank. Then, everything flows from here. (FinServices_SEA_Board)
>
> When we conduct the risk assessment, it is assessed against the business. In this way it is done with the business in mind and it is indirectly driven by strategy. (FinServices_SG_CISO)

d4 and d5 are bidirectional arrows ( ↔ ) to show the interactions and process flows between the sub-processes of “confirm risk appetite” and “define information security objectives to support business strategy” to “confirm information security strategy and objectives”. These show that the information security strategy and objectives are confirmed by consultation with outputs from the two preceding sub-processes:

> the board approves our information security risk profile when they [referring to the board] approve the overall organisation’s risk profile. Based on the risk profile, the CISO develops the security strategy and framework. These strategy and framework need to be presented and approved by the Board. (FinServices_SG_CIO)

> Based on the risk assessment and the willingness to accept the level of risks, we can define the required information security roadmap. (FinServices_MY_COO)

> always look at information as the asset of an organisation, then you classify the different type of information and consider the different level of protection to ensure the security of these information. (FinServices_MY_Board)

> we are able to plan out our investments and information security projects to align with our business strategy. (FinServices_MY_COO)

The final two interactions are d6 and d7, which are represented as bidirectional arrows ( ↔ ) to indicate the interactions between the sub-processes. d6 shows the interactions and sub-process flow between “confirm risk appetite” and “manage risk”. The bidirectional flow of inputs and outputs between these sub-processes shows that the output from “manage risk” which includes e.g. the results of risk assessment on the business operations are fed into “confirm risk appetite” for a decision on the risk level that an organisation is willing to accept. The output, i.e. the confirmed risk appetite will then be the input for the management to define the required actions to manage such risk:

> We then take this to the RMC and subsequently present to board for agreement and approval on the agreed level of risk acceptance. (FinServices_SG_CISO)

> The CRO team owns the risk assessment process and works with the CIO team on the information security risk assessment. (FinServices_SEA_IT-Architect)

> the bank conducts a detail risk assessment, understanding the various risk areas, its chances of happening and the impact of the loss … Based on the risk assessment and how much risk we are willing to accept, we define the actions to mitigate or manage the risks. (FinServices_MY_Board)

d7 shows the interactions and sub-process flows between “confirm information security strategy and objectives” and “implement information security standards, policies and controls” where the output e.g. information security strategy, objectives and standards are translated into detailed standards, policies and control for implementation:

> the CISO team develops the information security standards, policies and procedures. I as the CIO will be responsible to implement and comply to the standards from the information systems perspective. (FinServices_SG_CIO)

> The information security team helps in ensuring that the policies are implemented across the bank. (FinServices_SEA_IT-Architect)

> These will then be cascaded to the Exco, i.e. the executive committee or the senior management team who is responsible to make it happens on the ground. (FinServices_MY_COO)

#### *5.3.2.2 Monitor*

Analysis of data gathered from the case studies identified two second-order themes which were aggregated into the "monitor" dimension. These second-order themes are "measure and report performance" and "manage compliance", as shown in Figure 5-9.

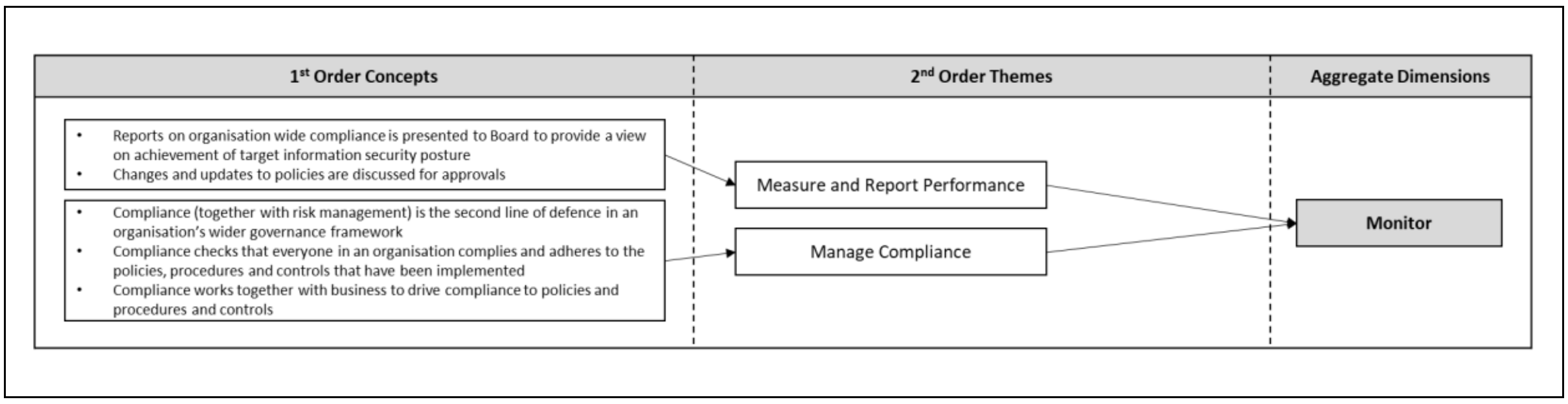

Figure 5-9: Data structure – ISG process (monitor)

For the purpose of ISG, interview data shows that board is responsible to ensure that information security objectives are achieved. Therefore, the board needs to be kept updated on the progress of the compliance. Performance measurements and compliance targets are set based on the "direct" process so that compliance reporting can be made against such measurements and targets:

> we present the ongoing monitoring of status, e.g. compliance check against the defined standards and policies. Reasons for noncompliance and remediation actions. Board is kept updated on the information security risk landscape. (FinServices_SG_CIO)

> We present to the board on our compliance status, a compliance status reporting by business, by areas, etc … Compliance is a non-negotiation matter in a bank. (FinServices_SEA_CIO)

The other second-order theme that emerged is "manage compliance". This refers to the process of monitoring the implementation of the standards, policies, procedures and controls, and assessing and ensuring that the implementation adheres to the intended objectives and standards defined by the board in "direct". "Manage compliance" is a sub-process that is executed through a dedicated compliance team within a compliance department in all case study organisations. The following interview excerpts highlight the importance of compliance management within the "monitor" dimension:

> On compliance, we have the compliance department who will conduct annual compliance. The Compliance team work closely with CISO and my information security team to conduct compliance checks. (FinServices_SG_CIO)
>
> In the bank, we have our compliance department that helps in ensuring the implementation … Based on the defined policies and standards, compliance department will conduct checks to ensure compliance are adhered. (FinServices_MY_Board)
>
> We have IT, risks and compliance involved in ensuring process, policies and procedures are implemented and adhered. (FinServices_MY_COO)
>
> we also have the compliance department that helps to make sure if everyone complies to the policies and procedures that have been defined. The Compliance team focuses on the monitoring to ensure that controls that have been recommended are actually implemented and adhered. (FinServices_MY_CRO)

"Manage compliance" is ongoing and is conducted regularly to ensure adherence to standards, policies, procedures and controls. Both FinServices_SEA_CFO and FinServices_MY_COO highlighted the need for regular compliance checks:

> We have an efficient compliance department that will conduct compliance check regularly. Our compliance team will check that every department comply to regulatory standards and policies (no excuses) and also to internal defined standards such as those for information/cybersecurity. (FinServices_SEA_CFO)

> In the bank, IT sets the information security standards, policies, and procedures. Compliance department will conduct their regular compliance audit on all the departments. This is done on a regular basis, i.e. annually. (FinServices_MY_COO)

In addition, documentation that was analysed reinforced the importance of compliance management as the case study organisations had standards and policies that adopted the requirements specified by regulatory guidelines. The following are excerpts from their policy documents on the role of compliance:

> Compliance is the responsibility of all officers within a financial institution … the compliance function is responsible for ensuring that controls to manage compliance risk are adequate and operating as intended. It is also responsible for assessing and monitoring of compliance risk faced by financial institutions. (Bank Negara Malaysia, 2015)

> Compliance processes should be implemented to verify that IT security standards and procedures are enforced. Follow-up processes should be implemented so that compliance deviations are addressed and remedied on a timely basis. (Monetary Authority of Singapore, 2013)

Further interview data that support the emergence of the "monitor" dimension and its associated second-order themes are found in Table 5-14 and Table 5-15.

Table 5-14: Data supporting "measure and report performance" (monitor).

| Dimension: **Monitor**; Theme: **Measure and Report Performance** | |
|---|---|
| **Case Study** | **Representative Quotes** |
| FinServices_SG | "Noncompliance will be reported and fixed. You can see the closed-loop management - risks agreed, standards and policies defined, compliance checked, remediation actions taken." (FinServices_SG_CIO) |
| FinServices_SEA | "The compliance department will present the compliance report to the CEO and RMC." (FinServices_SEA_CFO)<br>"It is held every two–three months and the focus is on reviewing critical audit findings or critical incidents. The board risks meeting will also cover any upcoming landscape items, potential issues and risks." (FinServices_SEA_CIO) |

| Dimension: **Monitor**; Theme: **Measure and Report Performance** | |
|---|---|
| **Case Study** | **Representative Quotes** |
| FinServices_MY | "Based on these compliance report, specific actions are taken to drive better compliance." (FinServices_MY_COO) |

Table 5-15: Data supporting "manage compliance" (monitor).

| Dimension: **Monitor**; Theme: **Manage Compliance** | |
|---|---|
| **Case Study** | **Representative Quotes** |
| FinServices_SG | "I as the CIO will be responsible to implement and comply to the standards from the information systems perspective, i.e. review and implement assess controls, systems protection be it networks, servers or applications." (FinServices_SG_CIO)<br>"The CISO team has a compliance team who will conduct a compliance check to ensure that the policies are adhered. We do this annually … compliance under CISO purview is to review and assist the business in ensuring compliance." (FinServices_SG_CISO)<br>"Oversight should include monitoring, compliance checks to ensure that we are actually doing what we are set out to do according to our policies and procedures." (FinServices_SG_Directo-InfoSecOfficer)<br>"The compliance team will be responsible to ensure that everyone complies to the standards and policies … Compliance will then come to check on the implementation against these standards, policies and procedures." (FinServices_SG_DeputyCIO) |
| FinServices_SEA | "Our compliance team will go around the bank to check against the policies and determine if they [referring to the business departments] comply." (FinServices_SEA_Board)<br>"The Risk and Prevention was more focused in terms of making sure that our overall risk is good and managed, and cybersecurity was nothing more than making sure we have firewalls and tight access controls!" (FinServices_SEA_CIO)<br>"The information security team helps in ensuring that the policies are implemented across the bank, everyone businesses and supporting functions, and we have the legal and compliance team that goes around making sure the bank complies." (FinServices_SEA_IT-Architect) |
| FinServices_MY | "we need to make sure we have all the required processes as required to ensure compliance." (FinServices_MY_Board)<br>"We have IT, risks and compliance involved in ensuring process, policies and procedures are implemented and adhered." (FinServices_MY_COO)<br>"While we must be compliant as we are regulated by BNM, it is important to have a strong risk and compliance to ensure a sound and prudent operations of a bank." (FinServices_MY_CRO) |

As discussed in this section, there are interactions between "measure and report performance" and "manage compliance". These interactions and process flows between the two second-order themes (m1) are shown with a bidirectional arrow (⟷) in Figure 5-10. The output from "manage compliance" e.g. compliance results against policies and standards are fed into "measure and report performance" where assessment is made and performance is measured to determine adherence and actions. The performance report is also fed back into the "manage compliance" sub-process:

> The compliance department will present the compliance report to the CEO and RMC. (FinServices_SEA_CFO)
>
> Based on these compliance report, specific actions are taken to drive better compliance. (FinServices_MY_COO)

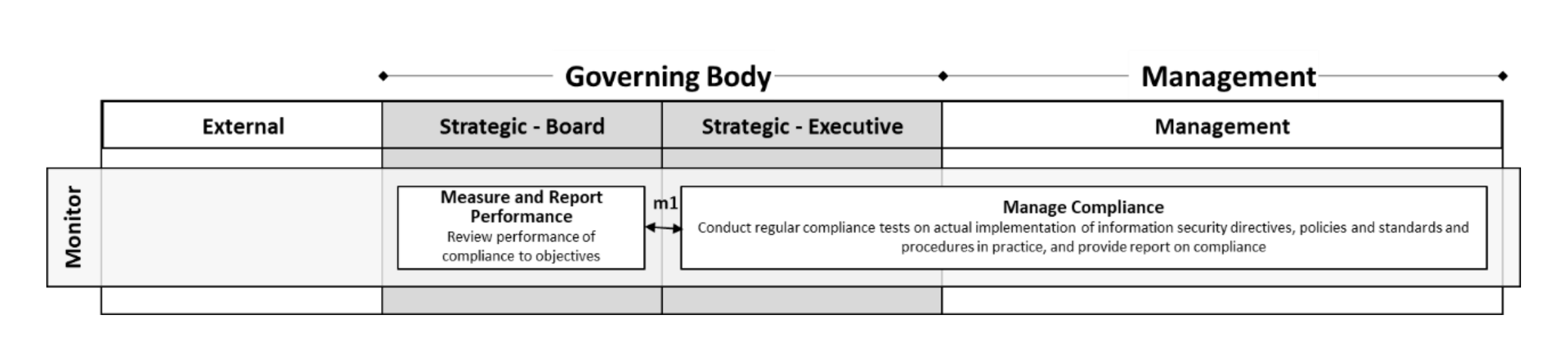

Figure 5-10: ISG second-order themes/sub-process flows and interactions (monitor)

"Monitor" comprises of two sub-processes i.e. "measure and report performance" and "manage compliance", which focuses on the continuous evaluation and assessment of information security strategy, objectives, policies, procedures and controls that have been implemented as per "direct" to understand the degree of compliance.

### *5.3.2.3 Evaluate*

In this research, the analysis of the case study data has identified two second-order themes, i.e. "evaluate and refine" and "collect and analyse" that are aggregated into "evaluate" as shown in Figure 5-11. Interview data also support the existence of a feedback-loop model where information security compliance data are collected, analysed and used to make the necessary adjustment which will be used as the input for the "direct" process.

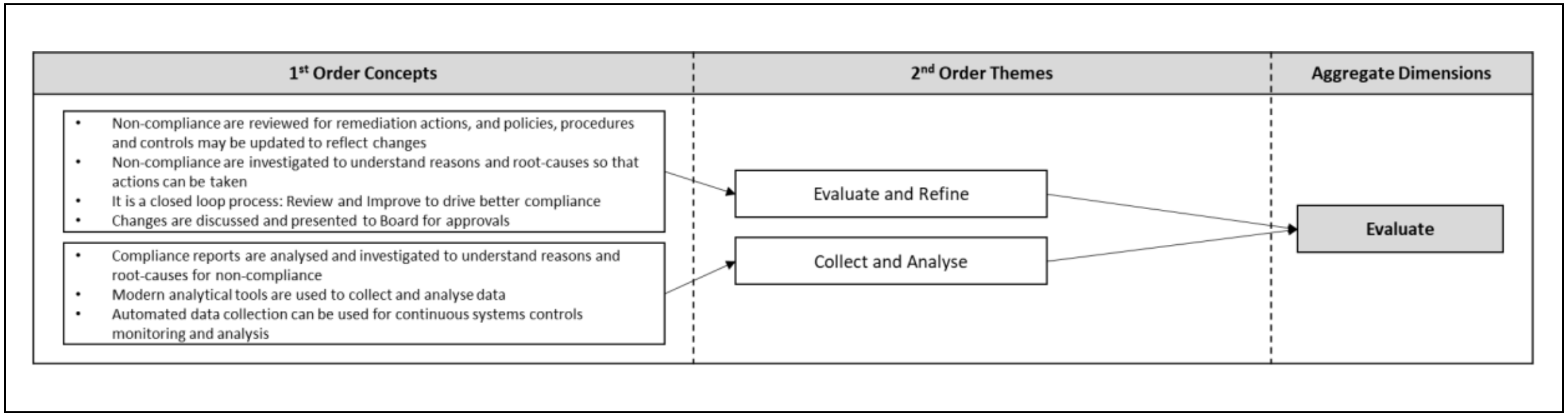

Figure 5-11: Data structure – ISG process (evaluate)

The following interview excerpts helped discover the "evaluate and refine" second-order themes:

> You can see the closed-loop management - risks agreed, standards and policies defined, compliance checked, remediation actions taken. This is repeated annually. (FinServices_SG_CIO)

> will study the gap and work with IT security to refine or update and required changes. (FinServices_SEA_Board)

> Changes are made if necessary based on the feedback of the compliance function. (FinServices_SEA_IT-Architect)

> We monitor the compliance and recommend changes that need to be done or changes to policies and standards to reflect changing requirements. (FinServices_MY_Board)

The following interview excerpts show that compliance data are gathered and analysed as part of "collect and analyse" prior to "evaluate and refine". Data are analysed to understand noncompliance or gaps that may be attributable to lack of awareness, changes in situation or information security requirements:

> If we find that there are big gaps in compliance, we need to study and understand the reasons behind the gap. Is it due to discipline of just not complying, is the policies are too difficult to comply or maybe even that our policies may be outdated! Maybe it is just due to training. We investigate the gaps thoroughly. (FinServices_SEA_Board)

There is a continuous monitoring on compliance and the evaluation on the applicability to determine if changes are required. (FinServices_MY_COO)

We have to explain why it happened, why it was not prevented, what went wrong and how to ensure it do not happen again. We also conduct a mid-year review on the risk profile to check if the environment or business have changed and if it needs to be updated. (FinServices_SG_DeputyCIO)

Additional interview data gathered from the various interviewees as shown in Table 5-16 and Table 5-17 supported the discovery of these second-order themes and the aggregated “evaluate” dimension.

Table 5-16: Data supporting “evaluate and refine” (evaluate).

| Dimension: **Evaluate**; Theme: **Evaluate and Refine** | |
|---|---|
| **Case Study** | **Representative Quotes** |
| FinServices_SG | “Reasons for noncompliance and remediation actions.” (FinServices_SG_CIO)<br>“If there are gaps, we need to fix or maybe even fix or modify our policies to address some of the gaps.” (FinServices_SG_Director-InfoSecOfficer)<br>“We review our standards, policies and procedures annually, will update them if required.” (FinServices_SG_Director-InfoSecOfficer) |
| FinServices_SEA | “If there are challenges, we will work through the challenges and decide if the noncompliance requires a change in standards or policies.” (FinServices_SEA_CFO) |
| FinServices_MY | “Priority for banks is to ensure compliance to regulatory, that is not negotiable, and board members will approve as long as it is for compliance to regulators!” (FinServices_MY_Board)<br>“we design the security approach, look at the investments required and evaluate against our risk appetite and make sure we comply with all the central bank guidelines.” (FinServices_MY_Board)<br>“Based on these compliance report, specific actions are taken to drive better compliance.” (FinServices_MY_COO) |

Table 5-17: Data supporting "collect and analyse" (evaluate).

| Dimension: **Evaluate**; Theme: **Collect and Analyse** | |
|---|---|
| **Case Study** | **Representative Quotes** |
| FinServices_SG | "They [referring to the compliance team] also help to highlight key gaps and noncompliance so that we can make a call on the actions, and these actions can sometimes be outdated standards and policies that we have to update. Therefore, compliance becomes a monitoring and feedback process for us." (FinServices_SG_DeputyCIO)<br>"Nowadays, we have a lot of smart tools that does the detailed work, e.g. automatic review of logs, etc." (FinServices_SG_Deputy-InfoSecOfficer) |
| FinServices_SEA | "They [referring to the compliance team] are the friendly parties that look at the noncompliance and work with us to take action on compliance." (FinServices_SEA_CFO) |

In addition to the identification of the two sub-processes of "evaluate and refine" and "collect and analyse", the interactions between these sub-processes (e1) are shown with a one-directional arrow (→) to indicate the interactions and process flows between them. "Collect and analyse" will collect and analyse information from compliance to determine gaps and reasons for the gaps as output to be presented to "evaluate and refine" where the information will be evaluated to determine changes/refinements that may be required to meet changes in information security objectives:

> Based on these compliance report, specific actions are taken to drive better compliance. (FinServices_MY_COO)
>
> There is a continuous monitoring on compliance and the evaluation on the applicability to determine if changes are required. (FinServices_MY_COO)

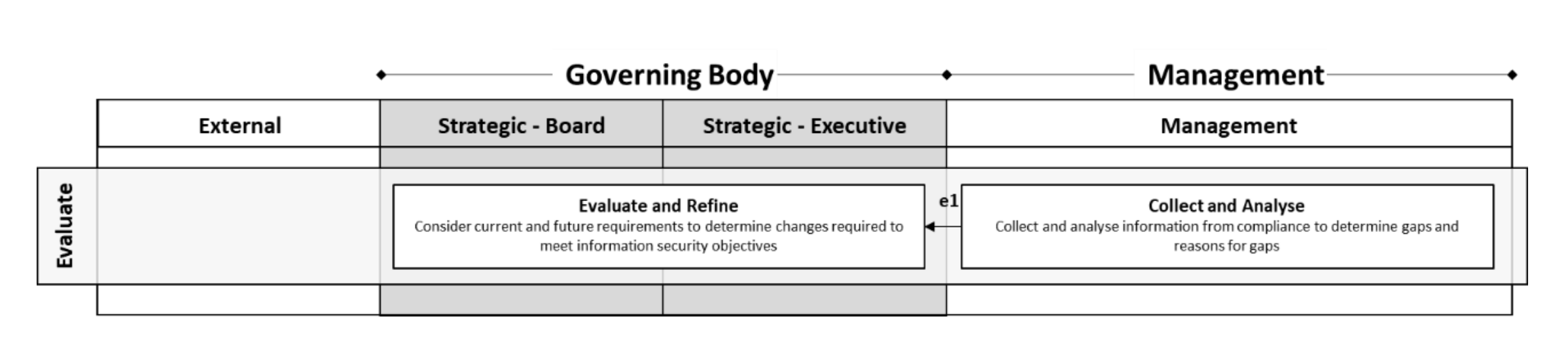

Figure 5-12: ISG second-order themes/sub-process flows and interactions (evaluate)

"Evaluate" is part of the direct-control-evaluate feedback-loop model. "Evaluate" refers to actions involved in the collection of data, analysis, evaluation and comparison to determine changes and adjustments required to meet current and future information security objectives:

> You can see the closed-loop management - risks agreed, standards and policies defined, compliance checked, remediation actions taken. This is repeated annually. (FinServices_SG_CIO)

#### 5.3.2.4 Communicate

"Communicate" is another aggregate dimension discovered together with "engage stakeholders" as a second-order theme in the case study analysis, as shown in the data structure in Figure 5-13.

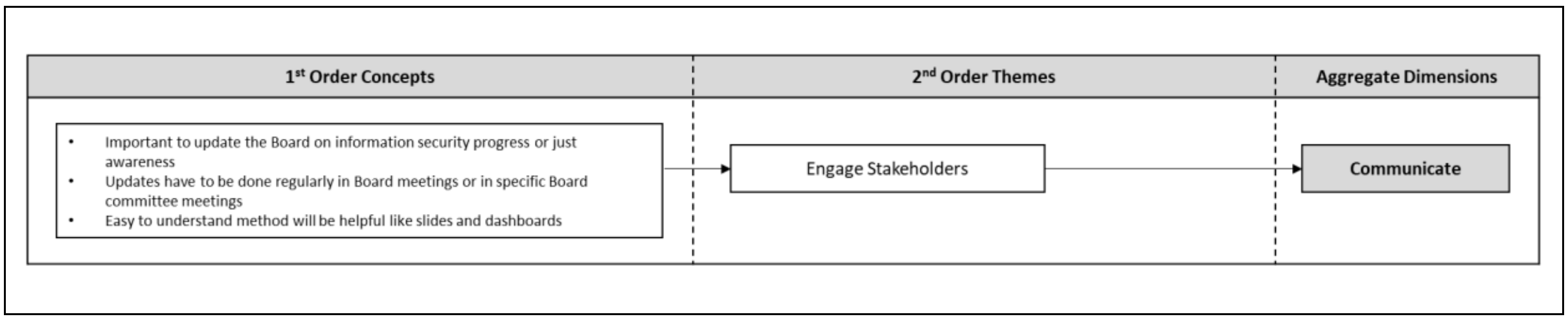

Figure 5-13: Data structure - ISG process (communicate)

All data emphasise the importance of engaging with the board as the board plays an important role in ISG. The engagement with the board covers organisation-specific information security topics and general topics in the industry as it is important that the board is kept aware of information security topics. FinServices_SG_CIO and FinServices_MY_COO confirmed that these are being done in their organisations:

> Board is kept updated on the information security risk landscape. (FinServices_SG_CIO)

> we have a reasonably good communication process to engage our board members with respect to information security. The board is kept aware. (FinServices_MY_COO)

FinServices_SG_DeputyCIO and FinServices_SEA_Board also emphasised the importance in engaging other stakeholders such as the regulators, shareholders and customers.

> It is important to have a process for communications, communicating to the board, communicating to the regulators, other stakeholders, to media, customers, etc. During an incident, communication is very important in managing the expectations. (FinServices_SG_DeputyCIO)

> It is also important to communicate outside of the bank, to manage expectations especially when an incident happens. We need to communicate to our shareholders, our customers, regulators, etc. I believe we have a proper communication process handled by our public relations. (FinServices_SEA_Board)

Table 5-18 provides further data that support the emergent “stakeholder engagement” theme and “communicate” dimension.

Table 5-18: Data supporting “communicate”.

| Dimension: **Communicate**; Theme: **Engage Stakeholder** | |
|---|---|
| **Case Study** | **Representative Quotes** |
| FinServices_SG | “Board is kept updated on the information security risk landscape. We also update the board on near misses of threats, actual incident, the incident response processes on whether we met our incident response KPIs, etc.” (FinServices_SG_CIO)<br>“The risk assessment will be reviewed every six months and updated to the board.” (FinServices_SG_CISO)<br>“We have separate information security updates to the board every other month, where we provide a summary of our RMC update to the board. Should there be any important update, e.g. a security breach or incident, that will be done as a special update during any of the monthly board meeting.” (FinServices_SG_CISO)<br>“we have a communication process in the event of an incident. Our communication process documents the steps required to design the message and get the approvals, and the actual communications to different audience via different means.” (FinServices_SG_CISO)<br>“we update the board on all areas of risks, covering financial, operations and information security. In the information security update, we present a snapshot of the situation, i.e. any threats detected, near misses and any actual incident that occurred. We update our information security compliance status of the |

| Dimension: **Communicate**; Theme: **Engage Stakeholder** | |
|---|---|
| **Case Study** | **Representative Quotes** |
| | different businesses and status of our information security projects, e.g. hardening of servers, development of our cybersecurity framework, etc. Therefore, the communication to the board is very important." (FinServices_SG_Director-InfoSecOfficer) |
| FinServices_SEA | "In every board meeting, we allocate a section on information security where the head of the IT board committee will update the board on cybersecurity incidents that happened in the industry locally (if any), update on any cybersecurity incidents or threats in the banks (if any), what has been done and an update on cybersecurity initiatives that the bank is embarking and its progress. This is a good start. We started this last year, when the board have more awareness about cybersecurity." (FinServices_SEA_Board)<br>"The board members are more cybersecurity savy as they read widely and one of the board member is in the IT industry. So the board engages in some effective dialogues and ask some good questions." (FinServices_SEA_CFO)<br>"all these monthly meetings and all these are internal IT and operations. Every month, there is also a monthly risk meeting." (FinServices_SEA_CIO)<br>"The CRO and the CIO have to present to the RMC, providing regular updates on the information security risks situation of the bank, highlighting the key information security projects and security incidents that may have happened. I will update my CIO, the progress of the information security architecture … The RMC will in turn updates the board during the board meetings." (FinServices_SEA_IT-Architect) |
| FinServices_MY | "quarterly update to board. Both the CIO and CRO will update the board on information security activities or projects that are undertaken and any breaches that have been detected or had happened. So, we have a reasonably good communication process to engage our board members with respect to information security. The board is kept aware." (FinServices_MY_COO)<br>"The board needs to work closely with the management and we need to be updated regularly. Keeping us updated on a regular basis is important, either through special board communique or during board meetings." (FinServices_MY_Board)<br>"And will report the monthly potential cyber attacks, the near misses and real incidents." (FinServices_MY_CRO) |

"Communicate" is a core ISG process that comprises "engage stakeholders" as the sub-process that demonstrates accountability and transparency through reporting and communication regarding the information security program undertaken to protect the organisation and respond to security incidents. Figure 5-14 shows the "engage stakeholders" sub-process mapped against the cross-functional process diagram.

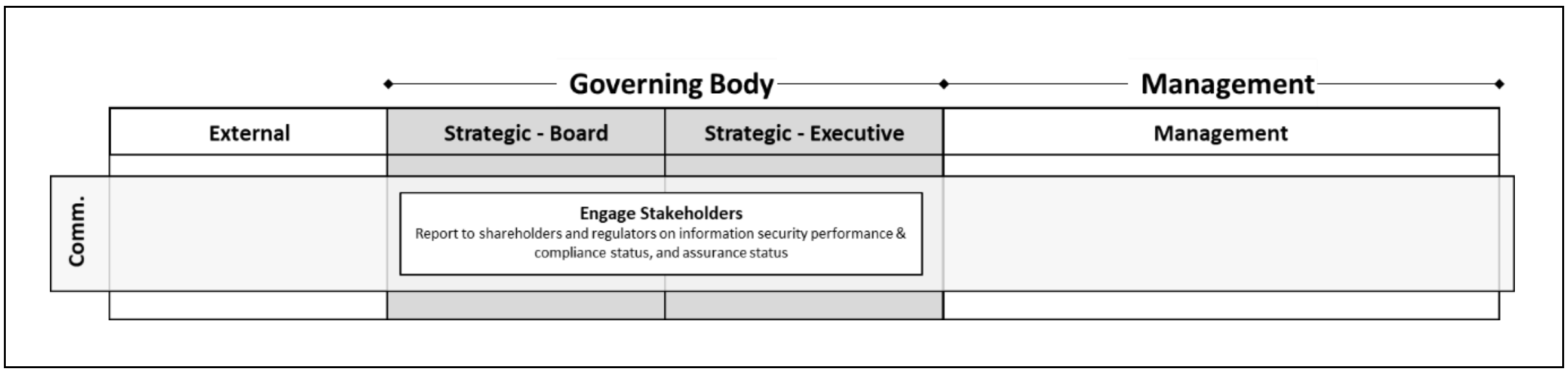

Figure 5-14: ISG second-order theme/sub-process (communicate)

#### *5.3.2.5 Assure*

"Assure" is the final component in ISG that has been identified from case study analysis. The analysis of the case study data identified 3 second-order themes, i.e. "conduct external audits and certifications", "provide oversight" and "conduct internal audit" as shown in the data structure in Figure 5-15.

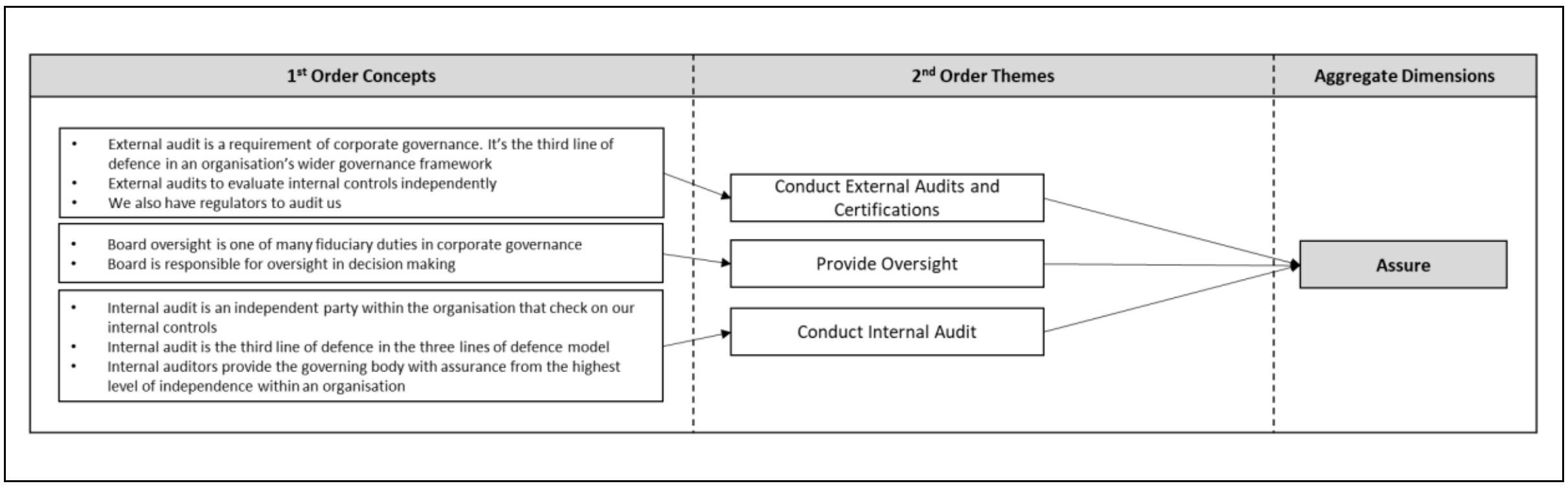

Figure 5-15: Data structure – ISG process (assure)

The following subsections provide analysis of the case study data in identifying the 3 sub-processes (i.e. second-order themes) in "assure".

##### 5.3.2.5.1 Conduct External Audits and Certifications

The "conduct external audits and certifications" theme refers to the independent assessment that is conducted by external parties, i.e. regulators, external auditors and consultants, that provides an independent assessment of the information security status of an organisation. While external audits are a statutory requirement that focus on financial information, these external audits have extended their scope to cover the auditing of the internal controls of the technology environment and controls in information security. Similarly, financial regulators have also expanded their audits to cover information security audits, as highlighted by FinServices_SEA_Board and shared by FinServices_SG_DeputyCIO and FinServices_SG_Director-InfoSecOfficer:

> Normally they [referring to the central bank regulator] will focus on financial audits, but with the internet banking and digital here and digital there, central bank comes to conduct information security audit and they are very detailed. (FinServices_SEA_Board)
>
> Similarly, we have the external audit done by our external auditors and also the regulator audit. (FinServices_SG_DeputyCIO)
>
> In the recent regulator technology risk audit, we assigned a team to work closely with regulator in supporting their audit. It went on for nearly two months as it was such a comprehensive audit. (FinServices_SG_Director-InfoSecOfficer)

In addition to statutory requirements, financial institutions have begun to seek the expertise of independent consultants to provide assessment and reviews of the information security posture of the organisations. This approach was shared by all the case study organisations, as shown in the following interview excerpts:

> we have an external assurance function where we engage external consultants to conduct a review annually just to ensure that we are in compliance with our internal policies and procedures. In addition, the external assurance ensures we are kept aligned with industry best practices and standards. (FinServices_SG_CISO)
>
> IT would like to hire consultants to conduct information security review. (FinServices_SEA_CFO)

> Board started to engage more and more consultants to review the general security posture of the bank. (FinServices_MY_COO)

Figure 5-15 shows the progressive data structure and Table 5-19 provides additional supporting data for this theme.

Table 5-19: Data supporting "conduct external audits and certifications" (assure).

| Dimension: **Assure**; Theme: **Conduct External Audits and Certifications** | |
|---|---|
| **Case Study** | **Representative Quotes** |
| FinServices_SEA | "there is also regulators doing assurance, our central bank. They will conduct scheduled audit, maybe every two years, but they can also conduct a surprise audit!" (FinServices_SEA_Board)<br>"all banks have excellent assurance processes as we always need a third party to provide assurance." (FinServices_SEA_CIO)<br>"we have our internal audit and external auditors and sometimes we get the BNM auditors. So we have ample assurance on information security risks." (FinServices_SEA_IT-Architect) |
| FinServices_MY | "In addition, you also have the external auditors that conduct the independent checks." (FinServices_MY_Board)<br>"Finally, for banks, assurance is important. We hire consultants to undertake our independent checks to ensure that we are doing the right thing. We also hire consultants to help us put in place proper processes and learn from best practices adopted by banks globally and regionally." (FinServices_MY_Board)<br>"The central bank mandated that all major enhancements and new internet-based systems to go through an external and independent review and attestation by a board member responsible for security before banks can launch such systems." (FinServices_MY_COO) |

##### 5.3.2.5.2 Provide Oversight

"Provide oversight" is another emergent theme discovered from case study analysis which is performed by the board in relation to ISG. The board receives information from both "conduct external audits and certifications" and "conduct internal audit" in discharging its oversight function. The oversight role of the board as part of assurance is reinforced in the following interview data gathered from the case studies:

> The role of governance is to ensure that there is a proper structure, there is segregation of duties, the need of oversight, and a check and balance to

> ensure that the business is doing the right thing for the shareholders. (FinServices_SG_CISO)

> Governance is important to be the third eye to provide an oversight, making sure information security initiatives are done right. (FinServices_SG_Director-InfoSecOfficer)

> We have a very onerous responsibility to provide oversight to ensure that the bank runs properly. (FinServices_SEA_Board)

> Board delegates the independent oversight over to the board committees to work with the executive management. However, the ultimate responsibility and the final decision rest with the board. (FinServices_MY_Board)

Further supporting data was gathered from documentation such as the following:

> The Audit Committee is established to provide independent oversight over internal and external audit functions, internal controls and ensuring checks and balances within the bank … Oversee management's implementation of the company's governance framework and internal control framework/policies. (FinServices_SEA Board Charter)

> reviewing the adequacy, effectiveness, independence, scope and results of the external audit and the company's internal audit function. (Monetary Authority of Singapore, 2018)

Figure 5-15 shows the progressive data structure and Table 5-20 provides additional supporting data for this theme.

Table 5-20: Data supporting "provide oversight" (assure).

| Dimension: **Assure**; Theme: **Provide Oversight** | |
|---|---|
| **Case Study** | **Representative Quotes** |
| FinServices_SG | "In my opinion, information security governance is the oversight that is ... By oversight, I mean the responsibilities to ask the right questions, check and get updates on the information security situation and to guide the organisation in the right direction." (FinServices_SG_CIO)<br>"The board is effective in providing the oversight as they ask the right questions and get involved in the right areas." (FinServices_SG_Director-InfoSecOfficer) |
| FinServices_SEA | "Most of the time, board members are selected from experienced industry practitioners … to the board to provide the oversight and contribute their experience and wisdom." (FinServices_SEA_Board)<br>"It is held every two–three months and the focus is on reviewing critical audit findings or critical incidents. The board risks meeting will also cover any upcoming landscape items, potential issues and risks." (FinServices_SEA_CIO) |
| FinServices_MY | "Some board members have specific experience and knowledge and they will ask the right questions on cybersecurity, how we derive the risk, have we have taken enough actions to manage and mitigate the risk." (FinServices_MY_CRO) |

#### 5.3.2.5.3 Conduct Internal Audit

"Conduct internal audit" is the third emergent theme discovered from the case study analysis, as shown in the data structure in Figure 5-15. This theme refers to the review and check of the adherence and compliance to defined standards, policies, procedures and controls, and is different from "manage compliance" as "conduct internal audit" acts as an independent party within an organisation. The board delegates the independent oversight role and responsibility to the internal audit committee, who works with the executive management in executing the internal audit functions. As highlighted by the interviewees, "conduct internal audit" is an important sub-process within "assure" in providing the overall assurance function in ISG:

> We also have the Internal Audit Committee where they will review internal audit findings. Information security audit is part of the company audit function and the result of an information security audit is also reported to the Internal Audit Committee. (FinServices_SG_CISO)

> we have the police that comes after us! That's the internal audit. The internal audit is a little different, they are like the police as they will come to check on you and send you a summon! Internal audit will check on us and will report our noncompliance too. (FinServices_SG_DeputyCIO)

Compliance is different from internal audit. Internal audit is the real police that will report noncompliance and we will have to explain and be penalised. It affects our KPIs. In a way, compliance provides the check and balance, providing the monitoring. Internal audit is strictly policing! (FinServices_SEA_CFO)

In the bank we have our internal audit and compliance team that will conduct regular audit and ensure that our policies and procedures are complied. (FinServices_SEA_Board)

Further supporting data for this theme is found in Table 5-21.

Table 5-21: Data supporting “conduct internal audit” (assure).

| Dimension: **Assure**; Theme: **Conduct Internal Audit** | |
|---|---|
| **Case Study** | **Representative Quotes** |
| FinServices_SEA | “Internal audit will conduct their audits to check on our compliance.” (FinServices_SEA_CIO)<br>“we have our internal audit and external auditors, and sometimes we get the BNM auditors. So we have ample assurance on information security risks.” (FinServices_SEA_IT-Architect) |
| FinServices_MY | “Internal audit will be the police that will check and make a report on any audit findings.” (FinServices_MY_Board)<br>“we have audit, both internal and external, to provide assurance.” (FinServices_MY_COO) |

The sub-processes within “assure” i.e. “conduct external audits and certifications”, “provide oversight” and “conduct internal audit” are mapped on the cross-functional process map shown in Figure 5-16.

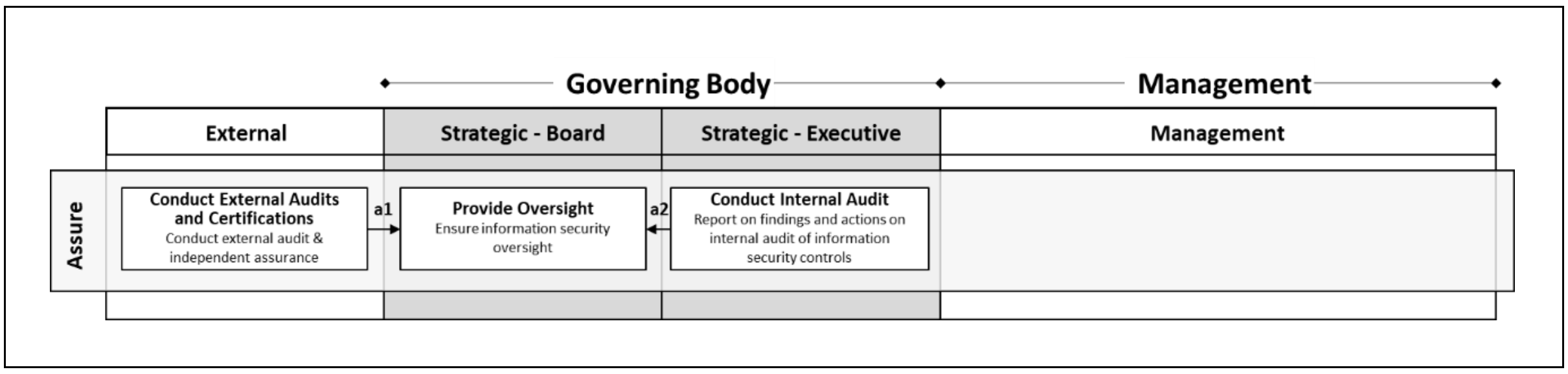

Figure 5-16: ISG second-order themes/sub-process flows and interactions (assure)

In order for the "provide oversight" sub-process to function, the sub-process receives information such as external audit reports/findings from external auditors or independent consultants. This interaction/sub-process flow from "conduct external audits and certifications" to "provide oversight" is represented with a one-directional arrow (→), a1:

> you have the audits, who provide the independent checks and gives a report card to the audit committee, the board. Don't forget the report from regulators too. (FinServices_SEA_Board)

Similarly, a one-directional arrow (→), a2 shows the interaction/sub-process flow from "conduct internal audit" to "provide oversight" where internal audit findings are presented to the "governing body":

> We also have the Internal Audit Committee where they will review internal audit findings. Information security audit is part of the company audit function and the result of an information security audit is also reported to the Internal Audit Committee. (FinServices_SG_CISO)

> Internal audit will conduct their own audit checks and report to us. (FinServices_MY_Board)

The "assure" process comprises checks and validations by independent parties (e.g. reviews, audits and certifications) to enable the board to exercise its oversight responsibility in ensuring compliance with the desired level of information security.

## 5.4 Towards a Refined ISG Process Model

This section brings together all the findings from the case study analysis discussed in the previous sections into a single ISG process model, as shown in Figure 5-17. This ISG process model, which is grounded on empirical data, is then compared against the conceptual ISG process model (Figure 4-15) developed based on the analysis of extant literature in Chapter 4 to confirm the theoretical concepts and to identify the differences in the refined ISG process model. The differences are shaded in Figure 5-17.

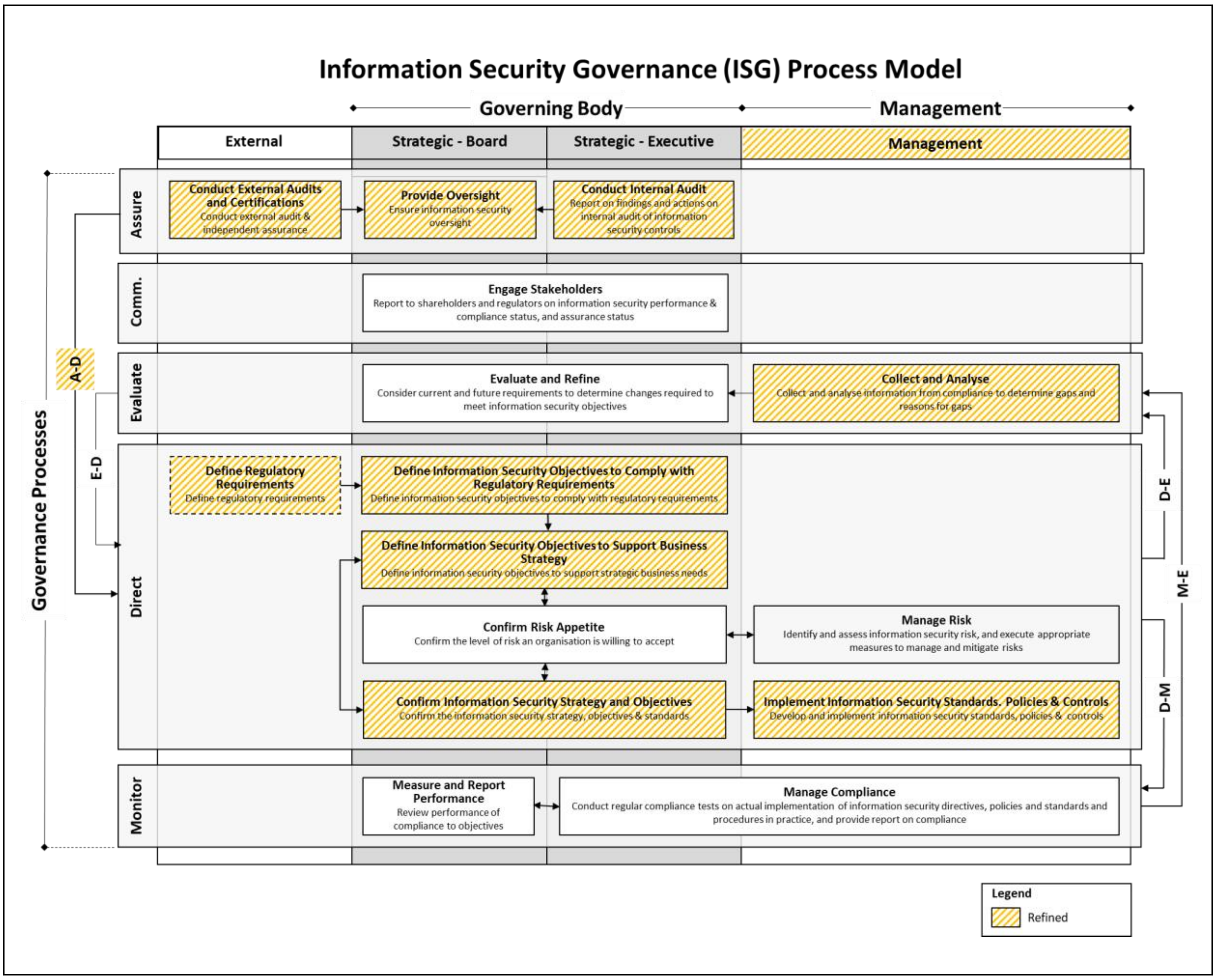

Figure 5-17: Development of refined ISG process model

### 5.4.1 ISG Stakeholder Groups

Similar to the conceptual ISG process model, the stakeholder groups are depicted on the top horizontal row of the refined ISG process model and comprise the following:

a. “External”
b. “Strategic - board” and “strategic - executive” collectively known as the “governing body”
c. “Management”

Data from the case studies led to the conclusion that there is a need for clear segregation of “governing body” and “management” but there were not enough data to confirm further segregation of “management” into “tactical management” and “operational management” as identified in the conceptual ISG process model. The reason behind the lack of data to justify this segregation could be the focus of data gathering, which was on the governance of information security, hence, governance structure and governance processes, and so very little was done to examine the structure required for the management and operations of information security. Although the conceptual model defined “management” to comprise “tactical management” and “operational management” as per the literature research, the refined ISG process model does not reflect the segregation of “tactical management” and “operational management”. The refined ISG process model shows only “management” which represents the management team which is responsible for implementing the policies and ensuring day-to-day operations.

### 5.4.2 Core ISG Processes

All 5 core ISG processes are shown on the left side of the refined ISG process model and comprise “direct”, “monitor”, “evaluate”, “communicate” and “assure”. These 5 core ISG processes are consistent with the core ISG processes identified in the conceptual ISG process model. “Direct-monitor-evaluate” represent a closed-loop process where objectives, standards, policies, procedures and controls are defined in “direct” and used as input to “monitor” to check for compliance and to “evaluate” for comparison to determine changes that are required for “direct”. The interaction between “direct” and “monitor” (D-

M) and the process flow from “direct” to “monitor” are shown with a one-directional arrow (→) connecting “direct” to “monitor”. As discussed in Subsection 5.3.2.2, these interactions and the information flow are confirmed by the interview data, as shared by FinServices_SG_CIO and FinServices_SEA_Board:

> we present the ongoing monitoring of status, e.g. compliance check against the defined standards and policies. Reasons for noncompliance and remediation actions. Board is kept updated on the information security risk landscape. (FinServices_SG_CIO)

> In the bank, we have our compliance department that helps in ensuring the implementation … Based on the defined policies and standards, compliance department will conduct checks to ensure compliance are adhered. (FinServices_MY_Board)

In a similar way, the information from “monitor” is used by “evaluate” to check against the objectives defined in “direct”. The interaction and process flow between “monitor” and “evaluate” (M-E) is shown with a one-directional arrow (→) connecting “monitor” to “evaluate” and the interaction and process flow between “evaluate” and “direct” (E-D) i.e. the comparison is also shown with a one-directional arrow (→) connecting “evaluate” to “direct”. These interactions and process flows were discussed in Subsection 5.3.2.3 and the findings were shared by all 3 case study organisations in interviews, as shown below:

> You can see the closed-loop management - risks agreed, standards and policies defined, compliance checked, remediation actions taken. This is repeated annually. (FinServices_SG_CIO)

> Changes are made if necessary based on the feedback of the compliance function. (FinServices_SEA_IT-Architect)

> will study the gap and work with IT security to refine or update and required changes. (FinServices_SEA_Board)

These interactions and process flows are consistent with the conceptual ISG process model and confirmed through the case study data.

One additional interaction and process flow (A-D) emerged from the case study data, i.e. the interaction and process flow from “assure” to “direct”. It is found that the information from “assure” is important and will be used as input to “direct” so that the relevant changes are made and adopted in “direct”. This interaction and process flow is depicted with a one-directional arrow ( → ) connecting “assure” to “direct”:

> the external assurance ensures we are kept aligned with industry best practices and standards. Assurance helps to ensure that we are adopting the right approach and that our policies and procedures are aligned to best practices in the industry. (FinServices_SG_CISO)

> We are learning and we also hired our external auditor to help us review and improve our information security process. (FinServices_SEA_Board)

> Look at the good side, the results of the audits will help us fix what are not right and put into implementations. (FinServices_SG_DeputyCIO)

There are further differences from the conceptual ISG process model that have been identified from the analysis of empirical data from the case studies. These differences as highlighted in Figure 5-17 are discussed below.

a. “Assure” has been broken down into 3 sub-processes, i.e. “conduct external audits and certifications”, “provide oversight” and “conduct internal audit” as the interview data provided more detail for the identification of these themes. The focus of the case study on governance helped in the discovery of these second-order themes.
b. “Collect and analyse” has replaced “collect and compare” in “evaluate”. “Collect and analyse” was discovered as an emergent theme as there was a focus on the analysis of information gathered from compliance to identify compliance gaps and determine the reasons for the gaps in meeting the information security objectives.

c. In relation to the “direct” process, it is apparent that ISG is regulatory, business and risk driven. The differences in the sub-processes within “direct” that have been discovered from the case study data include the following:
   - “Define regulatory requirements” was added as a sub-process that is undertaken by “external”.
   - “Define information security objectives to comply with regulatory requirements” is a sub-process that was not identified in the conceptual ISG process. The importance of complying with regulatory requirements was not discovered from extant literature in the development of the conceptual ISG process model, but the second-order themes on regulatory requirements were discovered in all the case study interviews. This can be attributed to the fact that all case study organisations were selected from the financial services industry, which is heavily regulated.
   - “Define information security objectives to support business strategy” replaced “align information security objectives with business strategy” to better reflect the need for the business-driven and actual activities involved in defining the information security objectives.
   - “Confirm information security strategy and objectives” and “implement information security standards, policies and controls” are used to better reflect the roles and responsibilities of the “governing body” and “management” as there were insufficient data to justify the granularity proposed in the conceptual ISG process model which comprised of 3 distinct sub-processes.

d. In “monitor”, “measure and report performance” was discovered as the emergent theme as there is a focus on reporting rather than purely measuring compliance.

## 5.5 Summary

This qualitative research is grounded on empirical data that was gathered primarily from interviews, documentation and to a lesser extent observations through on-site process walk-throughs. This chapter has provided the background to the 3 case study financial institutions that have been selected for the study based on the fact that these organisations have adopted good corporate governance and ISG practices. The chapter has also presented the empirical findings from the data collection and analysis. First-order concepts were coded, second-order themes were identified and visual representations via data structures have been provided throughout the chapter to illustrate the progressive identification of second-order themes and aggregated dimensions. Detailed analysis of the case study interview data facilitated the identification of the emerging themes and aggregated dimensions, which were subsequently used to refine the conceptual ISG process model. The refined ISG process model contains the stakeholders, the 5 core ISG processes and the relationships and process flows among these processes and sub-processes.

The refined ISG process model as shown in Figure 5-18 is a process-driven model that identifies the core ISG processes and sub-processes, the process interactions and the process flows. The refined ISG process model also maps these processes and sub-processes against the stakeholder groups to show the responsibilities of the stakeholder groups to these sub-processes.

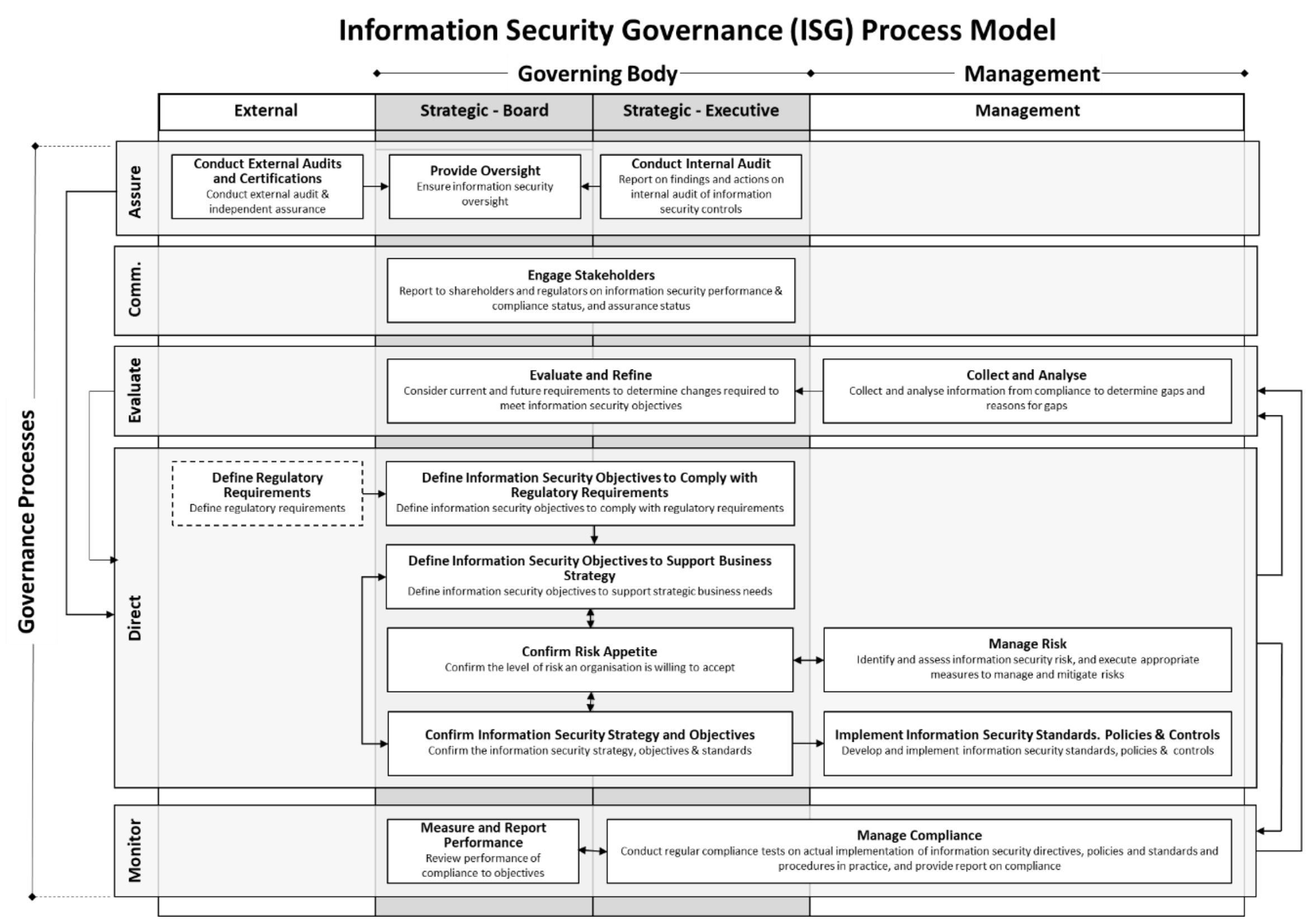

Figure 5-18: Refined ISG process model

The following chapter, Chapter 6, will present and discuss the findings and analysis of the expert interviews which validated this refined ISG process model.

# Chapter 6
# ISG Process Model Validation

Chapter 5 has described how the conceptual ISG process model was refined based on detailed analysis of empirical data from the case studies. This chapter presents the analysis and findings of Phase 3: Model validation (Figure 6-1) where expert interviews were used to further elicit the theory behind how ISG can be implemented in organisations and validate the ISG process model that was refined as described in the previous chapter.

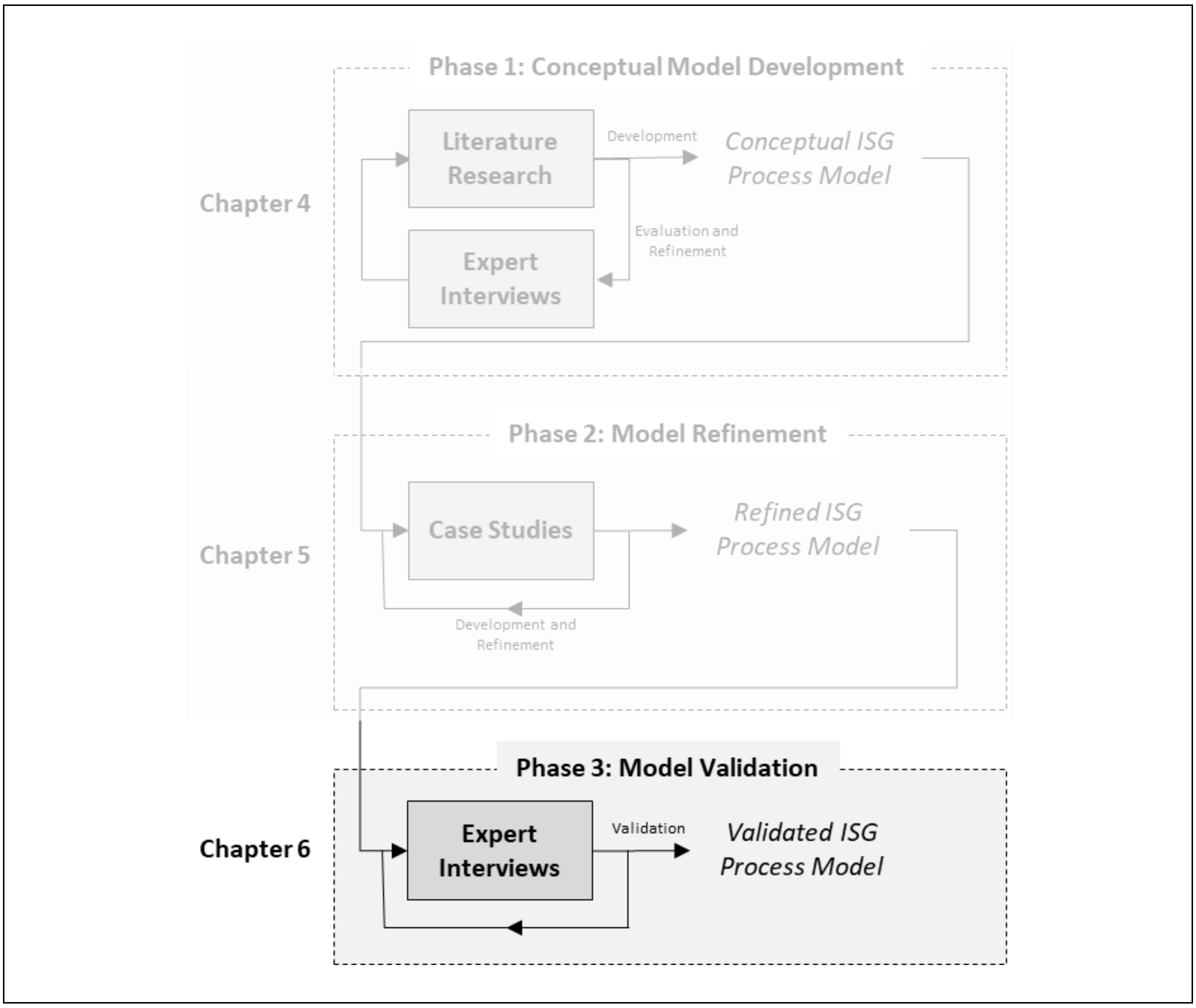

Figure 6-1: Phase 3: Model validation

This chapter is divided into 5 major sections. Section 6.1 provides a recap of the expert interview methodology adopted in Phase 3: Model validation, while Section 6.2 provides a profile of the experts, including a brief justification of the relevance of each interviewee's expertise and a summary of the interviews. Section 6.3 discusses the interview data and the analysis. Using these analyses, Sections 6.4 explains the validation of the ISG process model with expert interview quotes. A summary of the chapter is found in Section 6.5.

## 6.1 Expert Interviews

As discussed in Section 3.5, expert interviews have been adopted to validate the refined ISG process model. Bogner and Menz (2009) defined the expert interview as a method of data gathering via interview with an individual who possesses technical, process and interpretative knowledge in a specific field of action by virtue of the fact that the expert practises in that particular professional area. The purpose of expert interviews in Phase 3: Model validation was to gather further empirical data from these experts so that the research had more information relating to the practice of ISG that could be used to validate the refined ISG process model. The expert interviews were semi-structured interviews similar to the case study interviews with a list of open-ended questions that were prepared only as guiding topics to get the experts to discuss their insights freely and share their experience and activities in relation to ISG (Meuser & Nagel, 2009). All expert interviews were transcribed, coded and analysed. The approach used to analyse these expert interview transcripts was similar to the approach adopted for the analysis of the case study interviews that was discussed in detail in Section 3.8.

## 6.2 Profile of Experts and Summary of Interviews

Six experts were interviewed for Phase 3 between August 2018 and March 2019. The 6 experts, comprising 3 consultants and 3 industry professionals, were selected based on the specific criteria that were discussed in Section 3.5.1 and all experts were considered to be of equal status, i.e. all insights obtained from the experts are equally valuable without any bias towards a specific expert or group of experts.

a. Information security consultants

Three of the experts were experienced information security consultants who were working in the risk and information security area and had experience with a large number of companies during their careers. Consultants normally worked with several clients across different industries over time when conducting different projects. Their exposure to the different situations and challenges in their work over the years is seen as a particular advantage to the research as they were capable of identifying differences and similarities between companies and industries with different characteristics. In addition, these experienced consultants had expert knowledge and good theoretical background in their field of expertise (Schlegel, 2015). Two consultants from the Singapore offices of two global Big Four consulting firms were interviewed. These consultants were selected as they had advised customers on ISG and information security management, reviewed and developed ISG governance frameworks and conducted various information security audits. They were also Certified Information Systems Auditors (CISA) of ISACA. In addition, one consultant from the Singapore office of a global information security software company was interviewed as he had more than 25 years' experience in both financial and information risk consulting, having worked in several global risk consulting organisations and in an information security software company focusing on solutions to automate ISG, information security management and security operations processes.

b. Industry professionals

Three industry professionals were also interviewed as experts. These industry professionals were senior executives from the financial services industry who had extensive knowledge in information security and held positions directly involved in ISG and information security management. These 3 industry professionals were selected for the expert interviews as they had extensive experience in the topic of research, i.e. more than 25 years of experience and had worked across different countries. One of the experts had been involved in the information risk and security areas for over 35 years and worked in Europe and across the Asia-Pacific.

Table 6-1 shows a summary of the expert interviewees selected to validate the ISG process model and the following section provides profiles of the experts, a brief justification of the relevance of each interviewee's expertise and a summary of the interviews.

Table 6-1: Expert interviewees to validate ISG process model (extracted from Table 3-1).

| Interviewee | Organisation | Location & Role Coverage | Job Role | Involvement in Information Security | No. of Years in Industry |
|---|---|---|---|---|---|
| IS_ConsultingDirector | Consulting firm | Singapore & South-East Asia | Information Security Consultant | Direct | 20 years+ |
| IS_Consultant | Consulting firm | Singapore & South-East Asia | Information Security Consultant | Direct | 10 years+ |
| CISO-MY_Bank | Financial institution (bank) | Malaysia & South-East Asia | CISO | Direct | 25 years+ |
| CIO-SG_InvestmentCo | Financial institution (investment company) | Singapore | CIO | Direct | 25 years+ |
| ChiefInfoRiskOfficer-APAC_InsuranceCo | Financial institution (insurance company) | Malaysia & Asia-Pacific | Regional Chief Information Risk Officer | Direct | 35 years+ |
| IS_SoftwareConsultant | Information security software company | Singapore | Information Security Consultant | Direct | 25 years+ |
| | | | | **Total number of expert interviewees** | **6** |

### 6.2.1 IS_ConsultingDirector

The first expert interview was conducted with the information security consulting director (IS_ConsultingDirector) of a Big Four consulting firm based in Singapore. He was the information security consulting director for South-East Asia, responsible for leading a team of consultants in advising customers on ISG and information security management, conducting information security reviews, audits and forensic investigations, and assisting in the implementation of information security controls. He had more than 20 years of consulting experience and was a Certified Information Systems Auditor (CISA) of ISACA.

He started the interview by sharing with the researcher his most recent project where he led a review of a major incident that happened in a financial institution and brought down a nationwide infrastructure for a couple of hours. He highlighted that the project was at the request of the regulator and aimed to identify the root cause and improvements that should be adopted by the financial institution, and potential lessons learnt for the industry. He explained that the information security incident was not a malicious attack, but was due to an accidental error by a computer operator. He highlighted the importance of effective adherence to policies and controls, not just the implementation of policies and controls.

On ISG, he related it to the importance of ensuring correct and reliable information in corporate governance, and the duty of the board and senior executives of organisations:

> From information security perspective, it is not new any more as ensuring the CIA (confidentiality, integrity and availability) of information is part and parcel of good corporate governance. It is the duty of the board and senior executives to ensure correct and reliable information are maintained and disclosed as part of good corporate governance. While the focus is on financial information, it is not limited to financials, it relates to all information about the organisation to ensure external stakeholders can rely on. (IS_ConsultingDirector)

He continued to elaborate on the importance of ISG and management, and proceeded to explain that ISG is a set of structures and processes that drives oversight:

> Information security governance doesn't function in isolation. Information security governance, management and operations have very different functions and clarity among them is fundamental to the performance of each. We believe that besides structure, information security governance include the processes that ensure that reasonable and appropriate actions are taken to protect the organisation's information assets in the most effective and efficient manner in pursuit of its business goals. (IS_ConsultingDirector)

The interview continued for approximately 90 minutes with a detailed explanation of ISG, together with a framework from the view of his consulting firm. When the refined ISG process model was shared with him, he was very willing to provide his feedback. His insights were that the refined ISG process model has all the components that mirror the framework he shared although his model had different names and labels, and the proposed model has all the relevant ingredients of good ISG:

> You have it shown as a separation of governing body and management, which is good and provides a clear segregation of governance and management roles. It is good if you can incorporate the governance forums and committee. You have assure process, which we don't have. This is a good idea. (IS_ConsultingDirector)

The interview was very informative and the data was analysed as part of the analysis in this phase of the research. The insights from this interview served as good validation for this research.

### 6.2.2 IS_Consultant

The interview with IS_Consultant was the second expert interview. He was a consultant with another Big Four consulting firm based in Singapore and had extensive experience in working with clients to define information security frameworks, conduct technology security control reviews and develop control testing of applications. He had more than 10 years of consulting experience and was a Certified Information Systems Auditor (CISA) of ISACA. He had just taken up a new role in managing the security operations centre where the firm provided services in monitoring security threats for customers.

Leveraging his professional background in conducting information systems audits and controls reviews, he explained ISG from a control framework perspective, like how it was implemented in corporate governance. He explained the concept of the "three lines of defence" as proposed by the Internal Audit Association and its application in information security:

> We propose the three lines of defence concept, i.e. first line where the front line is responsible, then second line where we ensure that there is the check and balance, i.e. the risk and compliance functions, and finally the third line of defence where you have the assurance – the audits, both

> internal audit and external audit, and maybe even the regulatory audits. (IS_Consultant)

He continued to explain in detail how the 3 lines of defence were implemented to drive effective ISG, with examples of controls that were implemented in the front-end such as anti-virus software, password controls, etc., strong risk and compliance processes as the second line of defence, and the regulatory-driven internal and external audits in the third line of defence. He highlighted the importance of information security risk management in governance as he believed that the understanding of risk drives information security strategies and all its related actions and initiatives. He elaborated on the risk management processes and the role of the board and management in driving the risk management processes.

After the initial part of the interview and insights, the refined ISG process model was shared with IS_Consultant. When the researcher explained the various processes in the model, IS_Consultant was in strong agreement with the model that has been developed. He highlighted the fact that it is beneficial to segregate the generic compliance process into “monitor” and “evaluate” as this forces the activities of monitoring and evaluating changes to improve existing policies and controls. In his firm, this was treated as just standard management of compliance:

> Your model is more detailed as you separate “monitor” and “evaluate” as key processes when compared to our management of compliance process. (IS_Consultant)

In addition, he emphasised the importance of stakeholder engagement and was pleased that the researcher’s model incorporates the “communicate” process as one of the ISG processes:

> You have “communicate” as a process. We don’t, but we do have “communication and awareness” as a platform component that sits across all processes as we believe awareness is one of the most

> important. The customer's security posture is only as good as their awareness. (IS_Consultant)

The interview ended after a further discussion on the clear definition of roles and responsibilities of the governing body and management, and he had the following to say:

> I think the C-level executives are part of the governance process as they are delegated by the board to carry out the governance duties. I believe you reflected it well … the greyed part of the diagram. (IS_Consultant)

The second expert interview also provided insights that helped validate the refined ISG process model.

### 6.2.3 CISO-MY_Bank

The third expert interviewed was the CISO (CISO-MY_Bank) of a regional bank headquartered in Malaysia. CISO-MY_Bank was a seasoned IT professional who had assumed various roles in the bank over the last 20 years covering head of applications, head of data analytics and CIO for the consumer banking business before taking up the role as the CISO 5 years ago. In his role as the CISO, he was responsible for information security in the bank across South-East Asia covering Malaysia, Singapore, Indonesia, Thailand and Cambodia.

After formalities and a quick introduction to the research, CISO-MY_Bank had this to share when he was asked his understanding of ISG:

> In my view and my experience in the bank, information security governance is the oversight, the roles and responsibilities of the senior management and the board. The people who are responsible for decision making, setting the policies and monitoring the execution of the policies. That's my very high-level view. (CISO-MY_Bank)

CISO-MY_Bank also raised the issue that while responsibilities rested with the board, the challenge was that not all board members were technology or information security savvy.

Therefore, the bank had conducted many initiatives to raise awareness among the board members including scheduling special information and training sessions, and updating the board regularly on information security issues and trends in the financial services industry. CISO-MY_Bank believed that constant engagement of the board is important for effective ISG:

> the bank has established an Information Security Council. It is chaired by a board member and is driven by me ... In the Information Security Council, I have representation from CRO, CFO, CIO and the head of businesses (consumer banking, commercial banking, investment banking and operations) and internal audit. The role of this Council is to make decision on anything relating to information or cybersecurity. We have a monthly Council meeting and the Council will update the board every two months. Some of the areas we cover in the Council meeting include:
>
> - Update on security projects that is ongoing (costs against budgets)
> - Update on security situation, i.e. number of detected attacks, threats, etc.
> - Update on incidents (if any), actions required
> - Update on our assessment against BNM (central bank) requirements and compliance, e.g. update on our views of the new Technology Risk Management (TRM) guidelines to be introduced by BNM - challenges in adherence, what needs to be done, etc.
> - Get approval for new security projects (if any)
> - Get approval for new standards and policies. (CISO-MY_Bank)

CISO-MY_Bank also provided a detailed explanation of the information security risk management and compliance processes that had to comply with the regulator’s technology risk management guidelines. He emphasised that the information security risk management and compliance process was a subset of a very comprehensive bank-wide process.

After a detailed discussion on the processes adopted in the bank, the researcher shared the refined ISG process model with CISO-MY_Bank to solicit his comments. Some key comments that helped in validating the model include the following.

On the governance structure and defined roles and responsibilities:

> I like the organisation structure in a way that defines the governing body and management. We need to know who is driving governance and that is the board as I said earlier. You have included the senior management, which is correct. They work closely with the board. The board provides the oversight and approves the directions, and the senior management take the responsibilities to drive it through the organisation. (CISO-MY_Bank)

And on the core governance process:

> On the five key processes, yes, "direct" is definitely correct and is critical. I like the way you separate or break up "monitor" and "evaluate". It makes the processes a lot clearer. In practice we undertake both but is all part of one big process, i.e. the compliance and remediation part. We always treat it as one process, but by breaking into two discrete processes you can show the difference in focus. One is to check for compliance to defined policies and procedures, and the other focuses on the remediation or the changes required to reflect new requirements or changes in the organisations. I believe this also provide a feedback loop for the process. (CISO-MY_Bank)

The interview lasted 50 minutes. Many more details were discussed and shared in the interview. The interview was transcribed and analysed to identify the second-order themes and dimensions that were required to validate the refined ISG process model.

### 6.2.4 CIO-SG_InvestmentCo

The 4th expert interview was conducted with the CIO (CIO-SG_InvestmentCo) of a major investment company in Singapore. The CIO-SG_InvestmentCo was selected for the expert interview as he had nearly 30 years of experience, having worked for many years in the consulting industry, and had held many senior operational and technology roles in major organisations in South-East Asia. His experience in information security included the design of a payment security module that pioneered technologies that involved a two-factor authentication method during the dot-com era. His current role as the CIO included the responsibility for information security as the company's information security team reported directly to him.

CIO-SG_InvestmentCo's response to ISG was that it involves clear oversight by senior management, clear segregation of roles and responsibilities, and should be driven by business and risk. His precise comments on governance were as follows:

> Governance, like any governance, is similar be it corporate governance, IT governance and now information security governance. In my view, in governance it is important to have clear oversight by senior management, clear segregation of roles and responsibilities, and be driven by business and risk. In most cases, senior management, I mean the board, you get a board member to chair the board committee and be a champion for it … like board audit committee for corporate governance, board IT committee, board risk committee and maybe a board information security committee. (CIO-SG_InvestmentCo)

He shared that his company did not have a separate CISO but was in the process of hiring such a person as it was looking to implement a separate division focusing on information security. This was due to the growing importance and focus on information security, and at the request of the regulator.

He provided a detailed discussion of how ISG was implemented in his company. He also mentioned that the objective of ISG was to ensure that all information security initiatives were undertaken to help the company meet its business and risk objectives. He added regulatory objectives as his company was in the strictly regulated financial services industry.

He also brought up the importance of the fiduciary duties of the board and its increasing responsibilities in information security:

> The board committee focus on making sure the management is doing their job by making sure we have a regular, i.e. alternate month board committee updates. The committee helps to provide approvals for information security budgets and projects. In our regular updates, we provide the board committee with an update of our current information security situation - a health check on whether there are any breaches, any recorded near misses and an update of our information security projects. The updates are done by our head of information security. So the board committee aims to provide the oversight … an independent view, directions, comments and feedback. (CIO-SG_InvestmentCo)

On his feedback on the refined ISG process model that was shown to and discussed with him, he agreed with all 5 governance processes as this is consistent with what had been implemented in his company:

> Wow! This is interesting. You have defined five core processes in governance. Hmm … seems to be rather accurate. We adopt all the five processes, nothing more … Yes. Spot on. This is exactly what is practised. I like the way you illustrated them in a process model. The whole “direct”, “monitor” and “evaluate” provide a closed-loop process where you implement, check and improve. (CIO-SG_InvestmentCo)

CIO-SG_InvestmentCo also provided his feedback on the segregation of roles and responsibilities for all the governance processes in the ISG process model.

The interview ended after 50 minutes of discussion. Detailed information gathered during the interview was transcribed and analysed, and was used to validate the refined ISG process model.

### 6.2.5 ChiefInfoRiskOfficer-APAC_InsuranceCo

The 5th expert interview was conducted with the regional chief information risk officer (ChiefInfoRiskOfficer-APAC_InsuranceCo) who was responsible for information risk across the Asia-Pacific operations of a global insurance company. He was a very experienced individual who had over 35 years of experience in information security. He shared that he was involved from the early days of computer security to the current information security era covering technical, business and governance across many years working on technical security matters on IBM mainframes and now working on defining information security for start-ups that were working with the insurance company. He had spent 18 years working with the insurance company, starting with their head office in the UK and then across the Asia-Pacific.

In response to the researcher’s question on his understanding of ISG, he related his experience when he had been tasked to define the global corporate governance structure for the insurance company in 2004. While it had nothing to do with information security, the task involved the definition of the governance structure in risk management across the group in driving good governance practice. He learned the ways to drive a group-wide standard and the adaptation of such a global standard across various countries in the Asia-Pacific based on the different maturity levels of the countries. These lessons were used when he defined the insurance company’s group information security standards and policies. He highlighted the importance of change management, education and face-to-face meetings in driving acceptance among board members and executive management.

He emphasised the importance of clear roles and responsibilities in driving governance and how it was done in his company:

> The senior executives define the standards and the directions. For example, I will work with the team to define the standards and directions, understand what they want. From there, we will work out the detailed policies. Policies need to be simple for everyone to understand and implement to make sure staff don’t do the wrong thing … Each unit (business, technical or operations) have an owner and the owner is responsible for the execution. Training is conducted to ensure everyone is aware of the policies and is aware of the responsibilities. This is the execution or management part. (ChiefInfoRiskOfficer-APAC_InsuranceCo)

As part of governance, he reiterated the need to ensure policies are actually implemented via assessment and compliance checks. He also mentioned the need to have the right structure to drive governance:

> In addition to the processes and how information security governance and execution are implemented, we need to have the right structure to drive governance. We have an IT steering committee that help to drive the governance of information security governance. (ChiefInfoRiskOfficer-APAC_InsuranceCo)

Similar to other expert interviews, the researcher shared the refined ISG process model in the last part of the interview session to solicit feedback. His first reaction was to agree on the importance of the separation of the “governing body” and “management” structure as he believed that the structure is critical in driving clear roles and responsibilities:

> I like the structure in separating the roles of the governing body with the management who are responsible for the day-to-day operations and execution of the policies. It is important to understand the different roles and responsibilities … There is always a confusion and overlap between governance and management, and this will help identify the differences. (ChiefInfoRiskOfficer-APAC_InsuranceCo)

Further comments were solicited regarding the information governance processes that are depicted in the model. He agreed with all the process components and specifically highlighted the feature of “evaluate” with the following comments:

> Your “evaluate” process is just like our “review and improve” process where we look at the various dispensation and noncompliance, and review the need to change our policies or to review actions to drive better compliance. (ChiefInfoRiskOfficer-APAC_InsuranceCo)

The interview took 60 minutes and we continued with a casual discussion over coffee on the trends affecting the insurance industry. He also offered to provide further information should there be a need for follow-up discussion.

### 6.2.6 IS_SoftwareConsultant

This was the last expert interview that was conducted as the researcher believed that theoretical saturation had been achieved since no new themes and dimensions were identified beyond all the previous expert interviews. IS_SoftwareConsultant was a consultant based in Singapore from a global information security software company headquartered in the USA. IS_SoftwareConsultant was selected as he had more than 25 years of experience in financial and information risk consulting, having worked as a consultant with several global risk consulting organisations, with his latest role focused on providing information security solutions covering enterprise governance, risk and compliance (GRC), security operations (SecOps) and IT service management.

IS_SoftwareConsultant started the interview by using his software application to illustrate how the application supported GRC, SecOps and IT service management. He explained how the application could help in defining clear roles and responsibilities based on proper organisation structure and the automation of the processes from setting up to monitoring of policies and controls:

> From the GRC, governance, risk and compliance angle, we actually adopt salient points of corporate governance to ensure whatever we do even in IT and security, they are aligned to corporate governance expectations, e.g. SOX. We want our customers, for example, to have a proper organisation structure that drives the segregation of roles and responsibilities, ensuring check and balance for proper controls. As part of an implementation, we will work with customers to define their information security standards, policies and procedures, and will identify the standards that they may want to adopt, i.e. NIST, COBIT 5 or ISO. The application will capture all the policies, procedures and controls as the base checklist and compliance can do a compliance check against the list … automatically or manually. (IS_SoftwareConsultant)

While software helped in the automation, he emphasised the importance of a clear structure with defined roles and responsibilities, and an integrated process with defining and monitoring of security standards, policies and controls. He also highlighted that the software came with an auditing module that reinforced the need for assurance, providing a standard checklist and keeping track of all auditing information for traceability.

IS_SoftwareConsultant also discussed the GRC modules of the software that provided comprehensive functionalities that support risk identification, assessment and management. IS_SoftwareConsultant managed to successfully explain the complete ISG and management process by taking the researcher through the major functions of the software application.

In the latter part of the interview, IS_SoftwareConsultant was asked for his comments on the researcher's refined ISG process model. He was surprised and pleased to see an ISG process model as most published models do not focus on governance but on information security management:

> I have seen many information security models, NIST, COBIT, etc. covering information security management and operations, with some components on governance, but don't recall any model that is so specific to just information security governance. I agree that it is difficult to define governance clearly, but it is very good when you have such a model as it then drives clear definition of governance. (IS_SoftwareConsultant)

The researcher continued to discuss the model with IS_SoftwareConsultant, going through each process component and the relationships and interactions of the components. In his final remarks in validating the model, IS_SoftwareConsultant provided the following comment:

> I believe you've covered all areas. In our company's application terms we named it differently, but it's consistent. We have planning, risk management, compliance, implementation and assurance components, but it covers all areas from governance to management without really segregating governance and we didn't do it as a process model. Process model makes it easier to understand as most people are familiar with

> processes. It will definitely help senior executives to understand the model clearly in a process view. (IS_SoftwareConsultant)

This interview took approximately 45 minutes and IS_SoftwareConsultant also provided the researcher with some product documentation from the company.

## 6.3 Coding and Analysis

Each expert interview was transcribed and read to gain an overall understanding before the next expert interview. This was carried out throughout the expert interviews so that the researcher could add or modify the focus of the subsequent expert interviews to obtain specific information if required, akin to a theoretical sampling approach. All expert interview transcripts were coded and analysed using NVivo 11 in a similar way to how the case study interviews were coded and analysed. A sample screenshot of the coding in NVivo 11 is shown in Figure 6-2. The analysis of the expert interviews was based on previous knowledge from the analysis of the case study interviews, using the existing codes derived from Phase 2 as “a priori” codes as adopted in a template analysis approach (King et al., 2004, 2018). While existing codes, first-order concepts and second-order themes provided the basis for a quick start to coding in Phase 3, the analysis continued with an attempt to identify new codes and emergent first-order concepts.

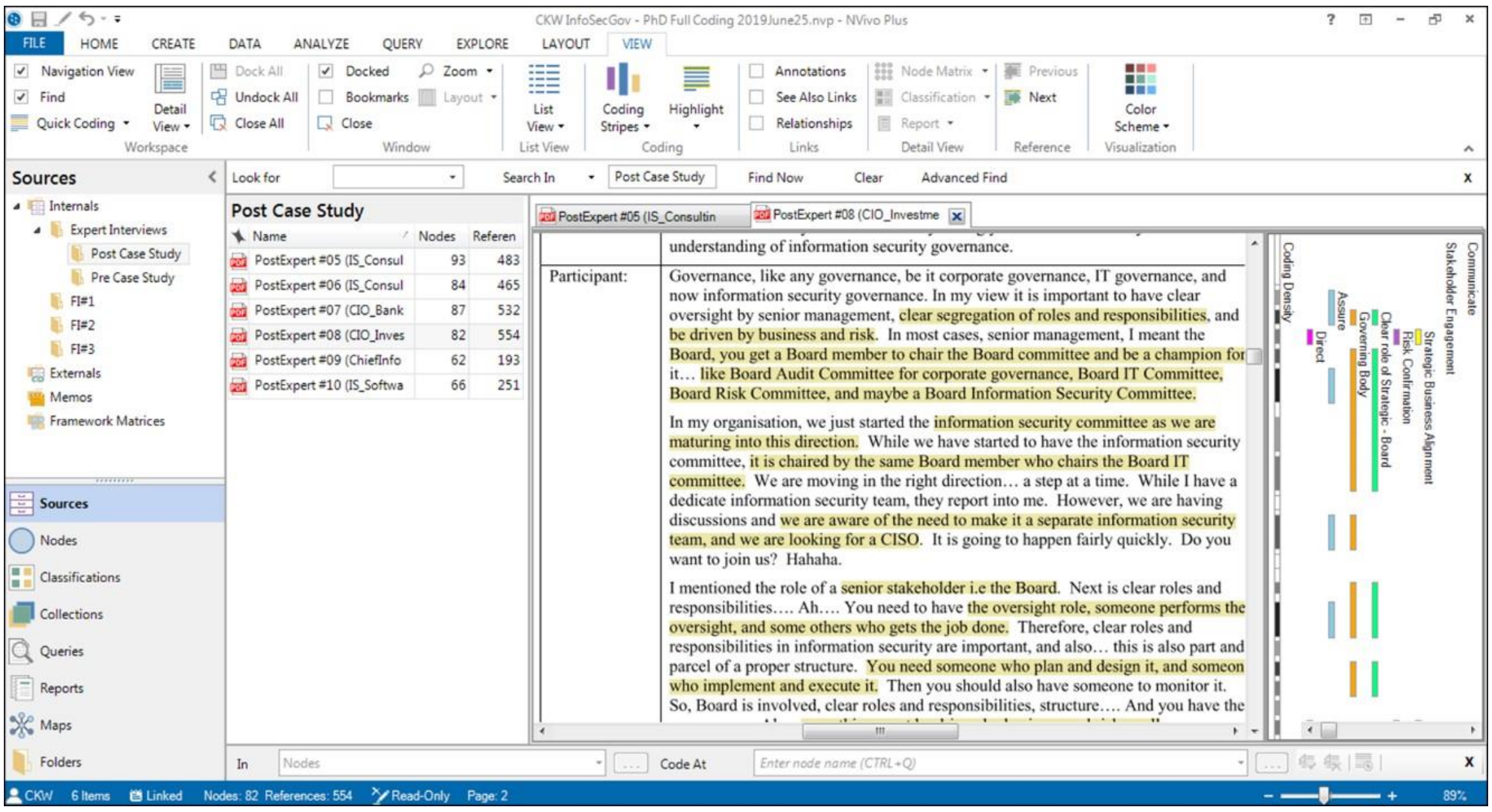

Figure 6-2: Coding and analysis of expert interview transcripts (sample)

The analysis of the 6 expert interview transcripts generated one new first-order concept relating to the “three lines of defence” as proposed by the Institute of Internal Auditors (Institute of Internal Auditors, 2013) which was brought up by IS_Consultant and CIO-SG_InvestmentCo:

> We propose the three lines of defence concept, i.e. first line where the front line is responsible, then second line where we ensure that there is the check and balance, i.e. the risk and compliance functions, and finally the third line of defence where you have the assurance - the audits, both internal audit and external audit, and maybe even the regulatory audits. (IS_Consultant)
>
> In line with this, we also adopt the three line of defence as proposed by our auditor. Clear segregation of duties in risk ownership and management, i.e. the operations are responsible for the first line of defence where they own the risk at the business. Then second line is our risk and compliance. The auditors, both internal and external, are our third line. I believe it is a universally accepted best practice for risk management and we adopt the same for information security risk. (CIO-SG_InvestmentCo)

However, further analysis of the relationships between codes and aggregation led to similar second-order themes and aggregated dimensions (i.e. assure, direct, monitor and evaluate). Therefore, after all 6 expert interview transcripts were coded and analysed, no new second-order themes or aggregated dimensions were discovered. This detailed process showed that the analysis had reached theoretical saturation and these expert interviews had also provided the validation of the process model.

Table 6-2 and Table 6-3 show a summary of the aggregated dimensions and themes as discovered in Phase 2: Model refinement against the expert interviews extracted from analysis in NVivo 11. The "intensity" represents the total number of individual statements in all interviews that relate to the particular dimension and theme, and provides the evidence for the validation of the themes and dimensions.

Table 6-2: Intensity by aggregated dimensions and themes (axial codes) against case studies for core governance stakeholder groups (extracted from analysis in NVivo 11).

| **Aggregated Dimension** | **Theme (Axial Code)** | **Intensity#** | | | | | | |
|---|---|---|---|---|---|---|---|---|
| | | **IS_ConsultingDirector** | **IS_Consultant** | **CISO-MY_Bank** | **CIO-SG_InvestmentCo** | **ChiefInfoRiskOfficer-APAC_InsuranceCo** | **IS_SoftwareConsultant** | **Total** |
| **External** | External | 1 | 3 | 6 | 6 | 1 | 2 | **19** |
| **Governing Body** | Strategic - board | 11 | 6 | 13 | 23 | 5 | 1 | **59** |
| | Strategic - executive | 6 | 4 | 10 | 9 | 3 | 2 | **34** |
| **Management** | Management | 6 | 2 | 3 | 5 | 2 | 2 | **20** |

#Intensity represents the total number of individual statements of all interviews that relate to a particular dimension and theme (or axial code)

Table 6-3: Intensity by aggregated dimensions and themes (axial codes) against case studies for core ISG processes (extracted from analysis in NVivo 11).

| Aggregated Dimension | Theme (Axial Code) | Intensity[#] | | | | | | |
|---|---|---|---|---|---|---|---|---|
| | | IS_ConsultingDirector | IS_Consultant | CISO-MY_Bank | CIO-SG_InvestmentCo | ChiefInfoRiskOfficer-APAC_InsuranceCo | IS_SoftwareConsultant | Total |
| **Assure** | Conduct external audits and certifications | 1 | 3 | 2 | 4 | 2 | 2 | **14** |
| | Provide oversight | 4 | 4 | 6 | 11 | 2 | 3 | **30** |
| | Conduct Internal Audit | 1 | 7 | 3 | 5 | 1 | 3 | **20** |
| **Communicate** | Engage stakeholders | 5 | 3 | 6 | 8 | 5 | 5 | **32** |
| **Evaluate** | Evaluate and refine | 2 | 2 | 4 | 1 | 3 | 2 | **14** |
| | Collect and analyse | 2 | 2 | 4 | 0 | 3 | 1 | **12** |
| **Monitor** | Measure and report performance | 1 | 0 | 3 | 2 | 2 | 3 | **11** |
| | Manage compliance | 2 | 11 | 2 | 4 | 7 | 5 | **31** |
| **Direct** | Define information security objectives to comply with regulatory requirements | 1 | 0 | 8 | 2 | 0 | 1 | **12** |
| | Define information security objectives to support business strategy | 17 | 5 | 12 | 11 | 2 | 2 | **49** |
| | Confirm risk appetite | 5 | 4 | 7 | 11 | 2 | 1 | **30** |
| | Manage risk | 9 | 27 | 15 | 12 | 4 | 8 | **75** |
| | Confirm information security strategy and objectives | 11 | 5 | 5 | 5 | 3 | 3 | **32** |
| | Implement information security standards, policies and controls | 13 | 8 | 6 | 6 | 3 | 7 | **43** |

[#]Intensity represents the total number of individual statements of all interviews that relate to a particular dimension and theme (or axial code)

The data source for the aggregated dimensions was the expert interviews and these represented the true perspectives of a set of experts who were consultants and practitioners based on their experiences in ISG. The discovery of the themes as indicated by the intensity in Table 6-2 and Table 6-3 has validated the ISG process model. The following sections of this chapter discuss the validation of the process model with evidence provided in the form of interview excerpts.

## 6.4 Validation of ISG Process Model through Expert Interviews

The following sections aim to describe the validation of the ISG process model through the 6 expert interviews. In each subsection, a table summarises the validation of each component of the ISG process model with representative quotes that corroborate the results.

### 6.4.1 ISG Stakeholder Groups and Structure

All experts had similar viewpoints that the structure drives defined roles and responsibilities and a clearly defined ISG structure is critical in improving ISG. All interview data confirm the segregation of “governing body” and “management” to clearly define the roles and responsibilities of the board, C-level executives and operational management. In addition, the expert interviews confirm the existence of “external”, which is comprised of external parties such as regulators, external auditors and independent consultants.

The ChiefInfoRiskOfficer-APAC_InsuranceCo reiterated the importance of clear segregation and identification of the roles involved in governance and management with the following remark:

> Good effort and I believe it will be very useful. There is always a confusion and overlap between governance and management, and this ISG process model will help identify the differences. (ChiefInfoRiskOfficer-APAC_InsuranceCo)

CIO-SG_InvestmentCo highlighted the clear roles and responsibilities and proper structure in ISG. He also emphasised the importance of the board within the structure:

> You need to have the oversight role, someone performs the oversight and some others who gets the job done. Therefore, clear roles and responsibilities in information security are important and also … this is also part and parcel of a proper structure. You need someone who plan and design it and someone who implement and execute it. Then you should also have someone to monitor it. So, board is involved, clear roles and responsibilities, structure … And you have the governance. (CIO-SG_InvestmentCo)

His views on governance structure were supported by IS_SoftwareConsultant, who stated that his global GRC and ITSM software application required such structure in the setup:

> In my company, we offer applications to support the overall GRC and IT service management and security operations. From the GRC, governance, risk and compliance angle, we actually adopt salient points of corporate governance to ensure whatever we do even in IT and security, they are aligned to corporate governance expectations, e.g. SOX. We want our customers, for example, to have a proper organisation structure that drives the segregation of roles and responsibilities, ensuring check and balance for proper controls. (IS_SoftwareConsultant)

IS_ConsultingDirector and CISO-MY_Bank also attempted to define the roles and responsibilities of the various stakeholders, namely, the board and the management:

> Members of governance committees must understand the differences between them in order to avoid dysfunction and meet the business risks and technology objectives. Very broadly, I believe we can define them as follows:
>
> - Information security governance – exists to ensure that the security program adequately meets the strategic needs of the business
> - Information security management – implements that program

- Information security operations – executes or manages security-related processes relating to current infrastructure on a day-to-day basis. (IS_ConsultingDirector)

> The board provides the oversight and approves the directions, and the senior management take the responsibilities to drive it through the organisation. (CISO-MY_Bank)

The following interview excerpts validate the fact that there are various stakeholders as depicted in the ISG process model which include "external" besides "governing body" and "management":

> The board is definitely a key player in the governance, as they are the one responsible for the oversights as you depicted in your model. Board works with external assurance providers to get their professional help in driving the assurance and the governance … I think the C-level executives are part of the governance process as they are the delegated by the board to carry out the governance duties. I believe you reflected it well … the greyed part of the diagram. (IS_Consultant)

> I like the organisation structure in a way that defines the governing body and management. We need to know who is driving governance and that is the board as I said earlier. You have included the senior management, which is correct. They will closely with board. The board provides the oversight and approves the directions, and the senior management take the responsibilities to drive it through the organisation. (CISO-MY_Bank)

> You have shown it well in your model. Actually, you have "external" and I believe this should include external auditors, regulators or anyone else that is outside the organisation! (IS_Consultant)

Your "external" should include every stakeholder, for example, the regulator like Monetary Authority of Singapore. (CIO-SG_InvestmentCo)

The interview excerpts and detailed analysis have validated the stakeholders and structure in ISG. Table 6-4, Table 6-5, Table 6-6 and Table 6-7 show the validation summary together with representative quotes.

Table 6-4: Validation data - "external" (external).

| Dimension: **External**; Theme: **External** | | |
|---|---|---|
| **Expert Interviewees** | **Validation** | **Representative Quotes** |
| IS_ConsultingDirector | Yes | "Singapore is very strict with new regulations set by regulators which are external to the organisations.“ (IS_ConsultingDirector) |
| IS_Consultant | Yes | "Board works with external assurance providers to get their professional help in driving the assurance and the governance" (IS_Consultant)<br>"You have shown it well in your model. Actually, you have 'external' and I believe this should include external auditors, regulators or any one else that is outside the organisation!" (IS_Consultant) |
| CISO-MY_Bank | Yes | "such as the external auditors and BNM (Central Bank)." (CISO-MY_Bank) |
| CIO-SG_InvestmentCo | Yes | "There are also the auditors. Both Internal and external." (CIO-SG_InvestmentCo)<br>"Your external should include every stakeholder, for example, the regulator like Monetary Authority of Singapore." (CIO-SG_InvestmentCo) |
| ChiefInfoRiskOfficer-APAC_InsuranceCo | Yes | "On the assurance, we have both internal and external assurance. External are by regulators and our external auditors." (ChiefInfoRiskOfficer-APAC_InsuranceCo) |
| IS_SoftwareConsultant | Yes | "Our auditing modules support the use by external auditors." (IS_SoftwareConsutant) |

Table 6-5: Validation data - "strategic - board" (governing body).

| Dimension: **Governing Body**; Theme: **Strategic-Board** | | |
|---|---|---|
| **Expert Interviewees** | **Validation** | **Representative Quotes** |
| IS_ConsultingDirector | Yes | "It is the duty of board … to ensure correct and reliable information are maintained and disclosed as part of good governance." (IS_ConsultingDirector) |

| Dimension: **Governing Body**; Theme: **Strategic-Board** | | |
|---|---|---|
| **Expert Interviewees** | **Validation** | **Representative Quotes** |
| IS_Consultant | Yes | “The board is definitely a key player in the governance, as they are the one responsible for the oversight as you depicted in your model.” (IS_Consultant) |
| CISO-MY_Bank | Yes | “The board provides the oversight and approves the directions, and the senior management take the responsibilities to drive it through the organisation.” (CISO-MY_Bank) |
| CIO-SG_InvestmentCo | Yes | “You need to have the oversight role, someone performs the oversight and some others who gets the job done. Therefore, clear roles and responsibilities in information security are important and also … this is also part and parcel of a proper structure. You need someone who plan and design it and someone who implement and execute it. Then you should also have someone to monitor it. So, board is involved, clear roles and responsibilities, structure … And you have the governance.” (CIO-SG_InvestmentCo) |
| ChiefInfoRiskOfficer-APAC_InsuranceCo | Yes | “The board risk committee approves policies and dispensation, and review and approve the audit reports.” (ChiefInfoRiskOfficer-APAC_InsuranceCo)<br>“A board member chairs the information security risk committee.” (ChiefInfoRiskOfficer-APAC_InsuranceCo) |
| IS_SoftwareConsultant | Yes | “I like the way you have segregated governing body and management, and the two groups in governing body, the board and executives.” (IS_SoftwareConsultant) |

Table 6-6: Validation data - “strategic - executive” (governing body).

| Dimension: **Governing Body**; Theme: **Strategic-Executive** | | |
|---|---|---|
| **Expert Interviewees** | **Validation** | **Representative Quotes** |
| IS_ConsultingDirector | Yes | “The board chairs the information security committee which is represented by the C-level executive.” (IS_ConsultingDirector) |
| IS_Consultant | Yes | “I think the C-level executives are part of the governance process as they are the delegated by the board to carry out the governance duties.“(IS_Consultant) |
| CISO-MY_Bank | Yes | “The board provides the oversight and approves the directions and the senior management take the responsibilities to drive it through the organisation.” (CISO-MY_Bank) |
| CIO-SG_InvestmentCo | Yes | “Board is responsible for oversight. Executives for reporting on internal findings.” (CIO-S_InvestmentCo) |

| Dimension: **Governing Body**; Theme: **Strategic-Executive** | | |
|---|---|---|
| **Expert Interviewees** | **Validation** | **Representative Quotes** |
| ChiefInfoRiskOfficer-APAC_InsuranceCo | Yes | "The senior executives define the standards and the directions." ((ChiefInfoRiskOfficer-APAC_InsuranceCo) |
| IS_SoftwareConsultant | Yes | "This structure is consistent with how our information security can be configured – separate governance comprising of executives." (IS_SoftwareConsultant) |

Table 6-7: Validation data - "management" (management).

| Dimension: **Management**; Theme: **Management** | | |
|---|---|---|
| **Expert Interviewees** | **Validation** | **Representative Quotes** |
| IS_ConsultingDirector | Yes | "Information security management – implements that program." (IS_ConsultingDirector) |
| IS_Consultant | Yes | "Management component where we focus on the management of information security." (IS_Consultant) |
| CISO-MY_Bank | Yes | "It is useful and practical. Good to separate governing body and management. They have different focus." (CISO-MY_Bank) |
| CIO-SG_InvestmentCo | Yes | "Clear segregation of duties in setting risk appetite and risk management." (CIO-SG_InvestmentCo) |
| ChiefInfoRiskOfficer-APAC_InsuranceCo | Yes | "This is the execution or management part." (ChiefInfoRiskOfficer-APAC_InsuranceCo) |
| IS_SoftwareConsultant | Yes | "It shows clear segregation of duties at least from the high level, it shows that there are differences in governing and management." (IS_SoftwareConsultant) |

The following section describes the validation of the ISG processes.

### 6.4.2 ISG Processes

All 6 expert interviews confirmed that the 5 information governance processes shown in the ISG process model are complete in governing ISG. Both IS_Consultant and CIO-SG_InvestmentCo also drew similarities with the "three lines of defence" model as proposed by the Institute of Internal Auditors (Institute of Internal Auditors, 2013) to demonstrate the relevance and validate the processes in the proposed ISG processes:

> We propose the three lines of defence concept, i.e. first line where the front line is responsible, then second line where we ensure that there is the check and balance, i.e. the risk and compliance functions, and finally

> the third line of defence where you have the assurance – the audits, both internal audit and external audit, and maybe even the regulatory audits. (IS_Consultant)

> In line with this, we also adopt the three line of defence as proposed by our auditor. Clear segregation of duties in risk ownership and management, i.e. the operations are responsible for the first line of defence where they own the risk at the business. Then second line is our risk and compliance. The auditors, both internal and external, are our third line. I believe it is a universally accepted best practice for risk management and we adopt the same for information security risk. (CIO-SG_InvestmentCo)

Experts such as IS_ConsultingDirector and CIO_Bank also supported the idea of an ISG process model:

> I like the way it is designed as a process model. It shows the dependencies. (IS_ConsultingDirector)

> I think this will help (organisations in implementing information security governance). It can be a process model or framework that we can assign individuals, names to the processes, and track them doing the job. (CISO-MY_Bank)
> Process model makes it easier to understand as most people are familiar with processes. It will definitely help senior executives to understand the model clearly in a process view. (IS_SoftwareConsultant)

Further supporting data for each of the processes are discussed in the following sections.

#### *6.4.2.1 Direct-Monitor-Evaluate*

The “direct”, “monitor” and “evaluate” processes work together as a closed-loop process where directions are defined, monitored for implementation and compliance, and evaluated for updates and changes to reflect the changing information security requirements. The interactions and relationships of these processes as depicted in the ISG process model were confirmed by IS_ConsultingDirector, CISO-MY_Bank and CIO-SG_InvestmentCo:

> These work in a closed-loop manner to provide an overall oversight. (IS_ConsultingDirector)
>
> On the five key processes, yes, "direct" is definitely correct and is critical. I like the way you separate or break up "monitor" and "evaluate". It makes the processes a lot clearer. In practice we undertake both but is all part of one big process, i.e. the compliance and remediation part. We always treat is as one process, but by breaking into two discrete processes you can show the difference in focus. One is to check for compliance to defined policies and procedures, and the other focuses on the remediation or the changes required to reflect new requirements or changes in the organisations. I believe this also provide a feedback loop for the process. (CISO-MY_Bank)
>
> Yes. Spot on. This is exactly what is practised. I like the way you illustrated them in a process model. The whole direct, monitor and evaluate provide a closed loop process where you implement, check and improve. (CIO-SG_InvestmentCo)

These closed loop interactions between the core governance processes are shown with the one-directional arrow (→) connecting "direct" to "monitor" and "evaluate", from "monitor" to "evaluate" and back from "evaluate" to "direct" providing the closed feedback loop.

#### *6.4.2.2 Direct*

In the expert interviews, it was also confirmed that within "direct" information security is driven by 3 key requirements, i.e. regulatory requirements, strategic business requirements and the need to manage information security risks. The second-order themes discovered from the expert interviews are consistent with the second-order themes identified in Subsection 5.3.2.1. which are shown as sub-processes within "direct" in the ISG process model, thus validating the ISG process model.

Only two second-order themes relating to "define information security objectives to comply with regulatory requirements" were not discovered in the expert interviews with IS_Consultant and ChiefInfoRiskOfficer-APAC_InsuranceCo. This non-discovery does not in any way show that these sub-processes do not exist. All 4 other expert interviews confirmed and validated these sub-processes:

> Spot on. This is exactly what is practised. I like the way you illustrated them in a process model. The whole direct, monitor and evaluate provide a closed loop process where you implement, check and improve. (CIO-SG_InvestmentCo)
>
> The "direct" process covers all the relevant processes involved. This is what we do in our organisation. (ChiefInfoRiskOfficer-APAC_InsuranceCo)

The following is evidence in the form of interview quotes from the expert interviews that validated the sub-processes in "direct".

a. Define information security objectives to comply with regulatory requirements

> So our information security processes (governance or not) is closely tied to risk and central bank compliance. Firstly, compliance to central bank. Central bank has a set of information security guidelines and the TRM, technology risk management guidelines that all banks must comply with. Both the risk management team and the information security or IT security team will study these guidelines and ensure that we develop our internal bank policies cover all the required guidelines. We will develop our standards and policies, and the required security controls required.

All IT controls must meet that of central bank. This is the first part in ensuring that our policies are at the minimum standard. (CISO-MY_Bank)

Don’t forget the MAS (Monetary Authority of Singapore), the regulator with a big stick. (CIO-SG_InvestmentCo)

Table 6-8 shows representative quotes from the expert interviews that validated this theme.

Table 6-8: Validation data for “define information security objectives to comply with regulatory requirements” (direct).

| Dimension: **Direct**; Theme: **Define Information Security Objectives to Comply with Regulatory Requirements** | | |
|---|---|---|
| **Expert Interviewees** | **Validation** | **Representative Quotes** |
| IS_ConsultingDirector | Yes | “PDPA (Personal Data Protection Act) laws and now many global/multinational are affected by the GDPR (General Data Protection Regulation). We need to comply to all these requirements.” (IS_ConsultingDirector) |
| IS_Consultant | No | No data was discovered from interview. |
| CISO-MY_Bank | Yes | “BNM (Central Bank) has a set of information security guidelines and the TRM, technology risk management guidelines that all banks must comply.” (CISO-MY_Bank) |
| CIO-SG_InvestmentCo | Yes | “Don’t forget the MAS (Monetary Authority of Singapore), the regulator with a big stick.” (CIO-SG_InvestmentCo) |
| ChiefInfoRiskOfficer-APAC_InsuranceCo | No | No data was discovered from interview. |
| IS_SoftwareConsultant | Yes | “the software supports regulatory requirements, like SOX, and … your information security standards must comply.” (IS SoftwareCosultant) |

b. Define information security objectives to support business strategy

information security governance includes the processes that ensure that reasonable and appropriate actions are taken to protect the organisation’s information assets, in the most effective and efficient manner in pursuit of its business goals … information security governance - exists to ensure that the security program adequately meets the strategic needs of the business. (IS_ConsultingDirector)

> That's where the information security strategy is defined based on the requirements of the organisation business objectives. (IS_Consultant)

> My view is that information security governance is similar - driven by business objectives, needs to meet risk management objectives … In organisations, you have the organisation information security strategy framework that is designed based on the organisation's business strategy or objectives. (CIO-SG_InvestmentCo)

> Regulatory requirements and risks together with business strategy drive our information security investments and standards … strategic alignment to business strategy and business initiatives which is more specific, just like we align to our digital transformation programs, etc. (CISO-MY_Bank)

Table 6-9 shows representative quotes from the expert interviews that validated this theme.

Table 6-9: Validation data for "define information security objectives to support business strategy" (direct).

| Dimension: **Direct**; Theme: **Define Information Security Objectives to Support Business Strategy** | | |
|---|---|---|
| **Expert Interviewees** | **Validation** | **Representative Quotes** |
| IS_ConsultingDirector | Yes | "information security governance includes the processes that ensure that reasonable and appropriate actions are taken to protect the organisation's information assets, in the most effective and efficient manner in pursuit of its business goals … information security governance – exists to ensure that the security program adequately meets the strategic needs of the business." (IS_ConsultingDirector) |
| IS_Consultant | Yes | "That's where the information security strategy is defined based on the requirements of the organisation business objectives." (IS_Consultant) |
| CISO-MY_Bank | Yes | "Regulatory requirements and risks together with business strategy drive our information security investments, and standards … strategic alignment to business strategy and business initiatives which is more specific, just like we align to our digital transformation programs, etc." (CISO-MY_Bank) |
| CIO-SG_InvestmentCo | Yes | "My view is that information security governance is similar – driven by business objectives, needs to meet risk |

| Dimension: **Direct**; Theme: **Define Information Security Objectives to Support Business Strategy** | | |
|---|---|---|
| **Expert Interviewees** | **Validation** | **Representative Quotes** |
| | | management objectives … In organisations, you have the organisation information security strategy framework that is designed based on the organisations business strategy or objectives.” (CIO-SG_InvestmentCo) |
| ChiefInfoRiskOfficer-APAC_InsuranceCo | Yes | “In the recent years, with the digital eco-systems, we are also putting in a lot of effort into privacy. All the mobile, cloud, etc. We need to review and develop a governance structure that focus on privacy, embedding the concept early such as privacy by design. Our information security objectives are driven by these business intiatives.” (ChiefInfoRiskOfficer-APAC_InsuranceCo) |
| IS_SoftwareConsultant | Yes | “work with customers to define their information security standards, policies and procedures which are driven by their business requirements” (IS_SoftwareConsultant) |

c. Confirm risk appetite

“Confirm risk appetite” and “manage risk” form the complete risk management process where the different stakeholders focus on the different activities and work together in addressing information security risks:

> Based on the risk assessment, we work with the senior executives and board, and the organisation will decide on an accepted risk profile. (IS_ConsultingDirector)
>
> Then, finalise it with the CRO and the senior management. Then we will bring this up the board risk committee and have a discussion with the board to seek their approval on the agreed risk level that the organisation can accept. This is critical as the board will decide on the level of residual risk that they are willing to take. Balance between what can be done, how much to invest in mitigation and the probability of happening, we agree on the risk appetite. (IS_Consultant) to an agreement on the risk appetite and the risk that we believe the bank will agree to by balancing the investment required to minimise or eliminate the risks. (CISO-MY_Bank)

Table 6-10 shows representative quotes from the expert interviews that validated this theme.

Table 6-10: Validation data for "confirm risk appetite" (direct).

| Dimension: **Direct**; Theme: **Confirm Risk Appetite** | | |
|---|---|---|
| **Expert Interviewees** | **Validation** | **Representative Quotes** |
| IS_ConsultingDirector | Yes | "Based on the risk assessment, we work with the senior executives and board, and the organisation will decide on an accepted risk profile." (IS_ConsultingDirector) |
| IS_Consultant | Yes | "Then we will bring this up the board risk committee and have a discussion with the board to seek their approval on the agreed risk level that the organisation can accept. This is critical as the board will decide on the level of residual risk that they are willing to take." (IS_Consultant) |
| CISO-MY_Bank | Yes | "to an agreement on the risk appetite and the risk that we believe the bank will agree to by balancing the investment required to minimise or eliminate the risks." (CISO-MY_Bank) |
| CIO-SG_InvestmentCo | Yes | "present our risk assessment to the risk board committee, get their blessing on the level of risks that they are willing to accept based on informed discussions." (CIO-SG_InvestmentCo) |
| ChiefInfoRiskOfficer-APAC_InsuranceCo | Yes | "The risk committee approves policies and sets the standards and the risk level." (ChiefInfoRiskOfficer-APAC_InsuranceCo) |
| IS_SoftwareConsultant | Yes | "Strategic alignment also drives risk appetite confirmation and in turn risk appetite confirmation drives infosec strategy, objectives and standards." (IS_SoftwareConsultant) |

d. Manage risk

This is done as part of an organisation-wide risk assessment. We will normally do this as part of a major risk assessment project (or it can be a specific information security risk assessment project). We work with all business divisions and the supporting divisions in assessing and identifying the information security risks and evaluate the potential impact. (IS_ConsultingDirector)

We conduct a risk assessment of the organisations from an information security angle, identify all the risk factors, and their chance of happening and the impact. We will summarise the risks into a risk heat map with the Red Yellow Green map, to show the risk profile. This is done with all the departments leaders. (IS_Consultant) in risk, we do an annual information security risk assessment, identify key weak points and understand our information security threats, and do an assessment of the potential impact should a breach happens. We look at potential weak points from the IT infrastructure in the network, server, customer portal and user access controls. Different business operations have different level of sophistication in protection and controls - we evaluate all that. We will do an assessment and identify risk mitigation strategies, and ensure we take proper actions on this. (CISO-MY_Bank)

Risk management is a key function of the software application. The module supports the end-to-end process of risk management. Beginning with the identification of all risk factors and risk areas, the application allows you to capture all identified risk factors. (IS_SoftwareConsultant)

Table 6-11 shows representative quotes from the expert interviews that validated this theme.

Table 6-11: Validation data for “manage risk” (direct).

| Dimension: **Direct**; Theme: **Manage Risk** | | |
|---|---|---|
| **Expert Interviewees** | **Validation** | **Representative Quotes** |
| IS_ConsultingDirector | Yes | “We work with all business divisions and the supporting divisions in assessing and identifying the information |

| Dimension: **Direct**; Theme: **Manage Risk** | | |
|---|---|---|
| **Expert Interviewees** | **Validation** | **Representative Quotes** |
| | | security risks, and evaluate the potential impact.” (IS_ConsultingDirector) |
| IS_Consultant | Yes | “We conduct a risk assessment of the organisations from an information security angle, identify all the risk factors, and their chance of happening and the impact. We will summarise the risks into a risk heat map with the Red Yellow Green map, to show the risk profile. This is done with all the departments leaders.” (IS_Consultant) |
| CISO-MY_Bank | Yes | “identify all the potential risks to business operations and identify both the probability of happening and the potential impact, be in financials, operations or reputation impact and implement actions to mitigate them…” (CISO-MY_Bank) |
| CIO-SG_InvestmentCo | Yes | “risk management processes where information security risks are identified and risk mitigation strategies are defined and actions are implemented.” (CIO-SG_InvestmentCo) |
| ChiefInfoRiskOfficer-APAC_InsuranceCo | Yes | “Risk management is a basic process where our risk management team conduct periodic assessment to ensure we are aware of our risk posture.” (ChiefInfoRiskOfficer-APAC_InsuranceCo) |
| IS_SoftwareConsultant | Yes | “Risk management is a key function of the software application. The module supports the end-to-end process of risk management. Beginning with the identification of all risk factors and risk areas, the application allows you to capture all identified risk factors.” (IS_SoftwareConsultant) |

e. Confirm information security strategy and objectives

As discussed earlier, information security strategy and initiatives in organisations are driven by regulatory requirements, business requirements and risks. Information security strategy and objectives are defined and approved by the senior executives and board, then detailed standards, policies, procedures and controls will be designed and implemented. Therefore, “confirm information security strategy and objectives” and “implement information security standards, policies and controls” are closely inter-related in practice:

> Then, the information security strategy will be defined to support the accepted and agreed risk profile. (IS_ConsultingDirector)

> Information security is defined and confirmed based on the organisation’s business objectives ...we worked with the business executives to define the security frameworks and the projects for the management to implement. (IS_Consultant)

The board committee focus on making sure the management is doing their job by making sure we have a regular, an alternate month board committee updates. The committee helps to provide approvals for information security budgets and projects, confirming the directions of information security for the organisation. (CIO-SG_InvestmentCo)

The senior executives define the standards and the directions. For example, I will work with the team to define and confirm the standards and directions, understand what they want. (ChiefInfoRiskOfficer-APAC_InsuranceCo)

Table 6-12 shows representative quotes from the expert interviews that validated this theme.

Table 6-12: Validation data for "confirm information security strategy and objectives" (direct).

| Dimension: **Direct**; Theme: **Confirm Information Security Strategy and Objectives** | | |
|---|---|---|
| **Expert Interviewees** | **Validation** | **Representative Quotes** |
| IS_ConsultingDirector | Yes | "Then, the information security strategy will be defined to support the accepted and agreed risk profile." (IS_ConsultingDirector) |
| IS_Consultant | Yes | "Information security is defined and confirmed based on the organisation's business objectives ...we worked with the business executives to define the security frameworks and the projects for the management to implement." (IS_Consultant) |
| CISO-MY_Bank | Yes | "We will develop our standards and policies, and the required security controls required." (CISO-MY_Bank) |
| CIO-SG_InvestmentCo | Yes | "we will define a set of initiatives as part of the larger information security plan. From here, we also confirm the information security strategy framework that drives the security architecture, hardware and application designs, standards, policies, procedures and finally the actual controls that will be implemented." (CIO-SG_InvestmentCo) |

| Dimension: **Direct**; Theme: **Confirm Information Security Strategy and Objectives** | | |
|---|---|---|
| **Expert Interviewees** | **Validation** | **Representative Quotes** |
| ChiefInfoRiskOfficer-APAC_InsuranceCo | Yes | "The senior executives define the standards and the directions. For example, I will work with the team to define and confirm the standards and directions, understand what they want." (ChiefInfoRiskOfficer-APAC_InsuranceCo) |
| IS_SoftwareConsultant | Yes | "Strategic alignment also drives risk appetite and in turn helps confirm the infosec strategy, objectives and standards." (IS_SoftwareConsultant) |

f. Implement information security standards, policies and controls

> Similarly, this will also drive the information security standards, which will then be translated into actual controls, procedures, etc. for implementation. (IS_Consultant)

> we also define the information security strategy framework that drives the security architecture, hardware and application designs, standards, policies, procedures and finally the actual controls that will be implemented. (CIO-SG_InvestmentCo)

> From there (confirm strategy and standards), we will work out the detailed policies. Policies need to be simple for everyone to understand and implement to make sure staff don't do the wrong thing. We have one policy, with one action – simple, separated by business, technical and operations. Each policy comes with an action and the reason why it is required or done. (ChiefInfoRiskOfficer-APAC_InsuranceCo)

Table 6-13 shows representative quotes from the expert interviews that validated this theme.

Table 6-13: Validation data for “implement information security standards, policies and controls” (direct).

| Dimension: **Direct**; Theme: **Implement Information Security Standards, Policies & Controls** | | |
|---|---|---|
| **Expert Interviewees** | **Validation** | **Representative Quotes** |
| IS_ConsultingDirector | Yes | “These (information security and objectives) are used to define the projects and initiatives that are required to meet the strategy, and to develop and implement the policies and procedures at the management and operational levels.” (IS_ConsultingDirector) |
| IS_Consultant | Yes | “Similarly, this will also drive the information security standards, which will then be translated into actual controls, procedures, etc. for implementation.” (IS_Consultant) |
| CISO-MY_Bank | Yes | “My technology team will define the architecture and the information security standards, and then the policy and procedures for implementation.” (CISO-MY_Bank) |
| CIO-SG_InvestmentCo | Yes | “we also define the information security strategy framework that drives the security architecture, hardware and application designs, standards, policies, procedures and finally the actual controls that will be implemented.” (CIO-SG_InvestmentCo) |
| ChiefInfoRiskOfficer-APAC_InsuranceCo | Yes | “From there (confirm strategy and standards), we will work out the detailed policies. Policies need to be simple for everyone to understand and implement to make sure staff don’t do the wrong thing. We have one policy, with one action – simple, separated by business, technical and operations. Each policy comes with an action and the reason why it is required or done.” (ChiefInfoRiskOfficer-APAC_InsuranceCo) |
| IS_SoftwareConsultant | Yes | “work with customers to define and implement their information security standards, policies and procedures.” (IS_SoftwareConsultant) |

#### *6.4.2.3 Monitor*

“Direct” sets the directions and implements the policies, procedures and controls, while “monitor” ensures that these policies, procedures and controls are implemented according to defined standards and are adhered to in the operations. In practice, “monitor” encompasses the compliance and reporting processes. In this research, the second-order themes within “monitor” are “measure and report performance” and “manage compliance”. Analysis of expert interview data shows 5 out of the 6 expert interviews validated the "measure and report performance" emergent theme, while all 6 expert interviews validated the "manage compliance" emergent theme.

a. Measure and report performance

> Assess the value of information security investments to gauge if organisation is receiving benefits as anticipated. Ensure that the execution of the information security program, and all its associated processes and activities, are done within the parameters set out by the program strategy, architecture and policy strategy. It is also a mechanism to measure and report to the governance committee. (IS_ConsultingDirector)

> Some of the areas we cover in the council meeting include:
>
> - Update on security projects that is ongoing (costs against budgets)
> - Update on security situation, i.e. number of detected attacks, threats, etc.
> - Update on incidents (if any), actions required
> - Update on our assessment against BNM (central bank) requirements and compliance, e.g. update on our views of the new technology risk management (TRM) guidelines to be introduced by BNM – challenges in adherence, what needs to be done, etc. (CISO-MY_Bank)

> In this area, we report the dispensation and noncompliance, and their actions to the risk committee for updates and approvals. (ChiefInfoRiskOfficer-APAC_InsuranceCo)

b. Manage compliance

> This covers the compliance process where we ensure that our customer have the relevant compliance process. Making sure they have a proper checklist to check against the actual adherence to policies and controls. Compliance will check and validate and report on the compliance matrix. (IS_Consultant)

> One is to check for compliance to defined policies and procedures, and the other focuses on the remediation or the changes required to reflect new requirements or changes in the organisations. (CISO-MY_Bank)

> On the governance part to ensure policies are actually implemented, we conduct assessments twice a year. One is a self-assessment and another is an assessment together with IT/risks. (ChiefInfoRiskOfficer-APAC_InsuranceCo)

Table 6-14 and Table 6-15 show representative quotes from the expert interviews that validated the themes for “monitor”.

Table 6-14: Validation data for “measure and report performance” (monitor).

| Dimension: **Monitor**; Theme: **Measure and Report Performance** | | |
|---|---|---|
| **Expert Interviewees** | **Validation** | **Representative Quotes** |
| IS_ConsultingDirector | Yes | “Assess the value of information security investments to gauge if organisation is receiving benefits as anticipated … it is also a mechanism to measure and report to the governance committee” (IS_ConsultingDirector) |
| IS_Consultant | No | No data was discovered from interview. |
| CISO-MY_Bank | Yes | “Update on security projects that is ongoing (costs against budgets) … Update on security situation, i.e. number of detected attacks, threats, etc … Update on incidents (if any).” (CISO-MY_Bank) |
| CIO-SG_InvestmentCo | Yes | “regular updates, we provide the board committee with an update of our current information security situation - a health check on whether there are any breaches, any |

| Dimension: **Monitor**; Theme: **Measure and Report Performance** | | |
|---|---|---|
| **Expert Interviewees** | **Validation** | **Representative Quotes** |
| | | recorded near misses and an update of our information security projects." (CIO-SG_InvestmentCo) |
| ChiefInfoRiskOfficer-APAC_InsuranceCo | Yes | "we report the dispensation and noncompliance, and their actions to the risk committee for updates and approvals." (ChiefInfoRiskOfficer-APAC_InsuranceCo) |
| IS_SoftwareConsultant | Yes | "In addition, there is an intelligent dashboard that shows the compliance status - can further drill down towards more granular details." (IS_SoftwareConsultant) |

Table 6-15: Validation data for "manage compliance" (monitor).

| Dimension: **Monitor**; Theme: **Manage Compliance** | | |
|---|---|---|
| **Expert Interviewees** | **Validation** | **Representative Quotes** |
| IS_ConsultingDirector | Yes | "Ensure that the execution of the information security program, and all its associated processes and activities, are done within the parameters set out by the program strategy, architecture and policy strategy … it is a management process to ensure everything works as intended." (IS_ConsultingDirector) |
| IS_Consultant | Yes | "This covers the compliance process where we ensure that our customer have the relevant compliance process. Making sure they have a proper checklist to check against the actual adherence to policies and controls. Compliance will check and validate and report on the compliance matrix." (IS_Consultant) |
| CISO-MY_Bank | Yes | "One is to check for compliance to defined policies and procedures, and the other focuses on the remediation or the changes required to reflect new requirements or changes in the organisations." (CISO-MY_Bank) |
| CIO-SG_InvestmentCo | Yes | "Management of compliance is done is done by both executives and management - and is part of our second line of defence." (CIO-SG_InvestmentCo) |
| ChiefInfoRiskOfficer-APAC_InsuranceCo | Yes | "On the governance part to ensure policies are actually implemented, we conduct assessments twice a year. One is a self-assessment and another is an assessment together with IT/risks" (ChiefInfoRiskOfficer) |
| IS_SoftwareConsultant | Yes | "The application will capture all the policies, procedures and controls as the base checklist and compliance can do a compliance check against the list … automatically or manually." (IS_SoftwareConsultant) |

#### 6.4.2.4 Evaluate

Together with "monitor", the "evaluate" process involves collection and analysis of data from "monitor" to determine if changes or updates are required to better improve the information security standards, policies, procedures or controls that have been implemented. This also covers the evaluation of whether the budget allocated to information security initiatives is required to be changed to reflect the required information security profile of the organisation. The analysis of the expert interviews data showed that all 6 expert interviews validated the "evaluate and refine" emergent theme, while 5 out of the 6 expert interviews validated the "collect and analyse" emergent theme.

a. Evaluate and refine

> Noncompliance are investigated to understand the reason and root cause so that actions can be taken. Actions can be taken against the users or maybe … the controls or procedures may need to be updated. (IS_Consultant)

> One is to check for compliance to defined policies and procedures, and the other focuses on the remediation or the changes required to reflect new requirements or changes in the organisations. I believe this also provide a feedback loop for the process. (CISO-MY_Bank)

> Your "evaluate" process is just like our "review and improve" process where we look at the various dispensation and noncompliance, and review the need to change our policies or to review actions to drive better compliance. (ChiefInfoRiskOfficer-APAC_InsuranceCo)

> It is an important closed loop, feedback loop process. It is of no use if people check for compliance but didn't take the effort to redesign their controls. (IS_SoftwareConsultant)

b. Collect and analyse

> where we look at the various dispensation and noncompliance, and review the need to change our policies or to review actions to drive better compliance. (ChiefInfoRiskOfficer-APAC_InsuranceCo)

> where we propose to assess the value of information security investments … many have been investing without knowing if they are doing it right. It is a good measure as a feedback to the governance committee or the board. (IS_ConsultingDirector)

> Noncompliance are investigated to understand the reason and root cause so that actions can be taken. (IS_Consultant)

Table 6-16 and Table 6-17 show additional representative quotes from the expert interviews that validated the themes for "evaluate".

Table 6-16: Validation data for "evaluate and refine" (evaluate).

| Dimension: **Evaluate**; Theme: **Evaluate and Refine** | | |
|---|---|---|
| **Expert Interviewees** | **Validation** | **Representative Quotes** |
| IS_ConsultingDirector | Yes | "Assess the value of information security investments to gauge if organisation is receiving benefits as anticipated … amend accordingly." (IS_ConsultingDirector) |
| IS_Consultant | Yes | "Noncompliance are investigated to understand the reason and root cause so that actions can be taken. Actions can be taken against the users or maybe … the controls or procedures may need to be updated." (IS_Consultant) |
| CISO-MY_Bank | Yes | "One is to check for compliance to defined policies and procedures, and the other focuses on the remediation or the changes required to reflect new requirements or changes in the organisations. I believe this also provide a feedback loop for the process." (CISO-MY_Bank) |
| CIO-SG_InvestmentCo | Yes | "I like the way you illustrated them in a process model. The whole 'direct, monitor and evaluate' provide a closed loop process where you implement, check and improve." (CIO-SG_InvestmentCo) |

| Dimension: **Evaluate**; Theme: **Evaluate and Refine** | | |
|---|---|---|
| **Expert Interviewees** | **Validation** | **Representative Quotes** |
| ChiefInfoRiskOfficer-APAC_InsuranceCo | Yes | "Your 'evaluate' process is just like our 'review and improve' process where we look at the various dispensation and noncompliance, and review the need to change our policies or to review actions to drive better compliance." (ChiefInfoRiskOfficer-APAC_InsuranceCo) |
| IS_SoftwareConsultant | Yes | "It is an important closed loop, feedback loop process. It is of no use if people check for compliance but didn't take the effort to redesign their controls." (IS_SoftwareConsultant) |

Table 6-17: Validation data for "collect and analyse" (evaluate).

| Dimension: **Evaluate**; Theme: **Collect and Analyse** | | |
|---|---|---|
| **Expert Interviewees** | **Validation** | **Representative Quotes** |
| IS_ConsultingDirector | Yes | "where we propose to assess the value of information security investments … many have been investing without knowing if they are doing it right. It is a good measure as a feedback to the governance committee or the board." (IS_ConsultingDirector) |
| IS_Consultant | Yes | "Noncompliance are investigated to understand the reason and root cause so that actions can be taken." (IS_Consultant) |
| CISO-MY_Bank | Yes | "We always treat is as one process, but by breaking into two discrete processes you can show the difference in focus. One focus on collecting data and the other focuses on the remediation or the changes required to reflect new requirements or changes in the organisations." (CISO-MY_Bank) |
| CIO-SG_InvestmentCo | No | No data was discovered from interview. |
| ChiefInfoRiskOfficer-APAC_InsuranceCo | Yes | "where we look at the various dispensation and noncompliance, and review the need to change our policies or to review actions to drive better compliance." (ChiefInfoRiskOfficer-APAC_InsuranceCo) |
| IS_SoftwareConsultant | Yes | "data is collected either manually or automatically through the system for analysis so next actions can be planned or implemented." (IS_SoftwareConsultant) |

#### *6.4.2.5 Communicate*

"Communicate" has been identified as a key dimension in the analysis of all 6 expert interviews, validating the dimension and the "engage stakeholder" emergent theme in the case study analysis. "Communicate" in ISG focuses on the engagement of stakeholders, primarily undertaken by C-level executives with the board, and between the board and C-level executives with external stakeholders, e.g. regulators, shareholders and customers. The following quotes on communications were shared by the experts:

> reporting to internal stakeholders and as part of public relation (PR) to external stakeholders like regulators and customers. It is seen as a PR exercise if it is to outside. For example, a system down or breach that impacts the customers, then it will be treated as a PR exercise to ensure we communicate the right message at the right time. (CIO-SG_InvestmentCo)

> They [referring to CIO and IT security head] will provide the regular updates scheduled every other month … We will also define a reporting framework driving engagement between the various stakeholders, e.g. the board and executive management. (IS_ConsultingDirector)

> We have monthly council meetings and the council will update the board every two months. (CISO-MY_Bank)

> In our regular updates, we provide the board committee with an update of our current information security situation - a health check on whether there are any breaches, any recorded near misses and an update of our information security projects. The updates are done by our head of information security. So the board committee aims to provide the oversight … an independent view, directions, comments and feedback. (CIO-SG_InvestmentCo)

> Communication is very important. It is required to ensure all management are updated and aware of new policies, noncompliance, actions taken, etc. It is communications that drive transparency in the overall governance, awareness for better compliance. This is extended to the wider organisation through emails, memos and training. As you have highlighted, upward communications to senior stakeholders and board are as critical. (ChiefInfoRiskOfficer-APAC_InsuranceCo)

The management dashboard provides attractive statistics that can be used to update the board committee. It is a very effective means of communications with board members. (IS_SoftwareConsultant)

Table 6-18 shows additional representative quotes from the expert interviews that validated the themes for "communicate".

Table 6-18: Validation data for "engage stakeholders" (communicate).

| Dimension: **Communicate**; Theme: **Engage Stakeholders** | | |
|---|---|---|
| **Expert Interviewees** | **Validation** | **Representative Quotes** |
| IS_ConsultingDirector | Yes | "They [referring to CIO and IT security head] will provide the regular updates scheduled every other month …We will also define a reporting framework driving engagement between the various stakeholders, e.g. the board and executive management." (IS_ConsultingDirector) |
| IS_Consultant | Yes | "We will also help them table this to the Exco meeting and the board." (IS_Consultant) |
| CISO-MY_Bank | Yes | "We have monthly council meetings and the council will update the board every two months." (CISO-MY_Bank) |
| CIO-SG_InvestmentCo | Yes | "reporting to internal stakeholders and as part of public relation (PR) to external stakeholders like regulators and customers. It is seen as a PR exercise if it is to outside. For example, a system down or breach that impacts the customers, then it will be treated as a PR exercise to ensure we communicate the right message at the right time." (CIO-SG_InvestmentCo)<br>"In our regular updates, we provide the board committee with an update of our current information security situation – a health check on whether there are any breaches, any recorded near misses and an update of our information security projects. The updates are done by our head of information security. So the board committee aims to provide the oversight … an independent view, directions, comments and feedback." (CIO-SG_InvestmentCo) |
| ChiefInfoRiskOfficer-APAC_InsuranceCo | Yes | "Communication is very important. It is required to ensure all management are updated and aware of new policies, |

| Dimension: **Communicate**; Theme: **Engage Stakeholders** | | |
|---|---|---|
| **Expert Interviewees** | **Validation** | **Representative Quotes** |
| | | noncompliance, actions taken, etc." (ChiefInfoRiskOfficer-APAC_InsuranceCo)<br>"Our experience has been a lot of effort in change management i.e. face-to-face meetings and discussions with board members, C-level executives, etc." (ChiefInfoRiskOfficer-APAC_InsuranceCo) |
| IS_SoftwareConsultant | Yes | "The management dashboard provides attractive statistics that can be used to update the board committee. It is a very effective means of communications with board members." (IS_SoftwareConsultant)<br>"All these are part of the overall governance process and the simple information in a nice dashboard facilitates discussions with the board committee - driving an effective oversight process." (IS_SoftwareConsultant) |

#### *6.4.2.6 Assure*

"Assure" provides "governing body" with comprehensive independent assessments and validations as part of ISG which generally comprises reviews, audits or certifications by independent parties such as internal and external auditors, and consultants. The data from all 6 expert interviews validated all the second-order themes, i.e. "conduct external audits and certifications", "provide oversight" and "conduct internal audits".

a. Conduct external audits and certifications

> an important component in our model is the assurance. The audits, both internal and external audits, to ensure that the customer has done the right things. And these components are important as part of corporate governance and the buyer is the board. You have shown it well in your model. Actually, you have "external" and I believe this should include external auditors, regulators or anyone else that is outside the organisation! (IS_Consultant)

> Just like corporate governance where auditors play a critical role in ensuring an independent assurance, it also applies to information security. When our external auditors conduct their audits, information security is a component of their audit. They conduct an assessment of our internal controls and also a check on our user access management. They provide a view on the strength of our internal controls. (CIO-SG_InvestmentCo)

There are also the auditors. Both internal and external. Don't forget the monetary authority, the regulator with a big stick. (CIO-SG_InvestmentCo)

And finally, audit is non-negotiable … On the assurance, we have both internal and external assurance. Based on the assurance findings, we will work on the actions and if required, redo the policies in December every year. (ChiefInfoRiskOfficer-APAC_InsuranceCo)

Table 6-19 shows representative quotes from the expert interviews that validated this theme.

Table 6-19: Validation data for "conduct external audits and certifications" (assure).

| Dimension: **Assure**; Theme: **Conduct External Audits and Certifications** | | |
|---|---|---|
| **Expert Interviewees** | **Validation** | **Representative Quotes** |
| IS_ConsultingDirector | Yes | "We provide advice to help drive effective corporate governance and audit is also a key component in ensuring good corporate governance." (IS_ConsultingDirector) |
| IS_Consultant | Yes | "Then you have the audits, who provide the independent checks and gives a report card to the audit committee, the board." (IS_Consultant) |
| CISO-MY_Bank | Yes | "they are definitely involved and in great extent, such as the external auditors and BNM." (CISO-MY_Bank) |
| CIO-SG_InvestmentCo | Yes | "Just like corporate governance where auditors play a critical role in ensuring an independent assurance, it also applies to information security. When our external auditors conduct their audits, information security is a component of their audit." (CIO-SG_InvestmentCo) |
| ChiefInfoRiskOfficer-APAC_InsuranceCo | Yes | "On the assurance, we have both internal and external assurance. Based on the assurance findings, we will work on |

| Dimension: **Assure**; Theme: **Conduct External Audits and Certifications** | | |
|---|---|---|
| **Expert Interviewees** | **Validation** | **Representative Quotes** |
| | | the actions and if required, redo the policies in December every year." (ChiefInfoRiskOfficer-APAC_InsuranceCo) |
| IS_SoftwareConsultant | Yes | "there is also an auditing module. The auditing module in the application facilitates external and internal auditors to review and check all internal information security controls implementation." (IS_SoftwareConsultant) |

b. Provide oversight

> If you look at the corporate governance model, audits are important to support board oversight. (IS_SoftwareConsultant)

> The board is definitely a key player in the governance, as they are the one responsible for the oversights as you depicted in your model. (IS_Consultant)

> The board committee focus on making sure the management is doing their job by making sure we have a regular, an alternate month board committee updates … Board committee aims to provide the oversight … an independent view, directions, comments and feedback. (CIO-SG_InvestmentCo)

> information security governance ensures that proper oversight is there to make sure all information security initiatives are defined accordingly to support business objectives and are defined to meet the objectives of risk management. (CIO-SG_InvestmentCo)

> The board provides the oversight. (CISO-MY_Bank)

Table 6-20 shows representative quotes from the expert interviews that validated this theme.

Table 6-20: Validation data for "provide oversight" (assure).

| Dimension: **Assure**; Theme: **Provide Oversight** | | |
|---|---|---|
| **Expert Interviewees** | **Validation** | **Representative Quotes** |
| IS_ConsultingDirector | Yes | "information security governance as a set of structure and processes that drives oversight … with right level of information … independent oversight." (IS_ConsultingDirector)<br>"We work with customers to … define the roles and responsibilities to ensure oversight." (IS_ConsultingDirector) |
| IS_Consultant | Yes | "The board is definitely a key player in the governance, as they are the one responsible for the oversights as you depicted in your model." (IS_Consultant) |
| CISO-MY_Bank | Yes | "The board provides the oversight." (CISO-MY_Bank) |
| CIO-SG_InvestmentCo | Yes | "Board is responsible for oversight. Executives for reporting on internal findings." (CIO-SG_InvestmentCo)<br>"In my view it is important to have clear oversight … like board audit committee for corporate governance, board IT committee, board risk committee and maybe a board information security committee." (CIO-SG_InvestmentCo) |
| ChiefInfoRiskOfficer-APAC_InsuranceCo | Yes | "We define the structure that is required to govern and make sure the processes are in place to support it, e.g. the self-assessment and independent assessment … the board committee approves policies and dispensation, and review and approve the audit reports (with actions). They provide the oversight." (ChiefInfoRiskOfficer-APAC_InsuranceCo) |
| IS_SoftwareConsultant | Yes | "If you look at the corporate governance model, audits are important to support board oversight." (IS_SoftwareConsultant) |

c. Conduct internal audit

> Then you have the audits, who provide the independent checks and gives a report card to the audit committee, the board ... The audits, both internal and external audits, to ensure that the customer has done the right things. (IS_Consultant)

> On the assurance, we have both internal and external assurance. Based on the assurance findings, we will work on the actions and if required, redo the policies in December every year. (ChiefInfoRiskOfficer-APAC_InsuranceCo)

For the internal assurance, we conduct both desktop reviews and on-site reviews, together with risk management and business teams. We will go through the findings, actions and will conduct trainings. (ChiefInfoRiskOfficer-APAC_InsuranceCo)

The auditing module in the application facilitates internal auditors to review and check all internal information security controls implementation. It tracks all the audit checklists and the audit findings, and it will also generate a simple-to-use charts for senior management discussions. (IS_SoftwareConsultant)

Table 6-21 shows representative quotes from the expert interviews that validated this theme.

Table 6-21: Validation data for "conduct internal audit" (assure).

| Dimension: **Assure**; Theme: **Conduct Internal Audit** | | |
|---|---|---|
| **Expert Interviewees** | **Validation** | **Representative Quotes** |
| IS_ConsultingDirector | Yes | "We provide advice to help drive effective corporate governance and audit is also a key component in ensuring good corporate governance. We also have a separate division focusing on risk, i.e. risk advisory which focuses on internal audit … helps customer in conducting assessment of their internal controls covering information security." (IS_ConsultingDirector) |
| IS_Consultant | Yes | "Then you have the audits, who provide the independent checks and gives a report card to the audit committee, the board ... The audits, both internal and external audits, to ensure that the customer has done the right things." (IS_Consultant) |
| CISO-MY_Bank | Yes | "with an independent reporting much like the risk and internal audit teams, where they can have better independence and authority in driving information security program." (CISO-MY_Bank) |
| CIO-SG_InvestmentCo | Yes | "we also adopt the three line of defence as proposed by our auditor … The auditors, both internal and external, are our third line. I believe it is a universally accepted best practice |

| Dimension: **Assure**; Theme: **Conduct Internal Audit** | | |
|---|---|---|
| **Expert Interviewees** | **Validation** | **Representative Quotes** |
| | | for risk management and we adopt the same for information security risk." (CIO-SG_InvestmentCo) |
| ChiefInfoRiskOfficer-APAC_InsuranceCo | Yes | "For the internal assurance, we conduct both desktop reviews and on-site reviews, together with risk management and business teams." (ChiefInfoRiskOfficer-APAC_InsuranceCo) |
| IS_SoftwareConsultant | Yes | "there is also an auditing module. The auditing module in the application facilitates external and internal auditors to review and check all internal information security controls implementation … It tracks all the audit checklists and the audit findings, and it will also generate a simple-to-use charts for senior management discussions." (IS_SoftwareConsultant) |

In addition to the validation of the second-order themes, the expert interviews validated the relationship between "assure" and "direct", which is represented with a one-directional arrow (⟶) connecting "assure" to "direct". Information and insights from "assure" are used to feed back into "direct" should improvements or changes be required. This was validated by ChiefInfoRiskOfficer-APAC_InsuranceCo and IS_SoftwareConsultant:

> On the assurance, we have both internal and external assurance. Based on the assurance findings, we will work on the actions and if required, redo the policies in December every year. (ChiefInfoRiskOfficer-APAC_InsuranceCo)

> In many cases, based on the audits, you may need to relook at your alignment, strategy or information security standards ... Either to meet new requirements or as part of improvements. (IS_SoftwareConsultant)

In addition to the data from the expert interviews, the ISG process together with the sub-processes identified in the ISG process model has been found to be consistent with the functionalities shown on an information security software application during the interview with IS_SoftwareConsultant.

The refined ISG process model developed in Chapter 5 has been validated in Phase 3 and this ISG process model is shown in Figure 6-3 as the validated ISG process model.

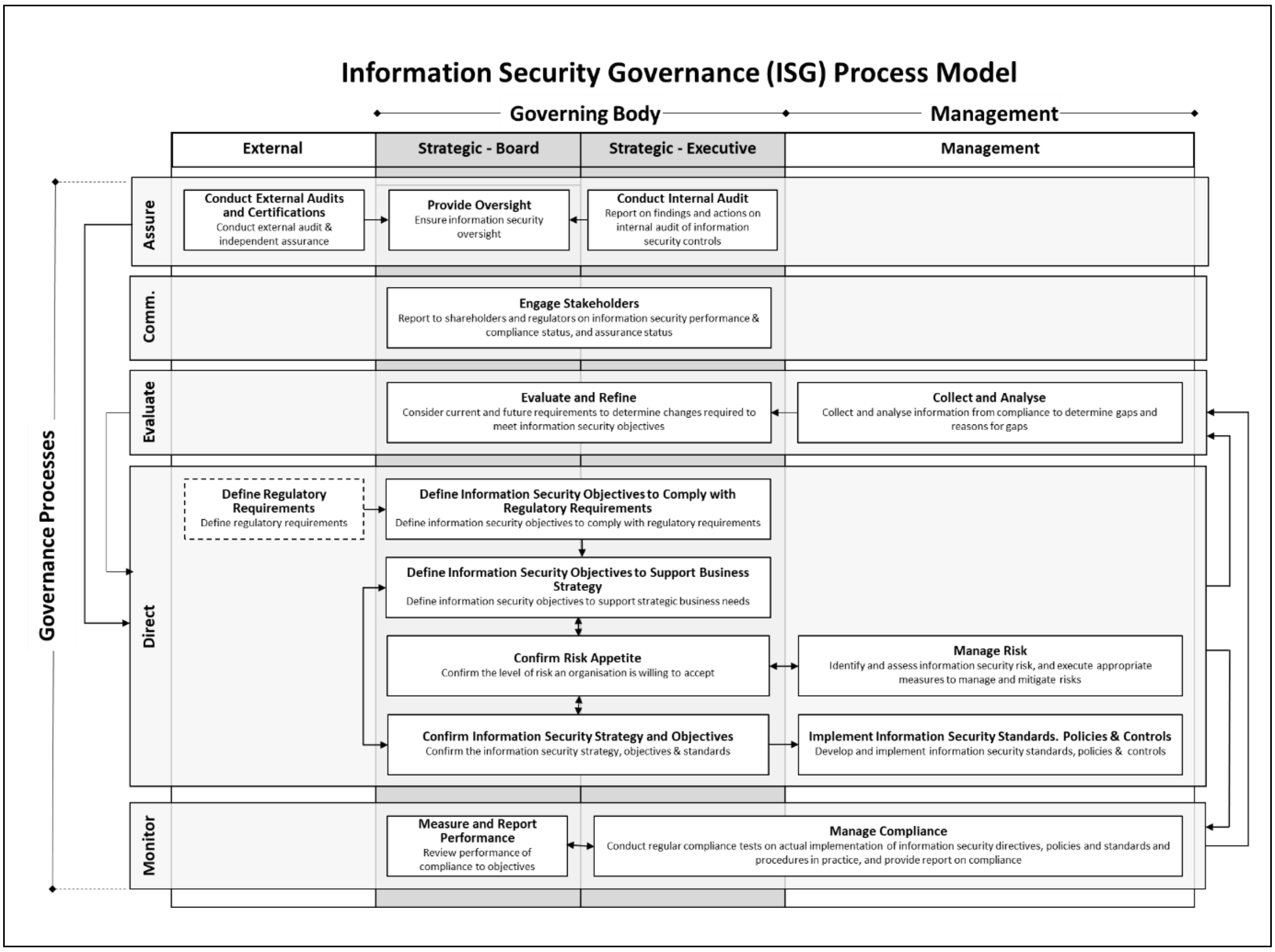

Figure 6-3: Validated ISG process model

## 6.5 Summary

The 6 expert interviewees expressed similar insights during the interviews and echoed the findings of the case study interview findings. The expert interview data were coded and analysed to validate the ISG process model. In the coding and analysis, a template analysis approach was adopted. Although new first-order concepts were identified, further analysis of the relationships between codes and aggregation led to similar second-order themes and aggregated dimensions. The analysis of the final 6 expert interviews served primarily as confirmation of the codes, first-order concepts, themes and aggregated dimensions already identified, and so confirmed that theoretical saturation was achieved in Phase 3. This validated the refined ISG process model as shown in Figure 6-3 as no further data emerged. Besides theoretical saturation, the expert interviews also served to address research validation as multiple sources of evidence and multiple approaches have been utilised to achieve consistent findings.

In essence, Phase 3 has validated that the stakeholder groups in ISG, namely, “external”, ”strategic - board” and “strategic - executive”, which form the “governing body”, and “management”. Phase 3 has also validated the ISG process model that comprises 5 core governance processes, i.e. “direct”, “monitor”, “evaluate”, “communicate” and “assure”. The aim of the validated ISG process model is to facilitate the implementation of ISG in organisations.

The next chapter will discuss the validated ISG process model in the context of existing literature. The implications of the results of this research towards information security research and practice will also be addressed.

# Chapter 7
# Discussion

This chapter discusses the proposed ISG process model and its theoretical and practice integration with extant literature, bringing in similarities and confirmations while explaining divergences from existing research where appropriate. This chapter is structured into 3 sections where Section 7.1 introduces the proposed ISG process model and Section 7.2 discusses the theoretical and practice integration and extension. Section 7.3 identifies the factors that have been discovered to influence the implementation of ISG that are beyond the original objective of the research.

## 7.1 Proposed ISG Process Model

ISG has been identified as a key area of concern in organisations as the board and executive management need to extend their fiduciary duties to include the protection of information because information is now a strategic asset for organisations. ISG models have been introduced by standards and professional bodies (International Organization for Standardization, 2013; National Institute of Standards and Technology, 2018b) with the objective of facilitating organisations to implement ISG, but these ISG models are generally normative models that focus only on the "what" of ISG and not "how" to implement ISG. While organisations have recognised the importance of ISG and there is an increasing need to implement ISG, organisations are continuously challenged as there are little relevant guidance available to help organisations to implement ISG.

Furthermore, the literature review (Section 2.5) has identified key gaps in the area of ISG research:

RG1: Lack of a holistic ISG framework or model that incorporates the broad areas of ISG

RG2: Lack of guidance on how to implement ISG

RG3: Limited ISG frameworks and models that are grounded in empirical studies

RG4: Lack of an ISG framework or model that easily identifies the processes required to be undertaken by various stakeholder groups involved in ISG

The motivation to address the practice problem and the key gaps in research has informed this research. The research has developed an ISG process model (Figure 7-1) that is informed by extant literature and grounded in empirical data that aims to address the research question:

“How can ISG be implemented in organisations?”

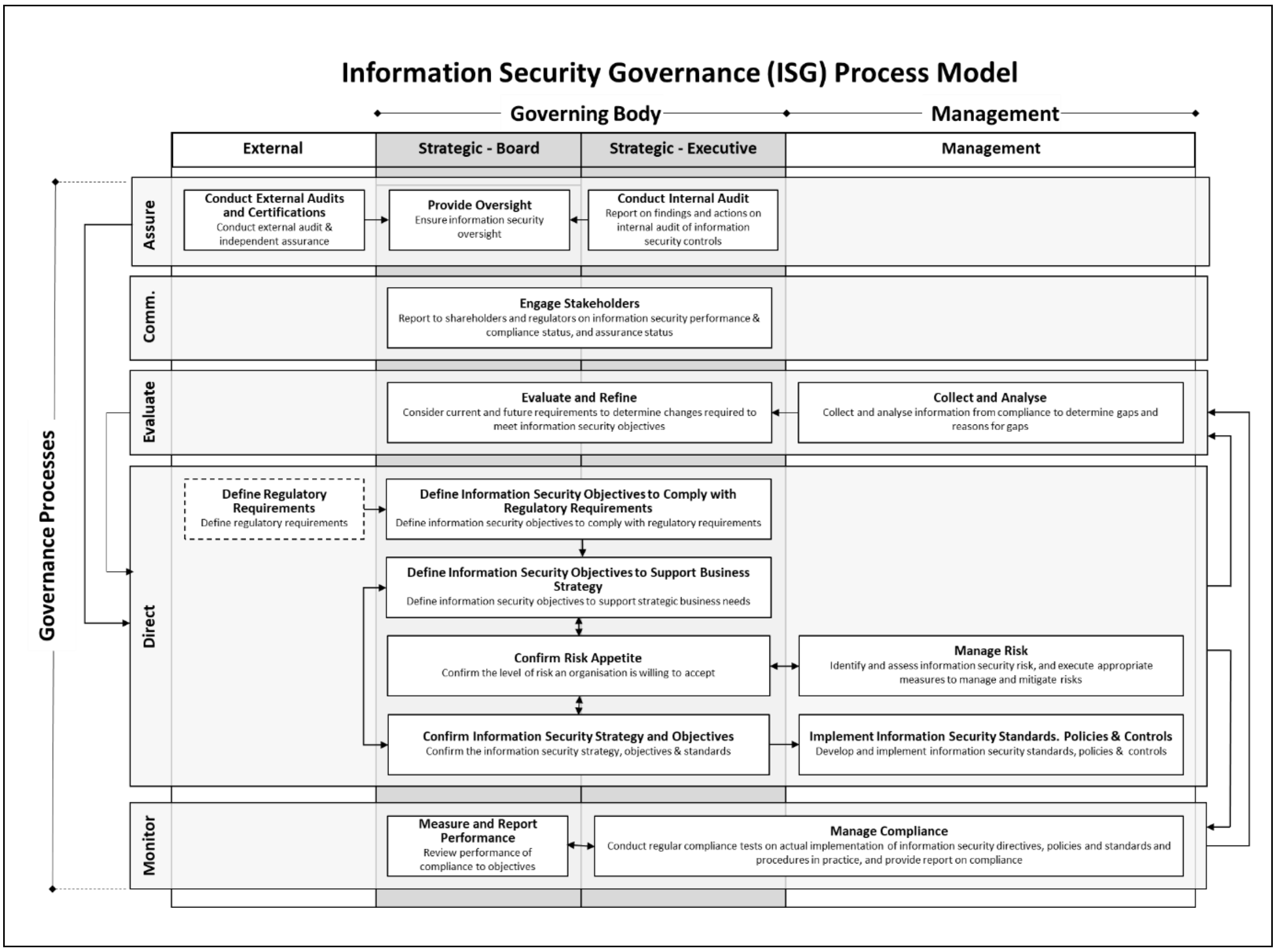

Figure 7-1: Proposed ISG process model

In this study, a conceptual ISG process model was first proposed based on synthesis of previous research, as discussed in Chapter 4. This conceptual model was subsequently refined based on empirical data gathered through case studies, as analysed in Chapter 5, and was finally validated with expert interviews, as presented in Chapter 6. The next section of this chapter provides a discussion of the key components of the ISG process model.

## 7.2 Theoretical and Practice Integration and Extension

The design of the proposed ISG process model incorporates key ISG components that were extracted from extant literature and enhanced with actual practices discovered in the case study research. The key ISG components comprise the stakeholder groups and the core process and sub-processes. Table 7-1 summarises the theoretical integration with extant literature and confirmation and extension from practice.

Table 7-1: Proposed ISG process model – theoretical and practice integration.

| Key ISG Components | Theoretical Integration | Practice Integration |
|---|---|---|
| a. ISG process model design principles | ISG process model design principles have been extracted from extant literature and are consistent with literature. | ISG process model design principles are adopted in case study organisations and were validated in expert interviews |
| • Organisation-wide and business driven | • (Allen, 2006; Buecker et al., 2013; Kayworth & Whitten, 2010; von Solms, 2006; Wu & Liu, 2019) | • Practised across organisations and driven by business requirements. Information security is on the agenda of board meetings and business discussions |
| • Risk management is fundamental to ISG as to corporate governance | • (du Plessis et al., 2011; Lindup, 1996; von Solms, 2001b; Webb, Maynard, et al., 2014) | • Adopted from a key principle of corporate governance (ASX Corporate Governance Council, 2019; Monetary Authority of Singapore, 2018; Securities Commission Malaysia, 2017) and practised in ISG |
| • Clearly identify governance processes with clear roles and responsibilities | • (Bart & Bontis, 2003; Ohki et al., 2009; von Solms & von Solms, 2006; Waitzer & Enrione, 2005; Williams, 2007a) | • ISG processes and sub-process are clearly defined and practised by different stakeholder groups |
| • Closed-loop processes that drive continuous improvement | • (Alqurashi et al., 2013; Mathew, 2018; Ohki et al., 2009; von Solms & von Solms, 2006) | • Closed-loop process of direct-monitor-evaluate is closely followed in practice |
| b. Stakeholder groups and structure | • Clear identification of stakeholder groups with clear responsibilities as identified in extant literature i.e. "external", "governing body" and "management" (Ohki et | • Clear definition of "governing body" comprising board members and C-level executives and management<br>• Insufficient evidence to segregate "management" into "tactical management" |

| Key ISG Components | Theoretical Integration | Practice Integration |
|---|---|---|
| | al., 2009; von Solms & von Solms, 2009, 2006). | and “operational management”<br>• “External” stakeholder group is confirmed as it comprises external auditors, regulators and external consultants |
| c. Core ISG processes / sub-processes | • The proposed ISG process model has extended the core ISG process with additional sub-processes (Ohki et al., 2009; von Solms & von Solms, 2006)<br>• Core ISG processes and sub-processes are consistent with multiple extant pieces of literature and research which focused on specific processes<br>• “Direct” (Alves et al., 2006; Conner & Coviello, 2004; Georg, 2017; Tan et al., 2017; Webb, Maynard, et al., 2014; Yaokumah & Brown, 2014b)<br>• “Monitor” (Da Veiga & Eloff, 2007; Ohki et al., 2009; Park et al., 2006; Tan et al., 2010; von Solms & von Solms, 2006)<br>• “Evaluate” (Ohki et al., 2009; von Solms & von Solms, 2006; Williams et al., 2013; Yaokumah & Brown, 2014b)<br>• “Assure” (Allen & Westby, 2007a; Farrell, 2015; Ohki et al., 2009; Pathak, 2004; Steinbart et al., 2018)<br>• “Communicate” (Alqurashi et al., 2013; Koh et al., 2005; Maynard et al., 2018; Mishra, 2015; Ohki et al., 2009; Sajko et al., 2011) | • The 5 core ISG process are confirmed as practised in the case study organisations<br>• Further sub-processes have been discovered for “assure” and “direct” reflecting detailed sub-processes that are practised in the case study organisations |

### 7.2.1 Theoretical Integration and Extension

This research has developed an ISG process model that incorporates all areas of ISG that have been discovered and consolidated from extant literature. These areas of ISG are encapsulated in the ISG definition as adopted in this research which has framed the scope of this research. In this research, ISG is defined as follows:

> ISG is the framework of rules, relationships, systems and processes by which the security objectives of the organisation are set and the means of attaining those objectives and monitoring performance are determined.

This research has addressed the research gaps in the following areas:

a. ISG process model with detailed processes and relationships/process flows

This research has developed an ISG process model that explains the governance of information security where the core ISG processes and sub-processes together with the relationships/process flows are identified, effectively addressing RG1 and RG4. ISG model research by von Solms and von Solms (2009, 2006) and Ohki et al. (2009) informed the development of the proposed ISG proposed model. The proposed ISG process model has extended ISG research by defining the sub-processes and their relationships/process flows together with the interactions among the stakeholder groups. The interactions among the stakeholder groups also represent the clear roles and responsibilities of these stakeholder groups in ISG. The proposed model is the first ISG process model that has incorporated the 5 core ISG processes and associated sub-processes, together with the relationships/process flows, illustrated in a cross-functional process map which also maps the processes/sub-processes against the stakeholder groups.

The von Solms direct-control cycle (von Solms & von Solms, 2006) defined the core processes at a high level, while Ohki et al. (2009) identified additional core processes such as “report” and “oversee”. In more recent research, Mathew (2018) also developed a high-level process model based on the plan–do–check–act cycle model of Deming but their ISG process model identified only the key processes and did not include any detailed sub-processes.

This research has addressed the lack of process models in both information systems and specifically information security research (Mathew, 2018; Shaw & Jarvenpaa, 1997) by developing an ISG process model that incorporates all core ISG processes and sub-processes, as well as the relationships/process flows and stakeholder groups involved in the governance of information security.

b. ISG process model which focuses on “how” to implement ISG

The literature review has found that the majority of ISG model research focused on defining the “what” in ISG. Among the ISG models that were developed to facilitate implementation of ISG, these were primarily normative models that describe and identify what *should* be done in governance of information security. Three ISG process models (Mathew, 2018; Ohki et al., 2009; von Solms & von Solms, 2006) attempted to address the “how” by identifying the core ISG processes but remained at a very conceptual level.

This research has developed an ISG process model from practice to explain how to implement ISG, addressing RG2 and RG4. The model illustrates all the core ISG processes together with the sub-processes that are required for ISG practice. Rather than just presenting these processes and sub-processes as identified components, as in many previous research and normative models (Da Veiga & Eloff, 2007; Kim, 2007; Mathew, 2018; Ohki et al., 2009; Park et al., 2006; Peggy et al., 2011; von Solms & von Solms, 2006), this model illustrates the relationships and process flows of all the core processes and sub-processes. In addition, the model illustrates how the various ISG sub-processes are mapped against the stakeholder groups in a cross-functional process map to identify the roles of the stakeholder groups.

The identification of the core ISG processes and sub-processes, and the relationships and process flows, together with the mapping of the stakeholder groups as illustrated in the process model will facilitate organisations in implementing ISG.

c. ISG process model grounded in empirical data

This model addresses RG3. The majority of ISG models in the literature are conceptual and have been derived from analysis of theoretical ISG concepts, principles and best practices. Among the 9 ISG models grounded in empirical data, only two ISG models (Mishra, 2015; Musa, 2018) were informed by case studies.

This research is the most comprehensive field study conducted to date that refines and validates an ISG process model using empirical evidence from real-world practice (17 interviews, complementary evidence such as documents as well as participant observations, validation by 6 expert interviews). This research is distinct from prior work in the following ways:

- It has adopted a case study research methodology where interviewees from the various ISG stakeholder groups from 3 case study organisations were interviewed about how ISG was practised in all 3 case study organisations.
- In addition to interviews, follow-up discussions and process walk-throughs were conducted to observe how ISG was practised in all 3 case study organisations where processes were reviewed and documented.
- The case study interviews and process walk-throughs focused on the ISG processes and activities related to the governance of information security, rather than management of information security.
- Interviews were conducted with interviewees across the different stakeholder groups involved in ISG within the same organisation, which facilitated the identification of segregation of responsibilities between governance and management of information security.

Previous research that adopted a multiple case study approach includes a study by Mishra (2015), who conducted 52 expert interviews across 9 different industries in an attempt to develop organisational security governance objectives, while Musa (2018) conducted interviews across 8 Malaysian public listed companies with the objective of validating the development of an IT security governance and internal controls framework. Neither research project explained how governance can be practised in organisations.

d. ISG process model which covers the broad scope of ISG

Besides being one of the most detailed ISG process models that have been developed, the proposed ISG process model is among the few ISG models that have incorporated multiple areas of ISG, which has normally been the focus of individual studies, thus addressing RG1 and RG4. Extant literature identified detailed research in specific areas of ISG, i.e. research in setting directions (Holgate et al., 2012; Maynard et al., 2018; Rastogi & von Solms, 2006; Yaokumah & Brown, 2014a), control and monitoring (Allen & Westby, 2007b; von Solms & von Solms, 2006), risk management (Ahmad et al., 2012; Antoniou, 2018; Maynard et al., 2018; von Solms & von Solms, 2009), assurance (Anhal et al., 2003; Fitzgerald, 2012; Holzinger, 2000) and communication (Allen, 2005; Bihari, 2008; Georg, 2017; von Solms, 2001b). This research has taken the initiative to bring together the insights developed by these scholars and extend the research to consolidated scholarship in ISG. The result of the research is an ISG process model that incorporates the broad scope of ISG.

e. Additional perspectives from corporate governance theory

The ISG process model has been developed with key inputs from corporate governance theories, which have been used to provide an additional theoretical perspective on the concept of governance. As information is treated as a key corporate asset, the understanding of the theories behind corporate governance has substantiated the process model. Three corporate governance theories, i.e. agency, stakeholder and stewardship theories, were analysed to identify the concepts to be incorporated into ISG. The ISG processes model has incorporated these corporate governance theories, thus extending the applicability of the ISG from a governance perspective. The ISG process model is aligned with key principles of agency theory, which focuses on risk management, controls and monitoring to safeguard the organisation (compliance/risk management model), key principles of stakeholder theory, which balances stakeholder needs by driving the alignment of information security with meeting strategic business requirements (stakeholder model), and finally key principles of stewardship theory, which focuses on performance improvement through strategic business alignment and coordinated decision-making such as execution of directions, decision on risk appetite, and evaluations and improvements across the governance and management stakeholder groups (partnership model). This addresses RG1.

### 7.2.2 Practice Integration

The conceptual ISG process model has been refined with data from 3 case study organisations. The insights on how ISG was practised in these 3 case study organisations, i.e. 3 financial institutions, have informed the following refinements of the conceptual ISG process model.

a. Clear identification of stakeholder groups involved in governance of ISG

There are significant similarities across the extant literature in relation to the need to separate the stakeholder groups for governance and management, but there was no clear identification of these stakeholder groups. The direct-control cycle ISG model (von Solms & von Solms, 2009, 2006) identified 3 level of management, i.e. strategic, tactical and operations, that are involved in ISG, while Ohki et al. (2009) identified corporate executives and auditors involved in ISG and managers involved in information security management. Data from the case study organisations on how ISG is practised has facilitated the validation of the ISG process model. In the proposed ISG model, there are 3 internal stakeholder groups, i.e. “strategic - board”, “strategic - executive” and “management”. Only “strategic - board” and “strategic - executive” are directly involved in ISG and are collectively known as the “governing body”. This “governing body” works closely with the “management”, which is responsible for information security management. There is another stakeholder group, i.e. “external” that is identified in the proposed ISG process model which comprises the external auditors, regulators and consultants that organisations engage to provide independent assurance and assessment.

The insights from practice can be explained accordingly.

Firstly, the proposed ISG process model developed in this study identifies 3 internal stakeholder groups, i.e. “strategic - board”, “strategic - executive” and “management”, which are different from the stakeholder categorisation in the direct-control cycle ISG model (von Solms & von Solms, 2009, 2006). The direct-control cycle ISG model identified two levels of management, i.e. tactical and operational. This difference can be explained by the fact that the direct-control cycle ISG model identified governance processes that involved all stakeholders in an organisation, while the proposed ISG process model in this study is more specific to the stakeholder groups who are responsible for the governance of ISG, i.e. the role of the “governing body”, which comprises “strategic - board” and “strategic - executive”. However, this “governing body” is collectively defined as the “strategic level” in the direct-control cycle model (von Solms & von Solms, 2009). This study did not proceed further in segregating “management”, which includes both tactical and operational management as identified in the direct-control cycle model, as insufficient empirical data on this theme were collected and the “management” stakeholder group is deemed to be responsible for the management of information security, i.e. the execution of the directives from the “governing body”.

Secondly, it is also interesting to draw attention to the differences in the definition of the “governing body” of the proposed ISG process model compared to the normative ISG model in ISO 27014 (International Organization for Standardization, 2013). In the proposed ISG model, the “governing body” includes both “strategic - board” and “strategic - executive”, while the ISO model may not include the executive management as part of the governing body. It is not exactly clear on the definition of the governing body as provided in the ISO 27014 model as the governing body was defined as part of the top management but does not include the executive management, which represents the C-level executives. Furthermore, there is no definition of who top management represents, hence the lack of clarity. The proposed ISG process model includes “strategic - executive” as part of the “governing body” as all empirical evidence support the fact that the executive management are deeply involved in governance processes, especially in working with the board in decision-making and in various oversight activities.

There is another stakeholder group, i.e. “external” that is identified in the proposed ISG process model which comprises the external auditors, regulators and consultants that organisations engage to provide independent assurance and assessment.
The empirical evidence in this study extends the earlier work of previous researchers by confirming and expanding the definitions of the various stakeholder groups who are involved in governing information security.

b. Clear segregation of “monitor” and “evaluate” processes

The “direct-monitor” cycle in ISG was introduced by von Solms and von Solms (2006) as the “direct-control” cycle adopting the governance principles of corporate governance. This was further developed into the “direct-monitor-evaluate” processes by Ohki et al. (2009) and ISO 27014 (2013). This study’s findings have shown that the ability to provide direction and subsequently monitor and adapt to changes are critical in ISG in driving a positive information security environment and culture. A fundamental difference in the study findings is that there is little segregation between “monitor” and “evaluate” in practice although these processes are both being executed, because it is natural that evaluation is done during monitoring to determine the appropriate next actions. However, further findings have concluded that the segregation of “monitor” and “evaluate” is beneficial and they should be segregated to provide clearer roles and responsibilities as these distinct processes may be undertaken by different teams.

## 7.3 Factors Influencing ISG Model Implementation

Hypothetical models that were developed purely on capability identification and synthesis of theoretical models were not able to identify some factors that influence the implementation of these models. The exploratory research design adopted in this study has provided the opportunity to discover additional insights beyond the original objectives of this research.

The following factors have been discovered from data gathered during case studies and expert interviews in the refinement and validation of the ISG model. While these factors have not been identified in previous ISG model research, some of these factors are consistent with research in other areas of information security (Commonwealth of Australia, 2006; Flores & Farnian, 2011; Mishra, 2007; Sajko et al., 2011) and corporate governance (Chhotray & Stoker, 2009; Christopher, 2010; Huse, 2008; Levrau & Van den Berghe, 2007) that considered effectiveness in the implementation of ISG and corporate governance.

While the proposed ISG process model as shown in Figure 7-1 may not change, the following factors influence the emphasis of the governance processes during implementation, adoption and acceptance of the ISG model, and ultimately influence the effectiveness of the governance of information security.

### 7.3.1 Regulatory Environment

Regulatory requirements influence governance and dictate the development of internal control structures and policies. Critical infrastructure industries such as financial services, utilities, telecommunications and utilities operate in heavily regulated environments and various regulatory compliances are required to ensure the continuous operations of these businesses. Regulatory requirements imposed by regulators in the financial services industry such as APRA (Australian Prudential Regulation Authority (APRA), 2019c), Monetary Authority of Singapore (Monetary Authority of Singapore, 2013) and Bank Negara Malaysia (Bank Negara Malaysia, 2018) have defined strict frameworks and requirements for information security. These requirements drive, and to a certain extent dictate, the governance approach of financial institutions to information security risk management, governance structures and internal information security policies. Additional penalties and fines for breaches of these regulations drive a strong regulatory compliance business environment. This regulatory compliance business environment in financial institutions was apparent in all 3 financial institutions in the case studies and confirmed many studies on this topic (Georg, 2017; Kim et al., 2008; Williams, 2014). Georg (2017) highlighted that further alignment with global regulations such as Sarbanes Oxley (Sarbanes-Oxley Act, 2002) and European Union Global Data Protection Regulation (European Parliament and Council of the European Union, 2016) has forced board members to become involved in stricter governance of regulatory compliance to avoid global sanctions.

### 7.3.2 Emphasis on Corporate Governance

Corporate governance promotes investors' confidence as it provides a solid foundation for management and oversight (ASX Corporate Governance Council, 2019). As information security risk is a key business risk in today's business environment, a similar perspective to that of corporate governance is applied to ISG.

Since corporate governance is associated with independence, board characteristics, audit committee, risk management, compliance, transparent disclosures and accounting and auditing with effective internal controls, strong corporate governance discipline helps in translating to a good practice of ISG as both adopt similar concepts and processes (Holzinger, 2000; Pathak, 2004; Rothrock et al., 2018; von Solms, 2001b). The influence of corporate governance on ISG was evident in the case studies, as organisations with good corporate governance aim to adopt ISG although ISG is not clearly defined. The discussions with board members and senior executives of financial institutions in the case studies always referred to their respective codes of corporate governance (Monetary Authority of Singapore, 2018; Securities Commission Malaysia, 2017) when the various ISG processes such as compliance, assurance and information disclosure were explored.

Thus, smaller organisations which are not subject to corporate governance may face challenges in implementing ISG as they may not be used to adopting such practices of good governance.

### 7.3.3 Power Distance Index

The power distance index (PDI) is one of the natural cultural dimensions identified by Hofstede (2001) that shows the degree of inequality among employees of organisations within a country and he argued that there is a correlation between a country's PDI and the readiness to accept power differences. This can be interpreted as that countries with higher PDI will drive a more compliant working environment where there is strong dependence of employees on their management and there may be less violation of standards, policies and procedures (Alshare & Lane, 2008). Based on the PDI country comparison for Malaysia and Singapore (Hofstede Insights, 2020), it is shown that both Malaysia (score of 100) and Singapore (74) have high PDI scores compared with countries with a Western culture such as the USA (40), Australia (38) and the UK (of 35). Both PDI scores for Malaysia and Singapore are high, which means that there are clear hierarchical structures in organisations, subordinates expect to adhere to rules and there is little challenge of authority. These hierarchical structures in organisations and compliant working environments are clearly evidenced in the study findings where ISG governance has a strong emphasis on compliance, i.e. operational compliance with standards, policies and procedures, and strict organisation compliance with regulators' requirements and heavy penalties for noncompliance. The strong "direct-monitor-evaluate" and "assurance" processes are diligently adopted and adhered to in the governance of information security, creating an information security-compliant culture.

While a high PDI translates to a strongly ISG compliant culture, the flipside is that this may discourage the voicing of ideas and feedback. The governance of information security may suffer from a lack of diversity in decision-making where insights on the operational levels are important for quick decision-making and feedback from the operational levels is required to drive innovations and continuous improvements in updating controls, procedures, policies and standards (Maynard et al., 2018).

PDI influences the implementation of ISG, hence a more detailed ISG model may be required to facilitate the implementation of ISG in high-PDI countries as their organisations need details and specifics to follow, while a high-level normative model may suffice for organisations in countries with lower PDIs.

### 7.3.4 Maturity Level of Information Security

The maturity level of information security indicates the level of rigour and sophistication in information security risk management practices (National Institute of Standards and Technology, 2011). Based on the 2017 *Cyber maturity in Asia-Pacific region* report published by the Australian Strategy Policy Institute International Cyber Policy Centre (Uren et al., 2017), Singapore (weighted score = 87.7) was ranked 4th, while Malaysia (weighted score = 73.2) was ranked 7th among 25 Asia-Pacific countries in terms of cyber maturity. The International Telecommunication Union also ranked Singapore (score = 0.898) above Malaysia (score = 0.893) in its 2018 Global Cybersecurity Index (International Telecommunication Union, 2019), which is a benchmark for the level of commitment of countries to cybersecurity covering 5 key pillars of legal, technical, organisational, capacity building and cooperation in building a national cybersecurity culture.

Based on the maturity level of information security, the empirical findings of this study may explain why FinServices_SG, which is located in Singapore which has a higher maturity level, focused more on risk management processes as compared to FinServices_SEA and FinServices_MY, which are located in Malaysia which emphasises compliance. Organisations with a higher maturity level of information security tend to focus on other ISG areas, beyond pure compliance to regulatory requirements.

The findings in this study allude to the potential influence of the maturity level of information security on ISG implementation in organisations, as organisations at different levels of maturity have different emphases on governance processes, as discussed in previous research (Brown & Nasuti, 2005; Da Veiga & Eloff, 2007; Maleh et al., 2018; Sajko et al., 2011).

### 7.3.5 Composition of Board Members

Studies have shown that an engaged board is critical in driving the right culture in governance, be it corporate governance or ISG, and this is true for driving the right information security culture (Anhal et al., 2003; Barker, 2015; Beretta, 2019; Bihari, 2008; Conner et al., 2003; Deloitte, 2015; Ernst & Young, 2018; Williams, 2007a). The right composition of board members, which consists of board members who have strong understanding of technology, information security and risk management, will set the right tone at the top. This is demonstrated by allocating adequate time to discuss information security matters as a business issue, having the right communication process in updating the board members on information security initiatives and incidents, and getting involved in oversight practices and disclosures by asking the right questions. These practices are evident in this study, where all 3 case study organisations showed the importance of appointing board members who were former technologists or consultants who understood technology, information security and risk management, and who could actively contribute to the oversight of information security. Empirical findings in the study have shown a shift in focus to appointing board members who are not just retired government bureaucrats, former accountants or lawyers, but IT professionals and consultants.

### 7.3.6 Structure and Responsibility of Information Security

The study has shown that responsibility for information security is assigned to various departments in organisations. While the proposed ISG process model may be the same, the focus and emphasis on the ISG processes vary depending on where ISG sits in an organisation. When information security responsibilities are assigned to a CISO who reports to a CIO, the focus of ISG can be very technology driven with a strong emphasis on IT security. When information security responsibilities are assigned to a CISO who reports to a CRO who has a larger responsibility for the total organisation risk, ISG will tend to be more holistic, with an organisation-wide perspective. In a similar manner, when information security responsibilities fall under the chief legal officer or CCO, the ISG will be compliance driven. Therefore, it is important to determine where information security responsibilities sit within an organisation in order to understand the emphasis and impact of ISG in driving compliance, risk management and innovation (Aguilar, 2014; Deloitte, 2019; Tan et al., 2010). A number of academic studies (Moulton & Coles, 2003; Rothrock et al., 2018; Williams, 2007a) and professional publications (Australian Prudential Regulation Authority (APRA), 2019c; Deloitte, 2015, 2018; Ernst & Young, 2018) have recommended that information security should be given higher priority as a business concern as information security risk is critical to business operations. When ISG is implemented as an organisation requirement similar to corporate governance, ISG will achieve the right focus in driving the required information security posture in an organisation, i.e. information security needs “a seat at the table” (Deloitte, 2019).

### 7.3.7 Awareness and Training

Awareness and training on information security risk and actions required to protect from information security breaches have been highlighted as key factors in driving a strong information security posture of an organisation (Holzinger, 2000; Peggy et al., 2011; von Solms, 2001a). Moreover, the awareness of information security of the board and senior executives of organisations strongly influences the governance of information security, as this drives a top-down information security awareness culture across everyone in the organisation (Georg, 2017; Williams, 2014, 2007a). A well-informed board ensures information security topics get a seat at the leadership table (Deloitte, 2017; Ernst & Young, 2019b). They encourage regular engagement in information security-related discussion, updates on information security initiatives and incident reporting, and general industry updates. The awareness and training provided to the board ensures that the board is kept abreast of information security trends and requirements, and facilitates decision-making such as providing approval for information security standards and policies, setting the information security risk appetite and discussion and approval of investments required in managing information security. Such training for the board is encouraged by various regulators in driving more effective ISG (Australian Prudential Regulation Authority (APRA), 2019b; Monetary Authority of Singapore, 2015; Securities Commission Malaysia, 2016). The interview data showed that the case study organisations were actively organising training and update sessions for their boards of directors to equip them with relevant knowledge to govern better.

While this research did not set out to focus on these influencing factors, the exploratory nature of the research has discovered these influencing factors from the case study and expert interviews conducted during the study. This study was not able to explore these influencing factors in more detail due to the study focus and limitations of the scope and resource availability; however, these influencing factors could be interesting areas of research in future study of ISG implementation, adoption and effectiveness, and ISG implementation across different industries and cultures.

## 7.4 Summary

This chapter has examined the proposed ISG process model in relation to existing research and discussed its theoretical integration and extension with extant literature. The discussion has confirmed that the proposed ISG process model is consistent with extant literature and research and expands on existing ISG model research including ISO 27014. In certain areas where divergences from existing research were identified, discussion has explained the divergences. Based on the empirical data obtained from the case study and expert interviews, this chapter has also identified other factors that influence the implementation of ISG. The study has found that while the ISG process model is similar, the focus on ISG may differ due to these influencing factors, contributing to differing emphasis on regulatory compliance, strategic business alignment and risk management.

The next chapter will conclude this study by summarising the research background, research methodology, how the research questions have been addressed, the proposed ISG process model, and the contributions and implications of this study. It will also discuss the limitations of this study and outline directions for future research.

# Chapter 8
# Conclusions and Future Directions

This final chapter of the thesis summarises the research and demonstrates how the objectives and the research question as introduced in Chapter 1 have been addressed. It then discusses the main contributions and implications of the study to academic research and practice. The chapter also highlights the limitations of this research and opportunities for future research.

## 8.1 Overview of Research

ISG has gained greater importance recently as businesses are becoming more digitalised and the roles of the board and executive management are receiving greater scrutiny in protecting their businesses from information security incidents. ISG has gained significant importance in both academic research and practice as organisations are looking to improve the ways to govern information security, because information is a strategic asset. As a response to this requirement, research in ISG covering areas such as information security strategy and policy definition, risk management, controls and compliance, incident response, communications and ISG frameworks and models has increased over the last two decades. However, the literature review has alluded to the fact that research in ISG frameworks and models specifically has been fragmented, not cumulative, and has adopted diverse interpretations of the concepts of ISG.

It is imperative that research helps provide a clear definition of ISG, i.e. what is ISG, why we need ISG and how ISG can be implemented to improve the overall governance of information security in driving towards a sound information security posture in organisations. In addition, the literature review has also confirmed the need for an ISG process model that is grounded on empirical findings that can be practically implemented in organisations to improve the governance of information security, hence the objective of this research.

The aim of this research has been to answer the research question:

"How can ISG be implemented in organisations?"

To answer this question, the research developed an ISG process model to explain how ISG can be implemented in an organisation. This research started with an initial literature review to understand the state of research in information security, corporate governance, ISG and specifically the development of ISG models. The literature review provided the underlying concepts for ISG and the stakeholders that are involved in the governance of information security, which informed the development of the conceptual ISG process model. An additional literature review on corporate governance theories provided an additional perspective to ensure that the conceptual ISG process model was also consistent with corporate governance requirements.

An exploratory research approach was adopted in this study. First, a conceptual ISG process model was proposed based on an assessment of existing hypothetical ISG frameworks and models, as well as theories of corporate governance. Subsequently, the conceptual ISG process model was refined through case studies of 3 financial institutions where 17 interviews with board of directors members and C-level executives were conducted. Finally, the refined ISG process model was validated with 6 expert interviews with experts comprising information security consultants, a CISO, a CIO and a chief information risk officer. The result is a validated ISG process model that is grounded on empirical data, as shown again in Figure 8-1 .

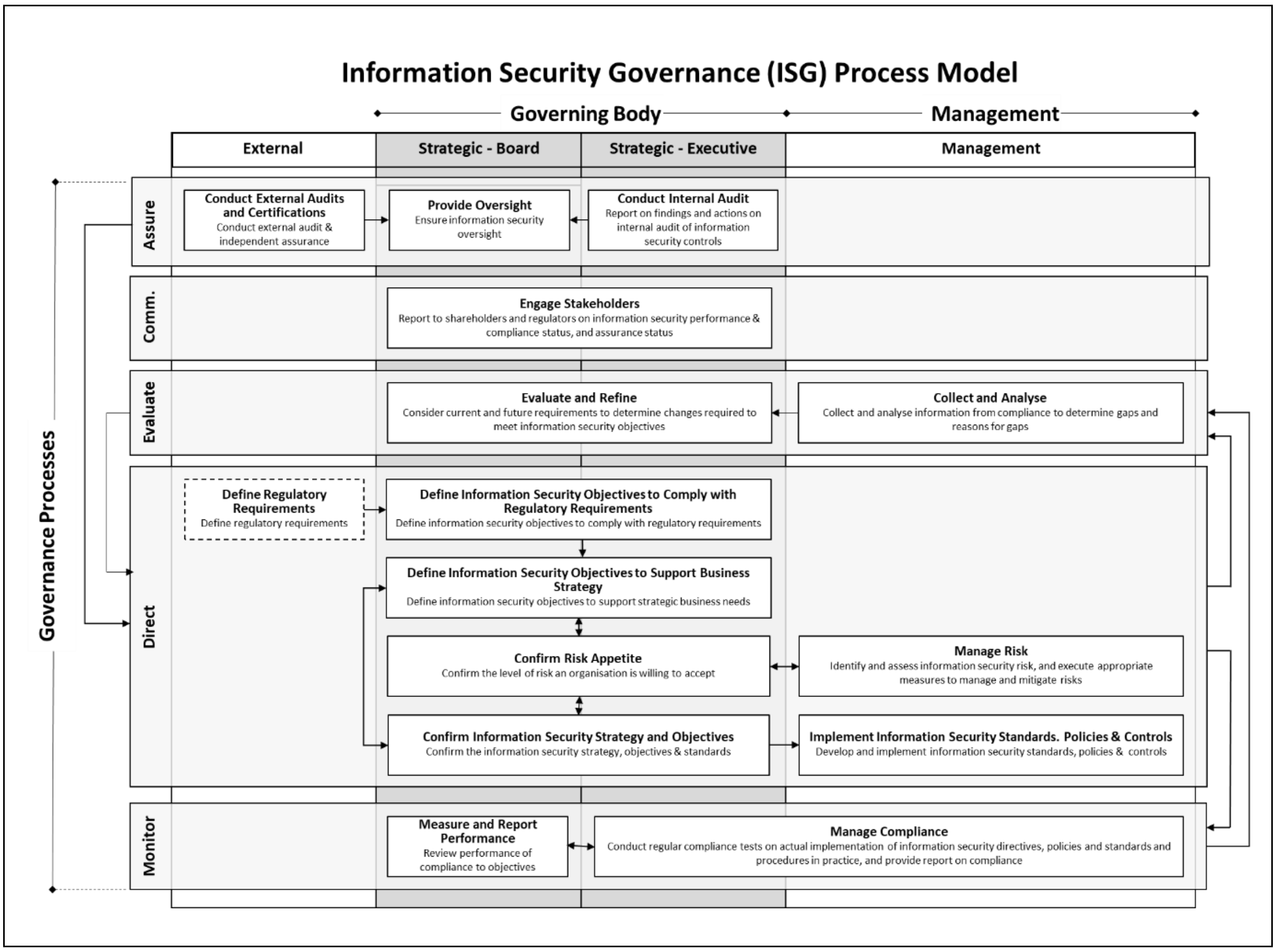

Figure 8-1: Proposed ISG process model

The proposed ISG process model shows the 5 key ISG processes and the related sub-processes, their interactions and the mapping against the ISG stakeholder groups, providing a clear illustration of the sequence of processes. The proposed ISG process model provides a practical reference model to facilitate the implementation of ISG in organisations. With the proper implementation of ISG, it is hoped that organisations will improve their governance of information security.

## 8.2 Contributions and Implications of Study

This exploratory research has taken a participatory form to study real-world problems and sought to bridge the theory-practice gap (Mathiassen, 2017). The primary contribution of this research is the development of an ISG process model to answer “how” to implement ISG in organisations by explaining how ISG is operationalised in organisations. Although limited to the scope of this specific study, this research contributes to solving the practical problem of improving ISG in organisations while also contributing to the theoretical foundation of ISG models.

### 8.2.1 Contributions and Implications for Theory

ISG in organisations is an under-researched area with little empirical backing or theoretical development. Although there are generic guidelines in industry standards on how to conduct ISG, there is very little guidance on how organisations can *practise* ISG (particularly through the integration of people, processes and technology). It is also apparent from the literature reviews that there have been many attempts to develop ISG frameworks and models, and each of these frameworks and models has a different focus or adopts different underlying theories and principles. There is also a gap where research on ISG frameworks is fragmented in that it does not build cumulatively on knowledge from prior research.

Firstly, from a theory perspective the ISG process model is a contribution to ISG process theory. Previous research on ISG frameworks and models mainly focused on identifying the key components of ISG and the critical success factors required for ISG, while the identification of processes involved in ISG received little attention. The justification for developing process theory is that: (1) ISG is a process theory, i.e. a series of events occurring within an organisational context; and (2) the ISG process is complex and multifaceted, so it is best studied through the interpretations of stakeholders. This research has developed an ISG process model that identifies all the ISG processes required to govern information security, namely, “direct”, “monitor”, “evaluate”, “communicate” and “assure”, and the sub-processes, the sequence of these processes and sub-processes, and the interactions among the various stakeholders in organisations. The mapping of the processes against the various stakeholder groups has also assisted in the definition of the roles and responsibilities of the stakeholder groups.

Secondly, in developing the ISG process model this research has also developed an information-processing perspective on ISG, as the process model identifies the sources of information, the requirements of the communication flows and the relationships between the stakeholder groups. The research has developed an information-processing network and explained how organisations can control information flow to increase the effectiveness of ISG. It is hoped that this perspective on an information-processing capability can attract experts in information theory to further develop our information-processing perspective towards improved effectiveness in ISG.

Thirdly, by adopting a multiple case study methodology and expert interviews, this research has empirically examined the implementation of ISG in 3 case study organisations where data were validated by expert interviews. Multiple sources of data which comprised stakeholder interviews, process walk-throughs and documentation were analysed in each case study to provide triangulation of data sources. Furthermore, cross-case analysis was conducted to identify process patterns across the cases to ensure theoretical replication and improved generalisability of the process model. This research is a comprehensive study where an empirically grounded ISG process model is developed and validated through a multiple case study methodology and expert interviews.

Fourthly, this research has studied key corporate governance theories, i.e. agency theory, stakeholder theory and stewardship theory, to provide the additional perspective of governance so that the proposed ISG process model is consistent with the expectations of corporate governance. This is an important contribution to research as businesses are becoming more digitalised in the new economy and information is now a critical asset to businesses, so that ISG is inseparable from corporate governance, thus bringing closer research on information systems and on business management. Information security and ISG research are primarily done as within the information systems discipline and very little in the business management discipline. It is hoped that these findings provide initial insights for future collaboration between information security and business management disciplines in ISG research.

The 5th contribution of this research is that it has identified some original findings in terms of factors influencing ISG that were not highlighted in previous ISG model development. These include, in particular, the impact of strict regulatory requirements and legislation, the maturity level of information security risk management and the influence of behavioural and cultural perspectives such as the profile of the board of directors, composition of the board and PDI. These influencing factors could be examined as part of detailed process theory research as these could represent the antecedent, contextual conditions or external constraints that influence the occurrence of events that shape the processes and the outcomes which affect the effectiveness of the model (Radeke, 2010).

The 6th and final contribution of this research is the development of one of the most comprehensive ISG process models that is grounded on empirical findings. Extant literature has indicated that most ISG frameworks and models have been developed as hypothetical models based on research on the key requirements and principles of ISG. This study has resulted in the development of an ISG process model that has been built on synthesis of cumulative knowledge from previous research and has been validated with empirical data providing further evidence on existing ISG theories and concepts. The research has expanded the seminal research on ISG by von Solms and von Solms (2009, 2006) and the ISG models of Ohki et al. (2009) and ISO 27014 (2013).

### 8.2.2 Contributions and Implications for Practice

Extant literature has confirmed the difficulty of understanding ISG and the challenges in implementing ISG in organisations, hence the motivation of this research in answering the research question. Moreover, there is a lack of empirical work in this area of ISG and ISG models. A few professional publications and standards (Gartner, 2010; International Organization for Standardization, 2013; National Institute of Standards and Technology, 2011) have introduced ISG models and practice guides to help organisations to implement and improve the governance of information security, but these are normative models based on expert opinions (and not necessarily empirically grounded) that specify what is required to govern information security, but do not inform how organisations can implement ISG. The need for unique ISG models to address specific technologies (e.g. artificial intelligence, the cloud, the Internet of Things, mobile applications) and business requirements (e.g. privacy, knowledge leakage) suggests that current ISG models are not flexible enough to adapt to the dynamic environment of information security risk (Lidster & Rahman, 2018), hampering the implementation of ISG in organisations. It is apparent from extant literature that there is still a substantial theory-practice gap in ISG and this research contributes to information security practice by introducing a practical ISG model that incorporates the key ISG components, i.e. 5 core governance processes and 4 stakeholder groups that can be adapted with modifications to be implemented by organisations across different industries implementing various technologies. The significant contribution is the introduction of a practical reference ISG process model that expands the normative model introduced by ISO 27014, expanding the “what” into “how” ISG can be implemented by practitioners.

The second contribution to information security practice is that the ISG process model identifies the core ISG processes together with their sub-processes using a process flow approach. This process flow illustrates how the ISG processes interact and integrate, translating ISG into a procedural approach that facilitates ISG implementation in organisations.

In addition to illustrating the core ISG processes, the proposed ISG model maps the processes against key stakeholder groups - defining the roles and responsibilities of the stakeholders in organisations, namely, the board of directors, the C-level executives and the rest of management. This is a significant contribution to practice as it confirms that information security is organisation-wide and the whole organisation is involved, albeit in different roles and capacities. This mapping of processes against stakeholder groups facilitates the implementation and tracking of ISG processes and sub-processes, and the translation to actual activities that are undertaken by the different stakeholder groups.

The final contribution of the ISG process model is that the model ensures that all ISG processes are aligned with corporate governance concepts, thus complying with corporate governance requirements and simplifying the understanding of ISG among non-information-systems and non-information-security practitioners. ISG is also identified as an organisation-wide responsibility and not delegated to IT. This can be illustrated by a few practical scenarios:

a. Budget for information security programs can be allocated based on business and risk management strategies and regulatory compliance requirements. This facilitates traceability and auditability in information security spending.
b. Information security programs are monitored and evaluated to ensure they adhere to original plans so that divergence from plans can be quickly identified and addressed.
c. There is a clear communication process that promotes transparency in disclosures to all internal and external stakeholders covering, e.g. management, employees, regulators, customers and shareholders.
d. There is an assurance component that helps the board of directors drive information security oversight, just like financial oversight in corporate governance. This assurance process aligns ISG with best practices in governance, risk and compliance, e.g. adopting the “three lines of defence” in assurance and risk management.

Ultimately, from a practice perspective: (1) process models are easier to visualise for practitioners than other types of models (e.g. variance models); (2) process models are easier to implement, as practitioners can structure their thinking according to the stages of the process model and change activities in their organisation; and (3) process models provide concrete *practice* insights rather than construct correlations (as is the case with variance models).

## 8.3 Limitations

While this research has taken all the required steps to ensure the validity and reliability of the findings, this study is not without limitations. The research focuses on interviewing key participants who are part of the governing body and this generally means personnel on the board and C-level executives who are directly or indirectly involved in the governance of information security. The purpose of this selection was to ensure that the research was able to obtain a real-world understanding of the practices involved in ISG. The first limitation in this research is that there was not enough time to expand the interviews to more participants covering personnel who were not involved in the governance of information security within each case study organisation, which could have provided insights from “outsiders” looking in to see if ISG was being performed as the participants claimed, e.g. the interplay between the governance of information security and the broader governance of the organisation.

Secondly, while information security is a highly researched area and a frequently discussed topic in practice, it is also a sensitive topic that is shrouded in confidentiality and many organisations would not be comfortable divulging details associated with their internal information security strategies and information security incidents as it might compromise their market-competitive positions. As the research was conducted in the financial services industry, which is highly regulated with strict legislation on privacy and confidentiality, there was difficulty in securing case study organisations and interviewees as not many were interested in discussing information security openly with external parties. In addition, interviewees who did agree to participate in interviews may have been hesitant to share specific data that would have enabled a more detailed extraction of information for analysis. This might have provided more specific information on recommendations for strategic priorities, criteria for prioritisation of budget allocations and risk management approaches in ISG.

There are also specific limitations regarding the case study organisations due to their concerns about confidentiality. Process walk-throughs by information security teams were undertaken in addition to participants' interviews to enable the researcher to better understand the processes that were involved in the governance of information security. While there were some process walk-throughs, many such discussions were done with PowerPoint slide presentations where the researcher was not able to observe the actual processes. Artefacts such as policy and control documents were shown but were not able to be collected as research evidence. However, these limitations are not expected to have any impact on the reliability, accuracy or completeness of the results.

The research is based on a multiple case study method conducted in 3 financial institutions in South-East Asia and validated with another 6 expert interviews with participants across different information security roles and functions to validate the ISG model. The methodology adopted has provided the required reliability and validity in the development of the ISG process model. While the financial services industry is the most mature in embracing ISG and corporate governance, it would have been the researcher's preference to conduct a similar set of case studies in another country such as Australia or another industry such as telecommunications or utilities to provide richer sets of empirical data in order to confirm if the ISG model was consistent across different countries and industries. Further research could be expanded to cover different countries and industries, as mentioned in more detail below in Section 8.4.

The last limitation of this research is potential research bias, as the researcher was the primary instrument for data collection and analysis in this qualitative research. The researcher has been working in the IT consulting and financial services industry, and therefore several of the participants in the research were known to the researcher. However, the topics of information security and ISG under investigation were new to the researcher and therefore the researcher had little influence on the behaviour of the participants and the interpretation of the data gathering and analysis processes. Furthermore, the researcher has regularly discussed the methodology and findings with supervisors during the data collection and analysis process, which has helped to minimise such researcher bias.

## 8.4 Future Research

This study has identified several interesting areas in ISG for future research. These future research areas have been identified from observations and the results of empirical findings and analysis, and some are derived from the limitations of this research.

In developing the ISG process model, this research has also developed an information-processing perspective on ISG. Further research could be conducted to study the function of information processing and understand how information processing can affect the implementation of ISG in organisations, and thus the effectiveness of ISG in organisations. With a similar intention, more research is needed to study ISG based on process theory to understand the antecedent conditions, contextual factors or external constraints that influence the occurrence of the processes and sub-processes, and their impacts on the outcomes and consequences. Research based on process theory would contribute to the identification of the various factors that impact on the effectiveness of ISG in organisations. Further, more research needs to be conducted around the management stakeholders, particularly on the segregation of tactical and operational governance aspects and the breakdown of processes for each.

This research has developed a proposed ISG process model based on empirical data from 3 case study organisations in the financial services industry. Subsequent research could be conducted in other critical infrastructure industries such as telecommunications, utilities and healthcare to further validate the model and confirm the ISG processes so that generalisations can be made to a standard reference model for ISG implementation across different industries. Furthermore, research could be undertaken across different industries with different levels of regulatory requirements as a comparative study to understand the impact of regulatory requirements and legislation on ISG processes and implementation. This would allow researchers to understand the role of regulatory requirements in driving ISG and the rationale behind the different emphases of ISG processes such as oversight, compliance and risk management.

Another critical area for future research would be to study the correlation of the proposed ISG process model with the maturity level of information security of an organisation. This research could be conducted in different organisations that are specifically selected based on their maturity level of information security as per the NIST framework (National Institute of Standards and Technology, 2018a) or the Cyber Maturity in Asia Pacific Region report published by the Australian Strategy Policy Institute International Cyber Policy Centre (Uren et al., 2017). Such research would provide insights into the different emphases of ISG processes or lack of certain processes in less mature organisations, allowing organisations to fine-tune ISG implementation according to the maturity level.

Future research on ISG could be undertaken from the organisation cultural perspective to study the evolving roles of the board of directors, C-level executives and management, and the influences of these stakeholders on the decision-making approaches in the governance of information security. The current study has alluded to initial findings that the composition and the background of board members influences ISG. Similar studies have been carried out on the influence of boards of directors on corporate governance (Huse, 2008; Ittner & Keusch, 2015; Kiel & Nicholson, 2003), therefore it would be important to undertake similar studies to provide insights into the composition of boards of directors in driving better governance of information security, especially in the areas of strategic business alignment and risk management.

This research on the development of an ISG process model has also been done on case study organisations in Singapore and Malaysia, where the PDI is high. Future research could be done to expand to case study organisations located in lower PDI countries such as Australia, the UK and the USA. Such research would help to identify the reasons behind the compliance-driven nature of ISG in different countries, the differing emphasis on compliance versus risk management and the decision-making ability of management and operational stakeholders. In addition, similar research in differing PDI countries could help validate the general applicability of various existing frameworks and models that have been developed based on research in Western countries, which generally have lower PDIs.

This research focused on the governance of information security which involves the processes that have been identified to cover “direct”, ”monitor”, ”evaluate”, “communicate” and “assure”. The proposed ISG process model shows that the “governing body” which comprises of the “strategic - board” and “strategic - executive” is responsible for the governance of information security while the “management” is responsible for the management of the information security. The ISG process model identifies the processes and sub-processes to be undertaken by the “governing body” in governing information security with the aim of making it simpler to operationalise ISG. The operations of the information security which is undertaken by the operational team as identified in von Solms (2006) is not shown in the proposed ISG process model as the focus of this research has been on ISG. A future research is recommended to extend the study to cover the management and operational aspects of ISG which are the responsibilities of the middle and lower management/administration (von Solms & von Solms, 2006). This future research can study the relationships between governance, management and operations of information security.

ISG is an important topic that continues to gain prominence in both academic research and practice as organisations continue to look for ways to govern information security, as information security risk is recognised as one of the key risks in the digital economy. The proposed future research in various areas of ISG would collectively contribute to both theory and practice in understanding the approaches to designing ISG processes that help improve ISG in organisations.

# Appendix A: Ethics Approval

Ethics approval from Human Ethics Advisory Group (HEAG) – Ethics ID: 1749890

THE UNIVERSITY OF MELBOURNE

27 September 2017

Dr A Ahmad
Computing and Information Systems
The University of Melbourne

Dear Dr Ahmad

I am pleased to advise that the Engineering Human Ethics Advisory Group has approved the following Minimal Risk Project.

| | |
|---|---|
| Project title: | **Improving Information Security Governance in Organizations** |
| Researchers: | **Dr A Ahmad, C Wong, Dr S B Maynard** |
| Ethics ID: | **1749890** |

The Project has been approved for the period: **9-Sep-2017 to 31-Dec-2017.**

It is your responsibility to ensure that all people associated with the Project are made aware of what has actually been approved.

Research projects are normally approved to 31 December of the year of approval. Projects may be renewed yearly for up to a total of five years upon receipt of a satisfactory annual report. If a project is to continue beyond five years a new application will normally need to be submitted.

Please note that the following conditions apply to your approval. Failure to abide by these conditions may result in suspension or discontinuation of approval and/or disciplinary action.

(a) **Limit of Approval:** Approval is limited strictly to the research as submitted in your Project application.

(b) **Amendments to Project:** Any subsequent variations or modifications you might wish to make to the Project must be notified formally to the Human Ethics Advisory Group for further consideration and approval before the revised Project can commence. If the Human Ethics Advisory Group considers that the proposed amendments are significant, you may be required to submit a new application for approval of the revised Project.

(c) **Incidents or adverse affects:** Researchers must report immediately to the Advisory Group and the relevant Sub-Committee anything which might affect the ethical acceptance of the protocol including adverse effects on participants or unforeseen events that might affect continued ethical acceptability of the Project. Failure to do so may result in suspension or cancellation of approval.

(d) **Monitoring:** All projects are subject to monitoring at any time by the Human Research Ethics Committee.

(e) **Annual Report:** Please be aware that the Human Research Ethics Committee requires that researchers submit an annual report on each of their projects at the end of the year, or at the conclusion of a project if it continues for less than this time. Failure to submit an annual report will mean that ethics approval will lapse.

(f) **Auditing:** All projects may be subject to audit by members of the Sub-Committee.

Please quote the ethics registration number and the name of the Project in any future correspondence.

On behalf of the Ethics Committee I wish you well in your research.

Yours sincerely

Dr Sherah Kurnia - Chair
Engineering Human Ethics Advisory Group

unimelb.edu.au

THE UNIVERSITY OF MELBOURNE

cc:
Dr S. B. Maynard
HEAG Administrator - Engineering
HEAG Chair - Engineering

unimelb.edu.au

# Appendix B: Invitation Letter, Plain Language Statement & Consent Form

THE UNIVERSITY OF MELBOURNE

## Invitation Letter

**Department of Computing and Information Systems**
**School of Engineering**

**< Insert Date>**

Dear <Insert Salutation / Name>

**Invitation to Participate in a Research on Improving Information Security Governance in Organizations – Case Study Interview**

Corporate Governance is critical to facilitate effective and prudent leadership that can deliver the strategic objectives of an organization where the Board of Directors are responsible for the governance of their organizations. In Corporate Governance, the Board of Directors and senior management have a fiduciary duty to protect the organization's assets and the value of corporation. As information is now considered a strategic asset for an organization to achieve its strategic business objectives, the fiduciary duty has extended to include the protection of such information. As a result, the governance of information security has become increasingly important for the Board of Directors and senior management and any failure in this duty can bring serious implications.

In view of the importance of Information Security Governance (ISG) and the associated challenges in implementing ISG in organizations, this research is motivated to explore ways to improve ISG in organizations. I am working on my PhD research with my supervisors to develop an ISG model & framework that can facilitate the implementation of ISG in organizations. As part of my research, I would like to invite you and selected individuals in your organization to participate in case study interviews where:

i. I would conduct interviews to understand how information security governance is undertaken in your organization, and how different people at different management levels are involved.

ii. I would also invite comments on the proposed ISG model. It will be a one-to-one interview and is estimated to take 60 minutes.

Attached to this invitation letter, you will find the following documents for your perusal:

i. Plain Language Statement
ii. Consent Form

**School of Computing and Information Systems**
The University of Melbourne, Victoria 3010 Australia
T: +61 3 8344 1501    F: +61 3 9349 4596
W: http://www.cis.unimelb.edu.au/

THE UNIVERSITY OF MELBOURNE

We recognise that your time is valuable and thank you in advance for your generous participation in this research project. We are confident that by sharing your experience, views and comments, we can contribute further to the research and development of an ISG model and framework that can facilitate the implementation of ISG in organizations, thus improving the governance of information security. Should you have any questions about this research, feel free to contact me for further clarifications.

Yours sincerely

Chee Kong WONG
PhD Student Researcher
Email: cheew2@student.unimelb.edu.au

**School of Computing and Information Systems**
The University of Melbourne, Victoria 3010 Australia
T: +61 3 8344 1501 F: +61 3 9349 4596
W: http://www.cis.unimelb.edu.au/

THE UNIVERSITY OF MELBOURNE

# Plain Language Statement

Department of Computing & Information Systems
School of Engineering

## *Project: Improving Information Systems Governance in Organizations*

Dr. Atif Ahmad (Responsible Researcher)
Tel: +61 3 8344 1396 Email: atif@unimelb.edu.au

Dr. Sean Maynard (Co-Researcher) Email: sean.maynard@unimelb.edu.au
Chee Kong Wong (PhD Student) Email: cheew2@student.unimelb.edu.au

## Introduction

Thank you for your interest in participating in this research project. The following few pages will provide you with further information about the project, so that you can decide if you would like to take part in this research.

Please take the time to read this information carefully. You may ask questions about anything you don't understand or want to know more about.

Your participation is voluntary. If you don't wish to take part, you don't have to. If you begin participating, you can also stop at any time.

## What is this research about?

Corporate Governance is critical to facilitate effective and prudent leadership that can deliver the strategic objectives of an organization where the Board of Directors are responsible for the governance of their organizations. In Corporate Governance, the Board of Directors and senior management have a fiduciary duty to protect the organization's assets and the value of corporation. As information is now considered a strategic asset for an organization to achieve its strategic business objectives, the fiduciary duty has extended to include the protection of such information. As a result, the governance of information security has become increasingly important for the Board of Directors and senior management and any failure in this duty can bring serious implications.

In view of the importance of Information Security Governance (ISG) and the associated challenges in implementing ISG in organizations, this research is motivated to explore ways to improve ISG in organizations. This research aims to develop an ISG model & framework that can facilitate the implementation of ISG in organizations.

School of Computing and Information Systems
The University of Melbourne, Victoria 3010 Australia
T: +61 3 8344 1501 F: +61 3 9349 4596
W: http://www.cis.unimelb.edu.au/

THE UNIVERSITY OF MELBOURNE

## What will I be asked to do?

Should you agree to participate in our case study research, we would like to conduct an interview with you to understand how information security governance is undertaken in your organization, and how different people at different management levels are involved. In addition, we will also invite your comments on our proposed ISG model. It will be a one-to-one interview and is estimated to take 60 minutes.

Should you agree to participate in the focus group discussion, we would like to gather your views and experience on how ISG is implemented in organizations, and invite your comments on our proposed ISG model. The focus groups will be a moderated session of 90 minutes discussion with a planned agenda.

With your permission, the discussion will be audio recorded and transcribed to further analyse the data.

## What are the possible benefits?

The participants in the case studies and focus groups can have an exchange of knowledge on ways to improve the governance of information security. The participants can obtain a copy of the publication of the research when the research is completed.

## What are the possible risks?

This project has received HREC clearance and involves minimal risk for the participants. We intend to protect your anonymity and the confidentiality of your responses to the fullest extent possible, within the limits of the law. Your name, contact details and affiliation will be kept in a separate, password-protected computer file away from any data that you supply or recordings we make. This can only be linked to your responses by the researchers. In any publications that may arise from this study, you will be referred to by a pseudonym. We will also remove any references to personal information that might allow someone to guess your identity or your affiliation. The data will be kept securely (using password protected access), in the Department of Computing Information System for five years following the date of last publication, before being destroyed.

## Do I have to take part?

No. Participation is completely voluntary. You are able to withdraw (quit) at any time.

## Will I hear about the results of this project?

Once the thesis arising from this research is completed, a brief summary of the findings will be available for you on request at the Department of Computing and Information Systems. It is also possible that the results will be published in academic journals and/or presented at academic conferences.

**School of Computing and Information Systems**
The University of Melbourne, Victoria 3010 Australia
T: +61 3 8344 1501  F: +61 3 9349 4596
W: http://www.cis.unimelb.edu.au/

THE UNIVERSITY OF MELBOURNE

### What will happen to information about me?

We intend to protect your anonymity and the confidentiality of your responses to the fullest extent possible, within the limits of the law. Your name, contact details and affiliation will be kept in a separate, password-protected computer file away from any data that you supply or recordings we make. This can only be linked to your responses by the researchers. In any publications that may arise from this study, you will be referred to by a pseudonym. We will also remove any references to personal information that might allow someone to guess your identity or your affiliation. The data will be kept securely (using password protected access), in the Department of Computing Information System for five years following the date of last publication, before being destroyed.

### Where can I get further information?

If you would like more information about the project, please contact the responsible researcher as identified in this Plain Language Statement.

### Who can I contact if I have any concerns about the project?

This research project has been approved by the Human Research Ethics Committee of The University of Melbourne. If you have any concerns or complaints about the conduct of this research project, which you do not wish to discuss with the research team, you should contact the Manager, Human Research Ethics, Research Ethics and Integrity, University of Melbourne, VIC 3010. Tel: +61 3 8344 2073 or Email: HumanEthics-complaints@unimelb.edu.au. All complaints will be treated confidentially. In any correspondence please provide the name of the research team or the name or ethics ID number of the research project.

**School of Computing and Information Systems**
The University of Melbourne, Victoria 3010 Australia
T: +61 3 8344 1501 F: +61 3 9349 4596
W: http://www.cis.unimelb.edu.au/

THE UNIVERSITY OF MELBOURNE

# Consent Form – Case Study Interview

**Department of Computing and Information Systems**
**School of Engineering**

## *Project: Improving Information Security Governance in Organizations*

**Primary Researcher:**

Dr. Atif Ahmad (Responsible Researcher) Email: atif@unimelb.edu.au

**Additional Researchers:**

Dr. Sean Maynard (Co-Researcher) Email: sean.maynard@unimelb.edu.au
Chee Kong Wong (PhD student) Email: cheew2@student.unimelb.edu.au

**Name of Participant:** ______________________________

1. I consent to participate in this project, the details of which have been explained to me, and I have been provided with a written plain language statement to keep.
2. I understand that the purpose of this research is to investigate an Information Security Governance model to improve governance of information security in organizations.
3. I understand that my participation in this project is for research purposes only.
4. I acknowledge that the possible effects of participating in this research project have been explained to my satisfaction.
5. In this project I will be required to be interviewed by your researcher to share with you how information security governance is undertaken in my organisation, and to share my comments on your proposed ISG model.
6. I understand that my interviews may be audio recorded for further analysis. However, I have an option to decide if I do not agree to my interviews being audio recorded.

   I agree to be audio recorded during the interviews ☐
7. I understand that my participation is voluntary and that I am free to withdraw from this project anytime without explanation or prejudice and to withdraw any unprocessed data that I have provided.
8. I understand that the data from this research will be stored at the University of Melbourne and will be destroyed after 5 years from the date of the last publication.
9. I have been informed that the confidentiality of the information I provide will be safeguarded subject to any legal requirements; my data will be password protected and accessible only by the named researchers.
10. I understand that after I sign and return this consent form, it will be retained by the researcher.

**Participant Signature:** ____________________ **Date:** ______________

**School of Computing and Information Systems**
The University of Melbourne, Victoria 3010 Australia
T: +61 3 8344 1501 F: +61 3 9349 4596
W: http://www.cis.unimelb.edu.au/

# Appendix C: Interview Guide

*Interview Guide*

## Interview Guide

This is not a survey. This is an interview guide that acts as a prompt, reminding the necessary topics to cover, questions to ask and areas to probe.

Estimated time for interview: 60 minutes.

### Introduction

Corporate Governance is critical to facilitate effective and prudent leadership that can deliver the strategic objectives of an organization where the Board of Directors are responsible for the governance of their organizations. In Corporate Governance, the Board of Directors and senior management have a fiduciary duty to protect the organization's assets and the value of corporation. As information is now considered a strategic asset for an organization to achieve its strategic business objectives, the fiduciary duty has extended to include the protection of such information. As a result, the governance of information security has become increasingly important for the Board of Directors and senior management and any failure in this duty can bring serious implications.

In view of the importance of Information Security Governance (ISG) and the associated challenges in implementing ISG in organizations, this research is motivated to explore ways to improve ISG in organizations.

**A. Case Study Interview Details**

Organization:

Location of Interview:

Date: Time:

**B. Participant's Background**

Gather background information about the participant.

1. What is the Participant's job title?

2. What is the Participant's daily role?

3. How long has Participant been in the role?

Ethics ID: 1749890.1
Interview Guide Version 1-0 07July2017 1

*Interview Guide*

## C. Participant's Understanding of Information Security & ISG

Gather the participant's understanding of ISG.

1. Based on your understanding, what is your current view on Information Security?

2. Do you believe organizations, especially Financial Services Organization are doing enough on the protection of Information Security?

3. In your opinion, what are some of the actions that should be taken by organization to better improved Information Security in organizations?

*Interview Guide*

4. Good Corporate Governance helps to facilitate effective and prudent leadership in organizations. New regulations and legislations e.g. Sarbanex-Oxley Act have been introduced to improve monitoring and disclosure, thereby, improving the governance of organizations.

   What are your views if similar governance is introduced to govern Information Security?

   What will be your understanding of a good ISG?

   Do you believe there ISG exists in organizations?

Ethics ID: 1749890.1
Interview Guide Version 1-0 07July2017 3

## D. ISG in Participant's Organization

1. Can you explain how ISG is implemented in your organization?

2. Who in your organization are involved in the governance of Information Security? Are you able to share with me an organization chart of people who are involved in Information Security?

3. Strategic Alignment. How do you know if Information Security activities are aligned to your organization's business requirements?

*Interview Guide*

4. Who are involved in ensuring Information Security activities are aligned with your strategic business requirements?

5. Are the Board of Directors involved? How?
Are the Senior Management involved? Who and How are they involved?

6. Does your organization has an Information Security committee represented by Board members (just like an Audit committee or other Board committees)? Who are represented in the committee? How often does the committee meet? What are the typical agenda and topics discussed in such committee?

Ethics ID: 1749890.1
Interview Guide Version 1-0 07July2017 5

*Interview Guide*

7. Risk Management. How does your organization handle risk management? Is Information Security part of the organization-wide risk management programme?

8. Investment. How does your organization decide on investment in Information Security activities? How $$$ and resources are allocated?

9. Is there a mechanism that is implemented to measure the effectiveness and compliance of the implementation of Information Security activities?

*Interview Guide*

10. Do you know if you are spending too little or too much on the protection of Information Security? Do you have any suggestions on how to improve the spending on the Information Security protection?

| |
|---|
| |

Ethics ID: 1749890.1
Interview Guide Version 1-0 07July2017 7

*Interview Guide*

## E. Comments on Proposed ISG Model

1. Show proposed ISG model (Attachment 1) to Participant.
   Explain the various components of the ISG model, and seek feedback and comments.

*Interview Guide*

2. Show ISG framework (Attachment 2) to Participant.
Explain the various processes within the ISG framework, and seek feedback and comments.

Ethics ID: 1749890.1
Interview Guide Version 1-0 07July2017 9

## F. Finishing the Interview

Ask Participant if there are any key points that should be considered in ISG.

Thanks the Participant for the time, and emphasize the adherence to confidentiality of information gathered.

*Interview Guide*

## Attachment 1.

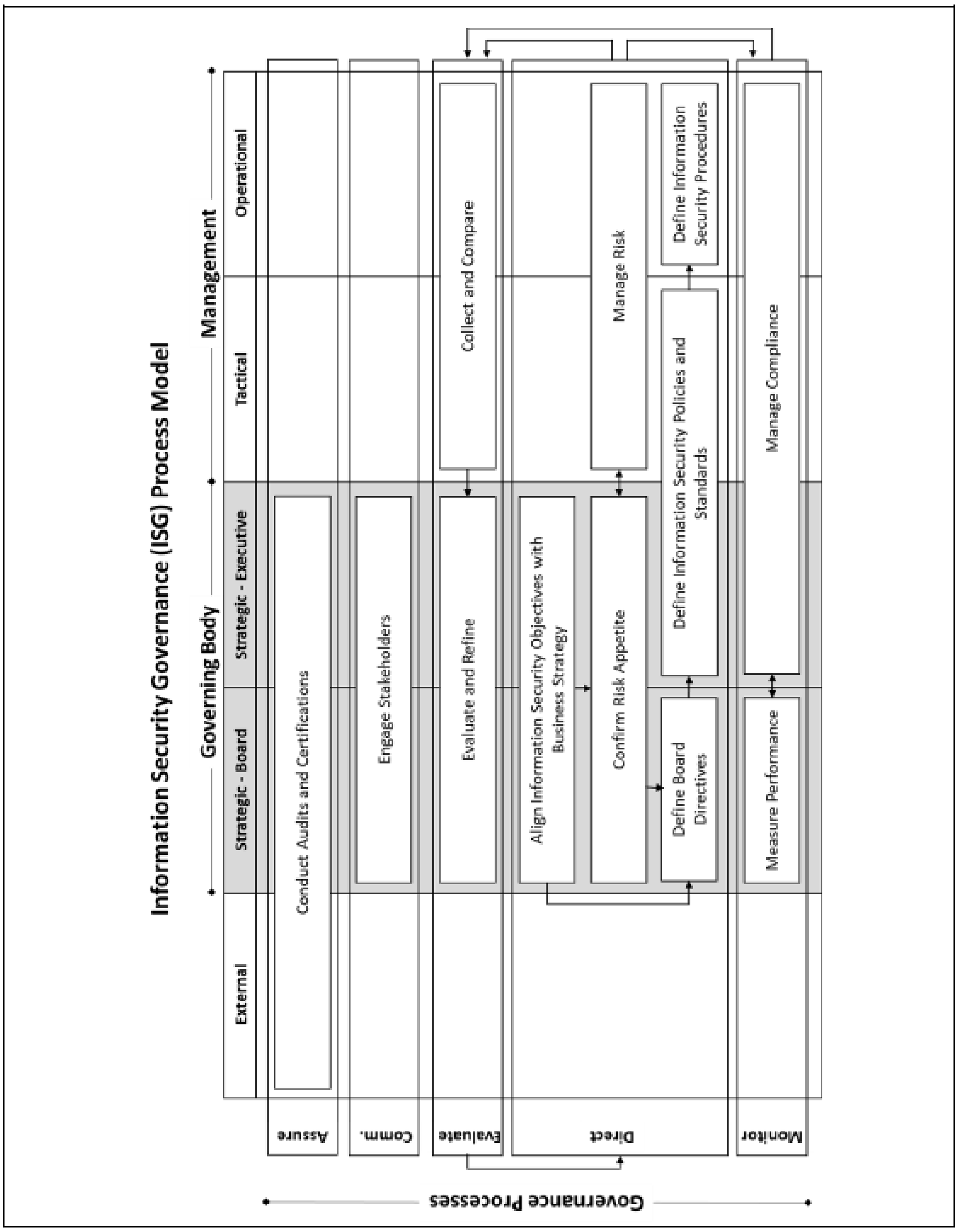

Ethics ID: 1749890.1
Interview Guide Version 1-0 07July2017 11